\documentclass{ws-rv961x669}

\usepackage{ws-rv-van}     
\usepackage{ws-rv-thm}     
\usepackage{subfigure}     
\usepackage{graphicx}
\usepackage{wrapfig,caption}
\usepackage{hyperref}

\makeindex

\newcommand{\taon}{$\tau$-lepton }
\newcommand{\taons}{$\tau$-leptons }
\newcommand{\Enu}{E$_\nu$}
\newcommand{\xmax}{\ensuremath{X_{\rm max}}}
\newcommand{\nmax}{\ensuremath{N_{\rm max}}}
\newcommand{\lsim}{\mathrel{\hbox{\rlap{\lower.75ex \hbox{$\sim$}} \kern-.3em \raise.4ex \hbox{$<$}}}}
\newcommand{\gsim}{\mathrel{\hbox{\rlap{\lower.75ex \hbox{$\sim$}} \kern-.3em \raise.4ex \hbox{$>$}}}}

\newcommand{\order}{\cal O}

\newcommand{\nubar}{\bar\nu}

\newcommand{\nutau}{$\nu_{\tau}$}
\newcommand{\nuebar}{{\bar \nu}_e}

\usepackage[usenames,dvipsnames]{xcolor}
\definecolor{orange}{cmyk}{0,0.5,1,0}
\definecolor{rossoCP3}{cmyk}{0,.88,.77,.40}
\definecolor{graa}{rgb}{0.8,0.8,0.8}
\definecolor{blaa}{rgb}{0.2,0.2,0.6}

\definecolor{pink1}{RGB}{226, 24, 166}

\begin{document}


\newpage

\chapter[Space-based Cosmic Neutrino Measurements]{Space-based Cosmic Neutrino Measurements} 

\author[John F. Krizmanic] {John F. Krizmanic\footnote{john.f.krizmanic@nasa.gov}}

\address{NASA/Goddard Space Flight Center\\
Laboratory for Astroparticle Physics  \\
Greenbelt, Maryland 21250 USA}

{\it Space, the final frontier $\cdots$} \newline
\hspace*{20mm} {Introductory line of Gene Roddenberry's {\it Star Trek}}
\vspace{1cm}

\begin{abstract}
    
Space-based neutrino detection employs celestial bodies, the Earth, moon, and the sun, as neutrino targets and particle cascade generators.  These cascades, or showers, have an immense content of charged particles, mainly electrons and positrons, that can form signals that can be detected by orbiting or balloon-borne (for the Earth) experiments. Here the focus will be on the using the Earth as the neutrino target and using the atmosphere to generate optical and radio signals from extensive air showers (EAS). Cascades in the Earth itself can also generate Askaryan radiation in ice, where the radio transparency allows the signal from in ice neutrino-induced showers to eventually refract out of the Earth's surface.  These different techniques also have different neutrino-flavor detection sensitivities and duty cycles, i.e. experimental live time fractions. For example, optical measurements need to be performed close to astronomical night and are also affected by the level of moonlight and distribution of clouds in the the viewing direction of the EAS \cite{2013APh....44...76A, 2019PhRvD..99f3011R}.  The radio technique can be operate with nearly 100\% efficiency, as long as sources of anthropogenic and other backgrounds are minimal \cite{2009APh....32...10A,Schoorlemmer_2016}. Given the distances for viewing the neutrino-induced signals from sub-orbital or low Earth orbit (LEO) altitudes, the neutrino energy thresholds for detection are currently above a PeV for optical Cherenkov detection \cite{2019PhRvD..99f3011R,2020PhRvD.101b3012A,2020PhRvD.102l3013V}and 100 PeV for radio detection \cite{2020arXiv201002892A}. In this chapter, the nature of the cosmic neutrino interactions, shower development, and optical and radio signal generation and detection will be detailed.  This will be done by discussing experiments that have used such signals to set limits on the very-high and ultra-high energy cosmic neutrino flux. We will also discuss the design and simulated performance of the next generation of space-based cosmic neutrino detection experiments.

\end{abstract}

\markboth{Neutrino Observations from Space}{J. F. Krizmanic} 

\body

\newpage

\tableofcontents

\newpage

\section{Introduction}\label{SNc_Intro}

The measurement of the spectrum of the Very-High Energy (VHE: \Enu $\gsim 1$ PeV) neutrino component of the cosmic radiation and the angular distribution of the sources on the sky provides a unique probe of high-energy astrophysical phenomena, in particular those involving hadronic processes, please see the theoretical chapter in this volume. The neutrino measurements from IceCube \citep{IceCube2015,IceCube2017,2018AdSpR..62.2902A} demonstrate the existence of an extra-solar system astrophysical neutrino flux from above 10 TeV to potentially 10 PeV, with one blazar, TXS 0506+056, so far identified as a source \cite{2018Sci...361.1378I,2018Sci...361..147I} and possibly a neutrino associated with the tidal disruption event AT2019dsg observed with a an optical counterpart \cite{2021NatAs.tmp...34S}.
The fundamental principle of space-based\footnote{Here we denote both Earth orbiting and sub-orbital experiments, such as balloon-borne, as space-based.}, neutrino and antineutrino\footnote{Neutrino will be used to denote both neutrino and antineutrino detection, since the neutrino-quark scattering kinematics are similar above a PeV.}  detection is to use either the Earth or atmosphere as a vast neutrino target whence some of the energetic products produce particle cascades in ice or in the atmosphere, the latter known as extensive air showers (EAS).

\begin{figure}
\begin{center}
\includegraphics[width=0.99\textwidth]{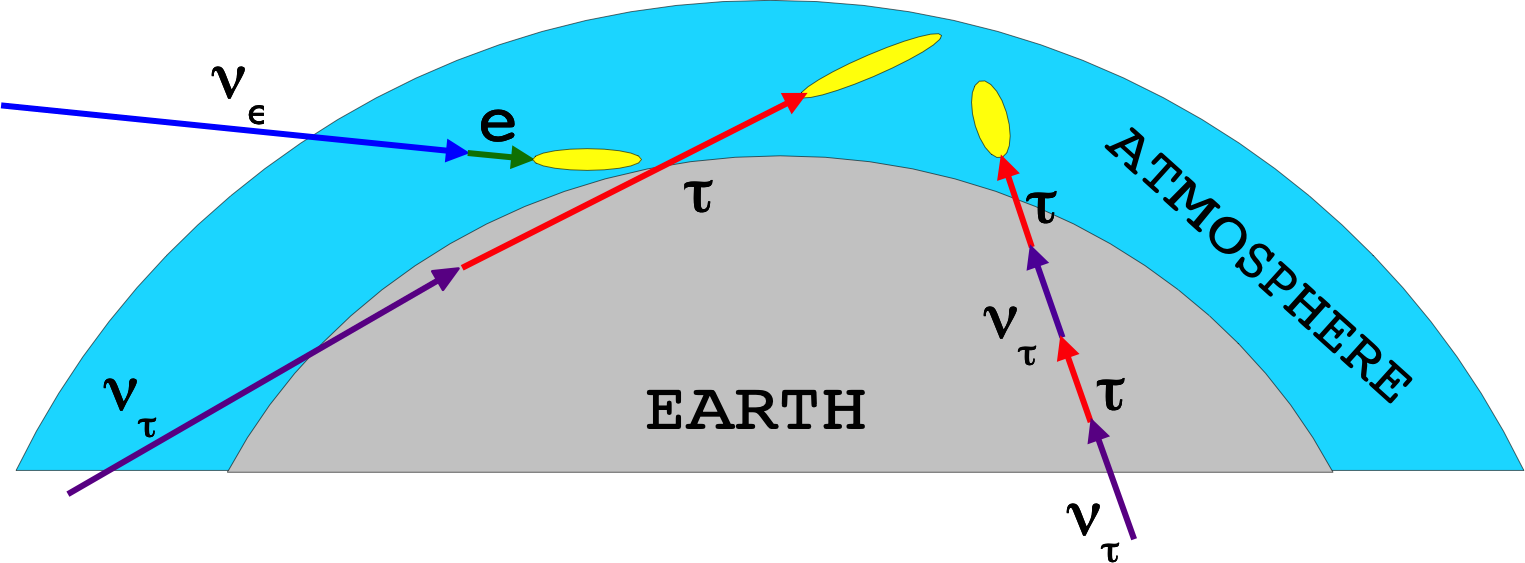}
\caption{Artist rendition of the possible neutrino interaction channels that produce extensive air showers that can be observed from space-based experiments.   From Ref. \citenum{2003APh....18..539B}.
DOI: \href{https://doi.org/10.1016/S0927-6505(02)00165-2}{10.1016/S0927-6505(02)00165-2}
\label{nuEASchannels}}
\end{center}
\end{figure}

Fig.~\ref{nuEASchannels} illustrated the various event types that can lead to a neutrino-induced EAS.  Neutrino interactions in the atmosphere of all neutrino flavors will produce an interaction that generates an EAS. Fig.~\ref{nuEASchannels} illustrates that for an electron charged-current neutrino interaction that places effectively  100\% of the neutrino energy into the EAS, while other lepton flavors produce EAS with a fraction of  \Enu in the EAS. Fig.~\ref{nuEASchannels} also shows the signals generated by tau neutrinos interacting in the Earth that lead to Earth-emergent upward-moving EAS.  For small Earth-emergence angles, the amount of material in the trajectory of generated \taon is such to put the majority of the neutrino energy into the Earth-emergent \taon.  For larger Earth-emergence angles, an effect known as tau neutrino regeneration \cite{Halzen:1998be} becomes increasingly important as the slant depth (in g/cm$^2$) of the \taon increases to allow decay and then regeneration of a lower energy tau neutrino that then interacts in the Earth.  These neutrino EAS signatures as well as those from radio emission of neutrino-induced particle cascades in ice will be discussed later in this chapter.

\begin{figure}[ht]
\begin{center}
\includegraphics[width=0.99\textwidth]{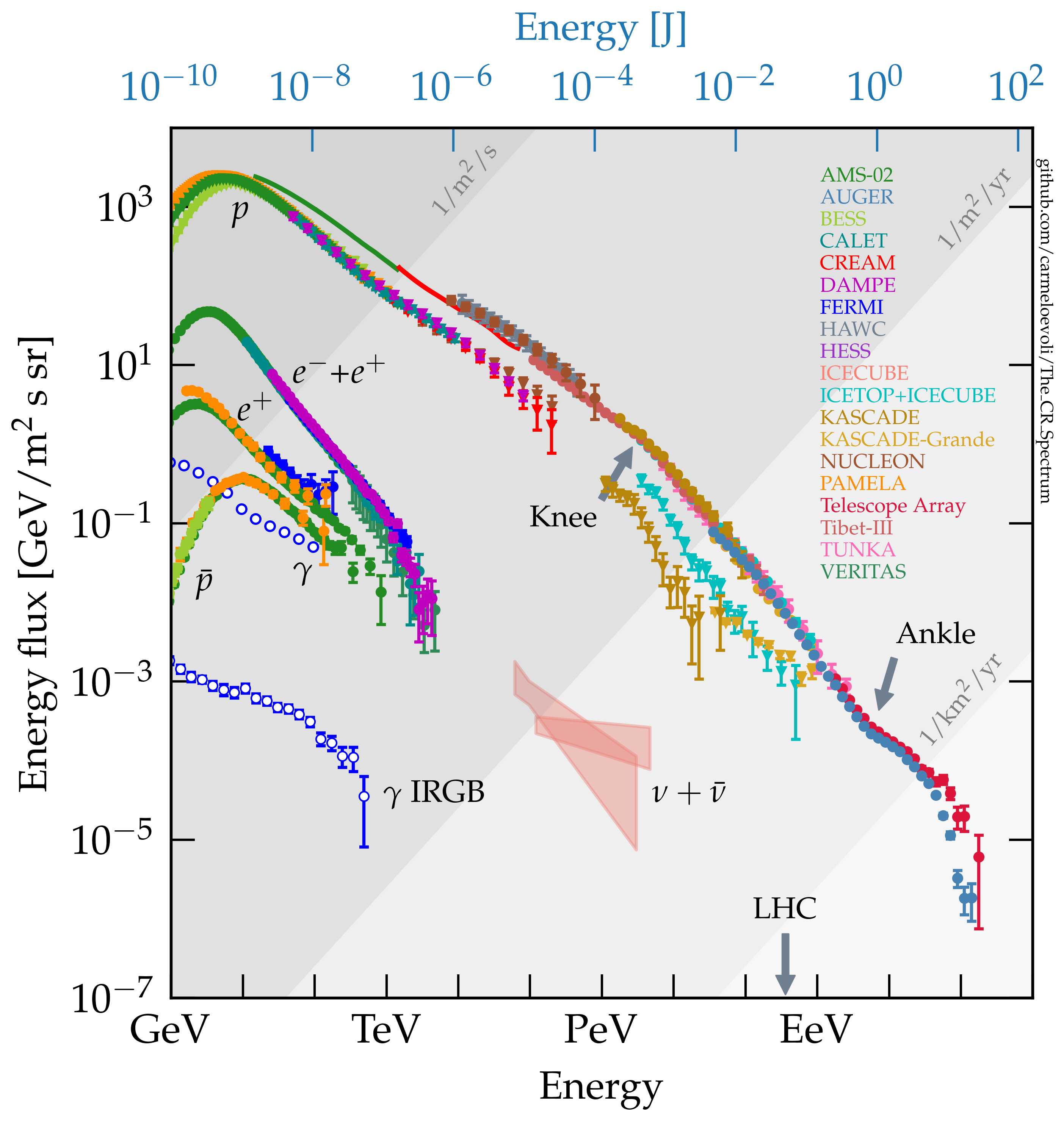}
\caption{The energy spectra of the various components of the cosmic radiation observed as Earth. Note the flux of neutrinos is well below that of hadronic cosmic rays.  From Ref.~\citenum{Evoli}.
DOI: \href{https://doi.org/10.5281/zenodo.4396125}{10.5281/zenodo.4396125}
\label{CRspectra}}
\end{center}
  \vspace{-5mm}
\end{figure}

Fig.~\ref{CRspectra} illustrates the flux of the various components of the cosmic radiation and demonstrates the need for multi-gigaton-year\footnote{Note that 1 km$^3$ of ice or water is 1 gigaton.} neutrino exposure to detect VHE and UHE cosmic neutrinos. Thus it is necessary to understand the properties of the EAS development induced by the different leptons and hadrons sourced from neutrino interactions and the particle, optical, and radio signals created by the EAS that can be used to measure these products of neutrino interactions.   The accuracy of the experimental measurements is intrinsically tied to the specific nature of the neutrino interaction and the EAS measurement technique. This ranges from determining to better than 20\% the incident energy via the air fluorescence technique of an EAS caused by an electron neutrino charged-current interaction in the atmosphere, to an energy resolution of several orders of magnitude when measuring an upward-moving \taon induced EAS. This is for both the radio and optical Cherenkov light measurement techniques and is due to the processes that induce variability in the EAS energies, which is largest for \taon decays generated from regenerative tau neutrino interactions in the Earth as shown in Figs.~\ref{nuEASchannels} and ~\ref{UpNuCher}.  The long propagation lengths for VHE \taons that can escape the Earth translates into gargantuan neutrino target, $\order$ $(10^{16}$t) if a large area of the Earth can be monitored for upward-moving EAS, for example see Ref.~\citenum{Reno:2021vf}. These along with the inherent angular resolution and neutrino flux sensitivity and astrophysics reach, in the context of a number of space-based experiments will be discussed in detail.  It will be shown that the experimental techniques have different accuracy of cosmic neutrino measurements and these techniques need to be fully understood, as well as the sources of backgrounds, which set the neutrino energy thresholds of the EAS measurements.
Using the atmosphere EAS signals for neutrino detection is not the only method for space-based neutrino detection, concepts include a space-based instrument to measure solar neutrinos as well as using the sun as a neutrino target \cite{2018LPICo2063.3002S} and using the lunar regolith as a neutrino target for the generation of sub-lunar cascades leading to an radio signal observed by an orbiting experiment \cite{2016AdSpR..58..464R,Romero-Wolf:2021ia}. These will be discussed in the final section of this chapter.


\section{History}\label{SNc_Hist}
Since many space-based neutrino observations are based on EAS measurements, it is worthwhile to understand these in the context of space-based measurements of ultra-high energy cosmic rays (UHECRs).  The space-based concept for observing UHECRs was suggested by Robert Benson and John Linsley in 1980 \cite{1980BAAS...12Q.818B, 1981ICRC....8..145B}
In 1995, Yoshi Takahasi proposed the concept of using an Earth-orbiting,  wide Field-of-View (FoV) telescope with meter-scale diameter to observe UHECRs using the dim EAS, near-UV fluorescence signal, called {\it Maximum-energy Auger (AIR)-Shower Satellite \newline
(MASS) \cite{1995ICRC....3..595T,1996SPIE.2806..102T}}. 


\begin{figure}[ht]
\begin{center}
{\includegraphics[width=0.99\textwidth]{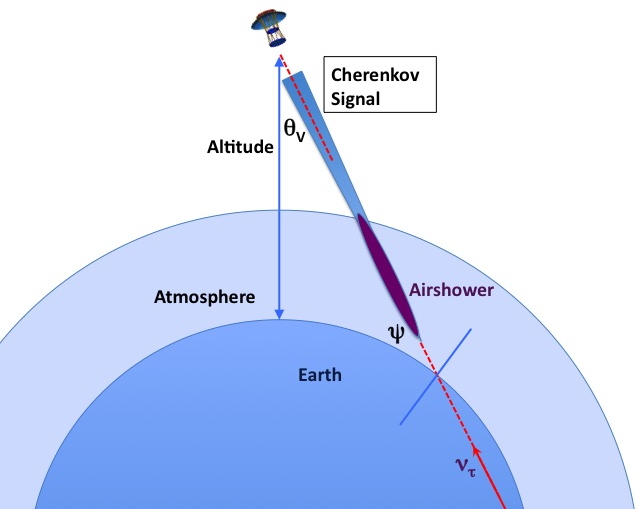}}
\caption{Schematic of the Cherenkov signal generated by
an upward-moving extensive air shower (EAS) generated by a the decay of a \taon induced by a tau neutrino interacting in the Earth. From Ref. \citenum{Krizmanic2011}.
 \href{https://arxiv.org/abs/1109.6544}{arXiv:1109.6544}
\label{UpNuCher}}
\end{center}
\end{figure}

\begin{figure}
\begin{center}
\includegraphics[width=0.99\textwidth]{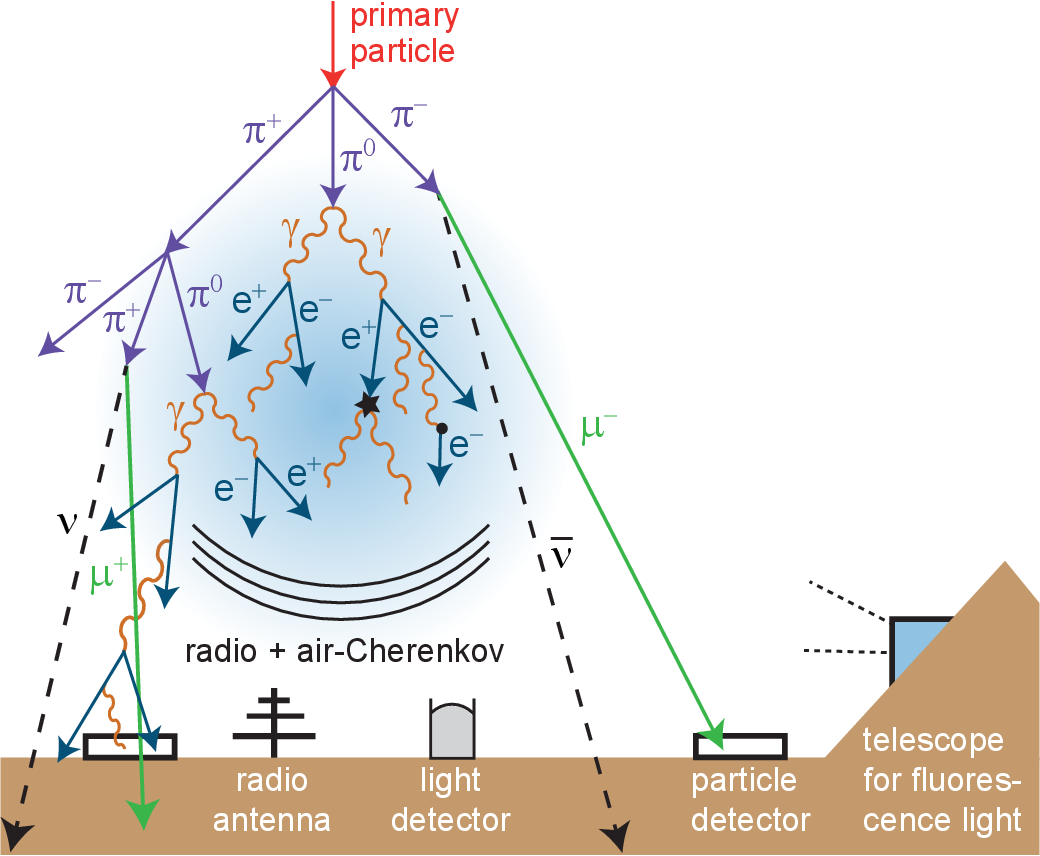}
\caption{Artist rendition of the particle development in an extensive air shower (EAS) and the various signals and ground-based detection techniques. Adapted from Ref.~\citenum{SCHRODER20171} DOI: \href{https://doi.org/10.1016/j.ppnp.2016.12.002}{10.1016/j.ppnp.2016.12.002}
Space-based EAS
detection uses similar experimental techniques as that on the ground but view the EAS from around 33 km altitude for
balloon-borne experiments and  $\sim$400 km or higher altitude from low Earth orbit.  However, space-based
measurement of EAS are more limited in the number of balloon-borne or spacecraft-based instruments/telescopes as
compared to that available for ground-based experiments.\label{EASdetectCartoon}}
\end{center}
  \vspace{-5mm}
\end{figure}

For space-based UHECR fluorescence measurements, the EAS spatial imaging requirements lead to a $\sim \mu$-second sampling resolution to accurately sample the waxing and waning of the EAS longitudinal profile. The EAS is initiated by a UHE particle interacting with matter that produces particle cascades in the atmosphere and this is shown schematically in Fig.~\ref{EASdetectCartoon} for a hadronic primary particle. These measurements measurements provide the data needed to reconstruct the EAS development to determine the energy and source direction.  and this leads A imaging size of around a km on the ground from low-Earth orbit, e.g. $400 -- 1000$ km altitude, which is very coarse resolution compared that for an optical telescope\footnote{At 400 nm, 0.1$^\circ$ resolution using a meter-diameter telescope is $\sim 10^4$ away from the diffraction limit.}.  In the early 2000's, this concept lead to the AirWatch mission concept \cite{1996SPIE.2806..102T} defined by a single orbiting telescope for monocular observation of EAS and the Orbiting Wide-angle Light collectors (OWL) mission designed to use the stereo fluorescence EAS technique.  Stereo reconstruction precisely measures the properties of the UHECRs and UHE\footnote{UHE is defined as $E_\nu \ge 1$ EeV.} neutrinos due to its superior angular resolution. 
Accurately reconstruction the 3-dimensional spatial EAS development provides high-resolution determination of UHECR nuclear composition, angular resolution, and $<20\%$ energy resolution.  
The AirWatch program led to the the Extreme Universe Space Observatory
on-board the Japanese Experiment Module of the International Space Station (JEM-EUSO) mission concept \cite{2013APh....44...76A} and several related balloon-borne missions, including EUSO-balloon \cite{2017ICRC...35..445B}, EUSO-SBP \cite{2019NIMPA.936..237O}, EUSO-SPB2 \cite{2020NIMPA.95862164S}, the CHerenkov from Astrophysical Neutrinos Telescope (CHANT) mission concept \cite{CHANT} and the Mini-EUSO experiment \cite{2021ApJS..253...36B} operating on the International Space Station starting in 2019. The OWL development also included using the air fluorescence technique to search for cosmic neutrinos interacting in the atmosphere \cite{OWL} and the beamed Cherenkov light from upward-moving EAS from \taon decay source from \nutau interactions in the Earth \cite{Krizmanic2003}. The OWL development also led to the concept of the neutrino
Balloon Air CHerenkov (neuBACH) ultra-long duration balloon (ULDB) optical Cherenkov neutrino detection instrument concept based on a design developed for a 2001 NASA/GSFC sub-orbital proposal \cite{NeuBACH}, which was not funded.

\begin{figure}
\vspace{-3mm}
 \begin{center}
    \includegraphics[width=0.48\textwidth]{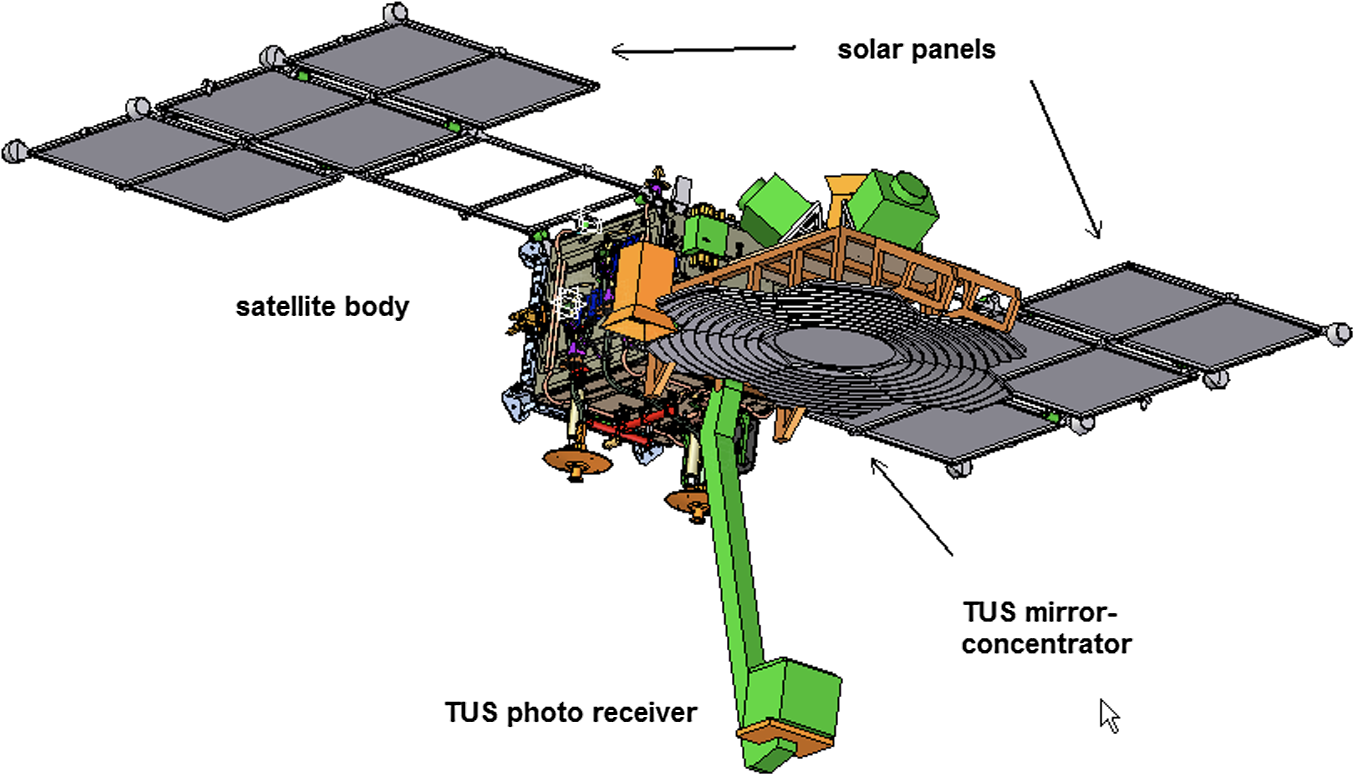}
    \hspace*{1 pt}
    \includegraphics[width=0.50\textwidth]{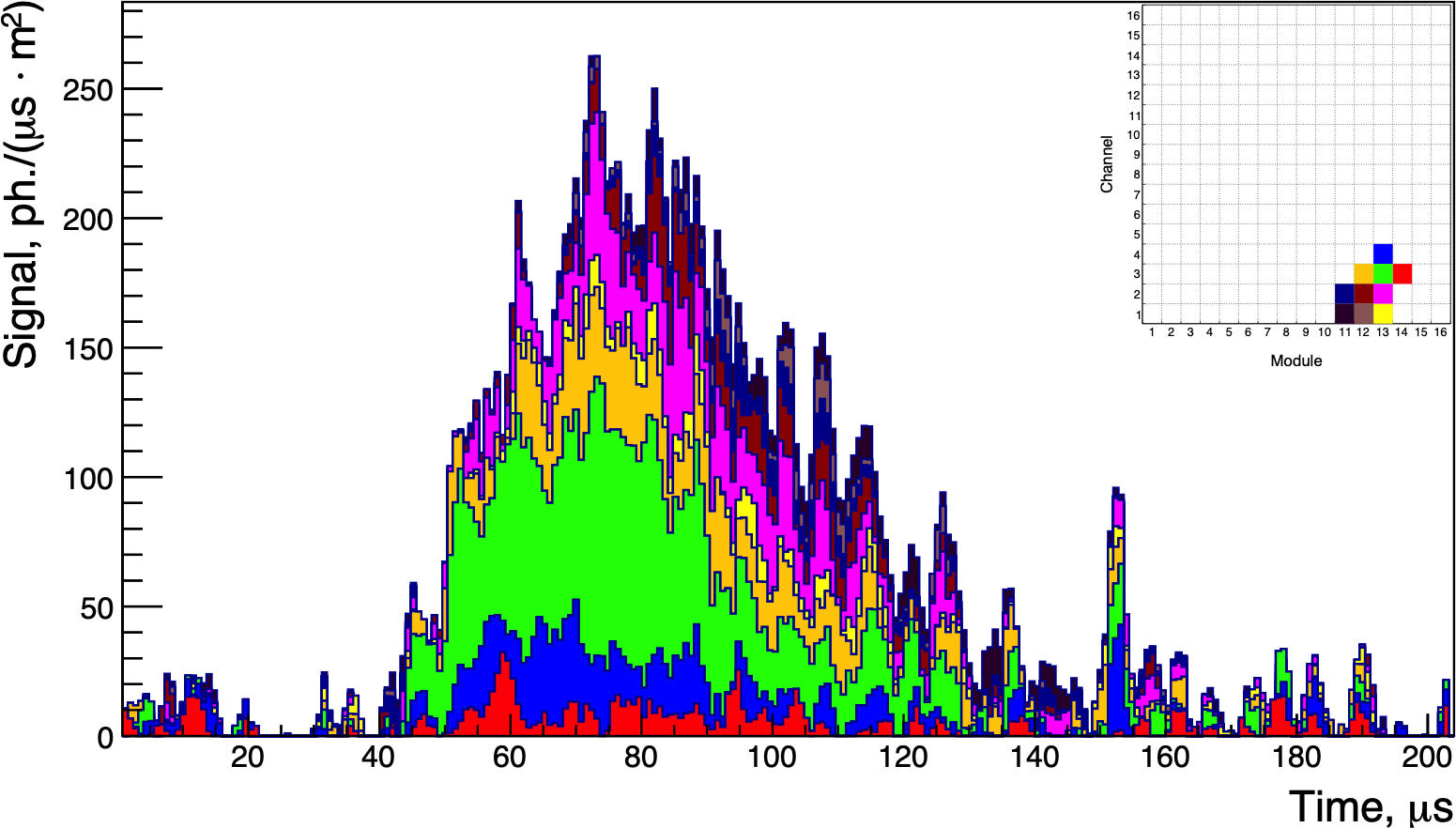}
  \end{center}
  \vspace{-1mm}
  \caption{Left: The schematic of the TUS experiment on the Lomonosov satellite. The Fresnel mirror, with $\sim$2 m$^2$ area, has a field of view of 9$^\circ$ imaging EAS in a focal plane with 256 pixels. Three years of operation at $\sim 500$ km altitude lead to an exposure of 12,000 km$^2$ sr yr for $E_{\rm UHECR} > 300$ EeV. From Ref.~\citenum{2017SSRv..212.1687K}.
  DOI: \href{https://doi.org/10.1007/s11214-017-0403-3}{10.1007/s11214-017-0403-3}
  Right: Results for a UHECR  fluorescence candidate event. The light curve stacks the 10 hit channels for the event that was observed over Minnesota in 2016  The inset shows the location of the 10 hit pixels for this event denoted TUS161003. From Ref.~\citenum{2020JCAP...03..033K}.
  DOI: \href{https://doi.org/10.1088/1475-7516/2020/03/033}{10.1088/1475-7516/2020/03/033}
  \label{TUS}}
  \vspace{-1mm}
\end{figure}
Somewhat independently, the Tracking Ultraviolet Set-up (TUS) experiment \cite{2017SSRv..212.1687K, Klimov:2017dul}, shown in the left panel of Fig.~\ref{TUS} was built, launched, and operated on the Lomonosov satellite from May 19, 2016 until May 30, 2019. The relatively modest size of the optical light collection for the near-UV air fluorescence signal led to a UHECR detection energy threshold of $\gsim 70$ EeV. TUS detected six UHECR EAS candidate events whose origin may be due to background \cite{2020JCAP...03..033K, 2021ICRC...37..316K}. The right panel in Fig.~\ref{TUS} show the measured light curve for one of these events (TUS161003) whose response is consistent with an EAS fluorescence observation for an event with $E_{CR} > 100$ EeV.
The TUS UHECR imaging system consisted of a $\sim$2 m$^2$ Fresnel mirror, a field-of-view of 9$^\circ$, and a focal plane with 256 pixels.  The optical system did not include a light shroud, which led to accepting more optical backgrounds during dark-sky operation. TUS also measured a wealth of transient atmospheric phenomena \cite{2019RemS...11.2449K}, including structures in the aurora \cite{2021JASTP.22005672K}, which are highly visible due to the Lomonosov satellite near-polar orbit. 

\begin{figure}[h]
\begin{center}
    \includegraphics[width=0.99\textwidth]{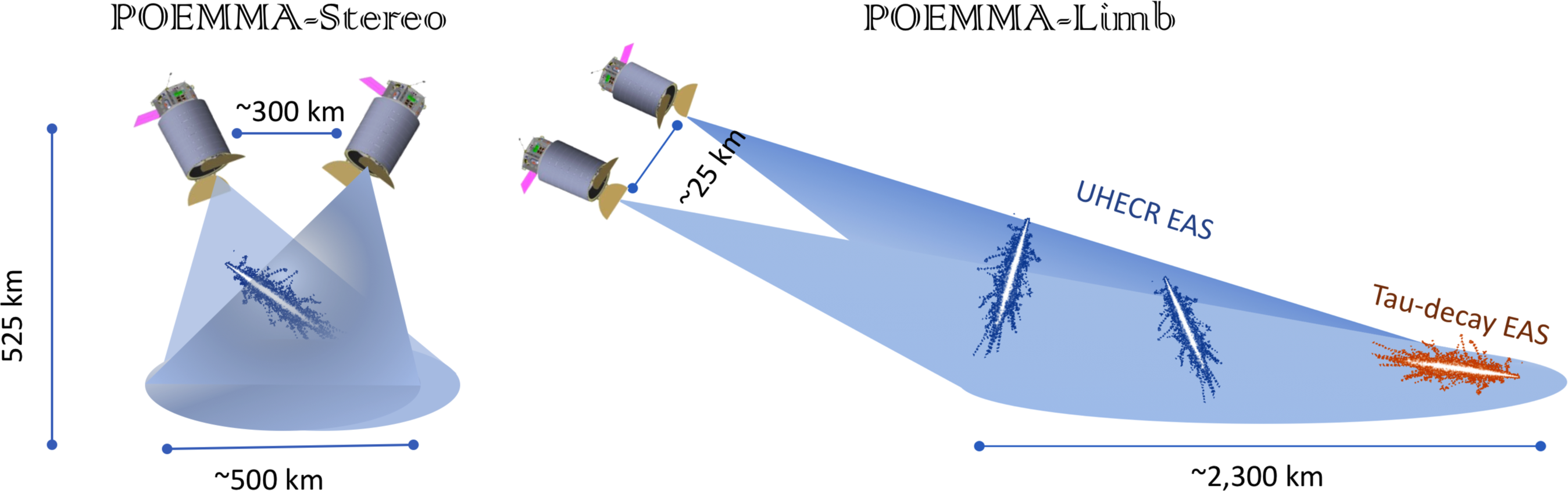}
 \end{center}
\vspace{-0.3cm}
  \caption{The POEMMA science modes. Left: POEMMA-Stereo where the spacecraft are separated and viewing a common atmospheric volume to measure the fluorescence emission from EAS. Right: POEMMA-Limb where the instruments are tilted to view near and below the limb of the Earth for optical Cherenkov EAS induced by tau neutrino events in the Earth. From Ref. \citenum{2021JCAP...06..007P}. 
  DOI:\href{https://doi.org/10.1088/1475-7516/2021/06/007}{10.1088/1475-7516/2021/06/007}
  }
  \label{POEMMAmodes}
\end{figure}

\begin{figure}[h]
\begin{center}
    \includegraphics[width=0.99\textwidth]{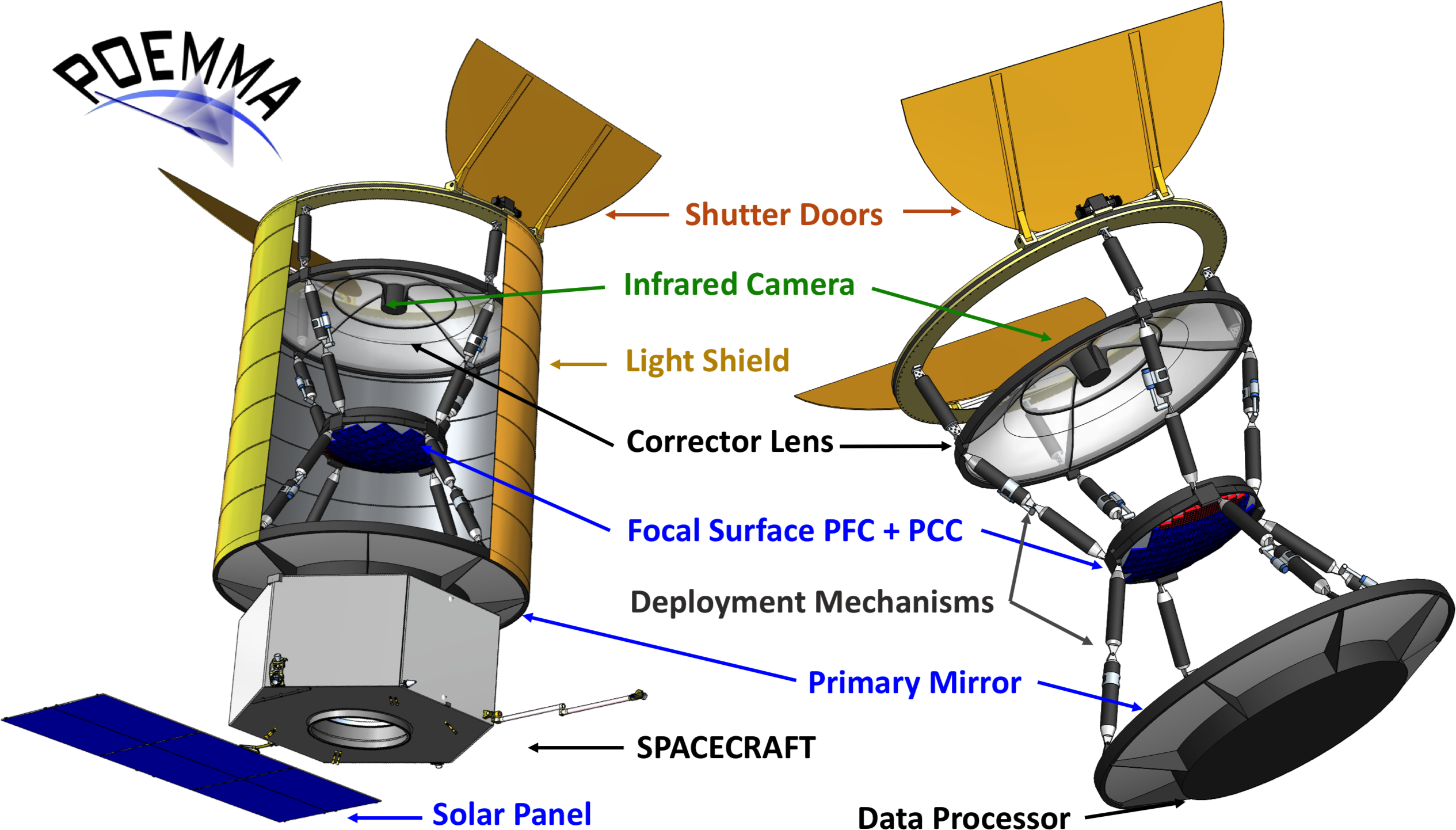}
 \end{center}
\vspace{-0.3cm}
  \caption{Schematics of a POEMMA satellite (left) and the Schmidt telescope (right) consisting of a 4-m diameter primary mirror, 3.3-m diameter corrector plate, and 1.6-m diameter focal surface comprised of 126,720 pixels in the POEMMA Flourescence Camera (PFC) and 15,360 pixels in the POEMMA Cherenkov Camera (PCC)  Several components are detailed in the schematic including infrared cameras which will measure cloud cover within the 45$\circ$ full FoV of each telescope during science observations. From Ref. \citenum{2021JCAP...06..007P}.
  DOI:\href{https://doi.org/10.1088/1475-7516/2021/06/007}{10.1088/1475-7516/2021/06/007}
 }
  \label{POEMMAinst}
\end{figure}

In 2018, these space-based experimental development efforts came together to leverage the work of each research program and led to the development of the Probe of MultiMessenger Astrophysics (POEMMA) mission \cite{2021JCAP...06..007P}. POEMMA was designed as a NASA Astrophysics Probe-class mission to perform precision high-statistics UHECR ($\ge 1,400$ events for a five-year mission) measurements above 20 EeV using the stereo air fluorescence technique, UHE cosmic neutrino searches using stereo air fluorescence measurements, and VHE Earth-interacting tau neutrinos using the beamed optical Cherenkov light generation from EAS for $E_\nu \gsim$ 20 PeV.
Fig.~\ref{POEMMAmodes} illustrates the two science modes of POEMMA.  The first is a Stereo mode where the telescope configuration is oriented to co-measure the EAS air fluorescence signal in a common atmospheric volume corresponding to nearly $10^{13}$ tons of atmosphere. The second had the telescopes pointed to view slightly below the limb of the Earth to be sensitive to the beamed optical Cherenkov signal from upward-moving EAS sourced from tau neutrino interactions in the Earth. Fig.~\ref{POEMMAinst} shows the schematics of each POEMMA satellite and Schmidt telescope, detailing the major components.  Assuming the guidance from the NASA probe study, the POEMMA mission would launch in 2030.



The use of the space-based measurements of the signals from EAS induced from the products of neutrino interactions either in the Earth's atmosphere \cite{1981BAAS...13R.883L,1982srca.conf..169L} or in the Earth itself \cite{Domokos1,Domokos2} has been recognized as a way to achieve incredibly large neutrino target mass, greater than $10^{13}$ metric tons for the atmosphere \cite{1981ICRC....7..125L,OWL}. The OWL \cite{OWL,Krizmanic2011} and POEMMA \cite{2021JCAP...06..007P, 2020PhRvD.101b3012A} instrument designs each used both the fluorescence detection of EAS at the highest energy scales, e.g. above 10 EeV for UHECRs, while using the beamed optical Cherenkov light from upward-moving EAS with neutrino energy thresholds around the PeV-energy scale \cite{2019PhRvD..99f3011R,2020PhRvD.102l3013V,2021PhRvD.103d3017C}. 

\begin{figure}
\vspace{-1mm}
 \begin{center}
    \includegraphics[width=1.0\textwidth]{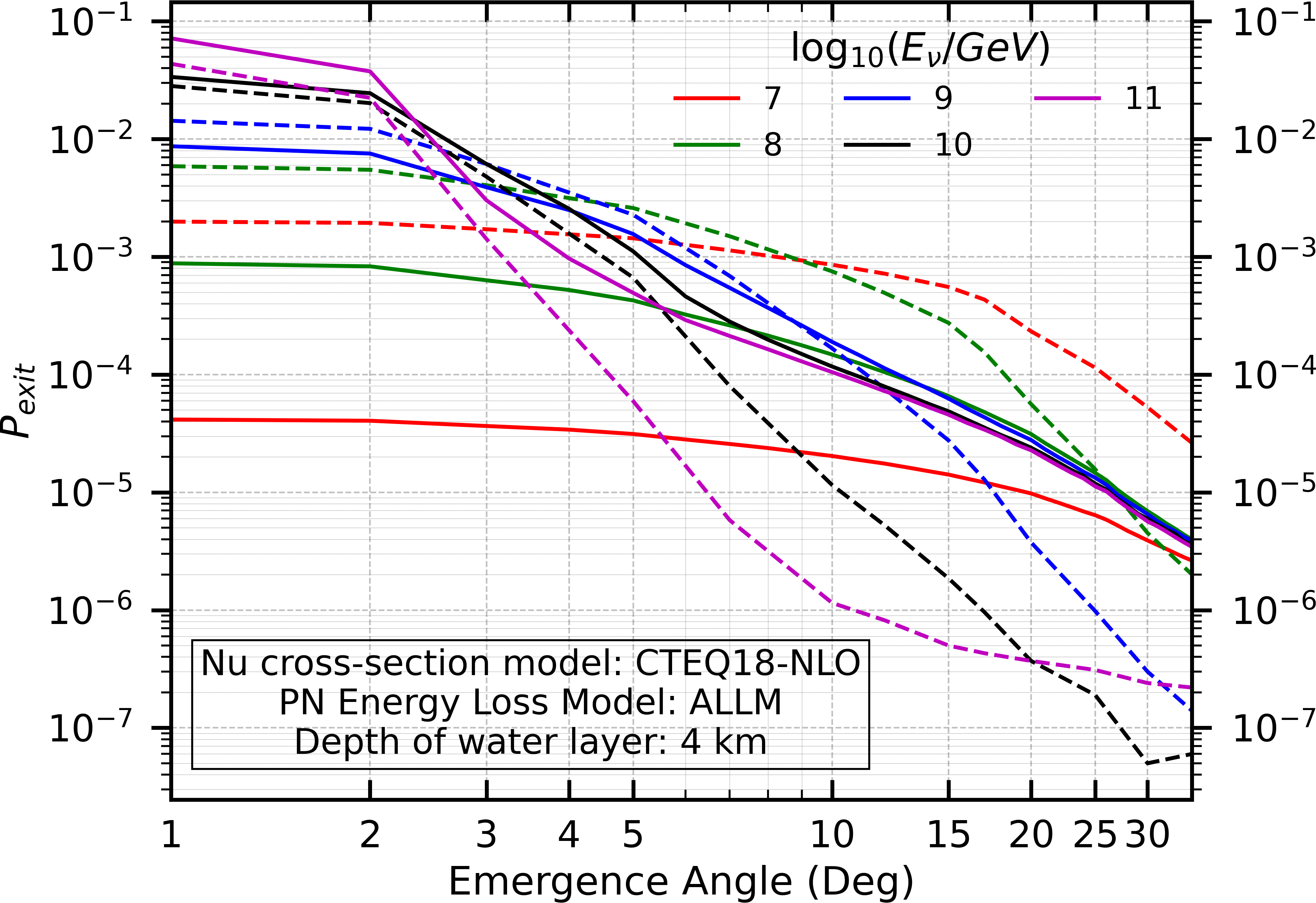}
  \end{center}
  \vspace{-1mm}
  \caption{\taon P$_{\rm Exit}$ versus tau neutrino energy and Earth-emergence angles using CTEQ18-NLO \cite{cteq18_nlo, hls_param} neutrino cross sections and  ALLM \cite{allm,reno_beta}  photonuclear \taon energy loss model: $\nu_\tau  \rightarrow \tau$ (solid); $\nu_\mu  \rightarrow \mu$ (dashed). From Refs.~\citenum{2021ICRC...37..1203P,2021ICRC...37..1205K}.
  DOI: \href{https://doi.org/10.22323/1.395.1205}{10.22323/1.395.1205}
  \label{Pexit}}
  \vspace{-1mm}
\end{figure}

When using the Earth as a tau neutrino converter,
the  \taon Earth-exit probabilities and energy spectra are strongly affected by the amount of material and composition, e.g. water/ice or rock, in which the \taon must propagate before exiting the Earth. At energies $\gsim$ PeV, \taons lose energy through ionization, bremsstrahlung, pair production, and photonuclear energy losses \cite{2019PhRvD..99l2001G,2021ICRC...37..1203P}. Furthermore, 
the \taon propagation through the Earth provides an additional \nutau flux from \taon decays, denoted as regeneration, which provides an enhancement of the lower energy \taon Earth-emergent flux \cite{Halzen:1998be}. The net effect of the \taon energy loss and regeneration is detailed in Fig.~\ref{Pexit} which shows the \taon P$_{\rm Exit}$ distributions as a function of incident tau neutrino energy and Earth-emergence angles.  Fig.~\ref{nuEASchannels} highlights two different regimes of tau neutrino interactions in the Earth. The first is denoted as ``Earth-skimming'' events where very high-energy \nutau interactions yield high exit probabilities and high-energy \taon leaving the Earth \cite{2021ICRC...37..1203P}, due to the \nutau interactions occurring in the thin outer water/ice layer of the Earth \cite{PREM} whose transient to rock is at an Earth-emergence angle $\psi \approx 2^\circ$ ($\psi$ defined in Fig.~\ref{UpNuCher}).  For larger Earth-emergence angles, the events are denoted as ``Earth-emergent'' since the \taon energy loss and decay process degrades the exiting \taon energy spectrum while also beginning the \nutau regeneration process.

The signals from upward-moving EAS sourced from tau neutrino interactions within the Earth provides a path to measure the astrophysical and the cosmogenic neutrino flux above one PeV \cite{Halzen:1998be,Feng:2001ue,2003APh....18..539B}, with a huge neutrino target volume of the order of $10^7$ km$^3$ \cite{Krizmanic2011, CHANT, 2021JCAP...06..007P}.
Simulation studies have shown that using the beamed optical Cherenkov light EAS signals can potentially yield EAS Cherenkov detection in the 1$-$10 PeV range with instruments with m$^2$-scale effective area \cite{2019PhRvD.100f3010R,2020PhRvD.102l3013V,2021PhRvD.103d3017C} while also having high sensitivity to measuring over-the-limb VHECRs \cite{2021arXiv210503255C}. While optical EAS measurements need to be performed near astronomical night, estimated to yield $\sim20\%$ duty cycle \cite{2019PhRvD.100f3010R, 2021NIMPA.98564614K}, they can be used to minimizing the energy threshold for EAS detection as compared to the radio detection technique, which currently has neutrino energy thresholds in the range of 0.1$-$1 EeV from balloon-borne instruments \cite{2009APh....32...10A,2020arXiv201002892A}. Both the optical Cherenkov and radio techniques have excellent angular resolution on the incident neutrino direction with the neutrino energy resolution dependent on the processes that lead to the showers producing the measured signals.

Pioneering searches for neutrino-induced Asrkayran radiation in the Greenland Ice Sheet was performed using the Fast On-orbit Recording of Transient Events (FORTE) satellite, but the sensitivity of the measurements corresponded to an energy threshold of around $10^{22}$ eV \cite{2004PhRvD..69a3008L}.
To date, the most successful balloon-borne cosmic neutrino measurements have been made during
the four long-duration balloon (LDB) flights of the Antarctic Impulsive Transient Antenna (ANITA) experiment \cite{2009APh....32...10A,2010PhRvD..82b2004G,2019EPJWC.21601009V,2019PhRvD..99l2001G}. ANITA has demonstrated the technique using beamed Askaryan radio emission \cite{Askaryan:1962hbi} from neutrino-induced particle cascades, e.g. showers, in the Antarctic ice. ANITA also has used the geomagnetic radio EAS signals to measure UHECRs and to search for upward-moving EAS source from Earth-emerging \taons from tau neutrino interactions in the Earth. The four flights of continuously updated versions of ANITA have provided sufficient sensitivity to place limits on the diffuse flux of UHE neutrinos \cite{2019PhRvD..99l2001G} as well as severe constraints \cite{ANITAtd} on speculative, so called `top-down' models that were postulated to provide high fluxes of UHECRs and UHE neutrinos.  ANITA-IV radio detection measurements \cite{2019PhRvD..99l2001G, 2020arXiv200805690A} included a class
of events consistent with that from Earth-emerging \taons observed near the Earth's limb from \Enu $\gsim$ 1 EeV \cite{2021PhRvL.126g1103G}.
Additionally, 
two anomalous EAS events from the ANITA-I and ANITA-III LDB flights have been measured \cite{2018PhRvL.121p1102G} consistent with the radio signal from EAS generated by $\nu_\tau$ interactions in the Earth, except that the reconstructed EAS energies ($\sim 500 PeV$) imply a tiny probability for these events given the projected material path lengths through the Earth \cite{2019PhRvD..99f3011R}. This has fueled a hypothesis that this event may be due beyond the Standard Model physics, such as due to an Earth-interacting sterile neutrino \cite{SterileNu}.  The ANITA-IV results include two over-the-limb directly observed UHECR events along with an analysis of the effects of atmospheric refraction for viewing just above the Earth limb.  Atmospheric refraction of optical and radio signals is a  potential VHECR/UHECR background to neutrino-induced upward-moving EAS \cite{2019PhRvD..99l2001G, 2020PhRvD.102l3013V}.
The technique using the beamed optical Cherenkov light EAS signals, searching for neutrino-induced EAS as well as measuring VHECRs from above the Earth's limb, will be demonstrated in the ULDB flight of EUSO-SPB2 \cite{2020NIMPA.95862164S} to be launched from Wanaka, New Zealand in the spring of 2023.

\section{Neutrino Interactions above 1 PeV}\label{SNc_NuInt}

\begin{figure}[ht]
\centerline{
\minifigure[Schematic diagram of neutrino-nucleon deep-inelastic scattering (DIS) where $\vec{p_1}$ is the momentum of the incident neutrino and $\vec{p_3}$ is the momentum of the outgoing lepton. The interactions are mediated by an appropriate vector boson which scatter off a parton with momentum $x \vec{p_2}$ where $x$ is the momentum fraction of the parton and $\vec{p_2}$ is the momentum of the incoming nucleon. The charged-current interaction is mediated by a W$^\pm$ boson and yields the charged-lepton ($\lambda$) associated with the neutrino flavor. The neutral-current is mediated by a Z$^0$ boson and the outgoing neutrino of the same flavor. The DIS process leads to an outgoing lepton with momentum fraction $y$ as well as a hadronic system with momentum $(1-y)$. The From Ref. \citenum{2019JHEP...09..083A}.
  DOI: \href{https://doi.org/10.1007/JHEP09(2019)083}{10.1007/JHEP09(2019)083}
]
{\includegraphics[width=2.5in]{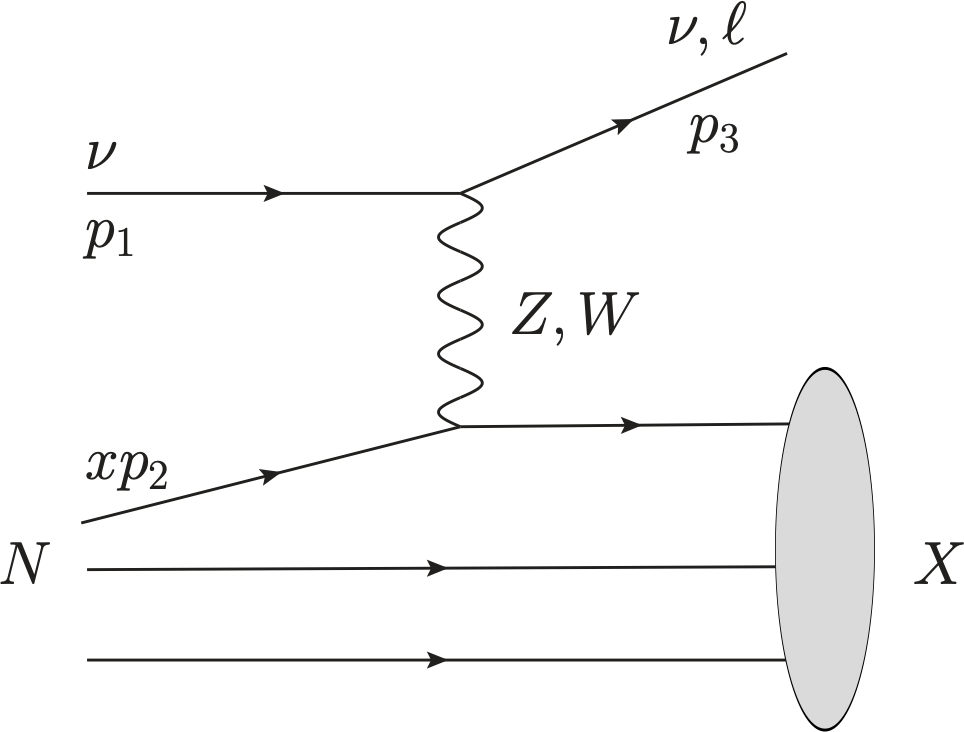}\label{nuScatter}}
\minifigure[Simplified schematic of an energetic neutrino scatter off a parton leading to two separate EAS: one from the generated lepton and the other from the generated hadronic ensemble. For the charged-current high-energy case, both the charged-lepton, with incident neutrino energy fraction $y$ and the hadronic system, with energy fraction $(1-y)$, will produce EAS, with the lepton flavor ($e,\mu,\tau$) determining the nature of the leptonic EAS. For the neutral-current neutrino interactions, only the hadronic system generates an EAS, with energy fraction $(1-y)$. The opening angle between these scales as $\approx 1/\gamma$, where $\gamma$ is the Lorentz boost of the lepton, and is exaggerated in the schematic.]
{\includegraphics[width=2.1in]{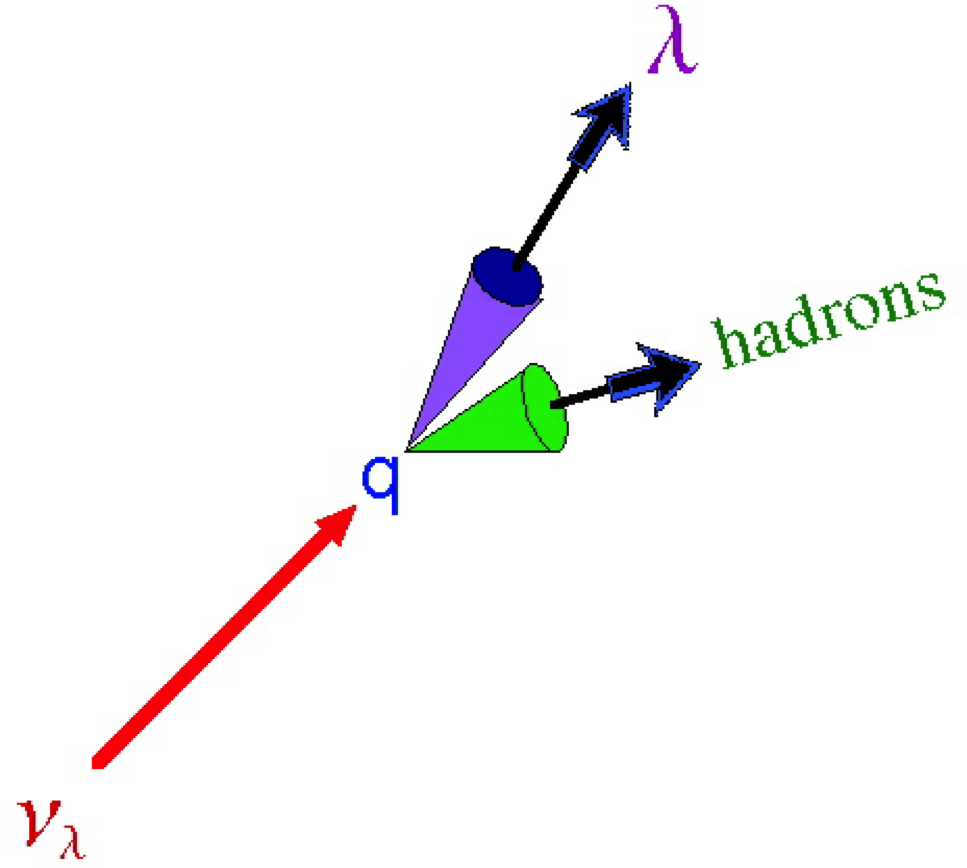}\label{nuIntCartoon}}
}
\end{figure}

\begin{figure}[ht]
\begin{center}
    
{\includegraphics[width=0.99\textwidth]{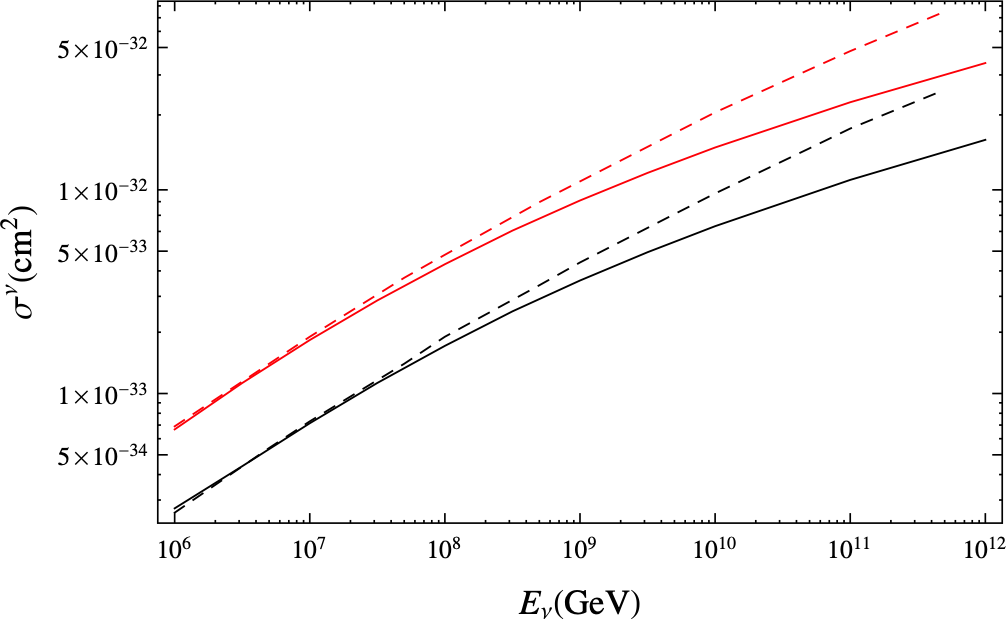}}
\caption{The charged-current (upper, Red) and neutral-current (lower, black) cross-sections as a function of neutrino energy for the DBH (solid curves) compared to that of the CSMS \cite{2011JHEP...08..042C} calculations (dashed curves). From Ref. \citenum{Block:2014kza}.
DOI: \href{https://doi.org/10.1103/PhysRevD.89.094027}{10.1103/PhysRevD.89.094027}
\label{BlockSigma}}
\end{center}
\end{figure}

\begin{figure}
\begin{center}
{\includegraphics[width=0.99\textwidth]{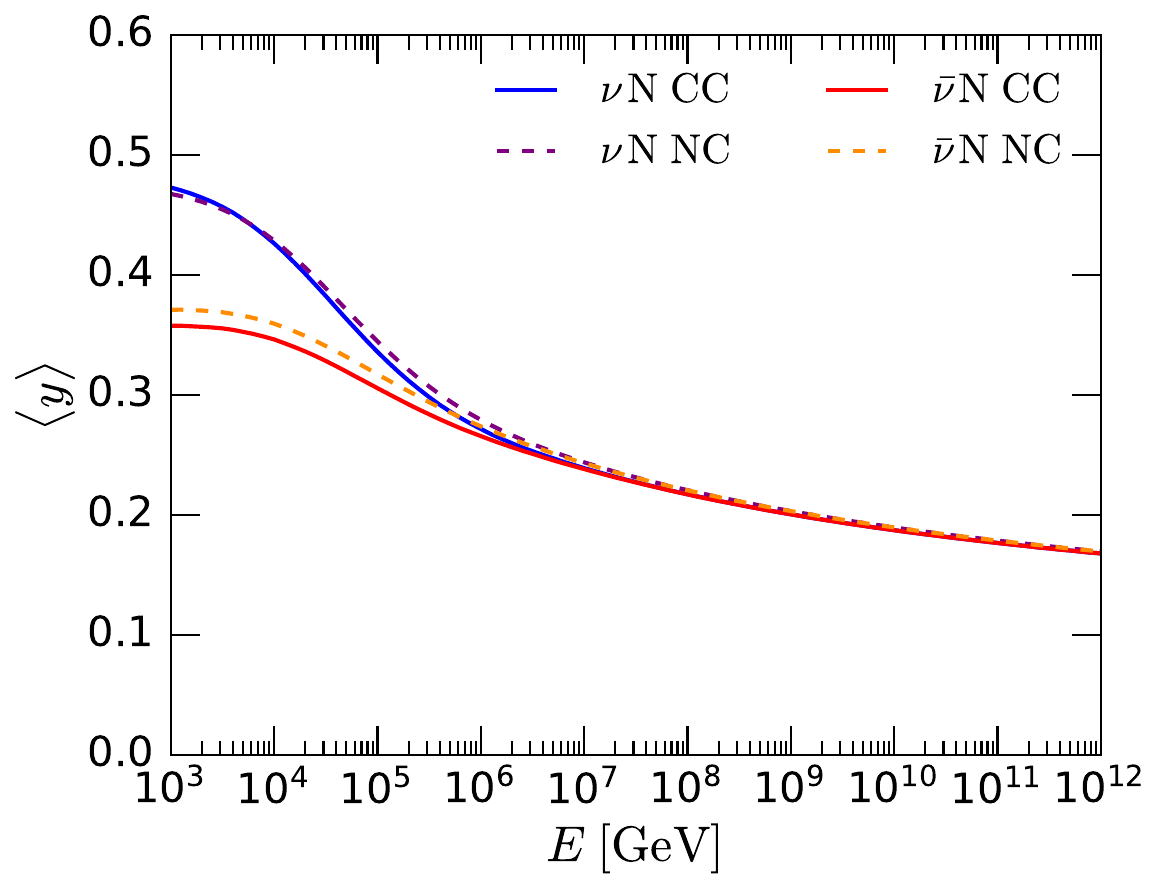}}
\caption{The average fractional energy loss  $<{\rm y}>$  as a function of energy for neutrinos and antineutrinos for charged-current and neutral-current scattering using CT18NLO PDFs \cite{2021PhRvD.103a4013H}. From Ref. \citenum{Reno21}.\label{nuYdep}}
\end{center}
\end{figure}

At high energies, \Enu $~\gsim 4$ GeV  both the charged-current (CC) and neutral-current (NC) neutrino and antineutrino cross sections are dominated by deep inelastic scattering off of nucleons\footnote{see Section 51 in Ref. \citenum{2020PTEP.2020h3C01P}}. This is schematically shown in Figs. \ref{nuScatter} and \ref{nuIntCartoon}. The process leads to multiple mesons in the final state as well as at least one nucleon\footnote{see Figs. 6.2$-$6.5 in Ref. \citenum{1990PhDT........94K}.}  One exception is the Glashow resonance \cite{PhysRev.118.316}, $\nuebar ~e^- \rightarrow W^- $ for $\nuebar$ energy of 6.3 PeV. Above around a PeV, the neutrino-nucleon and antineutrino-nucleon cross sections are essentially equal because the cross sections are dominated by scattering off of sea quarks in the nucleon \cite{PhysRevD.58.093009,PhysRevD.83.113009,2011JHEP...08..042C,Block:2013mia,Block:2014kza}. Since the neutrino energy scale for space-based observations is above $\sim$ 1 PeV, $\nu$ and $\nubar$ interactions will be referred to just neutrino interactions for simplicity.  At these high energies, the CC neutrino interaction process is given by $\nu_\lambda N \rightarrow \lambda X$ where $\lambda = \mbox{\{}e\mbox{, }\mu\mbox{, }\tau\mbox{\}}$, $N$ is the target nucleon, and $X$ is an excited hadronic state. For NC interactions, the process is $\nu_\lambda N \rightarrow \nu_\lambda X$, and $\sigma_{CC}/\sigma_{NC} \approx 0.41 - 0.43$ in the energy range $10^6 ~{\rm GeV} \le ~$\Enu$ ~ \lsim 10^{17}$ GeV based on the calculations by Block, Durand, and Ha (BDH) \cite{Block:2014kza}. However, these calculations extend the understanding of the particle physics, in particular the extrapolation of the measurements of parton distribution functions (PDFs) to very small Bjorken-x, to extreme energies and therefore lead to a theoretical systematic error, as illustrated in Fig.~\ref{BlockSigma}. At these energies, the average energy transfer for both CC and NC neutrino interactions to the hadronic system is $\langle y\rangle = (E_\nu - E_\lambda)/E_\nu\simeq 0.20$ for $10^6 ~{\rm GeV}\le ~$\Enu$ ~ \lsim 10^{11}$ GeV and decreasing to $y \simeq 0.08$ at $10^{16}$ GeV \cite{PhysRevD.58.093009,PhysRevD.83.113009,Block:2013mia}, as shown in Fig.~\ref{nuYdep}.  Thus for a CC neutrino  interaction, the charged lepton carries $\sim$ 0.80 of the initial energy with $\sim$ 0.20 going into the hadronic system, and each of these will yield appropriate EAS, dependent upon the flavor of the charged lepton, as shown in Fig.~\ref{nuIntCartoon}. For the NC interaction, an EAS is only created by the hadronic system with $\sim$ 0.20 of the initial neutrino energy and with a relatively reduced interaction cross section as compared to that for CC interactions.

High-energy particles interacting with matter produce particle cascades in the medium, and this is shown schematically in Fig.~\ref{EASdetectCartoon} for a hadronic primary particle interacting in the atmosphere. The depth of initial interaction is given by the nuclear interaction length for hadronic particles or by the radiation length of electromagnetic particles\footnote{For a detailed discussion see Ref. \citenum{GaisserBook}.}  The specific particle content of the EAS, even for hadronic primaries, is quickly (within several cascade interaction/decay steps) by the electromagnetic component: photons, electrons, and positrons as shown in Fig.~\ref{EASdetectCartoon}. The particle content grows to a maximum, $N_{MAX}$,  at some depth denoted as \xmax.  The longitudinal profile is shown in Figs.~\ref{EASprofile} and \ref{EAScomponents} as a function of slant depth in units of g/cm$^2$. Since electromagnetic particles dominate the EAS, the lateral width is given by the Moli{\'{e}}re radius, which a few hundred meter wide in the atmosphere at low altitudes, i.e. $\sim 10$ km. Thus, an EAS is effectively a pancake of particles a few meters thick with a width defined by the oli{\'{e}}re radius, traveling through the atmosphere at nearly the speed of light.


Fig.~\ref{EASprofile} shows the important properties of $E_{\rm CR} = 10^{11}$ GeV proton induced EAS cascades including the effects of initial interaction, or starting point fluctuations, and that the charged particle density at the peak of the profile, denoted as shower maximum (\nmax), is fairly constant for a fixed $E_{\rm CR}$ with only the {\it depth of the location of shower maximum}, denoted as \xmax in units of g/cm$^2$ showing the effects of starting point fluctuations. The particle cascades consist primarily of electromagnetic particles as shown in results from the CORSIKA/CONEX simulation code \cite{CORSIKA,CONEX} in Fig.~\ref{EAScomponents}.

An EAS is created when a high-energy particle interacts with an air molecule, or when a particle decays in the atmosphere. It consists of predominantly electromagnetic particles and grows in numbers to a maximum and then shrinks as the average energy of the particles in the shower decreases. The longitudinal size of the shower, the 
number of particle in the shower at a specific development stage, has a statistically determined standard shape in terms of the amount of material in the atmosphere through which the shower has passed, denoted as slant depth, measured in g/cm$^2$. Most of the uncertainty in shower development comes from the location of the first interaction point. The standard shape is the Gaisser-Hillas profile \citep{Gaisser-Hillas-1977-ICRC-15-8-353}: where \nmax\ is the size of the shower at maximum,
\begin{equation}
N(X) = \nmax \left(\frac{\xmax-X}{\xmax-X_0}\right)^\frac{\xmax-X_0}{\Lambda} \exp \left(\frac{\xmax-X}{\Lambda}\right) 
\end{equation}
\xmax\ is the depth of shower maximum, $X_0$ controls where the shower begins and $\Lambda$ controls the rate at which the shower grows and shrinks. For EAS created by downward-going cosmic rays, \nmax\ is roughly proportional to the primary energy, \xmax\ increases logarithmically with the energy with values in the range 500-1000 g/cm$^2$, and $\Lambda$ is in the range 50-100 g/cm$^2$ and is related to the width of the EAS. In practice,  $X_0$ is sometimes allowed to be negative to better mathematically fit the EAS profiles. A negative $X_0$ does not represent the physical first point of interaction). 

\section{Particle Cascade and Extensive Air Shower (EAS) Signal Development}

\begin{figure}[ht]
\centerline{
\minifigure[Example of the longitudinal profiles of simulated 100 EeV proton EAS \cite{1999ICRC....1..445M,1999ICRC....2..388K}.]
{\includegraphics[width=2.in]{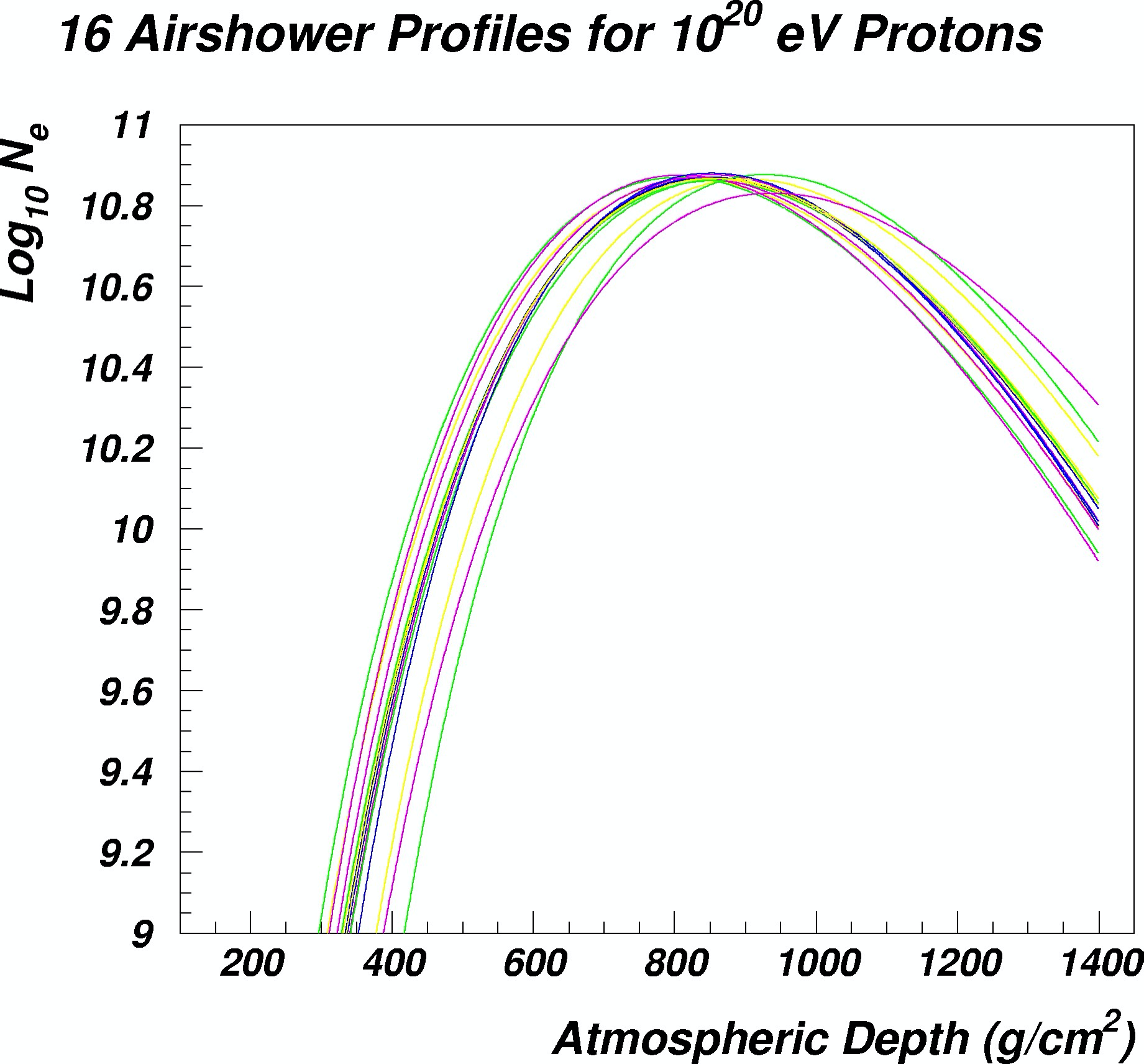}\label{EASprofile}}
\minifigure[The particle yield as a function of shower age for a 100 PeV pion EAS from the CONEX simulation as a function of starting altitude \cite{CONEX,2021ICRC...37..1205K}.
DOI: \href{https://doi.org/10.22323/1.395.1205}{10.22323/1.395.1205}
]
{\includegraphics[width=2.7in]{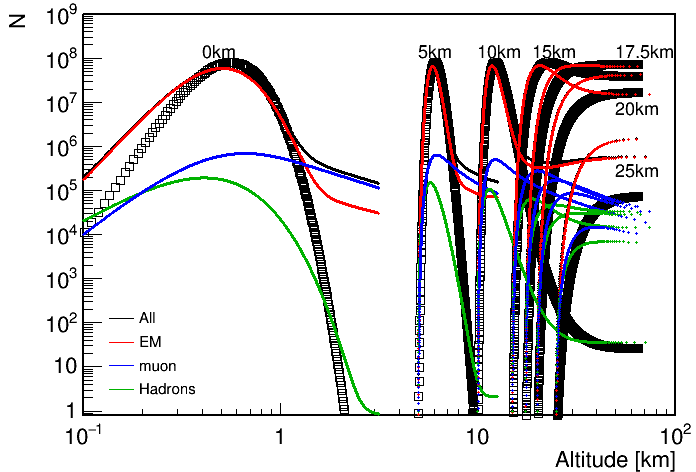}
\label{EAScomponents}}
}
\end{figure}

The EAS particles will excite nitrogen molecules in the atmosphere which will produce measurable mainly UV fluorescence light isotropically that can be detected remotely \cite{PDG}. The yield of fluorescence light (at STP) is of order of 10 photons per MeV of ionization energy deposited by EAS particles. Viewed from a distance of more than 10 km, the EAS can be well approximated as a line along which the shower progresses, at first brightening then diminishing. A parametric 1-D model using the Gaisser-Hillas profile \cite{Gaisser-Hillas-1977-ICRC-15-8-353} along the shower track combined with ray-tracing of the isotropic light produced suffices for fluorescence light simulation for showers \cite{2010APh....33..221B,2020PhRvD.102l3013V}.  After atmospheric attenuation of the signal to the instrument and applying an instrument response model, the ability of an experiment to measure cosmic particles can be modeled. The left schematic in Fig.~\ref{POEMMAmodes} illustrates stereo fluorescence detection using two Schmidt telescopes for POEMMA \cite{2021JCAP...06..007P} for UHECR and UHE neutrino measurements while the right panel shows the POEMMA configuration for measuring VHECRs and VHE neutrinos via the optical Cherenkov EAS emission.

The EAS also creates an Cherenkov light signal due to the fact the charged particles are moving faster than the speed of light in the medium, e.g. the atmosphere. The mean Cherenkov signal can be estimated using a 1-D longitudinal model of the EAS development, e.g. using the Greisen parameterization \cite{GaisserBook} or Gaisser-Hillas fits of simulated EAS, along with a model of the EAS electron and positron angular and energy distribution as a function of shower age {\cite{Hillas1, Hillas2}. Since the electrons and positrons dominate the fraction of charged particles in the EAS, their angular and energy distributions define the spatial and temporal extent of the EAS Cherenkov light signal \cite{PattersonHillas}. This modeling is based on the observation that the constituent EAS particle evolution have a universal, yet statistically determined, distributions after accounting for the starting point fluctuations of the EAS. More recent phenomenological models of these distributions have been developed by several groups \citep{Giller-2004-JPG-30-97,Nerling-2006-APP-24-421,Lafebre-2009-APP-31-243, Bergman-2013-ICRC-33-0983} which in turn have increased the fidelity of the modeling of the EAS Cherenkov light spatial and temporal distributions.

\subsection{EAS from \taon Decays}

The EAS formed from $\tau$-lepton decays depend on the decay channel and decay products \cite{PDG2}. $\sim 65
\%$ of the decays contain at least one hadron, such as a pion, while $\sim 17.5\%$ of the branching ratio leads to an electron with two neutrinos and $\sim 17.5\%$  with a muon and two neutrinos. \taon are produced polarized in the electroweak interaction and this has an impact on the \taon energy fraction put into the decay particles. The left panel in Fig.~\ref{tauDecays} shows the energy fraction for $\tau \rightarrow \pi\pi^{0}\nu_{\tau}$ decay channel assuming left-handed and unpolarized \taon decay obtained from the Pythia8 simulation package \cite{sjo15}.  Theoretical studies are investigating impact of \taon depolarization from the \taon energy loss processes in the Earth  \cite{2021arXiv211014662S,2021ICRC...37..1203P,2021ICRC...37..1205K}.

The right panel in Fig.~\ref{tauDecays} presents a sample of EAS simulated from several 100 PeV \taon decays. The results show an inherent variability in the charged-particle content of the EAS and with more minor variation in the location of \xmax. The example of the electron channel decay shows the effects of sampling the \taon decay energy fraction for the specific decay channel, sometimes only a small amount the the energy is in the emergent charged particle, in this case an electron.
The altitude of the \taon decay EAS development is a function of the Earth-emergence angle, $\beta_E = 90^\circ - \beta_{tr}$ shown in the left panel of Fig.~\ref{upGEOM}, and the energy of the emitted $\tau$-lepton. For small values of $\beta_E$, lower energy $\tau$-lepton decays will be constrained to low altitudes.  For example using $\beta_E=5^\circ$, 90\% (21\%) of $\tau$-leptons with $E_\tau=10^8~(10^9)$ GeV will decay below an altitude of 1 km. If an EAS completes its development at a distance of $\sim$10 km away from a detector, the EAS appears essentially as a 1-D quantity, assuming the  width of the EAS is $<$ 1 km.  However, for high energy \taon decays, the EAS particle development can extend above 20 km altitude where the development is stalled due to the lack of material in the rarified atmosphere at these altitudes.  This implies that a balloon-borne experiment may actually be illuminated by the EAS particles themselves if muon or \taon generated EAS occur at high altitudes. This variability in the altitude and angle away from zenith leads to a large variation in the strength of the atmospheric attenuation of the Cherenkov light, and this is quantified by plotting the optical depth as a function of wavelength and angle away from nadir.  The right panel of Fig.~\ref{upGEOM} presents the different optical depths defined by the optical path from the surface of the Earth to an 525 km observation altitude. For tau neutrino energies below 100 PeV and out to Earth-emergence angles $\sim 10^\circ$, the majority of tau decays in the atmosphere are near sea level, i.e. $\lsim 1$ km altitude, see Fig. 7 in Ref.~\citenum{2021NIMPA.98564614K}.

\begin{figure}{t}
\begin{center}
{\includegraphics[width=0.48\textwidth, trim = 0mm 0mm 0mm 0mm, clip]{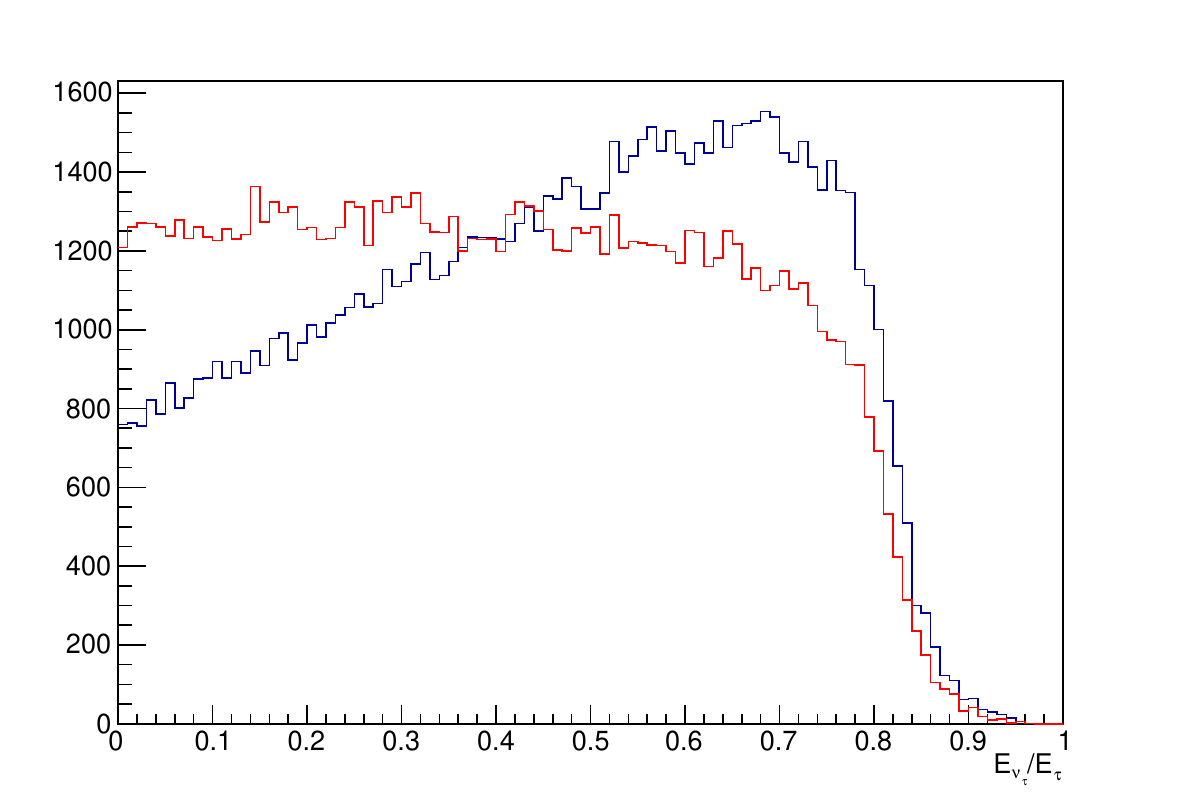}}
    \includegraphics[width=0.48\columnwidth,trim=0 1.5cm 0 1cm]{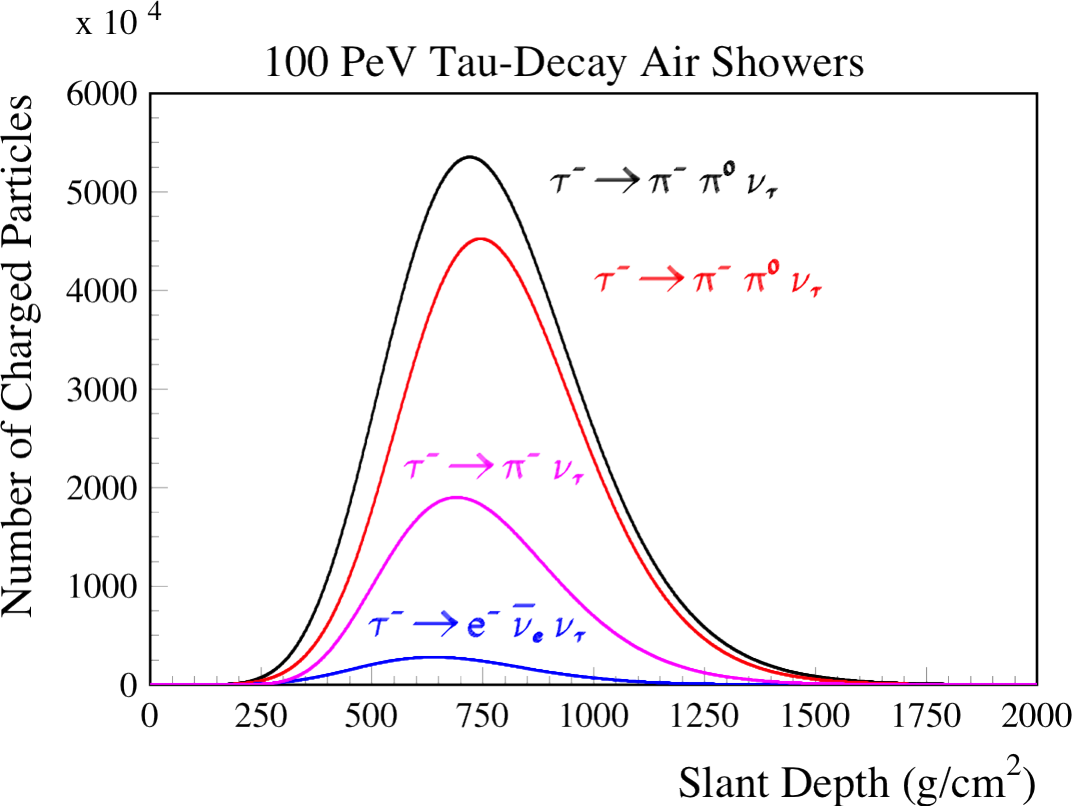}
\vspace{5mm}
\caption{Left: Histograms from Pythia 8 simulations of the decay mode $\tau \rightarrow \pi\pi^{0}\nu_{\tau}$. The blue histogram is the $\nu_{\tau}$ energy distribution for fully polarized left-handed $\tau$-leptons. The blue histogram is the distribution assuming unpolarized $\tau$-leptons decays. Right: The longitudinal profiles of four different simulated EAS from 100 PeV $\tau$-lepton decays. From top down, the 1$^{\rm st}$ decay has 82\% of the initial $\tau$-lepton energy in the hadronic system ($f_{\pi^-}$=49\%; $f_{\pi^0}$=33\% ) the 2$^{\rm nd}$ has 62\% ($f_{\pi^-}$=22\%; $f_{\pi^0}$=40\% ), the 3$^{\rm rd}$ has $f_{\pi^-}$=62\%, while the 4$^{\rm th}$ has $f_{e}$=3\%. Starting point fluctuations for the $\pi^-$ EASs vary from 12 to 180 g/cm$^2$. From Ref.~\citenum{2019ICRC...36..936K}.
  DOI:\href{https://doi.org/10.22323/1.358.0936}{10.22323/1.358.0936}
\label{tauDecays}}
\end{center}
\end{figure}

\begin{figure}[h]
\begin{center}
    \includegraphics[width=0.43\textwidth]{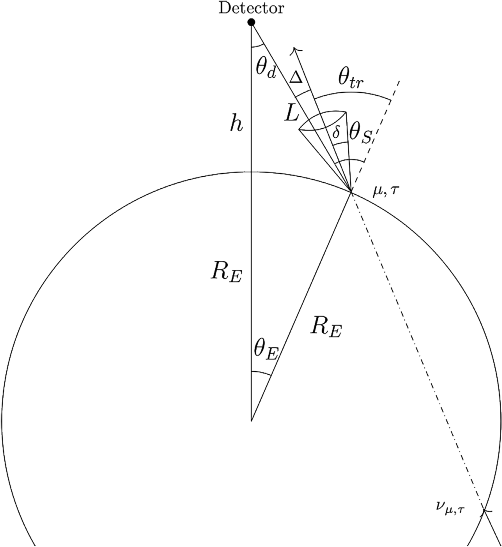}
   \includegraphics[width=0.55\textwidth]{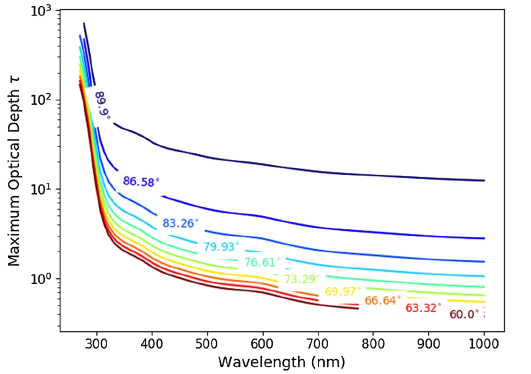}
 \end{center}
\vspace{-0.3cm}
  \caption{Left: The geometrical relationship between viewing angle ($\theta_d$) and Earth-emergence angle ($\beta_E = 90^\circ - \theta_{tr}$) for an observation at altitude $h$. The optical Cherenkov signal allows an EAS to be viewed at an angle $\theta_S - \theta_{tr}$ from the trajectory defined by $\theta_{tr}$. Below a maximum Earth-emergence angle defined by a maximum column depth in the Earth, the emerging \taon lepton flux becomes undetectable due to the Earth becoming opaque to these particles. Right: The total optical depth including aerosols, ozone, and Rayleigh scattering, as a function of zenith angle ($90^\circ - \beta_E$) from a point on the surface of the Earth. From Ref. \citenum{2021PhRvD.103d3017C}.
    DOI:\href{https://doi.org/10.1103/PhysRevD.103.043017}{110.1103/PhysRevD.103.043017}
    Right: The total optical depth, including the effects of aerosol, Rayleigh, and ozone attenuation, defined by the angle away from nadir to the surface of the Earth from a 525 km altitude. The limb of the Earth is located at an angle of $67.5^\circ$ in this geometry. From Ref.~\citenum{AustinPlot}.
    }
\vspace{-3 mm}
  \label{upGEOM}
\end{figure}

\section{Optical Cherenkov Light Emission from EAS}

In an EAS, the Cherenkov light is emitted in a narrow cone about the instantaneous direction of each electron with a width that depends on the local index of refraction. Since the electrons travel nearly parallel to the shower axis, the Cherenkov cones are all nearly centered on the shower axis. The transverse momentum of the electrons and the physical size of the shower causes the Cherenkov ring to become smeared. Fig.~\ref{Cspot} shows the Cherenkov profile at an altitude of 525 km for a upward-moving 100 PeV electron-induced EAS with an Earth-emergence angle $\beta_E=10^\circ$, simulated using the Hillas  parameterization of the EAS-electron angular and energy distributions as a function of EAS age \citep{Hillas2}. The Cherenkov signal was generated in the 200 - 1000 nm band and was attenuated using a baseline atmospheric model (along the lines discussed in a later section) that included the effects of Rayleigh scattering, Mie (aerosol) scattering, and ozone absorption \cite{2019PhRvD.100f3010R, 2021PhRvD.103d3017C}.  The structure of the Cherenkov photon density is approximately constant within the Cherenkov ring (horn-like structure in the profile), but for the Cherenkov light that falls outside the ring the flux drops off as a power law with radius. Within the Cherenkov ring, the temporal spread of the Cherenkov photons is of the order of 10 ns while the times spread increases into the microsecond range as the measurement movers further out into the power-law tail \cite{PattersonHillas, PhysRevD.104.063029}. Note that at 525 km altitude, the Cherenkov ring spans $\sim 60$ km in width and the photon density is quite substantial for a 100 PeV EAS. The Cherenkov ring is very pronounced in the profile due to fact that for small $\beta_E$ the EAS completes development at low altitudes, 3 km in this case, and the Cherenkov emission angle varies only by $\sim 0.1^\circ$ below 3 km \cite{2019PhRvD..99f3011R}.   However, at high altitudes where the EAS develops in a rarified atmosphere, the Cherenkov spatial structure is peaked along the axis of propagation due to several effects, including modifications to the EAS development, Cherenkov light angular and energy thresholds, and angular spread of the EAS charged-particles. These are mainly due to the fact that the physical length becomes much longer for a given slant depth (in g/cm$^2$), so much so that at altitudes $\gsim 10$ km the geomagnetic field (via the Lorentz force) significantly bends the trajectories of the Cherenkov-emitting electrons and positrons compared to the inherent spread due to Coulomb multiple scattering \cite{PhysRev.93.646.2,PhysRev.95.1705.4}.

\begin{figure}[ht]
\begin{center}
{\includegraphics[width=0.99\textwidth]{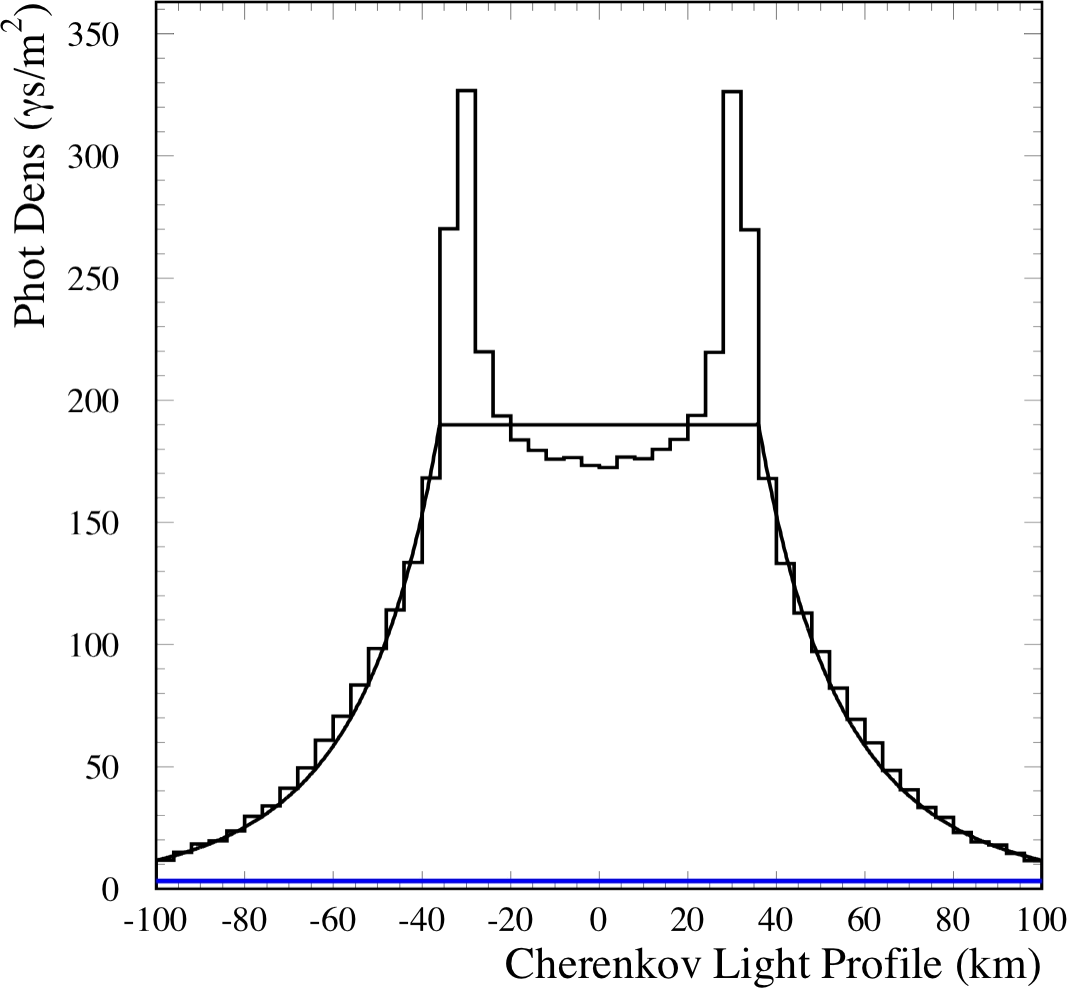}}
\caption{The spatial profile of the optical Cherenkov signal (photons/m2) at 525km altitude for a
100PeV upward EAS with a 15$^\circ$ Earth emergence angle ($\psi$ in Fig.~\ref{UpNuCher}.) From Ref.~\citenum{2021JCAP...06..007P}.
  DOI:\href{https://doi.org/10.1088/1475-7516/2021/06/007}{10.1088/1475-7516/2021/06/007}
\label{Cspot}}
\end{center}
\end{figure}

\begin{figure}
\begin{center}
{\includegraphics[width=0.99\textwidth]{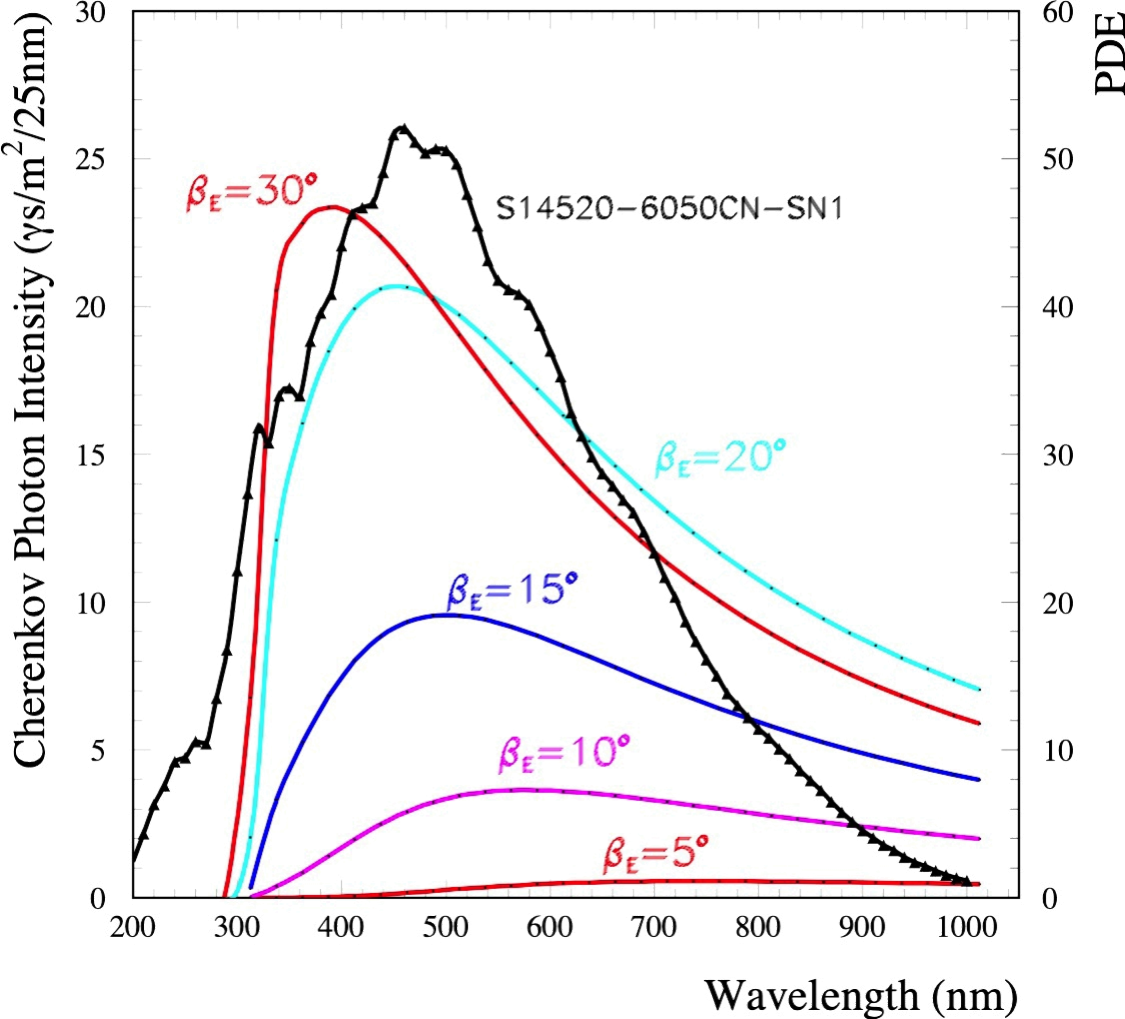}}
\caption{Cherenkov
signal intensity as function of wavelength for 100 PeV upward-moving EAS starting at sea level as a
function of Earth emergence angle ($\psi$) as observed at an altitude of 525 km. The measured photon detection efficiency (PDE) of a Hamamatsu
S14520 SiPM array \cite{OTTE2019283} is overlaid with the PDE scale given on the right horizontal axis. From Ref.~\citenum{2021JCAP...06..007P}.
  DOI:\href{https://doi.org/10.1088/1475-7516/2021/06/007}{10.1088/1475-7516/2021/06/007}
\label{CherSpectra}}
\end{center}
\end{figure}

The Cherenkov angle and threshold energy depends on the index of refraction of the medium, which is a function of altitude in the atmosphere. Using the model \cite{Hillas1} employed in Figs.~\ref{Cspot} and~\ref{CherSpectra}, the Cherenkov angle for relativistic charged particles (energy threshold) varies from $\sim$ 1.3$^\circ$ (22 MeV) at sea level to $\sim$ 0.2$^\circ$ (170 MeV) at an altitude of 30 km.  Thus the beaming of the Cherenkov signal gets stronger at higher altitudes while less of the electron energy spectrum in the EAS generates the Cherenkov signal for a fixed shower age. The temporal width of the Cherenkov signal is narrow within the Cherenkov ring but widens outside the ring \citep{PattersonHillas}. The temporal width of the Cherenkov signals will thus increase as the viewing angle away from the EAS propagation direction increases \cite{PhysRevD.104.063029}. The temporal spread can be accurately estimated using a geometric calculations for upward-moving 1-D EAS assuming \xmax = 750 g/cm$^2$ and a length of 500 g/cm$^2$, the temporal width  of the Cherenkov signal is calculated to be $\sim$6 ($\sim$70) ns for a viewing angles of 1.5$^\circ$ (5$^\circ$) for altitudes of observation of 33, 525, and 1000 km. This calculation does not include the effect of finite width of the EAS `pancake' that will add $\sim$ 10 ns.

The optical Cherenkov spectrum is strongly affected by scattering and absorption in the Earth's atmosphere \cite{2019PhRvD..99l2001G,2021PhRvD.103a4013H} and this is shown in Fig.~\ref{CherSpectra}. Included are the wavelength-dependent attenuation effects of Rayleigh scattering, Mie (aerosol) scattering, and ozone absorption on 100 PeV upward-moving EAS starting at sea level as a function of Earth-emergence angle ($\psi$ in Fig.~\ref{UpNuCher}). In particular ozone effectively eliminates transmission below 300 nm while the $\sim$1 km scale height aerosol layer near the Earth's surface strongly attenuate the optical Cherenkov signal, especially at the shortest wavelengths \cite{Elterman}.

\subsection{Optical Detection Instrumentation}

The most advanced current design for space-based UHECR and VHE and UHE cosmic neutrino detection experiment is Probe of MultiMessenger Astrophysics (POEMMA) \cite{2021JCAP...06..007P}. A NASA Astrophysics Probe-class mission, POEMMA is designed to observe cosmic neutrinos and to identify the sources of UHECRs with full-sky coverage. POEMMA consists of two spacecraft that co-view EAS while flying in a loose formation at 525 km altitudes with 300 km nominal spacecraft separation at 28.5$^\circ$ orbit inclination. Each spacecraft hosts a large Schmidt telescope with a novel focal plane, shown in Fig.~\ref{POEMMAfp} optimized to observe both the forward beamed, optical Cherenkov signals from extensive air showers (EAS) and the isotropic, near-UV fluorescence signal generated by EAS. In neutrino limb-viewing Cherenkov mode, POEMMA will be sensitive to cosmic tau neutrinos above 20 PeV by observing the upward-moving EAS induced from tau neutrino interactions in the Earth. The POEMMA spacecraft are designed to quickly re-orient to a Target-of-Opportunity (ToO) neutrino mode to view and follow transient astrophysical sources with unique flux sensitivity over the full sky. The spacecraft also include sufficient propulsion to adjust the satellite separation, especially reducing it to $\sim 25$km to optimize Cherenkov ToO observations \cite{2020PhRvD.102l3013V}.

\begin{figure}[h]
\begin{center}
    \includegraphics[width=0.99\textwidth]{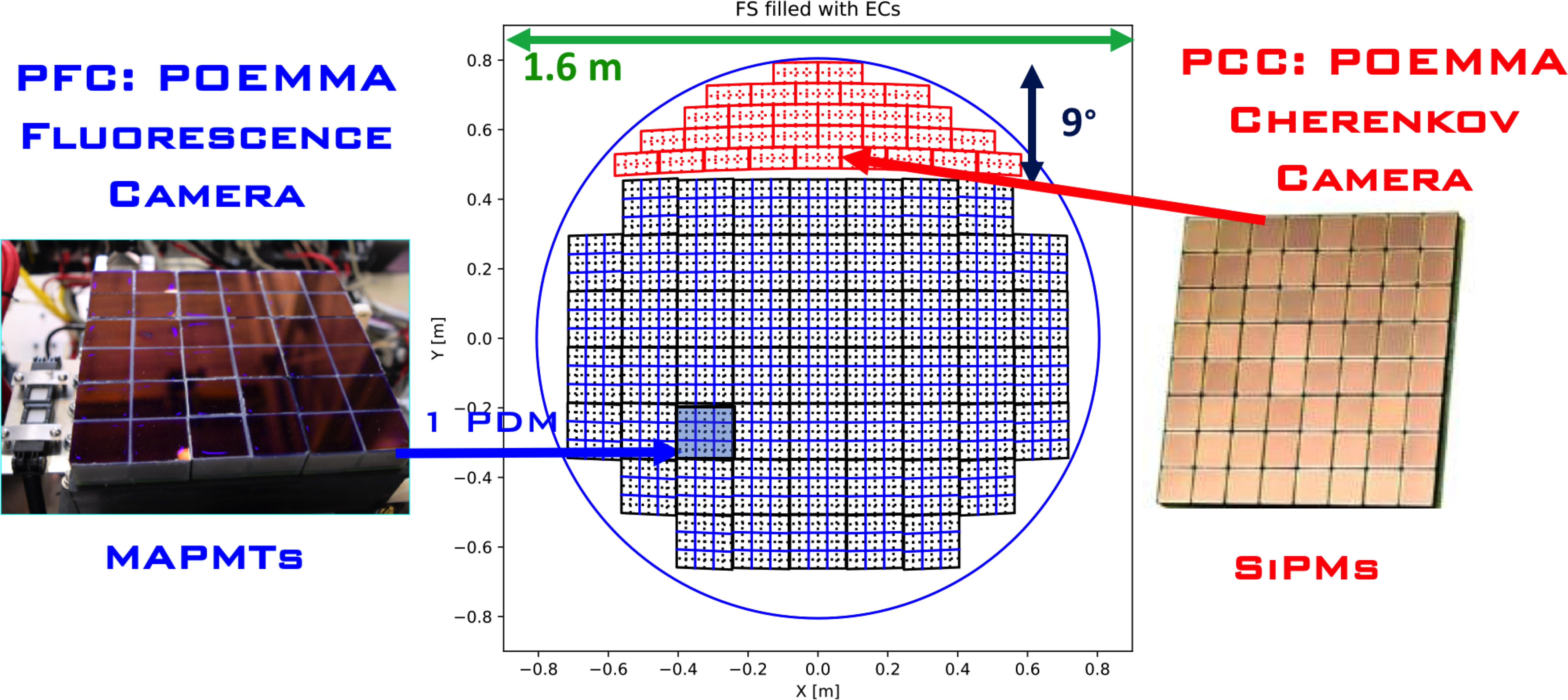}
 \end{center}
\vspace{-0.3cm}
  \caption{The layout of the hybrid focal plane of a POEMMA Schmidt telescope. The majority of the area is comprised of PFC MAPMT modules with a UV filter to record the 300$-$500 nm air fluorescence light in 1 $\mu$s snapshots. The PCC is comprised of SiPM pixels whose 300$-$1000 nm wavelength response is well-matched to that from the EAS optical Cherenkov signals and are recorded with 10 ns cadence. From Ref. \citenum{2021JCAP...06..007P}.
    DOI:\href{https://doi.org/10.1088/1475-7516/2021/06/007}{10.1088/1475-7516/2021/06/007}
  }
\vspace{-3 mm}
  \label{POEMMAfp}
\end{figure}

When the POEMMA telescopes are pointed towards the Earth's limb, a region corresponding to $9^\circ$ in elevation angle (defined from the nadir direction) and approximately $30^\circ$ in azimuth is sensitive to the Cherenkov signal from upward-moving EAS.  This angular range is illustrated by the SiPM (red) component in Fig.~\ref{POEMMAfp}.  This angular range constrains the measurement for diffuse neutrino flux measurements since a full $360^\circ$ azimuth range is desirable for an isotropic cosmic flux. The left panel in Fig.~\ref{upGEOM} defines the geometrical relationships between detector viewing angle ($\theta_d$) and the Earth-emergence angle ($90^\circ - \theta_{tr}$) of \taons.  In the case of POEMMA viewing EAS from 525 km altitude, a $\theta_d = 9^\circ$ corresponds to an Earth-emergence angle $90^\circ - \theta_{tr} \approx 26.5^\circ$, which then defines the Earth-emergence \taon flux versus incidence tau neutrino energy, as shown in Fig.~\ref{Pexit}. In practice, POEMMA would orient such that the telescopes view 2$^\circ$ band {\it above} the limb in order to measure the background from the Cherenkov signal from VHECRs refracting near the Earth's limb \cite{2020PhRvD.102l3013V,PhysRevD.104.063029}.  The 7$^\circ$ viewing below the limb corresponds to a maximum Earth-emergence angle of $\sim 19.6^\circ$. Thus a choice is made to sacrifice some low-energy \taon sensitivity in order to contemporaneously determine the extent of any VHECR over-the-limb background.  The choice for only using a $\Delta \alpha$ span of 9$^\circ$ for POEMMA was chosen to optimize both the limb-Cherenkov and stereo-Fluorescence measurement modes using the same telescopes.  The POEMMA hybrid focal plane configuration leads to a modest $\sim10\%$ reduction in the UHECR instantaneous geometry factor when in fluorescence-stereo mode, shown in the leftmost schematic in Fig.~\ref{POEMMAmodes}.  In this mode, POEMMA will have remarkable sensitivity to UHE neutrinos above 20 EeV, due to the required angular, energy, and EAS profile measurements imposed by the desire for accurate UHECR spectrum and composition measurements. The optimization of the optical wavelength and signal sampling performance of the hybrid focal plane defines the use of  SiPMs as the photo-detectors for viewing the optical Cherenkov EAS signal.

\begin{figure}[t]
\begin{center}
    \includegraphics[width=0.99\textwidth]{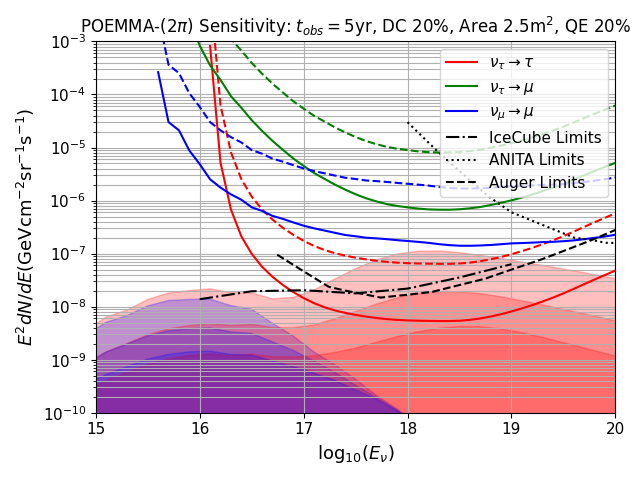}
 \end{center}
\vspace{-0.3cm}
  \caption{The 5 year neutrino diffuse flux sensitivity (90\% confidence level) for POEMMA (dashed lines) based on the simulated response for \taons $\rightarrow$ electron or hadrons, \taons $\rightarrow$ muons, and emergence muons from muon neutrino interactions in the Earth.  A 20\% duty cycle (live time fraction), an effective optical collection area of 2.5 m$^2$, and a 20\% photo-detection efficiency for the SiPMs in the focal planes were assumed.  The solid lines scale the sensitivity by  a factor of 12 to estimate the sensitivity for a POEMMA-like instrument with 360$^\circ$ azimuth coverage. Adapted from Ref. \citenum{2021PhRvD.103d3017C}.
    DOI:\href{https://doi.org/10.1103/PhysRevD.103.043017}{10.1103/PhysRevD.103.043017}
  }
\vspace{-3 mm}
  \label{diffuse}
\end{figure}

When calculating neutrino sensitivity limits, one procedure that provides results independent of the energy dependence of the incident neutrino spectrum is based on calculating the {\it neutrino aperture} or the geometry factor\footnote{See Refs. \citenum{1971NucIM..95....5S} \citenum{1972NucIM..98..187S} for the mathematical definition of the geometry factor of a particle telescope.} as a function of incident neutrino energy of accepting events that also includes the physics of neutrino signal generation. For Earth-emergent \taons, this would include the tau neutrino interactions in the Earth, \taon energy loss, EAS formation, signal generation, and signal detection over some threshold condition. Fig.~\ref{nuAperDiffuse} presents the diffuse-flux neutrino aperture for upward-moving EAS from Earth-emergent \taons for POEMMA compared to that for IceCube high-energy starting (HESE) events  \cite{PhysRevD.98.062003}\footnote{HESE events are defined by having the neutrino interactions occur in the active volume of the IceCube detector}. The neutrino apertures are corrected for the an average duty cycle for each experiment, 20\% for POEMMA and 100\%  for IceCube, to provide a more proper comparison of measurement capability on roughly yearly (and above) time scale.  When calculating the diffuse neutrino flux sensitivity, the experiment's response to an isotropic simulated flux is used for energies above a threshold energy, usually taken such to lead to a small probability of detecting a background event. An example of determining the threshold energy is given by that done for POEMMA. Assuming the dark-sky background dominates the optical background rate, a Poisson probability analysis was performed using a specific dark-sky background rate, the instrument response to the background spectrum compared to that for the signal.  For example, using the dark-sky background spectrum shown in Fig.~\ref{AirGlow} and the photo-detection-efficiency (PDE) response shown in Fig.~\ref{CherSpectra}, the dark-sky background weighted $<$PDE$>$ was determined to be $\sim 10\%$ while the $<$PDE$>$ for the optical Cherenkov signals was $\sim 20\%$ for the majority of the possible Cherenkov spectra seen by the POEMMA instruments \cite{2019PhRvD..99f3011R}. Assuming an average effective area of 2.5 m$^2$, 20\% duty cycle, and a {\it single telescope measurement}, a 20 photo-electron (PE) threshold leads to a false positive rate of $\lsim 0.01$/year from the dark-sky background.  A 20-PE threshold corresponds to an optical Cherenkov intensity of 40 photons/m$^2$ for the accepted EAS.

Near threshold, the time span of the EAS Cherenkov signal is $\sim 10$ ns and if the two POEMMA satellites are within the Cherenkov light pool of an event, e.g. the satellites are separated by $\sim 25$ km for that shown in Fig.~\ref{Cspot}, then temporal coincidence can be used to reduce the effects of the random dark-sky background and thus lower the signal detection threshold\footnote{For an overview of time coincidence techniques, see Adrian C. Melissinos, Experiments in modern physics, Academic Press, 1966, Chapter 9, p. 108
 \citenum{alma991018707219703276} or Richard C. Fernow, Introduction to experimental particle physics, Cambridge University Press (1986), Chapter 6, p. 109. \citenum{alma991025288489703276}.} In general, if $N$ detectors each measure uncorrelated, random individual background pulses with rate $r_i$ then the Poisson probability of measuring one or more pulse in an individual detector within a time interval $\Delta_T$, is 
\begin{equation}
P_i(k \ge 1; \mu_i) = 1-e^{-r_i \Delta_T}
\end{equation}
where $\mu_i = r_i \Delta_T$ is the expected number of pulses within time $\Delta_T$.  An inherent assumption is that any time-of-flight corrections, e.g. due to different path lengths to different detectors, have been corrected when temporally comparing the signals. Assuming $\Delta_T \approx 10$ ns and one expects no more than an overall rate $\sim 1$ kHz based a modest data acquisition system throughput maximum, $r_i \Delta_T \ll 1$, then $P_i(k \ge 1; \mu_i) \approx r_i \Delta_T$. The overall background coincidence rate for $N$ detectors in coincidence is then
\begin{equation}
    r_{Coinc} = N \Delta_T^{N-1} \prod_{i=1}^N r_i
    \label{TcoinEqn}
\end{equation}
In the case for POEMMA, $N=2$, $r_i = r$, because the iFoV for each pixel is identical, and leads to the random dark-sky background pixel rates to be similar, while $\Delta_T = 30$ ns to take into account of the 1-meter knowledge of the absolute satellite position in orbit. These together lead to a factor of 2 reduction in the achievable threshold energy, e.g. the threshold Cherenkov light intensity is 20 photons/m$^2$.  This threshold is used in the POEMMA diffuse sensitivity limits shown in Fig.~\ref{diffuse}, which assume that satellites are separated by $\sim 25$ km and both Cherenkov telescopes simultaneously measure the upward, \taon EAS Cherenkov signal using temporal coincidence to achieve the 20 photons/m$^2$ Cherenkov light detection threshold. Detailed end-to-end neutrino $\rightarrow$ \taon $\rightarrow$ EAS $\rightarrow$ Cherenkov light $\rightarrow$ instrument detection simulations are then used assuming the $7^\circ$ span of viewing below the Earth's limb and $\sim 30^\circ$ span in azimuth are used to determine the neutrino aperture, as shown in Fig.~\ref{nuAperDiffuse} for POEMMA.

Once the neutrino aperture is calculated, sensitivity limits can be calculated using the Feldman and Cousins unified statistical analysis approach \cite{PhysRevD.57.3873} that Poisson processes with background and upper confidence limits for null results. Here the confidence level of $\chi$\% is defined by the range of a parameter of interest such that the parameter is within the prescribed range $\chi$\% of the time, where the prescribed range depends on the specific confidence level, i.e. 90\%, and whether or not there is background in the measurement, i.e. 0, 1, ... etc events. Mathematically, the procedure associates a number of signal events and background events with a specific confidence level (CL).  For example, for the case of 90\% CL on the sensitivity calculation and no background, the expectation is 2.44 events over some energy band. Assuming the energy band is defined by a band in energy, $\delta_E$ where a power-law form of the neutrino flux is valid \cite{PhysRevD.66.103002}, a conservative\footnote{In the calculation presented in Ref.~\citenum{PhysRevD.66.103002}, a $\frac{\sinh{\delta}}{\delta} > 1$ term is ignored, thus making the calculation conservative} determination of the sensitivity limits for $N_{CL}$ events are given by
\begin{equation}
    \phi_{\rm CL}(E_{\nu_\tau}) = \frac{N_{CL} \times N_\nu}{\ln{\delta_E} \times E_{\nu_\tau} \times <A\Omega>(E_{\nu_\tau}) \times t_{\rm obs}}
    \label{sensLimEqn}
\end{equation}
where $<A\Omega>(E_{\nu_\tau}$) is the neutrino aperture at energy $E_{\nu_\tau}$ and $t_{\rm obs}$ is the effective observation time that implicitly includes the duty cycle for the wall clock time period. For 90\% confidence limits assuming no background events and over a decade in energy, $N_{CL}=2.44$ and $\delta_E=10$.
For pure tau neutrino limits, $N_\nu =1$, and assuming that neutrino oscillations transform any cosmic cosmic neutrino flux to $\nu_e:\nu_\mu:\nu_\tau = 1:1:1$, $N_\nu =3$ for an all-flavor neutrino limits.  Note that the $\ln{10}$ term in the denominator of Eqn.~\ref{sensLimEqn} for $N_{CL}$ over a decade in energy is replace by an appropriate factor if the 90\% CL limits are taken over a different interval. For example, it would be replace with the term $\ln{10^{0.5}} = \frac{1}{2} \ln{10}$ for a half a decade in energy, due to the logarithmic interval used in the integral in the calculation, see Eqn. 12 in Ref.~\citenum{PhysRevD.66.103002}.
\begin{figure}[t]
\begin{center}
{\includegraphics[width=0.99\textwidth]{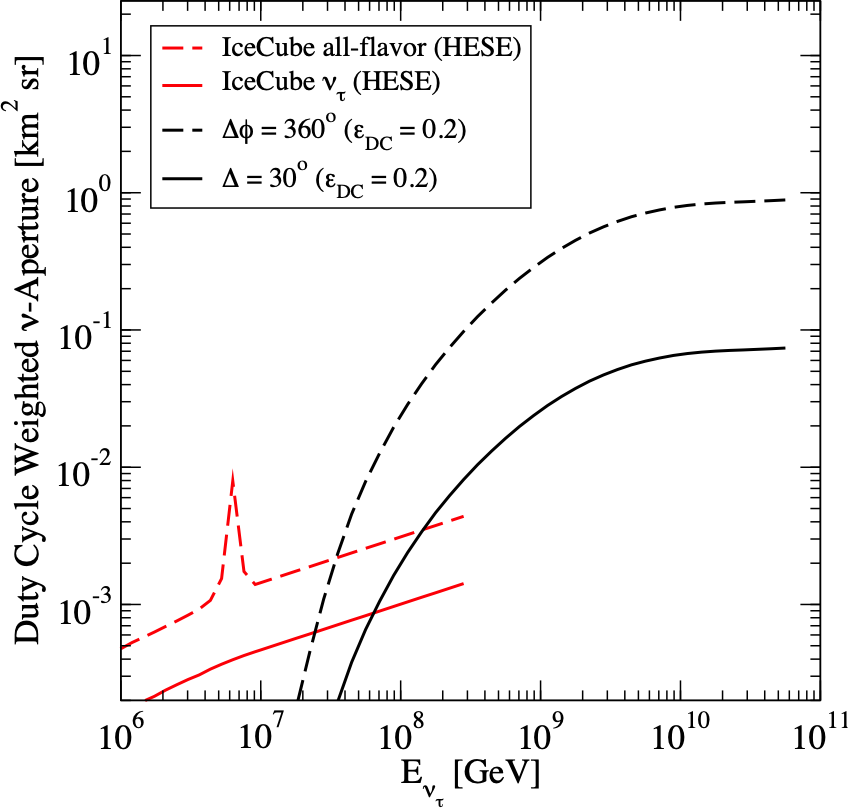}}
\caption{The POEMMA diffuse-flux, duty-cycle- weighted tau neutrino aperture as a function of neutrino energy for upward-moving EAS detected via optical Cherenkov light. The solid line is for POEMMA while the dashed line increases the POEMMA by multiplying by twelve to estimate the aperture for an instrument with $360^\circ$ azimuth field of view. Also shown is the IceCube all-flavor neutrino aperture (dashed) and tau neutrino neutrino aperture (solid) for high-energy starting events (HESE) in the IceCube detector, adapted from \citenum{PhysRevD.98.062003}. A 20\% duty cycle is used for the POEMMA results while a 100\% duty cycle is used for the IceCube results From Ref. \citenum{2021JCAP...06..007P}. 
  DOI:\href{https://doi.org/10.1088/1475-7516/2021/06/007}{10.1088/1475-7516/2021/06/007}
\label{nuAperDiffuse}}
\end{center}
\end{figure}

Fig.~\ref{diffuse} compares the POEMMA diffuse flux sensitivity, assuming 5 year operation and 20\% measurement duty cycle for the optical Cherenkov signal, for the detection of both tau neutrino $\rightarrow$ \taons and muon neutrinos $\rightarrow$ muons, compared to current IceCube, Pierre Auger Observatory, and ANITA all-flavor neutrino detection limits.  The calculation\cite{2021PhRvD.103d3017C} takes into account the different energy loss mechanisms for muons and \taons in the Earth and modeling of muon EAS formation from both Earth-emergent muons from $\nu_mu$ interactions and the muons from \taon decay in the atmosphere.  The threshold for EAS detection is based on assuming a constant level of the dark-sky airglow background in the 300 nm to 1000 nm wavelength band and applying a Poisson-based analysis to predict a false-positive rate of $\sim 1\%$ per year or operation due to this dominant background \cite{2019PhRvD.100f3010R,2021NIMPA.98564614K}. The dashed curves are for POEMMA ($\sim 30^\circ$ azimuth span) while the solid curves scale the sensitivities by a factor of twelve to estimate that for a full $360^\circ$ azimuth span Cherenkov detection POEMMA-like instrument. Note that this scaling assumes that the a mechanism exists to reduce the effects of the dark-sky airglow background, which implies better performance than that given by the current POEMMA design \cite{2021PhRvD.103d3017C}.  

\subsection{The Dark-sky Airglow Background}

\begin{figure}[h]
\centering
\includegraphics[width=0.99\columnwidth]{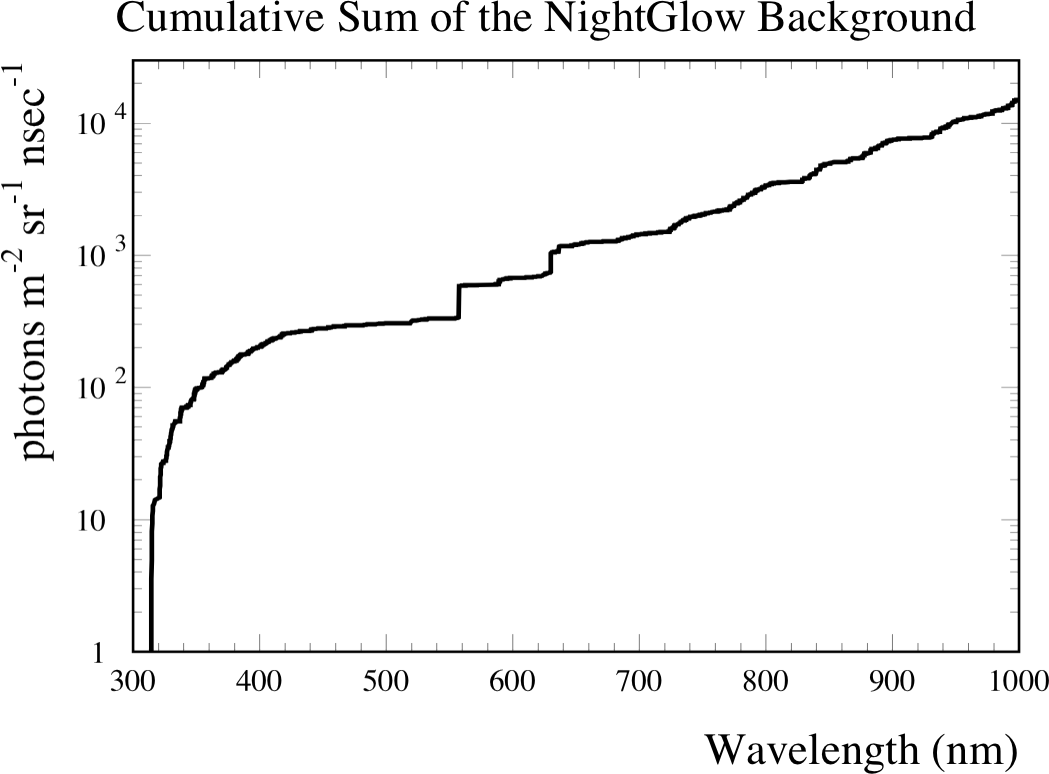}
\caption{The cumulative sum of the dark-sky airglow background using the measurements of Hanuschik \cite{2003A&A...407.1157H,2006JGRA..11112307C}. From Ref.~\citenum{2021NIMPA.98564614K}, Fig, 7.
  DOI: \href{https://doi.org/10.1016/j.nima.2020.164614}{0.1016/j.nima.2020.164614}
}
\label{AirGlow}       
\end{figure}

Optical measurements of EAS are made along with a continuous atmospheric background even during astronomical night. Chemiluminescent processes in the upper atmosphere of the Earth powered by solar and cosmic radiation ionization leads to production of faint light that is commonly known as an airglow \citep{Meier,Leinert}. These processes also cause aurora near the polar regions.  When views from LEO or balloon altitudes, the dark-sky airglow light together with the reflected component of the starlight and zodiacal light form a diffuse optical background that effectively sets the energy threshold for EAS detection. Additionally, the altitude of the observatory and its viewing direction in terms of angle away from nadir further yields to an increase in intensity, in part by viewing a longer path thru the dark-sky airglow layer, which is concentrated in a $\sim$10 km thick layer around 95 km in altitude  \cite{Roach}. Fig.~\ref{AirGlow} presents the dark-sky airglow measurements by Hanuschik at a time near high solar activity \cite{2003A&A...407.1157H,2006JGRA..11112307C} for the wavelength band 300$-$1000 nm in units of photons m$^{-2}$  ns$^{-1}$ sr$^{-1}$ as cumulative sum as a function of wavelength. The strong increase in intensity at longer wavelengths leads to $15 \times 10^3$ photons m$^{-2}$  ns$^{-1}$ sr$^{-1}$  in the 300$-$1000 nm band.
Given the strong wavelength dependence, the impact of dark-sky airglow on the measurements of EAS air fluorescence and optical Cherenkov needs to be considered and an analysis for both is detailed in Ref.~\citenum{2021NIMPA.98564614K}. The highly variable spectral variability of the EAS optical Cherenkov signals, with peak intensity spanning from $200-1000+$ nm, translates to dealing with the entire airglow spectrum shown in Fig.~\ref{AirGlow}.  Using SiPMs with response peaked at shorter wavelengths is desirable to span the EAS Cherenkov wavelength band.  However, these also have photo-detection-efficiency (PDE) $\lsim 20\%$ above 700 nm, which has sensitivity to the highest intensity of the airglow background. Fortunately, the 10-ns scale width of the optical Cherenkov signal in the maximum intensity region combined with modest-sized pixel instantaneous field-of-view (iFoV) the telescope help to reduce the impact \cite{2019PhRvD..99f3011R, 2020PhRvD.102l3013V,2020PhRvD.101b3012A,PhysRevD.104.063029}.
For $A_{Eff}= 1$ m$^2$, iFoV$=0.1^\circ$, $\Delta t = 10$ ns, and an effective PDE$= 30\%$ over the 200$-$1000 nm band yields a mean background of $\approx 0.14$ photo-electrons (PEs) using the airglow spectrum in Fig.~\ref{AirGlow}.

\subsection{Operational Duty Cycle and the Effects of Clouds}

A space-based optical signal EAS detection experiment's duty cycle, e.g. fraction of live time operation, depends on the fraction of sufficiently dark sky and cloud cover.  For optical Cherenkov EAS observations, the 200$-$1000 nm wavelength measurement band leads to a high intensity of the dark-sky airglow background in the measurements. Thus some fraction of moonlight can be tolerated if it is a fraction of the airglow background.  Studies performed for the JEM-EUSO mission \cite{2013APh....44...76A} show that a 30\% of observation time the condition of no sun and moonlight that is $\sim 50\%$ of the cumulative intensity of the dark-sky airglow.  At full moon, the moonlight intensity is approximately equal to that of the dark-sky background. This increases the 30\% fraction to 35\% but at the expense of raising the Cherenkov detection threshold to minimize the acceptance of background events. Ignoring difference in the moonlight, e.g. solar light spectrum, and airglow spectrum, the duty cycle including the effects of clouds can be estimated. Cloud faction analysis parsed into optical depth fraction versus height fractions was also performed for the JEM-EUSO mission \cite{2013APh....44...76A}.  The results show that $\sim 55\% (62\%)$ of the time the conditions are either cloud free or have clouds with $< 0.1 (1.0)$ optical depth.  This leads to a duty cycle estimate for optical Cherenkov EAS space-based measurements of 16.5\% to 18.6\%. Simulation studies have rounded this value to 20\% duty cycle which is used as the baseline.  It should be noted that the near-term VHECR optical Cherenkov measurements to be performed by EUSO-SPB2 will experimentally determine the achievable duty cycle as a function of Cherenkov intensity detection threshold.

The results of the simulated diffuse flux sensitivity shown in Fig.~\ref{diffuse} highlight several interesting aspects of space-based neutrino detection using the optical Cherenkov from upward-moving EAS.  
The first is that the optical signal from Earth-emergence muons from $\nu_mu \rightarrow \mu$ interactions dominant over that for $\nu_\tau \rightarrow \tau$ for POEMMA below $\lsim 10$ PeV and with the $\tau \rightarrow \mu$ decay channel. This provides a slight low-energy enhancement over that from the electron and hadron decay \taon decay channels.  Note that the these results use appropriate models of muon-induced EAS optical Cherenkov generation\cite{2021PhRvD.103d3017C}, which are intrinsically different\cite{Vankov} than that for electrons or hadrons. The second aspect is the fact that space-based experiments with a 5-year mission span and 20\% duty cycle for optical measurements have difficulty obtaining the exposure of large ground-based experiments, such as IceCube\cite{IceCube2015} and the Pierre Auger Observatory\cite{2019JCAP...10..022A}, unless the full potential of the optical Cherenkov measurements is realized, e.g. low-energy measurements with $\sim10$ PeV-scale neutrino thresholds with a $360^\circ$ azimuth field-of-view.

The distinction between diffuse, e.g. constant isotropic flux, and transient, e.g. variable planer flux from the location of an individual astrophysical source, neutrino sensitivities for space-based and ground-based experiments is evident in the comparison of that for IceCube and POEMMA. Fig.~\ref{nuAperDiffuse} compares the duty cycle (DC) weighted neutrino aperture for upward-moving EAS from Earth-emergent \taons compared to that for IceCube high-energy starting events (HESE) \cite{PhysRevD.98.062003} where the neutrino interactions occur in the active volume of the IceCube detector. Assuming 100\% duty cycle for IceCube and 20\% for POEMMA, the results present a comparison of the diffuse neutrino flux sensitivity as a function of energy. Above an EeV, POEMMA has around a factor of 10 larger DC-weighted neutrino aperture. However, this gain in performance becomes diluted since IceCube has been running for decades and POEMMA expects a 5-year mission.  If a neutrino aperture gain of twelve is available for a space-based experiment by increasing the azimuth FoV the full $360^\circ$, the space-based experiment becomes competitive. 

For neutrino fluxes from transient astrophysical sources that flare on timescales shorter that that for a space-based mission, the benefit of increasing the exposure by operating an experiment for longer times no longer holds, and a comparison to the neutrino effective area for short- and long-duration transient fluxes is the important measure of merit.  Fig.~\ref{nuToOarea} shows the relevant comparison between IceCube\cite{Aartsen_2017} and POEMMA\cite{2020PhRvD.102l3013V} where the sky-weighted effective areas are presented for long-duration, e.g. multiple month duration, events.  A 20\% duty cycle is applied for POEMMA while 100\% is assumed for IceCube. 
For short-duration events, the best case effective area is shown for POEMMA, where the duty cycle no longer factors in for transient time scales of $\sim 1000$ seconds. That is the ToO is either seen or not and the effects are on sky-coverage \cite{2020PhRvD.102l3013V}.
The benefit for space-base observation of transient neutrino events is highlighted in the Fig.~\ref{nuToOarea}. Gains of $\gsim 10^3$ in effective area at higher energies for short-duration observations are evident via comparison to the sky-averaged IceCube response.  These results demonstrate the benefits of space-based neutrino observations using the optical Cherenkov signals from upward-moving EAS generated by transient astrophysical sources.

\begin{figure}
\begin{center}
{\includegraphics[width=0.99\textwidth]{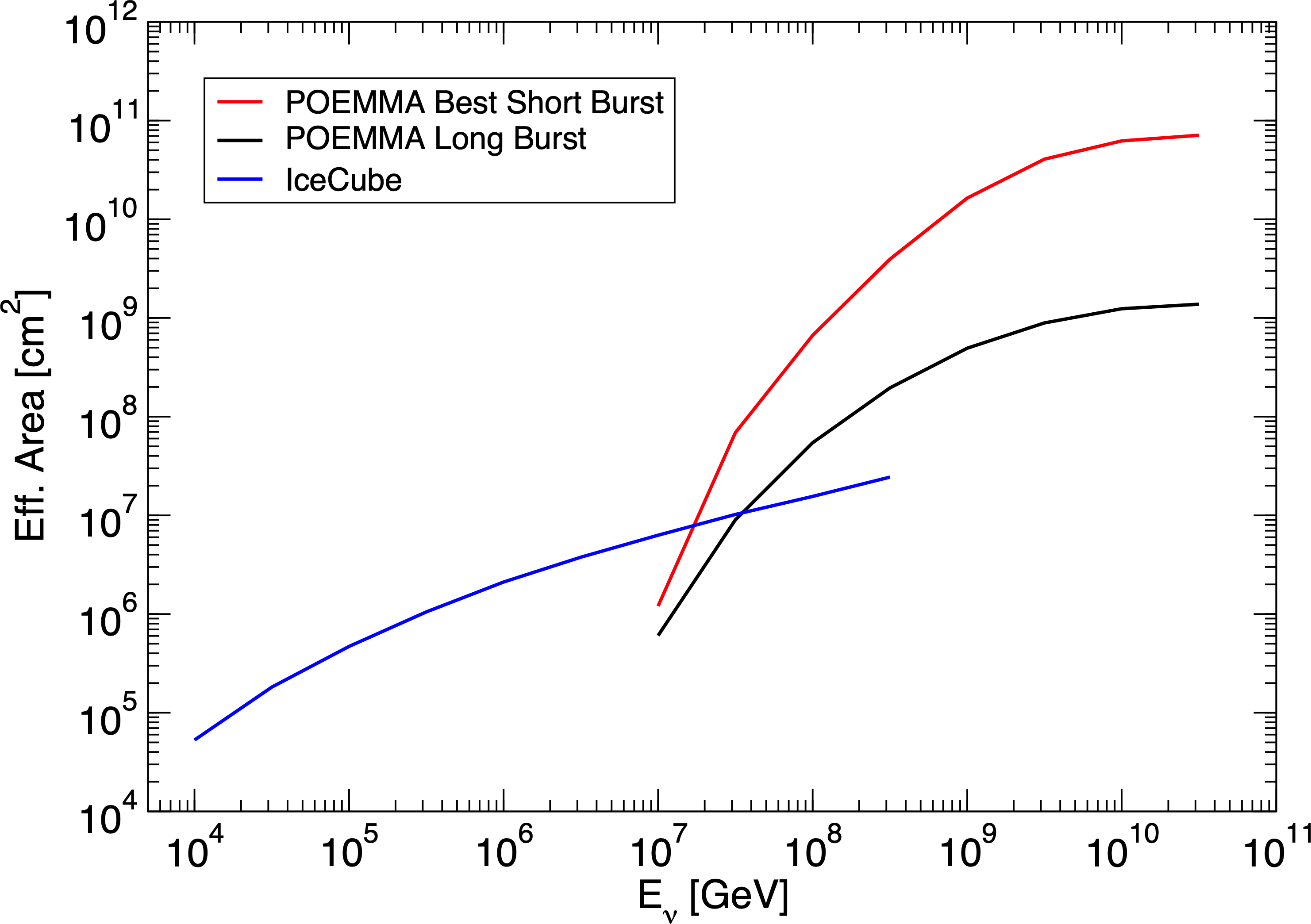}}
\caption{The effective area of POEMMA for long- and short-duration transient neutrino sources as a function of energy. For the long-duration events, a 20\% duty cycle was assumed. The IceCube\cite{Aartsen_2017} sky-averaged effective area is shown for comparison \cite{ToniV21}.\label{nuToOarea}}
\end{center}
\end{figure}

\begin{figure}[t]
\begin{center}
    \includegraphics[width=0.99\textwidth]{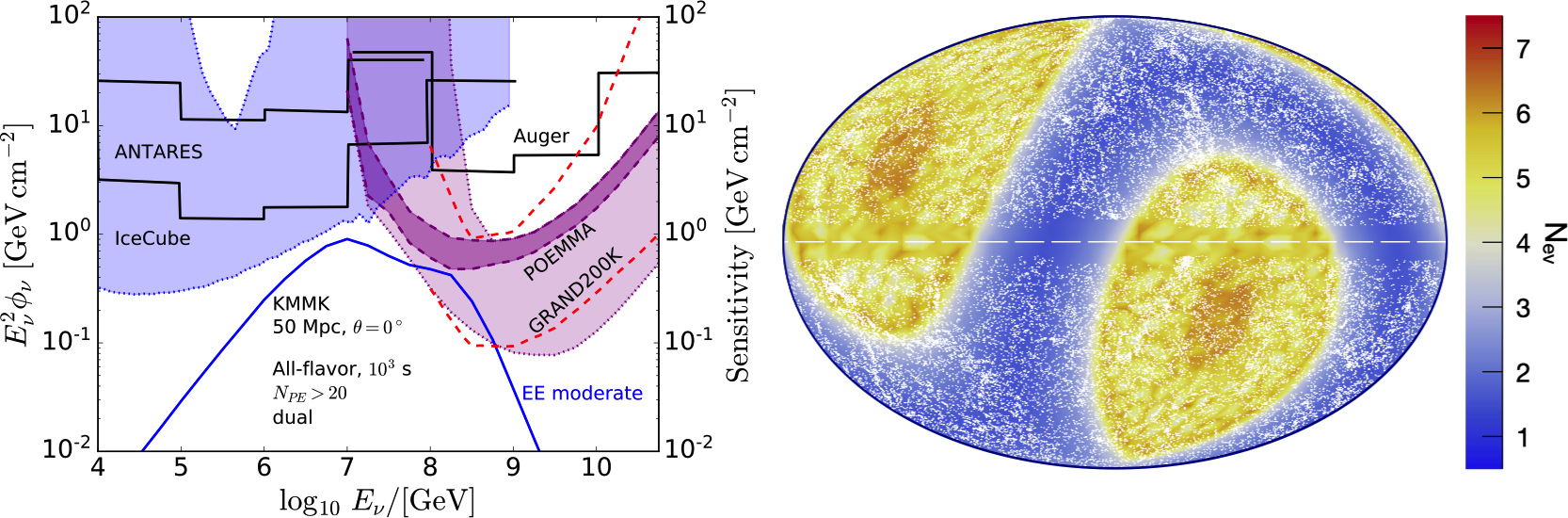}
 \end{center}
\vspace{-0.3cm}
  \caption{Left: POEMMA neutrino sensitivity to short, 1000-s neutrino transient bursts compared to that for the ground-based IceCube, Auger, and GRAND experiments. Right: the sky plot, in galactic coordinates, showing the number of POEMMA detected tau neutrino events using the Kimera, Murase, M{\'{e}}sz{\'{a}}ros, and Kiuchi (KMMK) sGRB model\cite{Kimura_2017} for a transient at 50 Mpc. From Ref. \citenum{2021JCAP...06..007P}.  DOI:\href{https://doi.org/10.1088/1475-7516/2021/06/007}{10.1088/1475-7516/2021/06/007}
}
\vspace{-3 mm}
  \label{POEMMAsb}
\end{figure}

\begin{figure}[h]
\begin{center}
    \includegraphics[width=0.99\textwidth]{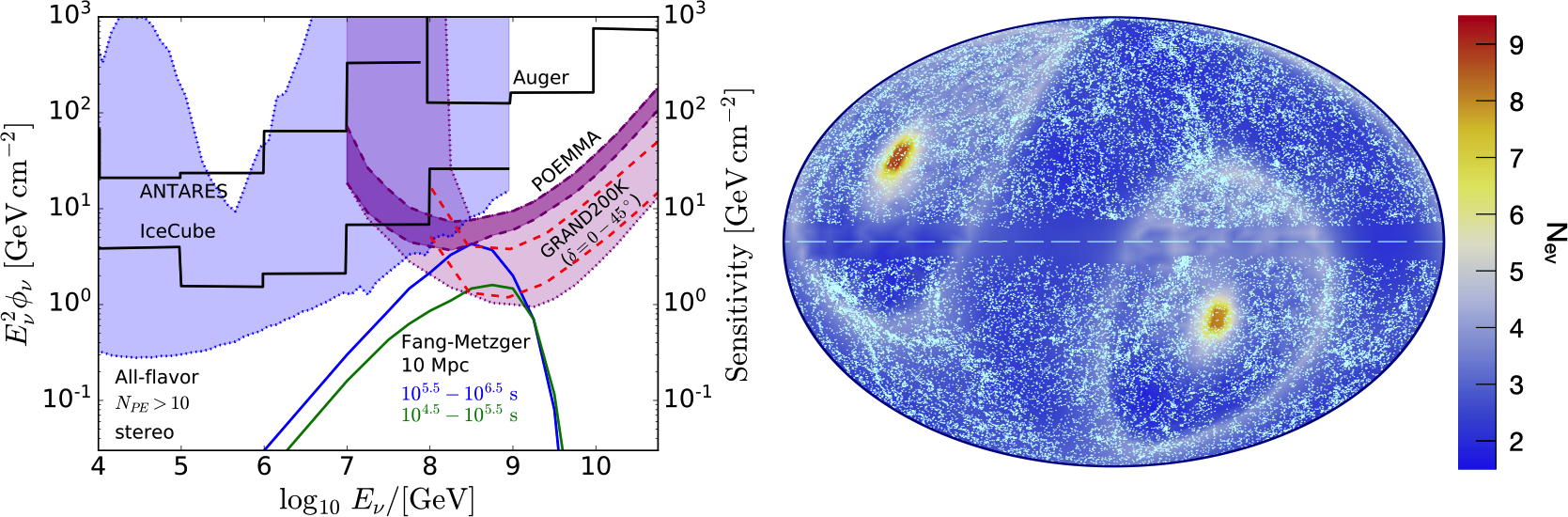}
 \end{center}
\vspace{-0.3cm}
  \caption{Left: POEMMA neutrino sensitivity to long-duration neutrino transient bursts compared to that for the ground-based IceCube, Auger, and GRAND experiments. Right: the sky plot, in galactic coordinates, showing the number of POEMMA detected tau neutrino events assuming the Fang and Metzger binary neutron merger (BNS) model\cite{FangMetzger_2017} for a source at 10 Mpc. From Ref. \citenum{2021JCAP...06..007P}.  DOI:\href{https://doi.org/10.1088/1475-7516/2021/06/007}{10.1088/1475-7516/2021/06/007}
}
\vspace{-3 mm}
  \label{POEMMAlb}
\end{figure}

\begin{figure}[t]
\begin{center}
    \includegraphics[width=0.99\textwidth]{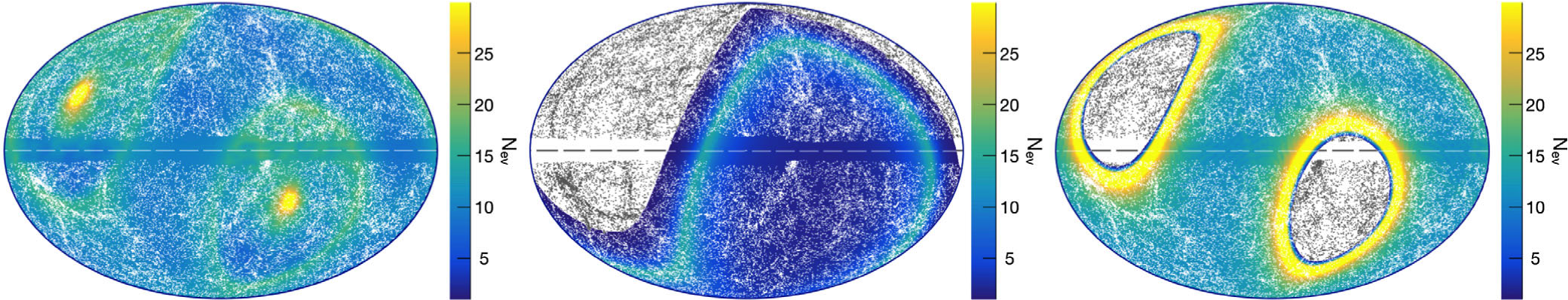}
 \end{center}
\vspace{-0.3cm}
  \caption{The POEMMA (left), IceCube (middle), and GRAND (right) neutrino event sky maps, in galactic coordinates, assuming the Fang and Metzger model\cite{FangMetzger_2017} with the BNS located at 5 Mpc distance.  From Ref. \citenum{2020PhRvD.102l3013V}.
  DOI:\href{https://doi.org/10.1103/PhysRevD.102.123013}{10.1103/PhysRevD.102.123013}
  }
\vspace{-3 mm}
  \label{POEMMAbnscomp}
\end{figure}

While Figs.~\ref{POEMMAsb}, \ref{POEMMAlb}, and \ref{POEMMAbnscomp} detail the neutrino detection reach for POEMMA in terms of cosmic distance to the transient sources, the next, nature question is what is the rate of detection for these transients based on the underlying theoretical cosmological distribution. Fig.~\ref{POEMMAToOrate} presents these rates as a function of mission time for POEMMA \citenum{2020PhRvD.102l3013V}, demonstrating that the detection is challenging yet promising for a space-based neutrino detection mission with POEMMA-like performance.

\begin{figure}[t]
\begin{center}
    \includegraphics[width=0.99\textwidth]{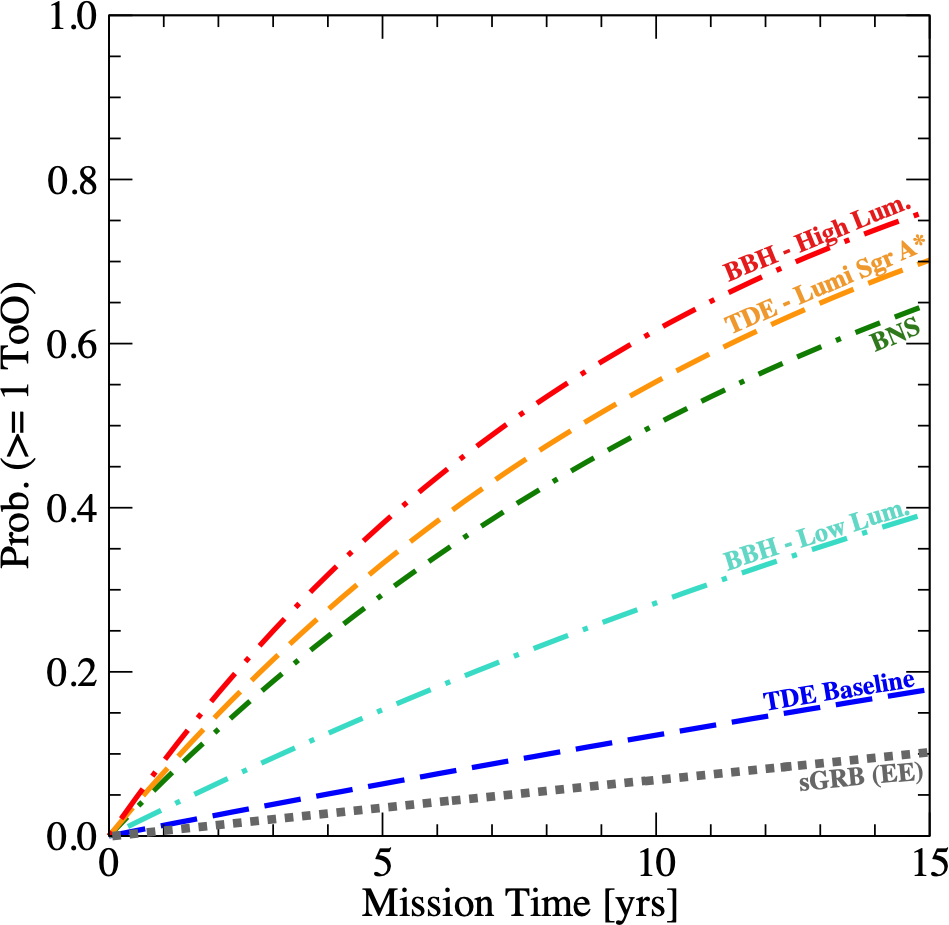}
 \end{center}
\vspace{-0.3cm}
  \caption{The probability for POEMMA to observe at least one transient versus mission operation time for several modeled source classes. Featured source models are TDEs from Lunardini and Winter \cite{PhysRevD.95.123001}, BNS mergers from Fang and Metzger \cite{FangMetzger_2017}, BBH mergers from Kotera and Silk \cite{Kotera_2016}, and sGRBs with moderate EE from KMMK \cite{Kimura_2017}.  From Ref. \citenum{2020PhRvD.102l3013V}.
    DOI:\href{https://doi.org/10.1103/PhysRevD.102.123013}{10.1103/PhysRevD.102.123013}
    }
\vspace{-3 mm}
  \label{POEMMAToOrate}
\end{figure}

\section{Neutrino EAS Measurements using Air Fluorescence Technique}

The charged particles in an EAS produce ionization that excites air fluorescence that is emitted isotropically  as the few hundred meter wide (determined by the Moli{\'{e}}re radius, here taken at altitudes below 10 km) and few meter deep 'pancake' of EAS particles moves through the atmosphere close to the speed of light.
While the wavelength band air fluorescence extends from below 200 nm to over 1000 nm \cite{1964JChPh..41.3946D}, the vast majority is in the band from 300 nm to 450 nm \cite{1964JChPh..41.3946D,1967PhDT........28B}. This allows for the use of UV filters to limit the effects of the dark-sky chemiluminescence airglow background to be  $\sim$ 500 photons m$^{-2}$ ns$^{-1}$ $sr^{-1}$ \cite{2005APh....22..439B}, as shown in Fig.~\ref{AirGlow}.
For experiments in low Earth orbit (LEO), if the optical system viewing the air fluorescence EAS signals has sufficiently large field-of-view (FoV) a vast amount of atmosphere can be  monitored, e.g.  $\sim 10^{13}$ metric tons of atmosphere for full FoV $= 45^\circ$ \cite{OWL,2021JCAP...06..007P}. The EAS imaging requirements of resolving approximately 1 km spatial lengths on the Earth's surface set the iFoV $=0.084^\circ$ for each pixel in the POEMMA focal plane shown, see Fig.~\ref{POEMMAfp}.  Thus the vast atmospheric volume viewed by POEMMA is effectively finely sampled in terms of the iFoV of the pixels.

An experiment that has high UHECR sensitivity also will have high sensitivity for UHE neutrino interactions that occur deep in the atmosphere that are well separated from UHECR EAS \cite{2020PhRvD.101b3012A}.  The EAS development
depends on the emergent charged-lepton from the neutrino interaction, but the hadronic system also initiates a EAS at the neutrino interaction point. At high energy scales, $\gsim~ 10 EeV$, EAS are subject to the Landau–Pomeranchuk–Migdal effect \cite{Landau:1953um,1956PhRv..103.1811M} which will effectively spatially extend the EAS, and models of the LPM effect are usually included in the EAS generators used in the UHE EAS simulations \cite{CORSIKA}.

\begin{figure}[t]
\begin{center}
    \includegraphics[width=0.99\textwidth]{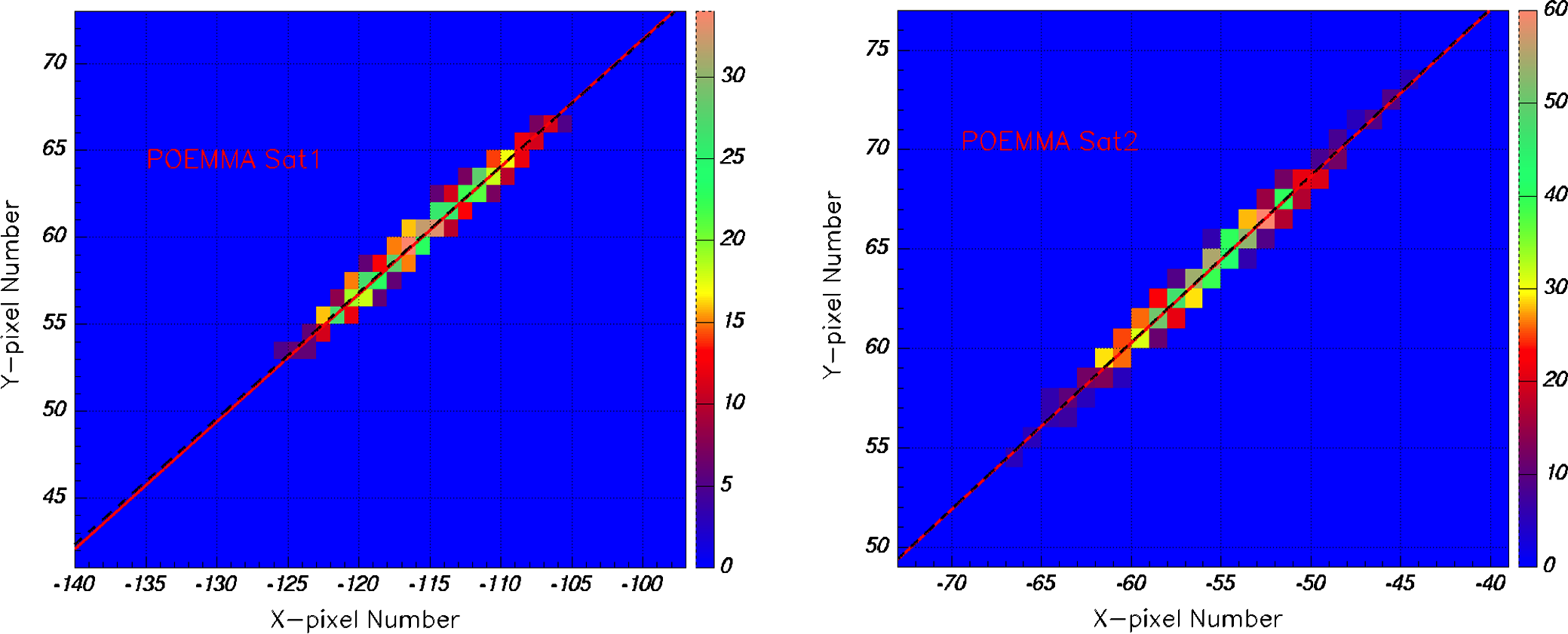}
 \end{center}
\vspace{-0.3cm}
  \caption{A co-viewed, stereo reconstructed 50 EeV simulated UHECR in the two POEMMA focal planes. The solid line shows the simulated trajectory while the dashed line is defined by the reconstructed trajectory. The color map provided the photoelectron statistics simulated for each pixel. From Ref. \citenum{2020PhRvD.101b3012A}.
      DOI:\href{https://doi.org/10.1103/PhysRevD.101.023012}{10.1103/PhysRevD.101.023012}
}
\vspace{-3 mm}
  \label{EASstereo}
\end{figure}

The stereo fluorescence technique, pioneered by the ground-based Fly's Eye experiment in Utah \cite{1985NIMPA.240..410B} provides a experimental technique to precisely measure the development of EAS and thus achieving superb angular, energy, and \xmax resolution. Applied from a space-based experiment, this technique allows for an immense amount of atmosphere to be used as a UHECR and UHE neutrino detector. Fig.~\ref{EASstereo} shows the simulation results for a 50 EeV UHECR proton event as seen in the focal plane of the POEMMA telescopes.  The accurate measurement using two well-separated EAS detector planes provides a straightforward geometrical method to determine the line where the two planes overlap, thus accurately determining the 3-dimensional spatial location of the EAS.  Even for energies above 100 EeV, neutrinos will interact nearly uniformly as a function of slant depth of the trajectory, providing a method to separate more horizontal neutrino events from the much more copious UHECR events.
For POEMMA, the selection of $X_{\rm Start} \gsim 2000 ~{\rm g/cm}^2$ has been shown, based on the analysis of the simulated $X_{\rm Start}$ distributions for UHECR protons for $E_{\rm CR} \ge 40 $EeV for the POEMMA mission with $\gsim 10^4$ rejection factor for UHECR background events\cite{2020PhRvD.101b3012A}.  These UHE neutrino measurements are performed whenever the POEMMA satellites are in Stereo-fluorescence UHECR operational mode, e.g. they come for `free'.
It is noted that Earth-emergent EAS from neutrino interactions in the Earth also provide a fluorescence signal that can be observed from space if the instruments are pointed to the Earth's limb  \cite{2015arXiv150905995V}. However, the tau neutrino energy thresholds are in the 10-EeV decade, e.g. similar observational threshold energies as UHECR EAS fluorescence measurements.  Unless the cosmic neutrino flux is such that it turns on in the energy range and given the usual power-law nature of astrophysical neutrino spectra, the sensitivity obtained using the optical Cherenkov measurements upward-moving EAS will provide better sensitivity due to the $\gsim 10$ PeV neutrino energy thresholds for space-based optical Cherenkov measurements based on modeling studies. These include the LEO POEMMA mission \cite{2019PhRvD..99f3011R} and for the ultra-long duration balloon (ULDB) EUSO-SPB2 experiment scheduled to fly in 2023 \cite{2020PhRvD.102l3013V, 2021PhRvD.103d3017C}.

\begin{figure}[h]
\begin{center}
    \includegraphics[width=0.99\textwidth]{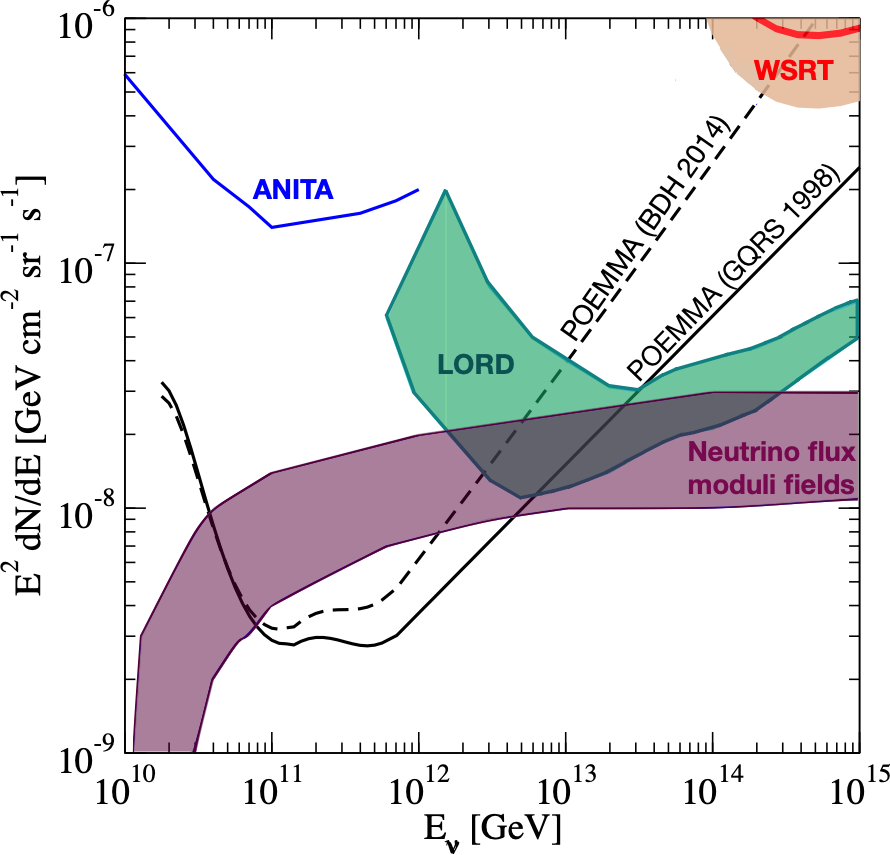}
 \end{center}
\vspace{-0.3cm}
  \caption{All-flavor sensitivity for detection of UHE neutrinos interacting in the atmosphere via the Stereo-fluorescence technique using POEMMA compared to the simulated response of the LORD experiment and theoretical predictions for the neutrino flux from strongly-coupled string moduli formed early in the Universe. From Ref. \citenum{2020PhRvD.101b3012A}.
        DOI:\href{https://doi.org/10.1103/PhysRevD.101.023012}{10.1103/PhysRevD.101.023012}
}
\vspace{-3 mm}
  \label{nuUHElims}
\end{figure}

Fig.~\ref{nuUHElims} presents the all-flavor 90\% confidence level (CL) limits for UHE neutrinos using POEMMA UHE stereo fluorescence measurement simulations. The sensitivity curves assume no background and 2.44 events per decade, see Ref.~\citenum{PhysRevD.57.3873}. At these extreme energies, the neutrino cross sections are obtained via extrapolation and thus contain theoretical uncertainties. Two different neutrino cross-section models are used to illustrate the uncertainty. The measurements are sensitive to all three flavors of neutrinos as well as both CC and NC interactions, with the nature of the flavor and interaction determining the properties of the EAS, as detailed in section \ref{SNc_NuInt}. Also shown is the simulated sensitivity of the Lunar Orbital Radio Detector (LORD) \cite{2016AdSpR..58..464R} mission that is being develop to detect the radio signal from neutrino-induced extensive particle cascades in the lunar regolith. Also shown are the theoretical prediction of a neutrino flux produced by decay of highly-boosted, strongly coupled, scalar field moduli, radiated by relic cosmic strings created early in the Universe\cite{PhysRevD.84.085006}. This unique all-flavor UHE neutrino sensitivity will be performed whenever the POEMMA satellites are in stereo-fluorescence UHECR operational mode, just requiring a separate analysis chain once the data is telemetered to the ground.

\begin{figure}[ht]
\begin{center}
    \includegraphics[width=0.98\textwidth]{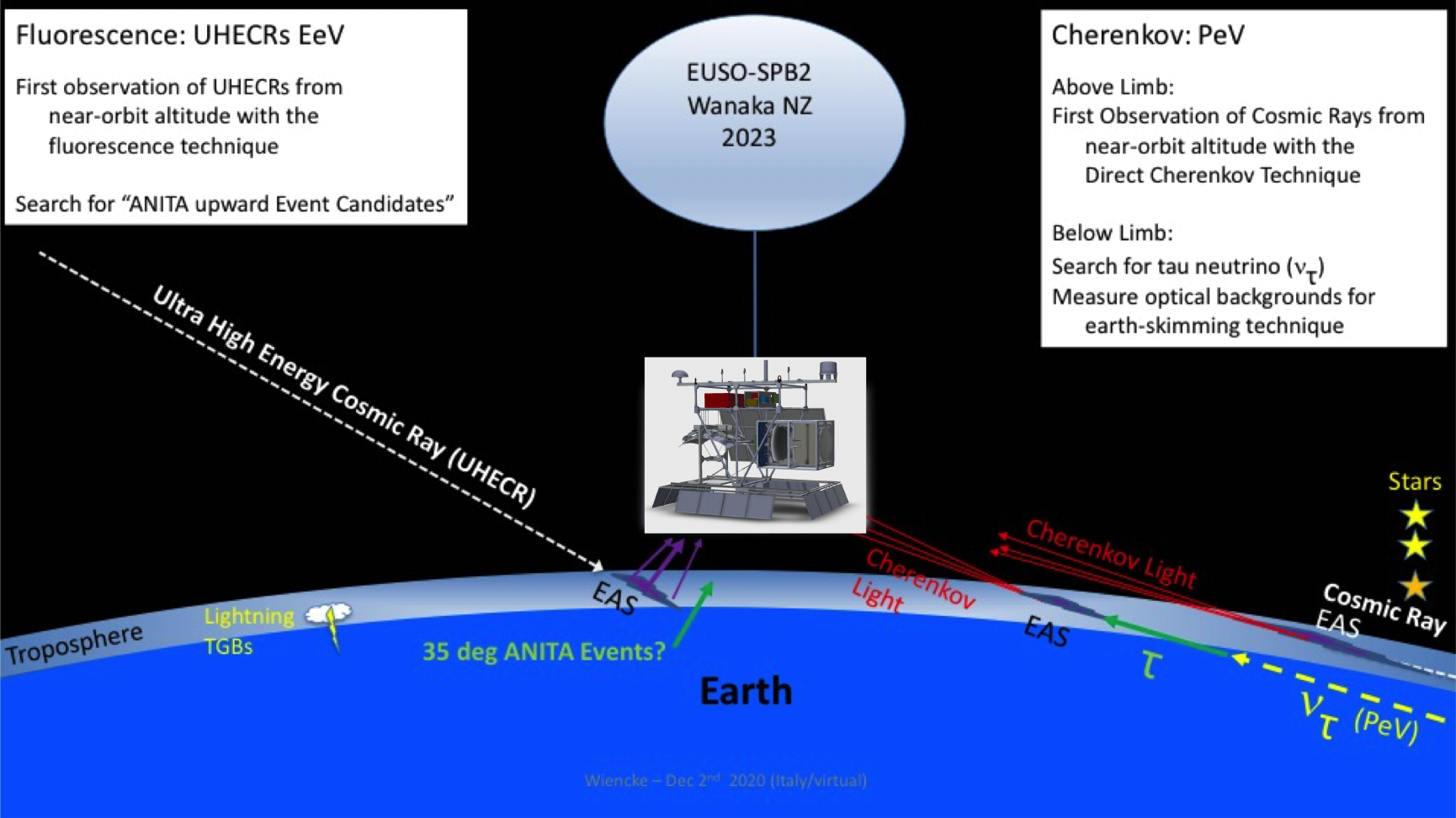}
 \end{center}
\vspace{-0.3cm}
  \caption{Artist's sketch of the EUSO-SPB2 payload and listing of science signals (fluorescence UHECR, Cherenkov from over-the-limb VHECRs, Cherenkov from upward EAS sourced from neutrino interactions in the Earth), background (dark-sky airglow, stars in the FoV of the CT), and other natural (lightning, transient luminous events (TLEs) optical light sources. Adapted from Ref. \citenum{Eser:2021H6}.
  DOIs: \href{https://doi.org/10.22323/1.395.0404 }{10.22323/1.395.0404 },   }
\vspace{-3 mm}
  \label{SPB2science}
\end{figure}

\section{Neutrino and UHECR EAS Optical Measurements using Balloon-borne Instruments}

While orbiting neutrino-detection missions using the EAS optical signals can be performed in multi-year mission time frames, the availability of ULDB, potentially of 100-day duration, and launched from mid latitudes, e.g. Wanaka, New Zealand, allow for astronomical night operation. This provides an environment to view the optical Cherenkov signal from above the limb VHECRs while also searching for the neutrino-induced, Earth-emergent upward-moving EAS as well as for the air fluorescence signal of downward-moving EAS from UHECRs. The EUSO-SB2 ULDB experiment, scheduled to launch in 2023, consists of a downward-looking fluorescence telescope (FT) to record, using $1 \mu$s snapshots, the waxing and waning of the fluorescence EAS signals from UHECRs.  Fig.~\ref{SPB2science} presents the various science and background optical signals the EUSO-SPB2 instrument will measure from $\sim$30 km altitude \cite{Eser:2021H6,Osteria:2021OJ,Filippatos:2021b5}. Fig.~\ref{SPB2schematics} shows the schematics of the EUSO-SPB2 payload, fluorescence telescope, and Cherenkov telescope. The FT is a 1-meter diameter optical aperture Schmidt telescope with a $37.4^\circ \times 11.4^\circ$ FoV.\cite{Osteria:2021OJ}. The FT focal plane is formed by 6912 pixels using three Photo Detection Modules (PMDs) developed for the JEM-EUSO mission \cite{2017NIMPA.866..150A}, and are planned to be used in the POEMMA fluorescence telescope. Each PDM is formed by 36 individual Multi-Anode Photomultipler tubes (MAPMTs) using the SPACIROC3 front-end electronic (FEE) readout with 1 $\mu$s sampling of the measured UHECR EAS development. Various systems based on this PDM design has been flown in the ULDB EUSO-SPB1 mission in 2017 \cite{2019NIMPA.936..237O} and EUSO-Balloon test flight mission \cite{2017ICRC...35..445B} flown in 2014, and Mini-EUSO instrument \cite{2021ApJS..253...36B} operating on the International Space Station starting in 2019. This PDM development is an example of the benefits of using relatively inexpensive balloon-borne missions to mature technologies for eventual LEO missions. The expected UHECR observation rate for EUSO-SPB2 for UHECR fluorescence events is about 1 in 8 hours for events slightly above 1 EeV, assuming optimum atmospheric conditions, e.g. cloud-free conditions during dark moonless nights \cite{Filippatos:2021b5}.  Assuming current neutrino flux limits, the potential for observing air fluorescence neutrino-induced EAS is small due to the relatively small atmospheric volume monitored by the FT. The FEE of the FT also includes a channel to register fast ($\sim 10$ ns), bright optical signals from the ground-reflected UHECR Cherenkov signal and searching for optical signals from anomalous upward-moving EAS signals observed in two of the ANITA flights \cite{2018PhRvL.121p1102G}.

\begin{figure}[ht]
\begin{center}
    \includegraphics[width=0.36\textwidth]{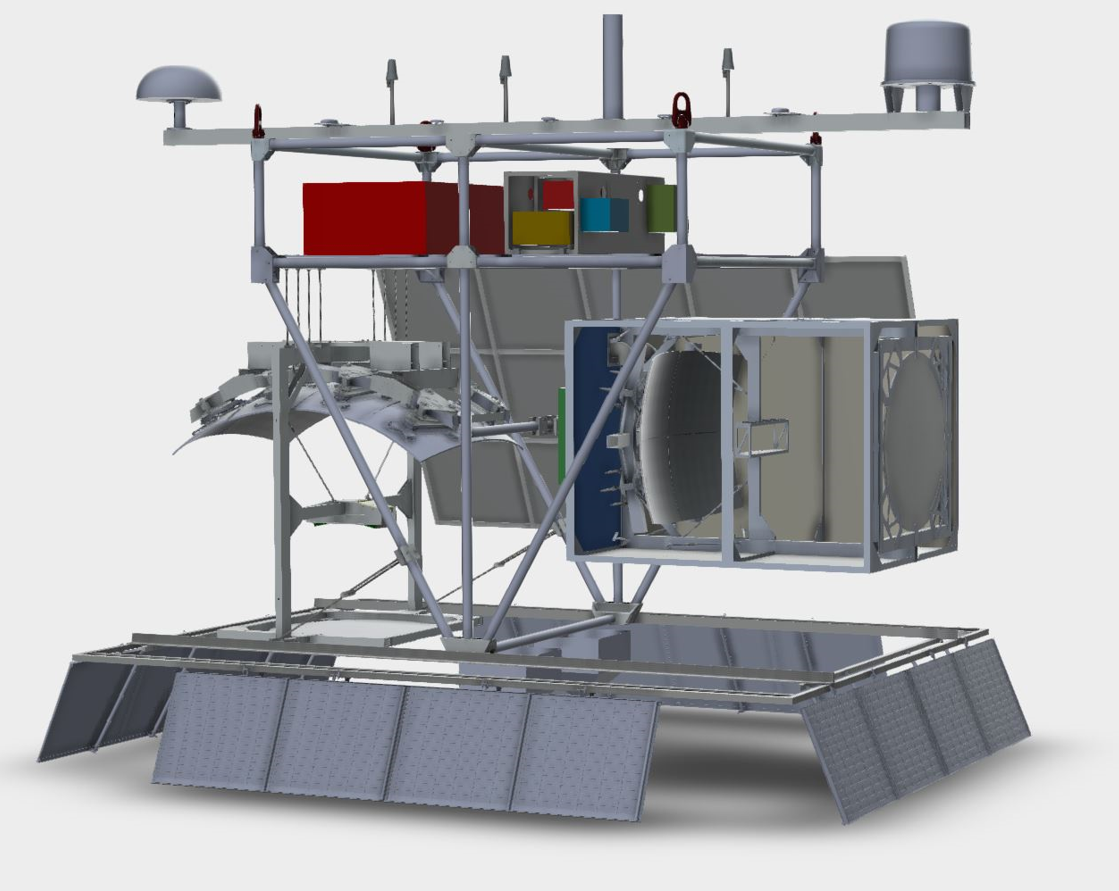}
    \includegraphics[width=0.32\textwidth]{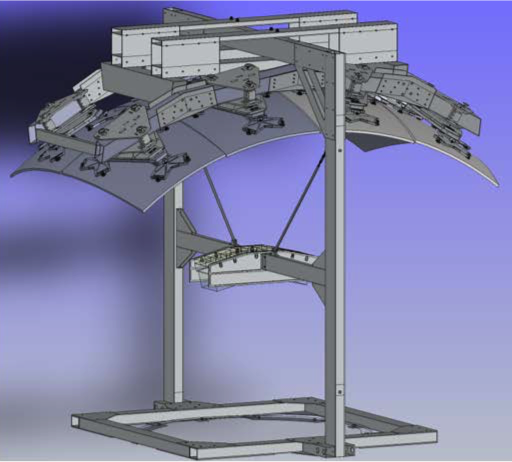}
    \includegraphics[width=0.29\textwidth]{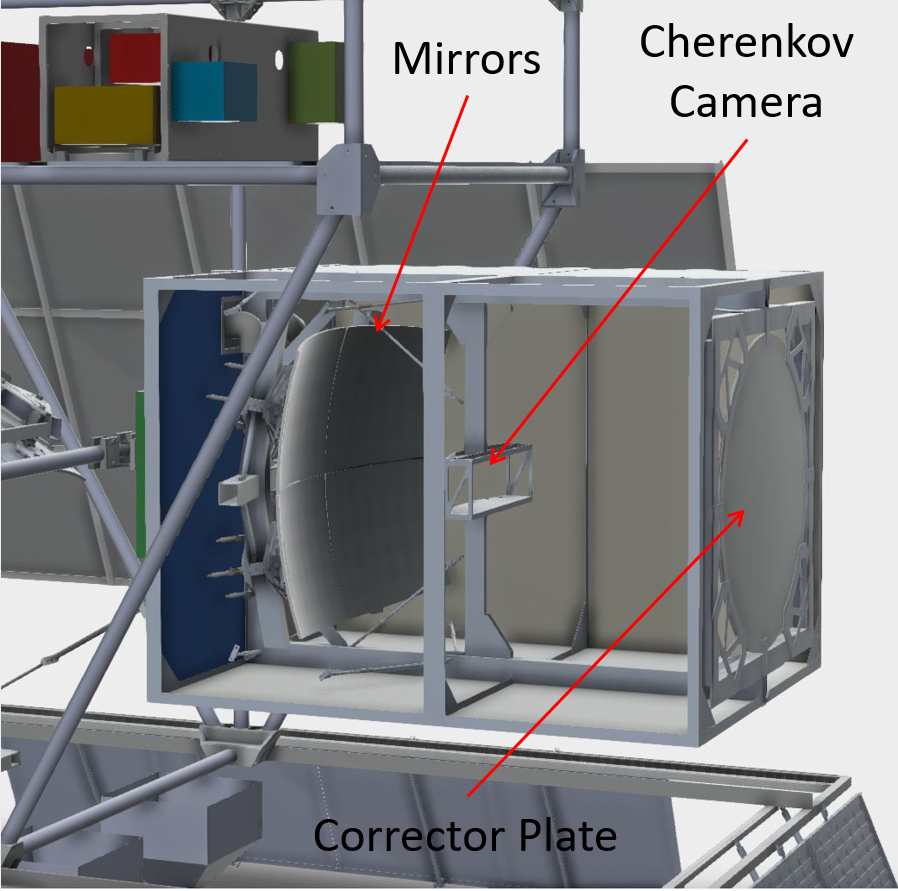}
 \end{center}
\vspace{-0.3cm}
  \caption{The EUSO-SPB2 payload configuration. Left: Schematic of the payload. Middle: Schematic of the fluorescence telescope. Right: Schematic of the Cherenkov telescopes. Not shown is the infrared cloud camera (see Ref.~\citenum{Diesing:2021fB} or the airglow monitoring sensor package (see Ref.~\citenum{2019NIMPA.922..150M}). From Refs. \citenum{Osteria:2021OJ,Filippatos:2021b5}.
  DOIs: \href{https://doi.org/10.22323/1.395.0206}{10.22323/1.395.0206}, \href{https://doi.org/10.22323/1.395.0405}{10.22323/1.395.0405}
  }
\vspace{-3 mm}
  \label{SPB2schematics}
\end{figure}

The EUSO-SPB2 Cherenkov telescope (CT) \cite{Eser:2021H6,Bagheri:2021s0} employs a Schmidt telescope with a $12.8^\circ \times 6.4^\circ$ horizontal by vertical (above or below the viewed Earth limb) FoV. The focal plane is comprised of a 512 pixel silicon photomultiplier (SiPM) camera using the MUSIC ASIC FEE readout \cite{10.1117/12.2231095} with a 100 MegaSample/second digitization. The choice of the SiPM is driven by the relatively high sensitivity in the 200$-$1000 nm wavelength band needed due to the inherent variability of the Cherenkov signal, e.g. see Fig.~\ref{CherSpectra}.  The 10 ns digitization span is optimized for the anticipated fast Cherenkov signal that is viewed at an angle close to from the EAS trajectory.   The optics in the CT design has a relatively large pixel instantaneous field-of-view, iFoV $=0.4^\circ$, which implies the count rate from the dark-sky background will be significant and dominate that due to dark-count rates in the SiPMs or other electronic noise \cite{Bagheri:2021s0}. The CT employs a novel technique that uses a bifocal mirror to split the $0.785$ m$^2$ effective area into two separate images on the focal plane.  This provides a coincidence trigger using spatially separated events, but at a cost of slightly increasing the effects of the dark-sky background. The goal of the bifocal design is to determine whether event backgrounds due to ionization in the SiPM pixels, i.e. due to traversing cosmic rays or other backgrounds, require the need of a such a bifocal trigger. 

\begin{figure}[ht]
\centerline{
\minifigure[The 100-day neutrino diffuse flux sensitivity (90\% CL) for EUSO-SPB2 (dashed lines) based on the simulated response for \taons $\rightarrow$ electron or hadrons, \taons $\rightarrow$ muons, and emergence muons from muon neutrino interactions in the Earth.  A 20\% duty cycle (live time fraction), an effective optical collection area of 1.0 m$^2$, and a 20\% photo-detection efficiency for the SiPMs in the focal planes were assumed.  The solid lines scale the sensitivity by  a factor of 28.125 to estimate the sensitivity for a EUSO-SPB2-like instrument with 360$^\circ$ azimuth coverage vs the 12.8$^\circ$ in the current design. Adapted from Ref. \citenum{2021PhRvD.103d3017C}.
DOI:\href{https://doi.org/10.1103/PhysRevD.103.043017}{10.1103/PhysRevD.103.043017}
]
{\includegraphics[width=0.75\textwidth]{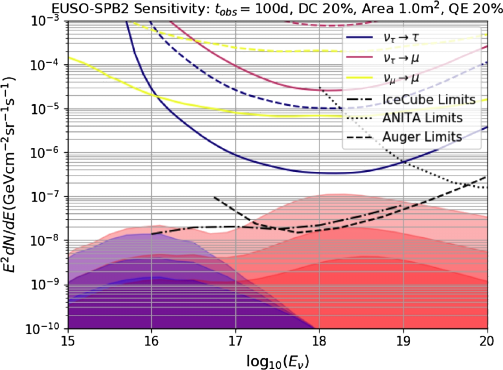}\label{SPB2diffuse}}
\hspace*{-10mm}
\minifigure[The conceptual design \cite{NeuBACH} for a balloon-borne instrument with two rings of optical Cherenkov light cameras to simultaneously measure the Cherenkov signals from upward-moving neutrino-induced EAS.]
{\includegraphics[width=0.21\textwidth]{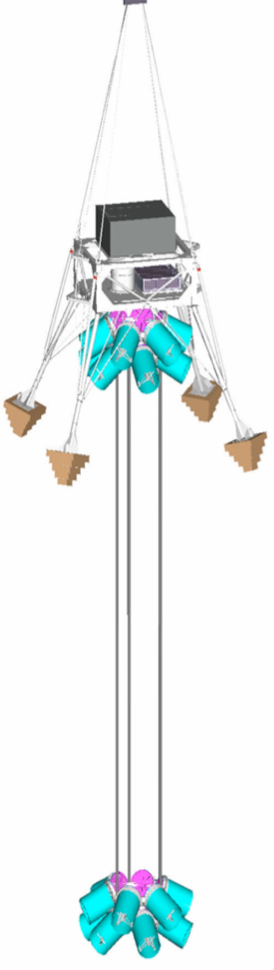}\label{NeuBACHfig}}
}
\end{figure}


The results of the simulated diffuse flux sensitivity for EUSO-SB2 is shown in Fig.~\ref{SPB2diffuse}. When compared to Fig.~\ref{diffuse}, several intriguing differences of optical Cherenkov measurements from upward-moving EAS for a ULDB mission versus that for one for LEO, such as POEMMA, are evident. As with that shown in Fig.~\ref{diffuse},  
the signal from Earth-emergence muons from $\nu_mu \rightarrow \mu$ interactions dominant over that for $\nu_\tau \rightarrow \tau$ for POEMMA below $\lsim 10$ PeV. However, if the azimuth acceptance of a balloon-borne instrument with similar performance could could be extended to 360$^\circ$, the sensitivity from the $\nu_mu \rightarrow \mu$ has significant sensitivity at and below 1 PeV.  An example of such a design is shown in Fig.~\ref{NeuBACHfig}, which was developed by a NASA/GSFC group in 2001. \cite{NeuBACH}. The design included two rings of Cherenkov telescope both to measure neutrino-induced upward-moving EAS in 360$^\circ$, with the lower ring of telescopes being lowered after launch. These provide two spatially separated measurements within the optical Cherenkov light pool on an event-by-event basis to reduce the threshold energy for observing neutrino events, e.g. by minimizing the effects of the dark-sky background. However, Fig,~\ref{SPB2diffuse} shows that the sensitivity curves for the hypothetical 360$^\circ$ azimuth balloon-borne version is orders of magnitude less sensitive than the results obtained by IceCube \cite{IceCube2015}, at lower neutrino energies, and the Pierre Auger Observatory \cite{2019JCAP...10..022A} amd ANITA \cite{2019PhRvD..99l2001G}, at higher energies.  Another factor is that the current length for ULDB missions is 100-days as compared with 5 or more years with LEO missions such as POEMMA, i.e. shorter duration missions will have lower exposure for a given instrument response, even when considering the distance to the EAS is shorter from a balloon altitude measurement, which increased the Cherenkov photon density for a given EAS energy.


Given the current limitations of around 100 days for an ULDB flight, there is a surprising sensitivity for transient cosmic astrophysical neutrino sources \cite{Venters:2021RV}, due to the fact that the instantaneous effective area for neutrino source detection (when duty cycle can be ignored) is significant even for EUSO-SPB2 with a CT with $12.8^\circ \times 6.4^\circ$ FoV and its optical Cherenkov light collection and signal digitization performance. Fig.~\ref{SPB2sky} shows the effective area for tau neutrino detection as a function of position in the sky assuming observation near Wanaka, New Zealand in May 2023 during full and near new moon. The EUSO-SPB2 payload is designed to rotate to point to a target-of-opportunity transient source, based on an external gravitional wave or electromagnetic detection alert, e.g. from the Gamma-ray Coordination Network (GCN) \cite{GCN} to facilitate ToO observations. The simulated neutrino  sensitivity to several astrophysics transient models indicates sensitivity to the transient neutrino signals at distances that include nearby galaxies, such as M31 \cite{Venters:2021RV}.  However, extrapolating the current theoretical rates shown Fig.~\ref{POEMMAToOrate}, to within a 100-day balloon mission implies a low probability for a transient astrophysical flare to occur within the smaller cosmological volume defined by the shorter balloon experiment transient detection distance.

\begin{figure}[h]
\begin{center}
    \includegraphics[width=0.99\textwidth]{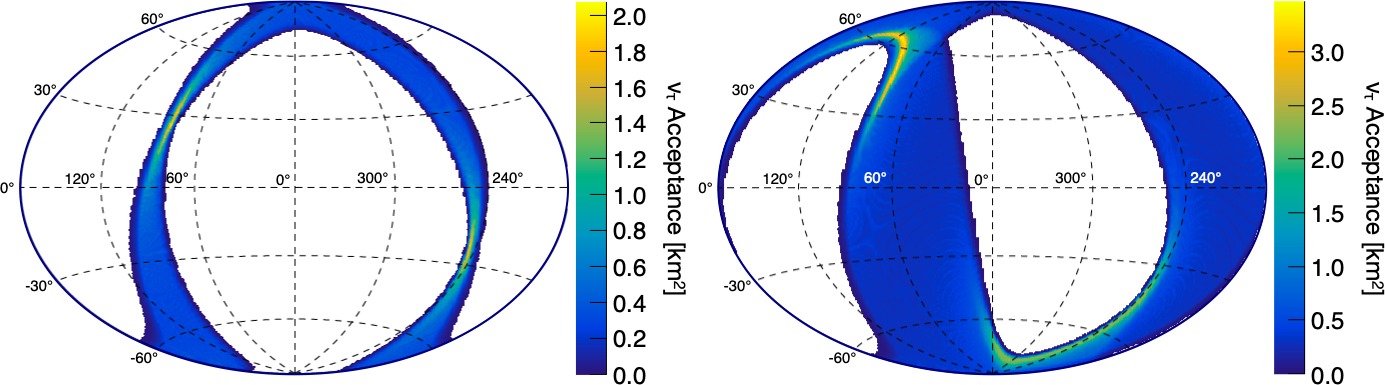}
 \end{center}
\vspace{-0.3cm}
  \caption{Sky plots in galactic coordinates detailing the effective area for $10^{8.5}$ GeV $\nu_\tau$ averaged over a day when the moon is full (left) or close to new (right) for EUSO-SPB2.  From Ref.~\citenum{Venters:2021RV}. 
          DOI:\href{https://doi.org/10.22323/1.395.0977  }{10.22323/1.395.0977}
}
\vspace{-3 mm}
  \label{SPB2sky}
\end{figure}

While it seems that the opportunity to detect a neutrino event from a balloon mission is challenging, the observation of the optical Cherenkov signals from very-high energy cosmic rays (VHECRs) observed slightly above the Earth's limb has been shown to be possible based on simulation studies \cite{PhysRevD.104.063029} and the results of the ANITA flights \cite{2021PhRvL.126g1103G}.  The left panel in Fig.~\ref{VHECR} shows the geometry of the an event that are viewed above the limb using the optical Cherenkov from EAS generated by VHECR. The atmospheric depth along the viewing direction quickly reduces to a negligible amount within a few degrees of the angle above the limb of the Earth, as shown in the right panel of Fig.~\ref{VHECR}. Viewing a few degress above the limb still provides a substantial atmospheric volume for detectable VHECRs even when the attenuation effects of atmospheric aerosols, ozone, and molecular scattering are considered, which are used to define the optical depth. A further complication is that the geomagnetic field will significantly increase the angular distribution of the electrons and positrons in the rarified atmosphere at high altitudes \cite{PhysRev.93.646.2,PhysRev.95.1705.4}. Fig.~\ref{VHECRrate} presents the simulated integral count rates above a VHECR threshold energy for the EUSO-SPB2 and POEMMA missions \cite{PhysRevD.104.063029}.  The results show that the EUSO-SPB2 ULDB experiment should have a VHECR threshold energy well below 1 PeV with a rate of several hundred observed events per hour just above threshold.  The rates based on three different assumptions on the geomagnetic field are shown to estimate the possible effects. In practice, setting different analysis thresholds on the Cherenkov signal should map out the VHECR rate.  The fact that the VHECR rate is experimentally fairly well-measured in this energy range provides an additional constraint to understand the EUSO-SPB2 CT response.  The simulated rates for POEMMA are also shown in Fig.~\ref{VHECRrate}. These over-the-limb VHECR events do form a background for upward-moving EAS neutrino searches due to the refraction of the VHECR optical Cherenkov signal when viewed near the limb \cite{2020PhRvD.102l3013V}, and therefore also provide a measure of this potential background in addition to an {\it in situ} calibration signal for a Cherenkov telescope.

\begin{figure}[h]
\begin{center}
    \includegraphics[width=0.49\textwidth]{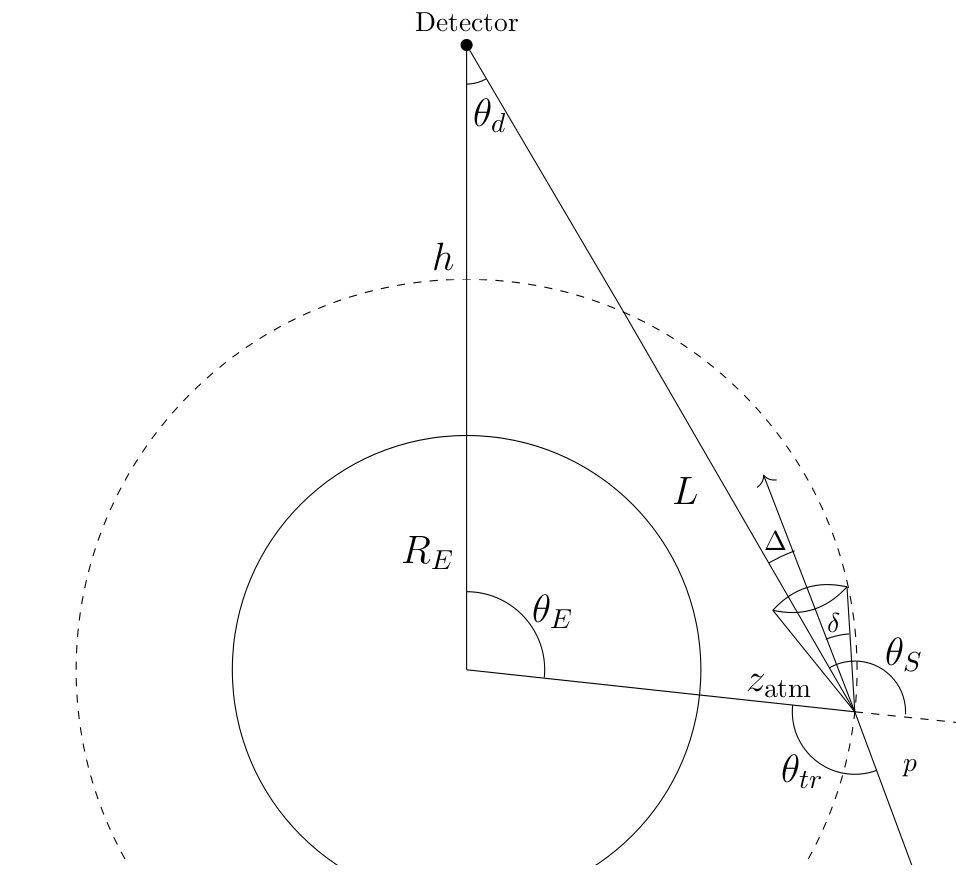}
   \includegraphics[width=0.49\textwidth]{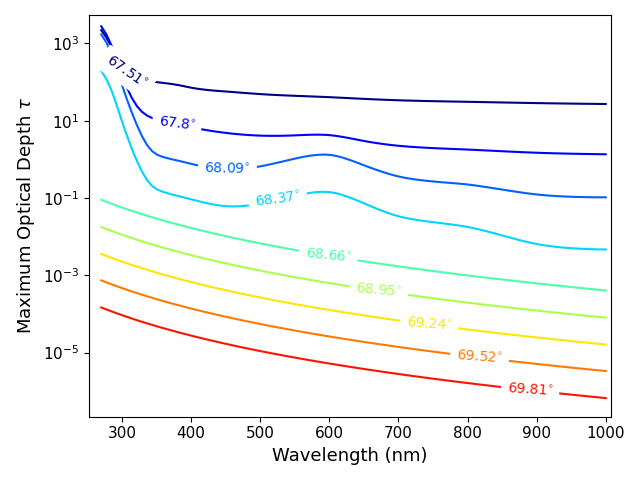} \end{center}
\vspace{-0.3cm}
  \caption{Left: The geometry for observing the beamed Cherekov signal from over-the-limb EAS generated by VHECR in the atmosphere.  Right: The total optical depth, including the effects of aerosol, Rayleigh, and ozone attenuation, defined by the angle away from nadir and obove the limb-of-the Earth, located at $67.5^\circ$ in this geometry, from a 525 km altitude. From Ref.~\citenum{PhysRevD.104.063029}. 
          DOI:\href{https://doi.org/110.1103/PhysRevD.104.063029}{10.1103/PhysRevD.104.063029}
}
\vspace{-3 mm}
  \label{VHECR}
\end{figure}

\begin{figure}[h]
\begin{center}
    \includegraphics[width=0.49\textwidth]{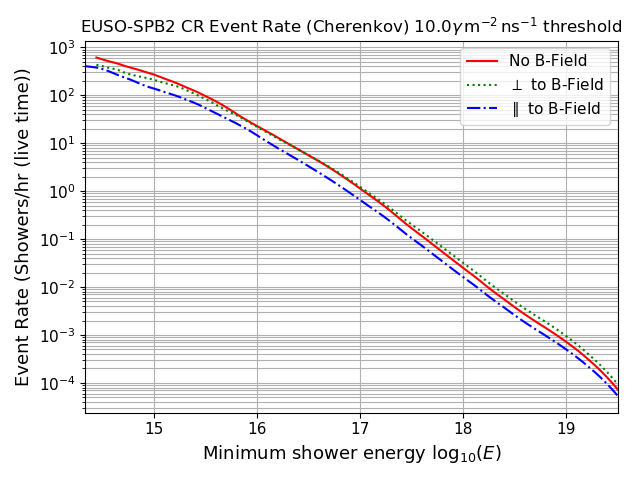}
       \includegraphics[width=0.49\textwidth]{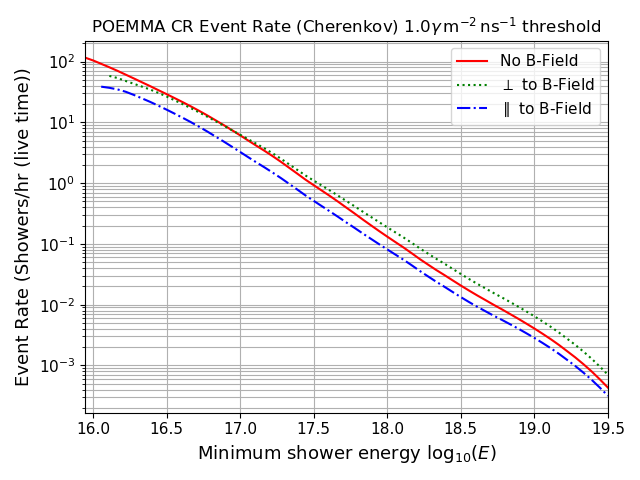}
 \end{center}
\vspace{-0.3cm}
  \caption{The integral event rates (above a threshold $E$) for observing over-the-limb VHECRs for EUSO-SPB2 (left) and POEMMA (right). The effects of the geomagnetic field are provided by assuming the acceptance for three different conditions. From Ref.~\citenum{PhysRevD.104.063029}. 
          DOI:\href{https://doi.org/110.1103/PhysRevD.104.063029}{10.1103/PhysRevD.104.063029} \label{VHECRrate}
}
\vspace{-3 mm}
  
\end{figure}

\begin{figure}[h]
\begin{center}
    \includegraphics[width=0.99\textwidth]{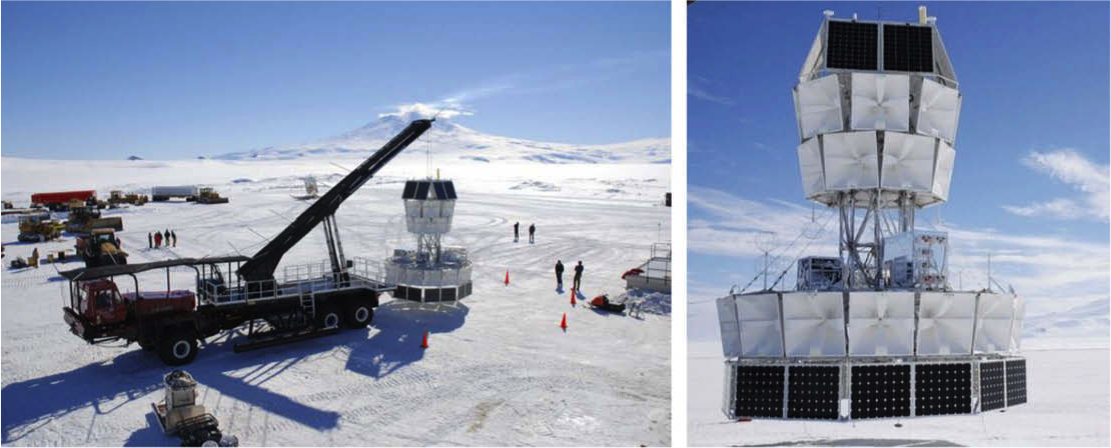}
 \end{center}
\vspace{-0.3cm}
  \caption{The ANITA-I payload configuration, with an upper 8-horn radio antenna array and lower 16-horn array, before its 35-day long-duration balloon (LDB) flight in December 2006 through January 2007. From Ref. \citenum{2009APh....32...10A}.
  DOI: \href{https://doi.org/10.1016/j.astropartphys.2009.05.003}{10.1016/j.astropartphys.2009.05.003}
  }
\vspace{-3 mm}
  \label{ANITApicts}
\end{figure}

\section{Radio Detection of Askaryan Radiation from Neutrino Interactions in Ice}

\begin{figure}[h]
\begin{center}
    \includegraphics[width=0.99\textwidth]{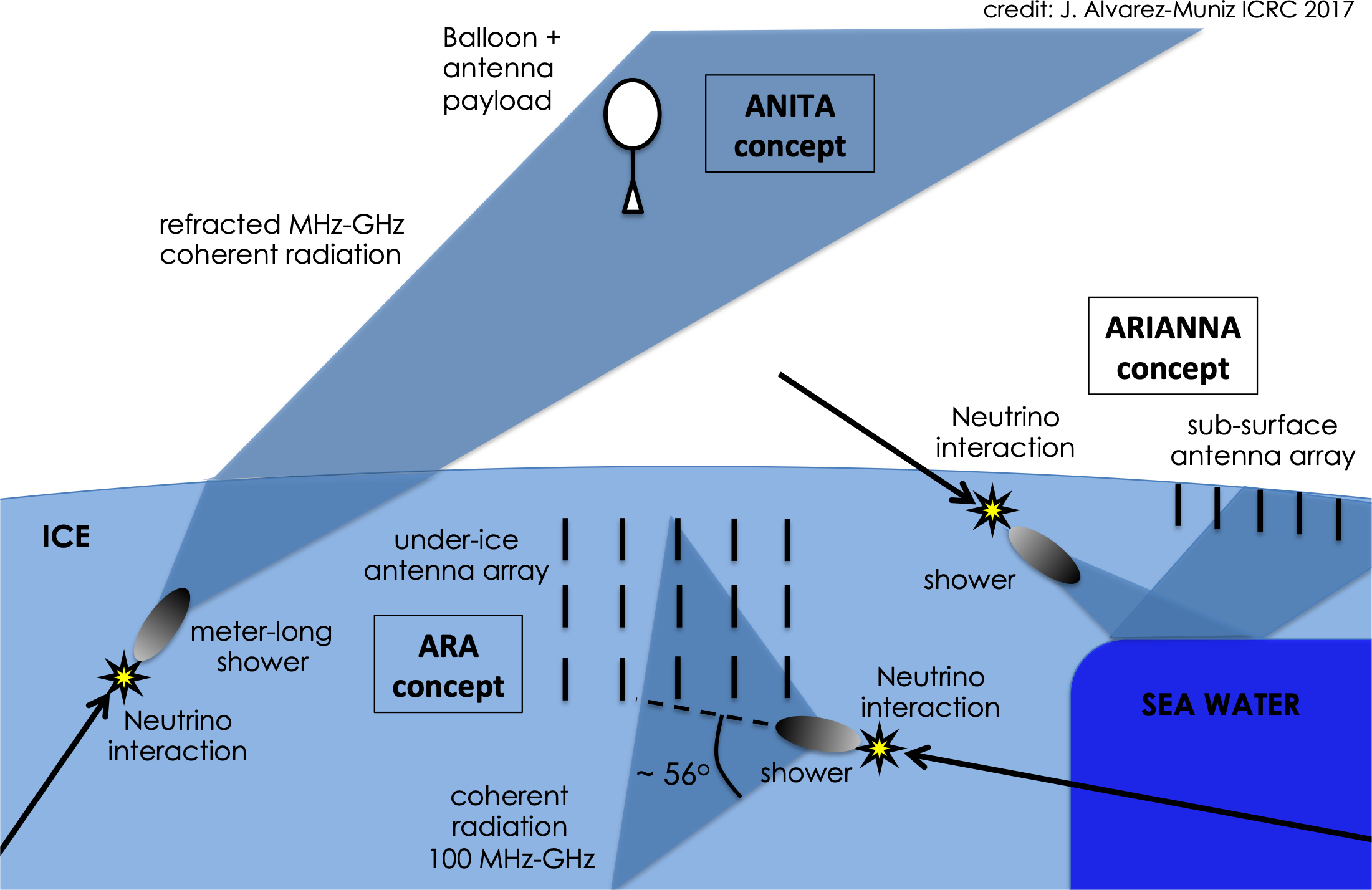}
 \end{center}
\vspace{-0.3cm}
  \caption{An artist rendition of the various methodologies to detect neutrino interactions in the Antarctic ice using the radio signal from the neutrino-induced cascades. From Ref.~ \citenum{2017ICRC...35.1111A}.
          DOI:\href{https://doi.org/10.22323/1.301.1111 }{10.22323/1.301.1111}
}
\vspace{-3 mm}
  \label{radioICEcartoon}
\end{figure}

In a relatively dense medium, such as ice or other materials, high energy particles initiate extensive particle cascades, see Fig. \ref{EASprofile}.  The cascade development is in principle similar to that for EAS, except that density of the medium is more uniform and larger, e.q. $\sim$ 1000 for ice/water, leading to meter-scale cascade longitudinal lengths.  As with EAS, the majority particle content of these cascades, even for hadronic primaries, quickly are dominated by electromagnetic components, with the charged particles being mainly electron and positrons. The latter have a finite lifetime a net negative charge imbalance develops with the evolution of the cascade. Since the electrons and positrons are moving faster than the phase velocity of photons in the dielectric, i.e. ice, the charge imbalance leads to the coherent emission of Askaryan radiation \cite{Askaryan:1962hbi} in the radio/microwave part of the electromagnetic spectrum (see the Chapter on {\it Radio Detection of High-Energy Neutrinos in Ice}.) Since the Askaryan effect is coherent, the power in the radio signal scales as the $N_{Cascade}^2$,  where $N_{Cascade}$, is the number of particles in the cascade that scales as $E_{Cascade}$,  The properties of ice, including that of Antarctic ice sheets, allow the propagation of these signals in the range of 10's of MHz to over a GHz with km-scale attenuation length \cite{2009APh....32...10A}, provide an immense amount of material for a radio-detection instrument with a large field-of-view that views the ice from high-altitude and thus a large cosmic neutrino target volume \cite{1992PhRvD..45..362Z,2010PhRvD..81l3009A}.  Fig.~\ref{ANITApicts} shows the ANITA-I payload and the arrays of 0.8-m wide horn antennae that measure the EAS radio signals from 200 MHz to 1200 MHz and provided a field-of-view of 360$^\circ$ in azimuth and in elevation angle from horizontal down to within 40$^\circ$ from nadir. For neutrino-induced Askaryan detection, this corresponds to a target volume 1$-$2 $\times 10^6$ km$^3$ \cite{2009APh....32...10A}. The differences between the index refraction of ice, $n_{ICE} \approx 1.35$ @ 300 MHz compared to air ($n_{AIR} \approx 1$) leads to significant refraction of the radio signal emerging from the ice. Fig. \ref{radioICEcartoon} provides a diagram of the generation of Askaryan radiation generated by neutrino-induced extensive cascades in the ice detectable by the ANITA experiment.  However, the strength of the radio signal even for a balloon-borne experiment leads to neutrino energy thresholds of the a significant fraction of an EeV, as shown in Fig.~\ref{PUEOsensitivity}.  

\begin{figure}[ht]
\begin{center}
{\includegraphics[width=0.99\textwidth]{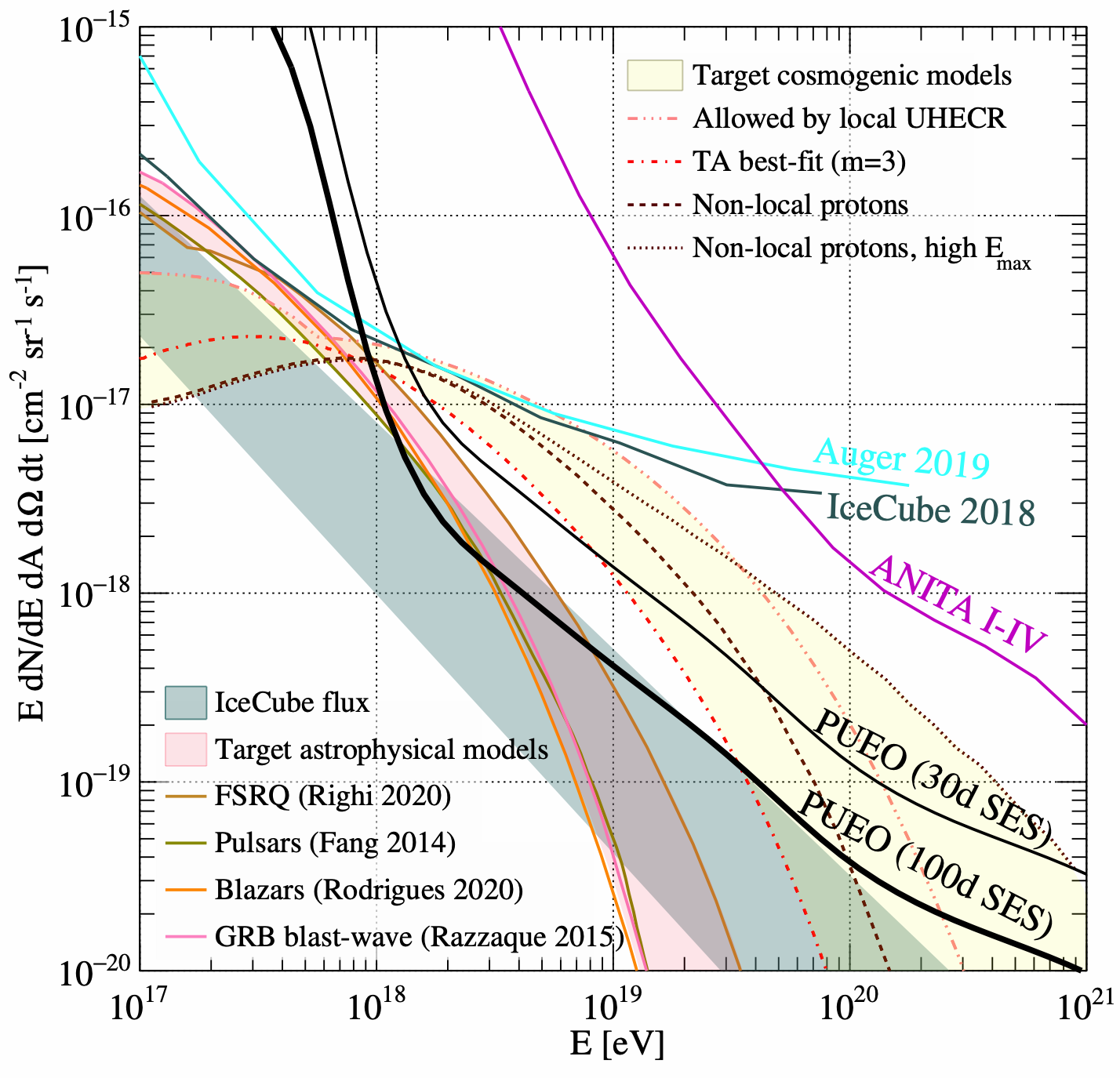}}
\caption{The PUEO simulated single-event sensitivity compared to the experimental limits from ANITA, IceCube, and Auger. From Ref. \citenum{2020arXiv201002892A}.  
DOI: \href{https://doi.org/10.1088/1748-0221/16/08/P08035}{ 	10.1088/1748-0221/16/08/P08035}
\label{PUEOsensitivity}}
\end{center}
\end{figure}

As with the optical Cherenkov signal, the temporal spread of the radio photons is of the order of 10 ns within the radio Akaryan ring,  while the times spread increases as the measurement samples the signal from further out into the power-law tail of the radio signal profile. Since Askaryan radiation is generated in any neutrino-induced particle cascade, in principle this technique is sensitive to all flavors of neutrinos in both the charged-current and neutral-current interactions, with the amount of neutrino energy put into the cascade dependent on the interaction and flavor.  For example, 100\% of an electron neutrino charged-current interaction is put into the cascade with around 80\% due to the generated electron and the remaining 20\% put into a hadronic cascade in the deep-inelastic scatter (see Section \ref{SNc_NuInt}.   Muons produced in a muon neutrino CC interaction can propagate over long slant depths and produce a different type of cascade, albeit with the average charged particle number lower by several orders of magnitude for a given energy \cite{Vankov}. While fluctuations in the muon-induced EAS formation can lead to an observable signal \cite{2021PhRvD.103d3017C}, they are effectively undetectable unless the muon energy is several orders above a given energy threshold for electromagnetic or hadronic EAS observation.  The experimental constraints on the ultra-high neutrino flux, see Fig.~\ref{PUEOsensitivity}, have made the CC muon EAS channel not competitive with the electron CC neutrino channel. The tau neutrino channel is unique in that the Lorentz-boosted decay length of the \taon can lead to these escaping the Earth, then decay to form upward-moving EAS detectable in the radio (as well as optical Cherenkov previously discussed), which is detailed in the next section. To summarize, the electron neutrino CC interaction channel provides a way to put 100\% of the neutrino energy into an EAS, while the muon and \taon lepton effectively only place around 20\% into the EAS, assuming these generated leptons with 80\% of \Enu are not observable. For the NC interactions, around 20\% of \Enu is put into the EAS, but with a smaller cross section than that for the CC, see Fig.~\ref{BlockSigma}.

\section{Radio Detection of Geomagnetic Radiation from EAS}

The measurement of UHECRs via detection of the EAS radio signals has been demonstrated to be a robust methodology as given by the LOFAR \cite{Buitink_2016}, the Auger Engineering Radio Array (AERA) \cite{Aab_2016_radio_CR_energy,2021ICRC...37..0387P}, and other ground-based UHECR experiments\footnote{For reviews see Refs. \citenum{SCHRODER20171, Schroeder:2019TB}}. However, the generation of the radio signal for upward-moving EAS exhibits unique characteristics, as opposed to the signal from downward-moving, UHECR-induced EAS. If the EAS develops in a low density (high altitude) part of the atmosphere, the magnetic field is able to deflect the particles enough so that  the tracks are no longer being amplified by the Cherenkov effect and leads to a saturation of the radio signal \cite{2021ICRC...37..1031P}. 
This effect is similar to that with the generation of the optical Cherenkov signal due to the small index of refraction during signal generation at high altitudes where the air index of refraction is closer to unity.

In the case that the particle cascade occurs in the atmosphere, an extensive air shower (EAS) is formed, with unique properties due to the lower density, $\rho_{AIR} \approx 1.2$ mg/cm$^3$ at STP.  Furthermore, the density profile of the atmospheric medium is exponential versus altitude and this translates into EAS developing over multiple km in length, i.e. the slant depth profiles shown in Fig.~\ref{EASprofile} corresponds to $\sim$ 10 km for a nearly vertical EAS. The EAS development and optical and radio signal generation differs whether the cascade is developing downward or upward, the latter having the atmosphere becoming less dense as a function of EAS age.

The radio emission from an EAS is due to the combination of several different radiation mechanisms and is denoted as geomagnetic radio emission \cite{Alvarez-Muniz_2012}. The evolving electron and positron distributions in the EAS are deflected by the Earth's magnetic field,
resulting in radio emission due to the time-varying transverse currents in the shower, which depend on the orientation to the geomagnetic field.  The small idenx of refraction in air in the radio wavelength band produces a Cherenkov-like radiation effect that focuses the radio emission forward along the propagation axis of the EAS.
The  geomagnetic field deflects the charged particles in the EAS in a plane that is orthogonal to the magnetic field vector and the direction of propagation via the Lorentz force. The EAS also produces an excess negative charge due to the dominance of electrons over positrons during the cascade development, leading to an additional Askaryan-like \cite{Askaryan_1962} 
contribution to the radio emission. Since the Earth's magnetic field is well known \cite{IGRF_2015}, and the beam is forward-directed, the impulse shape and polarization provide a signature for the radio measurement of EAS events, e.g. the EAS produces a characteristic radio impulse spectrum and polarization that make EAS-induced events straightforward to identify versus Askaryan signals generated in the ice \cite{ANITA2018}. The Cherenkov-like 
effects produce an impulsive electric field with a spectrum that increases linearly up to 5~MHz, is flat up to $\sim$100 MHz \cite{Alvarez-Muniz_2012}, and then falls exponentially at higher frequencies \cite{Alvarez-Muniz_2012b}. Fig. \ref{fig:RadioPulse} shows an example result of the ZHAireS simulations for one specific Earth-emergence angle of 10$^\circ$ as a function of starting, i.e. \taon decay, altitudes for upward-moving 1 EeV EAS \cite{2021ICRC...37..1205K,2021ICRC...37..1031P}. The radio emission beam for an EAS starting at sea level is aligned with the EAS propagation axis with a beam full-width half-maximum of $\sim$4$^\circ$ at $\nu \approx 50$ MHz, and produces a narrow cone peaking at $\sim$1.4$^\circ$ from the EAS axis with 0.5$^\circ$ width at $300$ MHz, becoming narrower as the radio frequency increases.  The location of Cherenkov-like ring occurs at angles closer to the EAS trajectory as the starting EAS altitude is increased, mimicking the effect that occurs with the optical Cherenkov generation due to the reduction of the index of refraction of air at higher altitudes. The backgrounds relevant to radio detection of EAS consist of sky noise, thermal emissions from the environment, i.e. ice, the atmosphere, the instrument, and anthropogenic backgrounds \cite{2009APh....32...10A,Thomson_Moran_Swenson_2017}. 

\begin{figure}

\begin{center}
     \includegraphics[width=0.99\columnwidth]{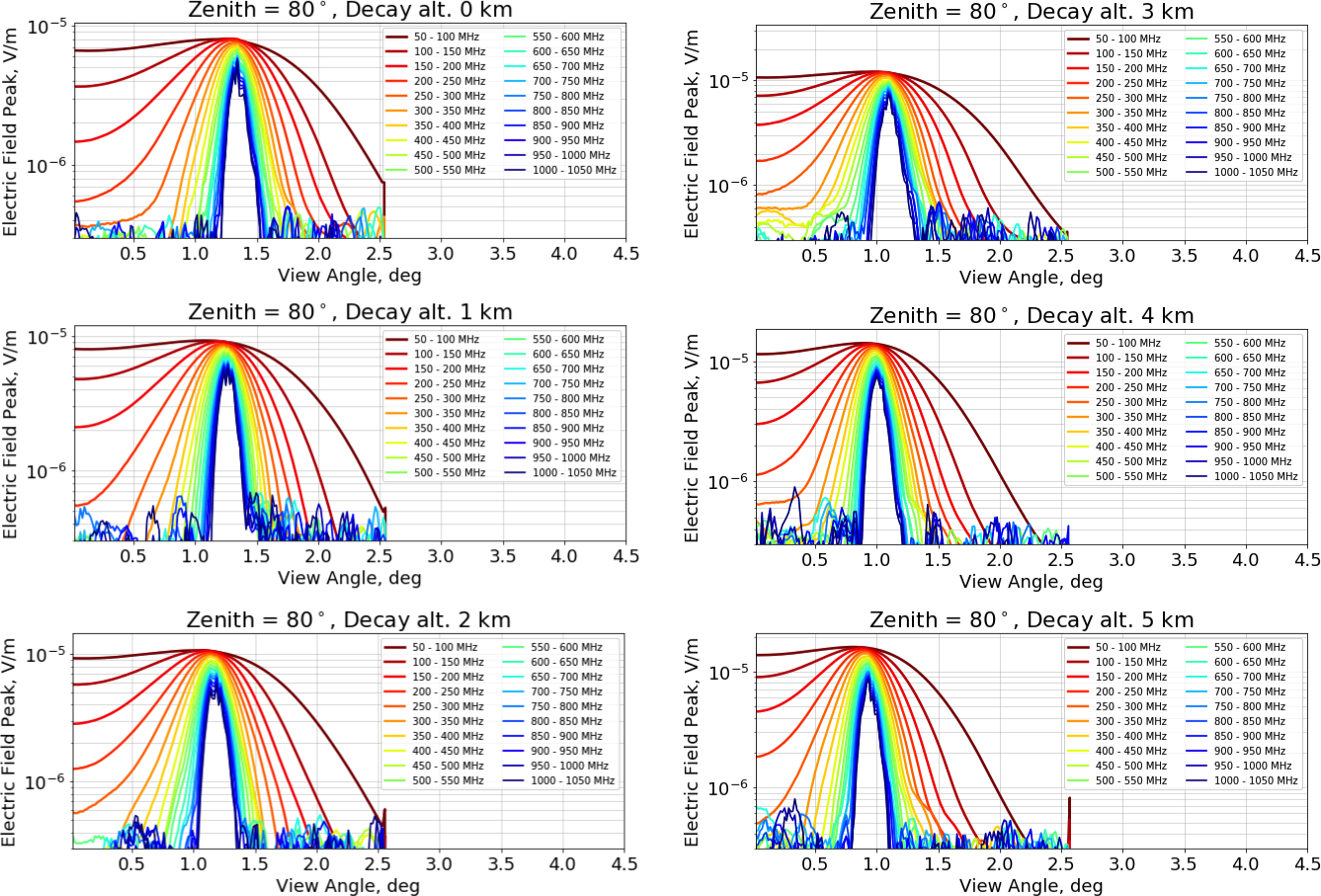}
\caption{\small {Results from the ZHAireS simulation showing the radio pulse spectra at 525 km altitude as a function of observer view angle of the shower for a zenith angle of 80 degrees and a \taon decay altitude of 4 km.  From Ref. \citenum{2021ICRC...37..1205K}.
  DOI: \href{https://doi.org/10.22323/1.395.1205}{10.22323/1.395.1205}
\label{fig:RadioPulse}}}
\end{center}
\vspace{-6mm}
\end{figure}

\begin{figure}[ht]
\begin{center}
{\includegraphics[width=0.99\textwidth]{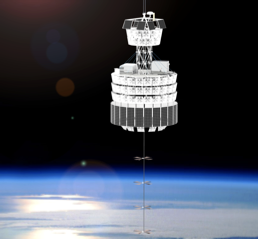}}
\caption{An artist rendering of the PUEO long-duration balloon payload, showing the various antennae including a low-frequency drop down antenna. From Ref. \citenum{2020arXiv201002892A}.
  DOI: \href{https://doi.org/10.1088/1748-0221/16/08/P08035}{ 	10.1088/1748-0221/16/08/P08035} \label{PUEOpict}
  }
\end{center}
\end{figure}

The four LDB flight of ANITA have detected and identified EAS events either reflected off the ground (with a flip in the signal polarity) or viewed directly arriving in directions above the limb of the Earth \cite{Hoover_2010}. These have yielded an independent measurements of the cosmic ray flux versus energy spectrum above 3 EeV \cite{Schoorlemmer_2016}.  ANITA has also found two upward-moving EAS whose phenomenology are consistent with $\tau$-leptons \cite{Gorham_2016, Gorham2018}, although the slant depth through the Earth strongly disfavors a $\nu_\tau$ origin \cite{Romero-Wolf_2017}. 
The results of the last, as to date, flight of ANITA, ANITA-IV  \cite{2019PhRvD..99l2001G, 2020arXiv200805690A} including a class
of events consistent with that from Earth-emerging \taons observed near the Earth's limb, which motivates further space-based and sub-orbital cosmic neutrino measurements. It should be noted that each ANITA instrument employed in the four different flights use different hardware, upgraded from the previous flight. Thus, one must take this consideration when comparing neutrino detection performance for each of the four ANITA flights.

The Payload for Ultrahigh Energy Observations (PUEO)  \cite{2020arXiv201002892A} is a NASA Pioneer-class mission for radio detection of neutrino events in the Antarctic ice,  \taon induced, upward-moving EAS from neutrino interactions in the Earth, and direct and reflected UHECR signals. The PUEO payload is shown in Fig.~\ref{PUEOpict}. While built on the ANITA development, flight, and analysis experience, PUEO employs a larger radio array in the 300$-$1200 MHz band, including drop-down low-frequency, 50$-$300 MHz radio antenna, enhancing the sensitivity to  UHECRs and neutrinos. PUEO also includes an interferometric phased-array trigger, that coherently sums radio waveforms from each antennae using the appropriate propagation time delays to provide an increase in the signal-to-noise $\approx \sqrt{N_{\rm antennae}}$ compared to using a single antenna. This phased-array trigger technique has been demonstrated in the Askaryan Radio Array (ARA) experiment \cite{ALLISON2019112}, and allows the reduction of the neutrino energy detection threshold by an order or magnitude, as shown in Fig.~\ref{PUEOsensitivity}.

Several instrument studies for future NASA sub-orbital, including the balloon-borne ExaVolt Antenna (EVA)\cite{2017EPJWC.13502002P}, Earth-orbiting observatories, such as synoptic wideband orbiting radio detector (SWORD) \cite{Romero-Wolf_2013}, and lunar-orbiting missions, including the Zettavolt Askaryan Polarimeter (ZAP) \cite{Romero-Wolf:2021ia} and the Lunar Orbital Radio Detector (LORD) \cite{2016AdSpR..58..464R} have been undertaken to develop EAS radio detection observatories for cosmic rays and neutrino-induced extensive particle cascades.

At frequencies $<10$~GHZ, well above the EAS radio detection frequency band, the atmosphere and clouds are radio transparent and do not affect the ability to observe radio emission of EAS. However, knowledge of the atmospheric properties is important for radio emission since the local index of refraction at the site of the EAS development determines the beam pattern similarly to the way the Cherenkov radiation beam pattern depends on the index of refraction of the medium \cite{Alvarez-Muniz_2012}. The atmospheric profile will refract the propagation path of radio emission, although this is typically $<~1^{\circ}$. 
For space-based radio measurements, the dispersion of radio impulses through the ionosphere is a major source of error. This was studied extensively for the FORTE analysis \cite{2004PhRvD..69a3008L} and the proposed space-based SWORD mission \cite{Romero-Wolf_2013}. In order to accurately trigger on these EAS radio events, the ionospheric dispersion has to be corrected in real time on the orbiting instrument to provide an appropriate event trigger \cite{Romero-Wolf_2013}.

\section{Neutrino EAS Measurements Ground-based Experiments}

Ground-based experiments such as the Pierre Auger Observatory \cite{ABRAHAM201029,ABRAHAM2010227} in Argentina and the Telescope Array \cite{ABUZAYYAD201287,TOKUNO201254} in Utah employ vast arrays, spanning areas of hundreds to thousands km$^2$, of ground-based particle detectors and several air fluorescence telescopes to measure the flux of UHECRs. These experiments also have sensitivity to cosmic neutrinos, primarily to tau neutrinos that interact in the Earth and produce Earth-emergence EAS \cite{2007APh....26..402G,2013ICRC...33.1579Y,2019JCAP...10..022A}.  The detection of the particles in the EAS cascades by the surface arrays has offered the best sensitivity due to their nearly 100\% duty cycle, compared to $\sim 10\%$ for fluorescence measurements \cite{2007APh....26..402G,2019JCAP...10..022A}.  The neutrino detection limits obtained by these techniques for Auger operation are shown in Figs.~\ref{diffuse} and ~\ref{PUEOsensitivity} for measurements of the cosmic diffuse neutrino flux and in Figs.~~\ref{POEMMAsb} and ~\ref{POEMMAlb} for astrophysical transient neutrino flux detection.

Several experiments focused on using ground-based detection of Earth-emergent EAS (EeEAS) from tau neutrino interaction in the Earth have been and are in development using both the optical Cherenkov and radio EAS detection technique.  Fig.~\ref{radioMountainCartoon}  artistically presents the concept of using ground-based radio detection arrays to detect Earth-emergent EAS. In principle optical Cherenkov telescopes \cite{2003ICRC....5.2991A,2019PhRvD..99h3012O} or even particle detector arrays \cite{2020arXiv200206475R} can be used to detect the EAS.  Separate development integrating coincidence electronics to the Owens Valley Radio Observatory telescopes for UHECR EAS radio detection \cite{2020NIMPA.95363086M} could in principle be also used for searches for neutrino-induced EeEAS.

\begin{figure}[h]
\begin{center}
    \includegraphics[width=0.99\textwidth]{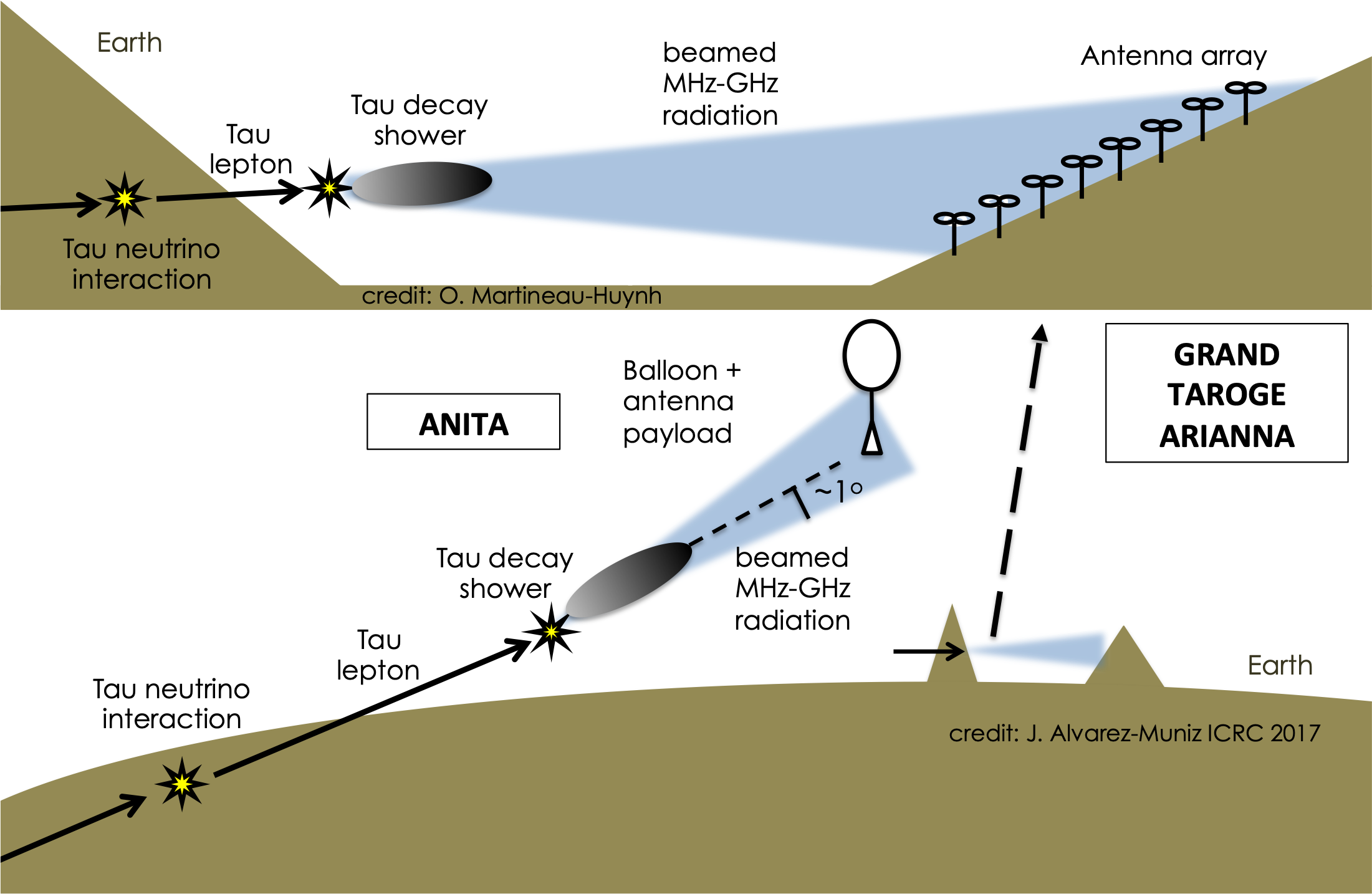}
 \end{center}
\vspace{-0.3cm}
  \caption{An artist rendition of the methodologies to detect neutrino interactions in the Earth leading to emergent \taon which decay to form EAS detectable using the radio signal from the neutrino-induced cascades, including both ground-based and balloon-borne detectors. From Ref. \citenum{2017ICRC...35.1111A}.
            DOI:\href{https://doi.org/10.22323/1.301.1111}{10.22323/1.301.1111}
  }
\vspace{-3 mm}
  \label{radioMountainCartoon}
\end{figure}

The ground-based optical Cherenkov EeEAS technique was pioneered by the All-sky Survey High Resolution Air-shower (Ashra) telescope experiment built and operated on Mauna Loa at 3300 m above sea level and faces aces to Mauna Kea in Hawaii \cite{2003ICRC....5.2991A,2013APh....41....7A,Ogawa:2019eB}. Fig.~\ref{ASHRA} shows the EAS detection concept in the left panel, the imaging Cherenkov camera in the middle panel, and the neutrino detection limit as a function of energy for $~1.2$ years of live observation time. d The camera in the telescope has a wide full FoV of 42$^\circ$ and uses a 20-inch photoelectric lens imaging tube
rather than an optical lens to produce a 1-inch diameter converging beam at the focal plane. The focal plane is  comprised of 58 photo-multiplier tubes that are part of the high
resolution imaging system with iFoV $\approx 0.05^\circ$ that also allowed for event triggers over different time scales \cite{2013APh....41....7A,Ogawa:2019eB}. The right panel of Fig.~\ref{ASHRA} shows that the Asrha-1 measurements have a neutrino detection threshold energy around 3 PeV.
Results from using the imaging atmospheric Cherenkov telescope (iACT) MAGIC to search for neutrino-induced EeEAS \cite{2018APh...102...77A} is also shown in the right panel of Fig.~\ref{ASHRA} and shows a slightly lower neutrino detection energy threshold due to the larger collecting power of MAGIC. The Ashra collaboration has reported that the group is working an extension of the experiment,the  Neutrino Telescope Array (NTA), which would also provide gamma-ray initiated EAS measurements \cite{Ogawa:2021dK}.

\begin{figure}[ht]
\begin{center}
    \includegraphics[width=0.53\textwidth]{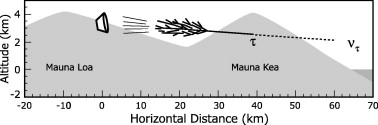}
    \hspace*{1 pt}
    \includegraphics[width=0.19\textwidth]{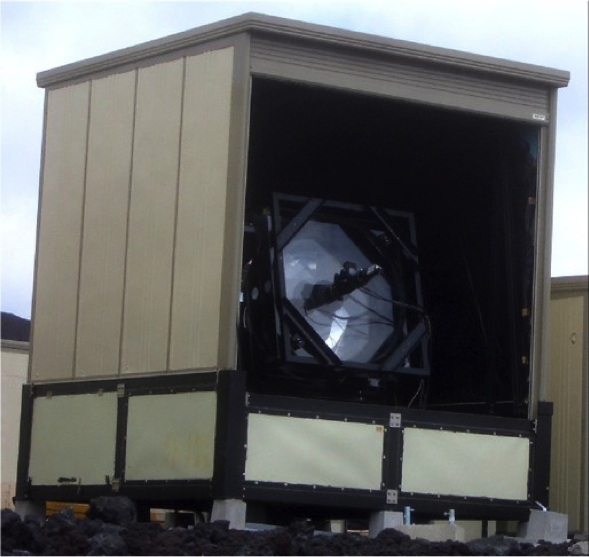}
    \hspace*{1 pt}
    \includegraphics[width=0.24\textwidth]{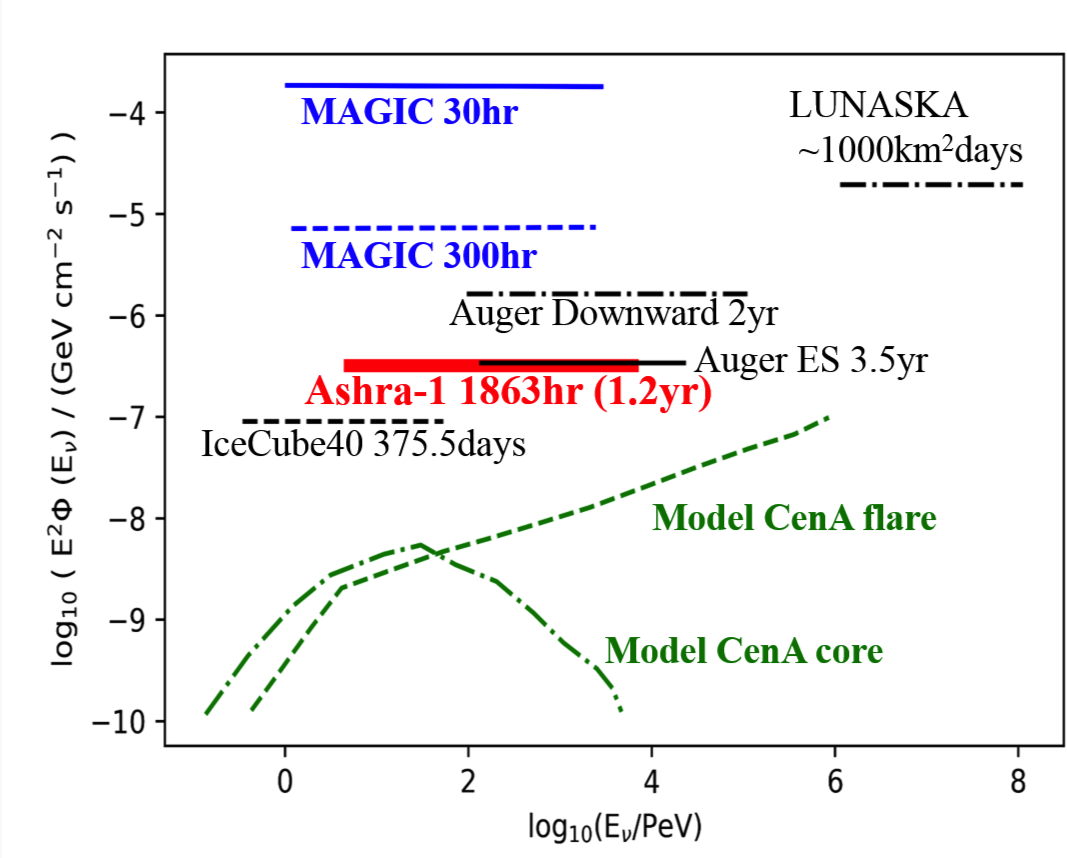}
 \end{center}
\vspace{-0.3cm}
  \caption{The Ashra-1 EeEAS experiment in Hawaii. Left: Schematic of the EAS detection technique. From Ref.~\citenum{2013APh....41....7A}; Middle: Picture of the Ashra-1 imaging Cherenkov telescope. Right: Ahsra-1 cosmic neutrino detection limits using 1863 hours of live time observations. From Ref.~\citenum{Ogawa:2019eB}. 
  DOIs: \href{https://doi.org/10.1016/j.astropartphys.2012.10.001}{10.1016/j.astropartphys.2012.10.001}, \href{https://doi.org/10.22323/1.358.0970}{10.22323/1.358.0970}
  }
\vspace{-3 mm}
  \label{ASHRA}
\end{figure}

To be located on a mountain top, Trinity \cite{2019PhRvD..99h3012O,2019BAAS...51g..67O} is a more ambitious ground-based Optical Cherenkov EeEAS neutrino detection experiment. The full configuration is proposed to be comprised of an array of 18 EAS imaging Cherenkov telescopes that provide a total 360$^\circ$ azimuth and 5$^\circ$ elevation viewing of EAS.  Each telescope will span $20^\circ$ in azimuth and be comprised of $\sim 3300$ SiPM pixels each with iFoV$ = 0.3^\circ$ with approximately 100 MHz readout defined by the $\sim 10$ ns Cherenkov signal width near threshold detection. A Trinity demonstration telescope is under development \cite{Brown:2021tf} and is shown along with the Trinity concept in Fig.~\ref{Trinity} and both will use appropriated sized SiPMs in the focal planes.

\begin{figure}[ht]
\begin{center}
    \includegraphics[width=0.67\textwidth]{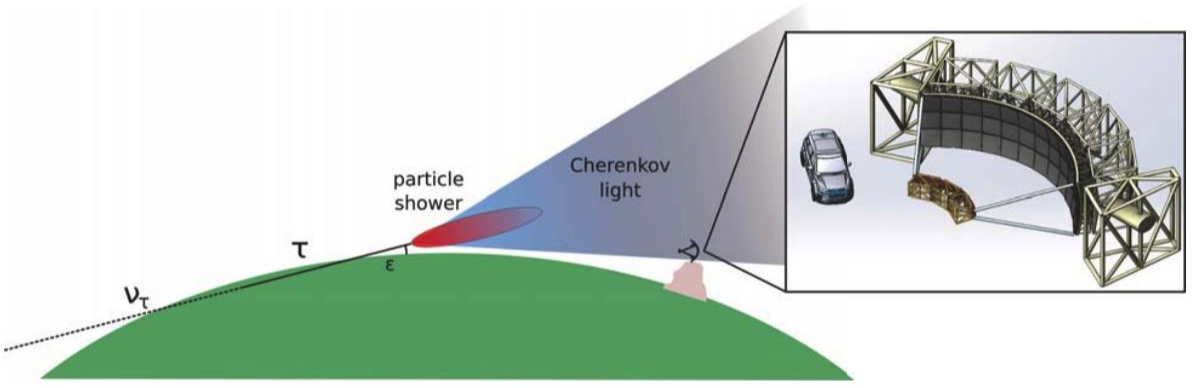}
    \hspace*{1 pt}
    \includegraphics[width=0.29\textwidth]{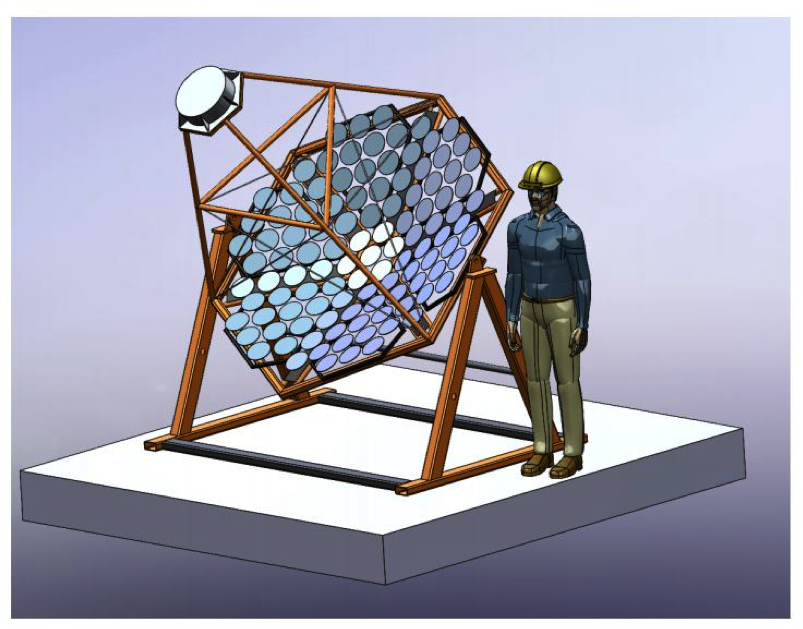}

 \end{center}
\vspace{-0.3cm}
  \caption{Left: Schematic of the Trinity experiment showing one of the 18 optical Cherenkov telescopes. 
  Right: Artist concept of the Trinity demonstration Cherenkov telescope. From Ref.~\citenum{Brown:2021tf}. 
  DOI: \href{https://doi.org/10.22323/1.395.1179}{10.22323/1.395.1179}
  }
\vspace{-3 mm}
  \label{Trinity}
\end{figure}

Both space- and ground-based EAS optical Cherenkov measurements need to be performed near astronomical night, e.g. duty cycle $\sim$20\%, and both are impacted by the airglow background and cloud cover.
A significant difference between space-based and ground-based optical Cherenkov EeEAS observations is the different viewing range around the Earth's limb and the different effects of atmospheric attenuation, especially due to aerosols. Using Trinity geometry as an example with a detector at 3 km altitude and assuming a spherical Earth, a 5$^\circ$ telescope elevation span corresponds to $\sim 6.5^\circ$ maximum Earth-emergence angle ($\beta_E = 90^\circ - \beta_{tr}$ in Fig.~\ref{upGEOM}.  Evaluation of the 100 PeV EAS spectra for these angles from Fig.~\ref{CherSpectra} shows an strong attenuation effect for Earth-emergent angles $\beta_E \lsim 10^\circ$, with the peak of the spectra at longer wavelengths. While the results shown in Fig.~\ref{CherSpectra} were calculated for a detector at 525 km altitude, the attenuation is dominated by the aerosol layer which has a scale height of $\sim$ 1 km above the Earth \cite{Elterman}, as implied by the optical depth profiles shown in Fig.~\ref{upGEOM}.  With the distance to the limb of $\sim$ 225 km for the Trinity example, EAS from higher energy Earth-interacting tau neutrinos that decay closer to the detector will reduce the viewed optical depth of aerosols to EAS maximum and thus reduce the effects of the atmospheric attenuation.  Similarly, the Cherenkov light intensity in terms of ${photons}/m^2$ within the Cherenkov cone, e.g. between the horns in Fig.~\ref{Cspot}, will be larger in the ground-observation case due to the shorter distance to the location of the EAS at shower maximum.  The net effect of these is that the EeEAS Cherenkov spectrum and intensity are both highly variable for both ground- and space-based observations.

\begin{figure}[ht]
\begin{center}
    \includegraphics[width=0.52\textwidth]{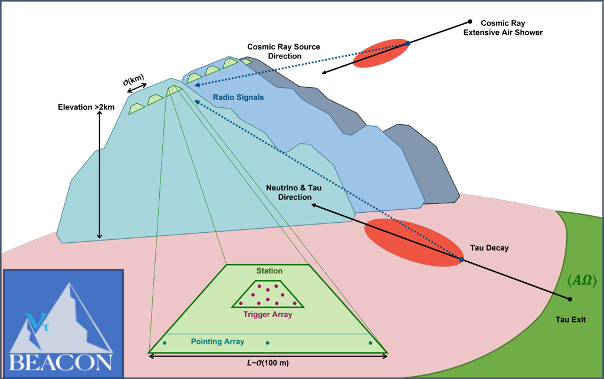}
    \hspace*{1 pt}
    \includegraphics[width=0.46\textwidth]{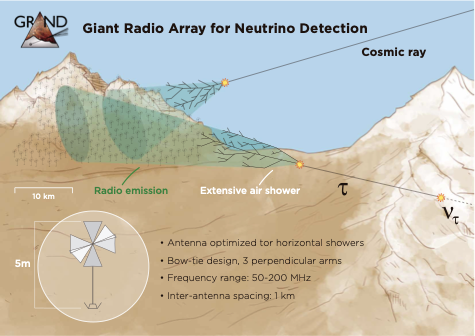}

 \end{center}
\vspace{-0.3cm}
  \caption{Left: Artist sketch of the BEACON ground-based radio EAS experiment. From Ref.~\citenum{2020JCAP...11..065W}.
  Right: Artist sketch of the ground-based EAS radio GRAND10k array. From Ref.~\citenum{2020SCPMA..6319501A}. 
  DOI: \href{https://doi.org/10.1007/s11433-018-9385-7}{10.1007/s11433-018-9385-7}
  }
\vspace{-3 mm}
  \label{NewGroundRadio}
\end{figure}

The ground-based radio EAS measurement technique has demonstrated its capability to accurately measure UHECR-induced EAS \cite{SCHRODER20171, Schroeder:2019TB}. As highlighted in Fig.~\ref{radioMountainCartoon}, radio EAS measurements can be used to search for EeEAS using radio detectors with the benefits inherent to the technique, nearly 100\% duty cycle and minimal atmospheric attenuation of the radio EAS signal. Several experiments currently under development are the TAiwan Radio Observatory for Geomagnetic Emission (TAROGE) and TAROGE-M, the Beamforming Elevated Array for COsmic Neutrinos (BEACON) \cite{2019BAAS...51g.191W,2020JCAP...11..065W}, and Giant Radio Array for Neutrino Detection (GRAND)\cite{2020SCPMA..6319501A} experiments.  The experiments employ arrays of radio antennae to have sensitivity sensitivity to both EeEAS from tau neutrinos and horizontal UHECRs. The left panel in Fig.~\ref{NewGroundRadio} illustrates the BEACON experiment, where sub-arrays of radio antennae form a phased-array using the beamforming technique similar to what is planned for the PUEO instrument \cite{2020arXiv201002892A}, but operating in the 30$-$80 MHz and the 200$-$1200 MHz bands. The BEACON arrays consist of a number of close packed 100-antenna stations separated by tens of meters and outrigger antenna separation by around a hundred meters to improve the angular resolution. A 30$-$80 Mhz prototype BEACON array has been operating at White Mountain Research Station in Bishop, California has been operating since 2019. The eventual full BEACON array is aimed to be completed around 2030 and will consist of 200,000 antennae covering and 200,000 km$^2$ area \cite{2020JCAP...11..065W}. The TAROGE-M experiment located on Mt. Melbourne in Antarctica has been operating using a similar high-elevation radio array operating in the 180$-$450 MHz band but does not use the beamforming technique \cite{Wang:20217k}. TAROGE-M is designed to detect downward-moving EAS from UHECRs as well as search for the EeEAS signals from tau neutrinos. TAROGE-M has heritage from the TAiwan Radio Observatory for Geomagnetic Emission (TAROGE) \cite{Chen:2021HZ} in the high coastal mountains of Taiwan and the ARIANNA-HCR (Horizontal Cosmic Ray) radio array \cite{Wang:2019qV}.

The Giant Radio Array for Neutrino Detection (GRAND) experiment is a ground-based radio array operating in the 50$-$200 MHz band that is to built in various stages with increasing numbers of antennae to an eventual number of 200,000 deployed over a 200,000 km$^2$ area \cite{2020SCPMA..6319501A}.  An initial 300-antenna prototype is to be started in 2021 in a radio quiet site of 200 km$^2$ near Lenghu China, with an altitude of $\sim 3$ km above sea level \cite{Zhang:2021Tk}. GRAND10k is anticipated to start in 2025 and GRAND200k to start after 2030.  The right panel in Fig.~\ref{NewGroundRadio} depicts the GRAND array at a single location, where EeEAS signals from tau neutrino interactions in nearby mountains can be measured. The simulated performance of GRAND200k predicts a UHECR exposure of $\sim 10^5$ km$^2$ sr year per year with good energy, angular and \xmax resolution. Since the footprint of the beamed EAS radio signal from UHECR EAS must be within the ground array, the geometry for GRAND200k accepts UHECRs in the zenith angular range $65^\circ$--$85^\circ$.

\section{Comparison of Future Space-based and Ground-based Cosmic Neutrino Measurements}

Both space-based and ground-based optical Cherenkov neutrino detection experiments employ imaging telescopes to measure the fast, $\sim$ 10 ns signal from EeEAS, but the space-based POEMMA implementation shows that the stereo-fluorescence technique can also be employed to achieve unique sensitivity to all flavors of neutrinos above 20 EeV.  However, an orbiting space-based EeAS optical Cherenkov telescope will scan the entire sky in less than a year's time, while a ground-based experiment at a single location will always have some areas of the sky that are not accessible, i.e. see the comparison of the BNS neutrino sky coverage for POEMMA, IceCube, and GRAND200k shown in Fig.~\ref{POEMMAbnscomp}.  As evidenced by IceCube, Auger, TA, and other experiments, ground-based experiments can be upgraded and operate over decades time frames. While space-based experiments are usually planned for 3$-$5 years of operation, the Fermi Gamma-ray Telescope have operated over 13 years since launch in 2008 \cite{2009ApJ...697.1071A,Fermi}. Besides the Hubble Space Telescope, free-flying missions usually do not have the capability to update hardware after launch.

Ground-based radio EeEAS experiments require over 100,000 separate, albeit inexpensive, antennae arranged over a large area to achieve the highest tau neutrino sensitivity proposed by these experiments. This is shown in Fig.~\ref{DiffuseCOmp} where the diffuse neutrino flux
sensitivity is presented to facilitate comparison to the space-based POEMMA results and in a context of several theoretical models of neutrino sources.
The left panel shows the 5-year all-flavor neutrino sensitivity for the space-based POEMMA mission, denoted as POEMMA30 where the 30 defines the azimuth span of Earth-limb Cherenkov observations, and a much more sensitive curve if an instrument could be developed that views the entire 360$^\circ$ azimuth with the same neutrino detection energy threshold. Comparison to current experimental limits as well that for several future experiments including GRAND10k. The right panel shows similar sensitivity limits for Trinity (10 years operation), BEACON (3 years operation), GRAND200k (3 years operation), and IceCube Gen2 with the radio extension \cite{2021JPhG...48f0501A}. 

In these comparisons of either measured or simulated diffuse neutrino flux sensitivity, two important aspects need to be discussed. The first is the assumptions regarding the confidence level and what energy span this is applied.  For example, all the sensitivity limits in Fig.~\ref{DiffuseCOmp} are 90\% CL per decade of energy, except for the BEACON results that are presented as 90\% CL per half decade of energy \cite{2020JCAP...11..065W}. As was discussed regarding the definition of Egn.~\ref{sensLimEqn}, there is a factor of two improvement if sensitivity limits for per half-decade versus per decade calculations. The second is the rigor of the underlying calculations of neutrino event acceptance that define the senisitivity.  For results that are based on actual measurements, i.e. that reported by IceCube, Pierre Auger Observatory, ANITA, and FORTE results, the assumption is that these are well-studying and if background events occur in the event sample, the limits use the appropriate $N_{CL}$, e.g. $N_{90}=3.28$ for an expected background on 1.0 event with no signal events, from Table XII in Ref.~\citenum{PhysRevD.57.3873}.  For simulated neutrino sensitivity determination, the robustness of the limits naturally are tied to the robustness of the assumptions in the neutrino $\rightarrow$ \taon $\rightarrow$ EAS $\rightarrow$ Cherenkov light or radio signal $\rightarrow$ instrument detection.  While an end-to-end neutrino simulation, $\nu$SpaceSim is in development \cite{2019ICRC...36..936K,2021ICRC...37..1205K} to provide a standard framework of optical Cherenkov and radio neutrino-induced EAS detection, currently each experiment use a variety of simulation software tailored to the specific experiment. In this context, the POEMMA simulated performance \cite{2019PhRvD.100f3010R,2020PhRvD.102l3013V,2020PhRvD.101b3012A} and that for EUSO-SPB2 \cite{2021PhRvD.103d3017C,PhysRevD.104.063029} are at a mature state (and provided the initial input for $\nu$SpaceSim development) as well as that for those experiments leveraging ANITA experience, including PUEO \cite{2020arXiv201002892A} and BEACON \cite{2020JCAP...11..065W}.  The new concept of Trinity, eg using mountain top optical Cherenkov detectors to detect \taon from Earth-interacting tau neutrinos, the simulations and detector response used to determine the sensitivities are at an early stage of development. Thus, the Trinity sensitivity presented in Fig.~\ref{DiffuseCOmp} are based on more fundamental assumptions on the performance, i.e. that the optical Cherenkov signal will be spread in four pixels, thus allowing the use of time coincidence (see Eqn.~\ref{TcoinEqn}, and background techniques used in iACTs allow for dark-sky background subtraction to yield a threshold of 24 PEs for the signal \cite{2019PhRvD..99h3012O}. The development of the Trinity prototype telescope \cite{Brown:2021tf} allows these assumptions to be experimentally validated while also providing information for optimization of the final design.  This highlights the importance of experimental prototypes, which is also being done for BEACON \cite{2020JCAP...11..065W,Zeolla:2021bl} and POEMMA, with EUSO-SPB2 to serve as the prototyp \cite{2020NIMPA.95862164S,Eser:2021H6}.

\begin{figure}[ht]
\begin{center}
    \includegraphics[width=0.44\textwidth]{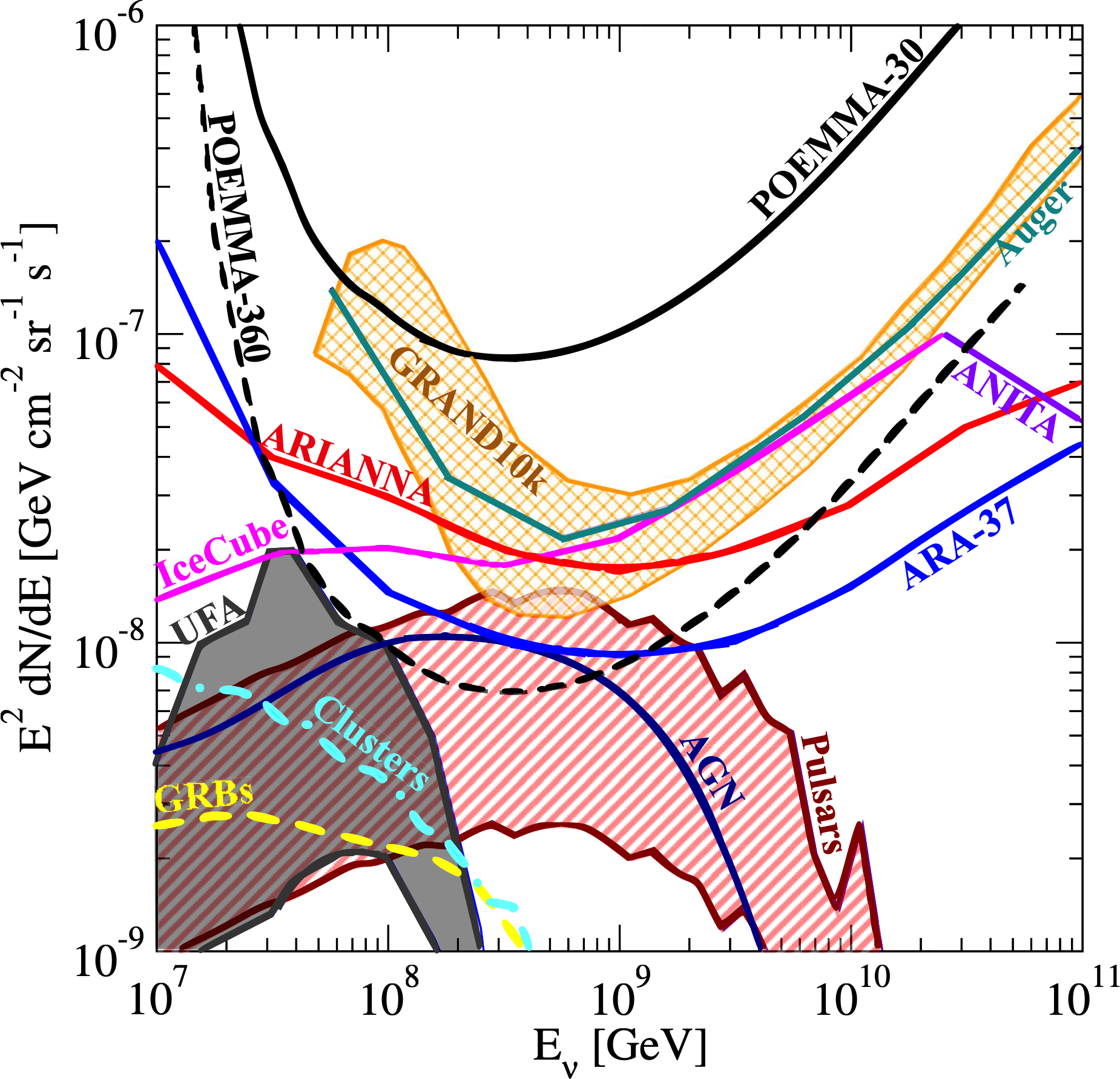}
    \hspace*{1 pt}
    \includegraphics[width=0.535\textwidth]{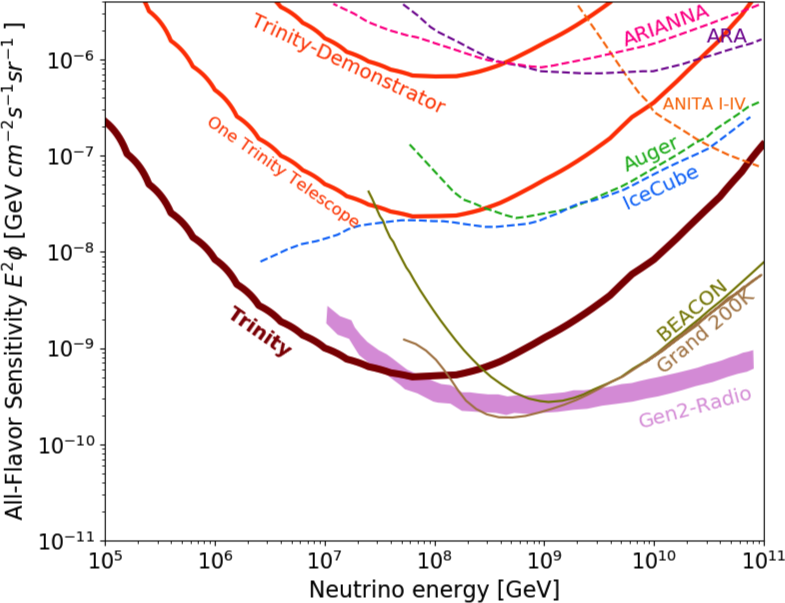}

 \end{center}
\vspace{-0.3cm}
  \caption{Left: All-flavor 90\% CL neutrino diffuse flux limits for POEMMA30 and POEMMA360 ($12 \times $POEMMA30) configurations compared to several measured (IceCube, Auger, ANITA) experimental limits and simulated limits for future experiments (GRAND10k, ARA-37). The theoretical neutrino flux predictions are for from newborn pulsars \cite{PhysRevD.90.103005}, active galactic nuclei (AGNs) \cite{2017nacs.book...15M}, clusters of galaxies with central sources \cite{Murase2008,2018NatPh..14..396F}, prompt and late flare emission from gamma-ray bursts (GRBs) \cite{PhysRevD.76.123001}, and from UHECR photodisintegration in the source itself (UFA) \cite{PhysRevD.92.123001}.
    DOI: \href{https://doi.org/110.1103/PhysRevD.100.063010}{10.1103/PhysRevD.100.063010}
  Right: All-flavor 90\% CL neutrino diffuse flux limits based on simulated response for Trinity, BEACON \cite{2020JCAP...11..065W}, GRAND200k \cite{2020SCPMA..6319501A}, and IceCube Gen2 with the radio array extension \cite{2021JPhG...48f0501A}. From Ref.~\citenum{Wang:2021/M}. 
  DOI: \href{https://doi.org/10.22323/1.395.1234}{10.22323/1.395.1234}
  }
\vspace{-3 mm}
  \label{DiffuseCOmp}
\end{figure}

\section{Summary}

Space-based cosmic neutrino detection offers a path to achieve unique sensitivity by using the Earth as a gargantuan neutrino target and converter and the atmosphere as an immense signal generator using the emitted optical and radio radiation from the generated extensive air showers.  While space-based neutrino missions are challenging, the benefits include full-sky coverage for astrophysical neutrino sources. In particular, the space-based cosmic neutrino sensitivity obtained using meter-scale area optics, based on the POEMMA studies, to measure the optical EeEAS Cherenkov signal lends itself for using a single SmallSat, constellation of SmallSats, or an International Space Station implementation that requires resources much smaller than a NASA Astrophysics probe-class mission and with only a modest degradation in performance. Furthermore, simulation development \cite{2019ICRC...36..936K,2021ICRC...37..1205K} is currently incorporating the optical Cherenkov and radio technique into a single framework to quantify the improvement in cosmic neutrino sensitivity if both techniques are used in tandem on the same experiment. The astrophysics that these measurements probe are highlighted in terms of time-domain multimessenger astrophysics, e.g. transient sources, but other fundamental physics includes unique sensitivity to detection of super-heavy dark matter (SHDM) decays or annihilation \cite{2021PhRvD.104h3002G}. The field is progressing in different areas of extraterrestrial neutrino detection: one technique proposes to use ground-based radar detection of neutrino-induced cascades in ice \cite{2021arXiv210400459P}, while another NASA Innovation Advanced Concept (NIAC) study investigated flying a neutrino detector close to the Sun to measure solar neutrinos in MeV energy decade as well as searching for gravitational-lensed neutrinos from the Galactic core region \cite{2019AAS...23412604S}. Thus, cosmic neutrino detection appears to have a bright and promising future as evidenced by the development and supporting research performed to define the future cosmic neutrino measurements, including the NASA Astrophysics Probe-class POEMMA mission and potentially smaller class missions.

\section{Acknowledgements}

The research presented in this chapter is meant to demonstrate the richness of the science and experimental techniques that underlie space-based and related cosmic neutrino measurements.  Over the past decades, many individual researchers and groups have made significant contributions to this field, and it is hoped that the breath of this effort is reflected in the materials presented in this chapter on space-based cosmic neutrino detection.

\newpage 

\bibliographystyle{ws-rv-van}
\bibliography{SpaceNeuBib-updated}

@INBOOK{1996SPIE.2806..102T,
       author = {{Takahashi}, Yoshiyuki and {Chipman}, Russell A. and {Dimmock}, John O. and
         {Hillmann}, Lloyd W. and {Lamb}, David J. and {Leslie}, Thomas M. and
         {Weimer}, Jeffrey J. and {Christl}, Mark J. and {Fishman}, Gerald J. and
         {Parnell}, Thomas A. and {Barbier}, Louis M. and {Boyce}, Kevin and
         {Christian}, Eric R. and {Krizmanic}, John F. and {Mitchell}, John W. and
         {Ormes}, Jonathan F. and {Stecker}, Floyd W. and {Stilwell}, Donald E. and
         {Streitmatter}, Robert E. and {Loh}, Eugene C. and {Sokolski}, Pierre and
         {Sommers}, Paul and {Cherry}, Michael L. and {Linsley}, John and
         {Scarsi}, Livio},
        title = "{Maximum-energy Auger (AIR)-shower satellite (MASS/AIRWATCH)}",
    booktitle = {Proceedings of the SPIE},
         year = 1996,
       editor = {{Ramsey}, Brian D. and {Parnell}, Thomas A.},
       volume = {2806},
       series = {Society of Photo-Optical Instrumentation Engineers (SPIE) Conference Series},
        pages = {102-112},
          doi = {10.1117/12.253970},
       adsurl = {https://ui.adsabs.harvard.edu/abs/1996SPIE.2806..102T}
}

@ARTICLE{2020PTEP.2020h3C01P,
       author = {{Particle Data Group} and {Zyla}, P.~A. and {Barnett}, R.~M. and {Beringer}, J. and {Dahl}, O. and {Dwyer}, D.~A. and {Groom}, D.~E. and {Lin}, C. -J. and {Lugovsky}, K.~S. and {Pianori}, E. and {Robinson}, D.~J. and {Wohl}, C.~G. and {Yao}, W. -M. and {Agashe}, K. and {Aielli}, G. and {Allanach}, B.~C. and {Amsler}, C. and {Antonelli}, M. and {Aschenauer}, E.~C. and {Asner}, D.~M. and {Baer}, H. and {Banerjee}, Sw and {Baudis}, L. and {Bauer}, C.~W. and {Beatty}, J.~J. and {Belousov}, V.~I. and {Bethke}, S. and {Bettini}, A. and {Biebel}, O. and {Black}, K.~M. and {Blucher}, E. and {Buchmuller}, O. and {Burkert}, V. and {Bychkov}, M.~A. and {Cahn}, R.~N. and {Carena}, M. and {Ceccucci}, A. and {Cerri}, A. and {Chakraborty}, D. and {Chivukula}, R. Sekhar and {Cowan}, G. and {D'Ambrosio}, G. and {Damour}, T. and {de Florian}, D. and {de Gouv{\^e}a}, A. and {DeGrand}, T. and {de Jong}, P. and {Dissertori}, G. and {Dobrescu}, B.~A. and {D'Onofrio}, M. and {Doser}, M. and {Drees}, M. and {Dreiner}, H.~K. and {Eerola}, P. and {Egede}, U. and {Eidelman}, S. and {Ellis}, J. and {Erler}, J. and {Ezhela}, V.~V. and {Fetscher}, W. and {Fields}, B.~D. and {Foster}, B. and {Freitas}, A. and {Gallagher}, H. and {Garren}, L. and {Gerber}, H. -J. and {Gerbier}, G. and {Gershon}, T. and {Gershtein}, Y. and {Gherghetta}, T. and {Godizov}, A.~A. and {Gonzalez-Garcia}, M.~C. and {Goodman}, M. and {Grab}, C. and {Gritsan}, A.~V. and {Grojean}, C. and {Gr{\"u}newald}, M. and {Gurtu}, A. and {Gutsche}, T. and {Haber}, H.~E. and {Hanhart}, C. and {Hashimoto}, S. and {Hayato}, Y. and {Hebecker}, A. and {Heinemeyer}, S. and {Heltsley}, B. and {Hern{\'a}ndez-Rey}, J.~J. and {Hikasa}, K. and {Hisano}, J. and {H{\"o}cker}, A. and {Holder}, J. and {Holtkamp}, A. and {Huston}, J. and {Hyodo}, T. and {Johnson}, K.~F. and {Kado}, M. and {Karliner}, M. and {Katz}, U.~F. and {Kenzie}, M. and {Khoze}, V.~A. and {Klein}, S.~R. and {Klempt}, E. and {Kowalewski}, R.~V. and {Krauss}, F. and {Kreps}, M. and {Krusche}, B. and {Kwon}, Y. and {Lahav}, O. and {Laiho}, J. and {Lellouch}, L.~P. and {Lesgourgues}, J. and {Liddle}, A.~R. and {Ligeti}, Z. and {Lippmann}, C. and {Liss}, T.~M. and {Littenberg}, L. and {Lourengo}, C. and {Lugovsky}, S.~B. and {Lusiani}, A. and {Makida}, Y. and {Maltoni}, F. and {Mannel}, T. and {Manohar}, A.~V. and {Marciano}, W.~J. and {Masoni}, A. and {Matthews}, J. and {Mei{\ss}ner}, U. -G. and {Mikhasenko}, M. and {Miller}, D.~J. and {Milstead}, D. and {Mitchell}, R.~E. and {M{\"o}nig}, K. and {Molaro}, P. and {Moortgat}, F. and {Moskovic}, M. and {Nakamura}, K. and {Narain}, M. and {Nason}, P. and {Navas}, S. and {Neubert}, M. and {Nevski}, P. and {Nir}, Y. and {Olive}, K.~A. and {Patrignani}, C. and {Peacock}, J.~A. and {Petcov}, S.~T. and {Petrov}, V.~A. and {Pich}, A. and {Piepke}, A. and {Pomarol}, A. and {Profumo}, S. and {Quadt}, A. and {Rabbertz}, K. and {Rademacker}, J. and {Raffelt}, G. and {Ramani}, H. and {Ramsey-Musolf}, M. and {Ratcliff}, B.~N. and {Richardson}, P. and {Ringwald}, A. and {Roesler}, S. and {Rolli}, S. and {Romaniouk}, A. and {Rosenberg}, L.~J. and {Rosner}, J.~L. and {Rybka}, G. and {Ryskin}, M. and {Ryutin}, R.~A. and {Sakai}, Y. and {Salam}, G.~P. and {Sarkar}, S. and {Sauli}, F. and {Schneider}, O. and {Scholberg}, K. and {Schwartz}, A.~J. and {Schwiening}, J. and {Scott}, D. and {Sharma}, V. and {Sharpe}, S.~R. and {Shutt}, T. and {Silari}, M. and {Sj{\"o}strand}, T. and {Skands}, P. and {Skwarnicki}, T. and {Smoot}, G.~F. and {Soffer}, A. and {Sozzi}, M.~S. and {Spanier}, S. and {Spiering}, C. and {Stahl}, A. and {Stone}, S.~L. and {Sumino}, Y. and {Sumiyoshi}, T. and {Syphers}, M.~J. and {Takahashi}, F. and {Tanabashi}, M. and {Tanaka}, J. and {Ta{\v{s}}evsk{\'y}}, M. and {Terashi}, K. and {Terning}, J. and {Thoma}, U. and {Thorne}, R.~S. and {Tiator}, L. and {Titov}, M. and {Tkachenko}, N.~P. and {Tovey}, D.~R. and {Trabelsi}, K. and {Urquijo}, P. and {Valencia}, G. and {Van de Water}, R. and {Varelas}, N. and {Venanzoni}, G. and {Verde}, L. and {Vincter}, M.~G. and {Vogel}, P. and {Vogelsang}, W. and {Vogt}, A. and {Vorobyev}, V. and {Wakely}, S.~P. and {Walkowiak}, W. and {Walter}, C.~W. and {Wands}, D. and {Wascko}, M.~O. and {Weinberg}, D.~H. and {Weinberg}, E.~J. and {White}, M. and {Wiencke}, L.~R. and {Willocq}, S. and {Woody}, C.~L. and {Workman}, R.~L. and {Yokoyama}, M. and {Yoshida}, R. and {Zanderighi}, G. and {Zeller}, G.~P. and {Zenin}, O.~V. and {Zhu}, R. -Y. and {Zhu}, S. -L. and {Zimmermann}, F. and {Anderson}, J. and {Basaglia}, T. and {Lugovsky}, V.~S. and {Schaffner}, P. and {Zheng}, W.},
        title = "{Review of Particle Physics}",
      journal = {Progress of Theoretical and Experimental Physics},
         year = 2020,
        month = aug,
       volume = {2020},
       number = {8},
          eid = {083C01},
        pages = {083C01},
          doi = {10.1093/ptep/ptaa104},
       adsurl = {https://ui.adsabs.harvard.edu/abs/2020PTEP.2020h3C01P}
}

@ARTICLE{Gorham2018,
       author = {{Gorham}, P.~W. and {Rotter}, B. and {Allison}, P. and {Banerjee}, O. and {Batten}, L. and {Beatty}, J.~J. and {Bechtol}, K. and {Belov}, K. and {Besson}, D.~Z. and {Binns}, W.~R. and {Bugaev}, V. and {Cao}, P. and {Chen}, C.~C. and {Chen}, C.~H. and {Chen}, P. and {Clem}, J.~M. and {Connolly}, A. and {Cremonesi}, L. and {Dailey}, B. and {Deaconu}, C. and {Dowkontt}, P.~F. and {Fox}, B.~D. and {Gordon}, J.~W.~H. and {Hast}, C. and {Hill}, B. and {Hughes}, K. and {Huang}, J.~J. and {Hupe}, R. and {Israel}, M.~H. and {Javaid}, A. and {Lam}, J. and {Liewer}, K.~M. and {Lin}, S.~Y. and {Liu}, T.~C. and {Ludwig}, A. and {Macchiarulo}, L. and {Matsuno}, S. and {Miki}, C. and {Mulrey}, K. and {Nam}, J. and {Naudet}, C.~J. and {Nichol}, R.~J. and {Novikov}, A. and {Oberla}, E. and {Olmedo}, M. and {Prechelt}, R. and {Prohira}, S. and {Rauch}, B.~F. and {Roberts}, J.~M. and {Romero-Wolf}, A. and {Russell}, J.~W. and {Saltzberg}, D. and {Seckel}, D. and {Schoorlemmer}, H. and {Shiao}, J. and {Stafford}, S. and {Stockham}, J. and {Stockham}, M. and {Strutt}, B. and {Varner}, G.~S. and {Vieregg}, A.~G. and {Wang}, S.~H. and {Wissel}, S.~A.},
        title = "{Observation of an Unusual Upward-Going Cosmic-Ray-like Event in the Third Flight of ANITA}",        journal = {Phys. Rev. Lett.},
         year = 2018,
        month = oct,
       volume = {121},
       number = {16},
          eid = {161102},
        pages = {161102},
          doi = {10.1103/PhysRevLett.121.161102},
archivePrefix = {arXiv},
       eprint = {1803.05088},
 primaryClass = {astro-ph.HE},
       adsurl = {https://ui.adsabs.harvard.edu/abs/2018PhRvL.121p1102G}
}

@ARTICLE{SterileNu,
   author = {{Cherry}, J.~F. and {Shoemaker}, I.},
    title = "{A Sterile Neutrino Origin for the Upward Directed Cosmic Ray Shower Detected by ANITA}",
  journal = {ArXiv e-prints},
archivePrefix = "arXiv",
   eprint = {1802.01611},
 primaryClass = "hep-ph",
     year = 2018,
    month = feb,
   adsurl = {http://adsabs.harvard.edu/abs/2018arXiv180201611C}
}

@ARTICLE{ANITA2018,
       author = {{Gorham}, P.~W. and {Allison}, P. and {Banerjee}, O. and {Batten}, L. and {Beatty}, J.~J. and {Bechtol}, K. and {Belov}, K. and {Besson}, D.~Z. and {Binns}, W.~R. and {Bugaev}, V. and {Cao}, P. and {Chen}, C.~C. and {Chen}, C.~H. and {Chen}, P. and {Clem}, J.~M. and {Connolly}, A. and {Cremonesi}, L. and {Dailey}, B. and {Deaconu}, C. and {Dowkontt}, P.~F. and {Fox}, B.~D. and {Gordon}, J.~W.~H. and {Hast}, C. and {Hill}, B. and {Hsu}, S.~Y. and {Huang}, J.~J. and {Hughes}, K. and {Hupe}, R. and {Israel}, M.~H. and {Liewer}, K.~M. and {Liu}, T.~C. and {Ludwig}, A.~B. and {Macchiarulo}, L. and {Matsuno}, S. and {Miki}, C. and {Mulrey}, K. and {Nam}, J. and {Naudet}, C. and {Nichol}, R.~J. and {Novikov}, A. and {Oberla}, E. and {Prohira}, S. and {Rauch}, B.~F. and {Roberts}, J.~M. and {Romero-Wolf}, A. and {Rotter}, B. and {Russell}, J.~W. and {Saltzberg}, D. and {Seckel}, D. and {Schoorlemmer}, H. and {Shiao}, J. and {Stafford}, S. and {Stockham}, J. and {Stockham}, M. and {Strutt}, B. and {Sutherland}, M.~S. and {Varner}, G.~S. and {Vieregg}, A.~G. and {Wang}, S.~H. and {Wissel}, S.~A. and {Anita Collaboration}},
        title = "{Constraints on the diffuse high-energy neutrino flux from the third flight of ANITA}",
      journal = {Phys. Rev. D},
         year = 2018,
        month = jul,
       volume = {98},
       number = {2},
          eid = {022001},
        pages = {022001},
          doi = {10.1103/PhysRevD.98.022001},
archivePrefix = {arXiv},
       eprint = {1803.02719},
 primaryClass = {astro-ph.HE},
       adsurl = {https://ui.adsabs.harvard.edu/abs/2018PhRvD..98b2001G}
}

@ARTICLE{PattersonHillas,
   author = {{Patterson}, J.~R. and {Hillas}, A.~M.},
    title = "{The relation of Cerenkov time profile widths to the distance to maximum of air showers}",
  journal = {Journal of Physics G Nuclear Physics},
     year = 1983,
    month = mar,
   volume = 9,
    pages = {323-337},
      doi = {10.1088/0305-4616/9/3/013},
   adsurl = {http://adsabs.harvard.edu/abs/1983JPhG....9..323P}
}

@ARTICLE{Krizmanic2003,
   author = {{Krizmanic}, J.~F. and {Sokolsky}, P. and {Streitmatter}, R.
	},
    title = "{Limitations on Space-Based Air Fluorescence Detector Apertures Obtained from IR Cloud Measurements}",
  journal = {Proceedings of the 28th ICRC (Tsukuba)},
   eprint = {astro-ph/0305404},
     year = 2003,
    month = jul,
   volume = 2,
    pages = {639},
   adsurl = {http://adsabs.harvard.edu/abs/2003ICRC....2..639K}
}

@ARTICLE{Vankov,
   author = {{Stanev}, T. and {Vankov}, H.~P.},
    title = "{Air-shower detection of ultrahigh-energy muons and neutrinos}",
      journal        = "Phys. Rev.",
      volume         = "D40",
     year = 1989,
    month = sep,
    pages = {1472-1476},
      doi = {10.1103/PhysRevD.40.1472},
   adsurl = {http://adsabs.harvard.edu/abs/1989PhRvD..40.1472S}
}

@MISC{CORSIKA,
   author = {{Heck}, D. and {Pierog}, T. and {Knapp}, J.},
    title = "{CORSIKA: An Air Shower Simulation Program}",
    howpublished = {Astrophysics Source Code Library},
     year = 2012,
    month = feb,
   archivePrefix = "ascl",
   eprint = {1202.006},
   adsurl = {http://adsabs.harvard.edu/abs/2012ascl.soft02006H}
}

@Book{GaisserBook,
  author =	 {Thomas K. Gaisser},
  title = 	 {Cosmic Rays and Particle Physics},
  publisher = 	 {Cambridge University Press},
  year = 	 1990
}

@ARTICLE{IceCube2015,
       author = {{Aartsen}, M.~G. and {Abraham}, K. and {Ackermann}, M. and {Adams}, J. and {Aguilar}, J.~A. and {Ahlers}, M. and {Ahrens}, M. and {Altmann}, D. and {Anderson}, T. and {Archinger}, M. and {Arguelles}, C. and {Arlen}, T.~C. and {Auffenberg}, J. and {Bai}, X. and {Barwick}, S.~W. and {Baum}, V. and {Bay}, R. and {Beatty}, J.~J. and {Becker Tjus}, J. and {Becker}, K. -H. and {Beiser}, E. and {BenZvi}, S. and {Berghaus}, P. and {Berley}, D. and {Bernardini}, E. and {Bernhard}, A. and {Besson}, D.~Z. and {Binder}, G. and {Bindig}, D. and {Bissok}, M. and {Blaufuss}, E. and {Blumenthal}, J. and {Boersma}, D.~J. and {Bohm}, C. and {B{\"o}rner}, M. and {Bos}, F. and {Bose}, D. and {B{\"o}ser}, S. and {Botner}, O. and {Braun}, J. and {Brayeur}, L. and {Bretz}, H. -P. and {Brown}, A.~M. and {Buzinsky}, N. and {Casey}, J. and {Casier}, M. and {Cheung}, E. and {Chirkin}, D. and {Christov}, A. and {Christy}, B. and {Clark}, K. and {Classen}, L. and {Coenders}, S. and {Cowen}, D.~F. and {Cruz Silva}, A.~H. and {Daughhetee}, J. and {Davis}, J.~C. and {Day}, M. and {de Andr{\'e}}, J.~P.~A.~M. and {De Clercq}, C. and {Dembinski}, H. and {De Ridder}, S. and {Desiati}, P. and {de Vries}, K.~D. and {de Wasseige}, G. and {de With}, M. and {DeYoung}, T. and {D{\'\i}az-V{\'e}lez}, J.~C. and {Dumm}, J.~P. and {Dunkman}, M. and {Eagan}, R. and {Eberhardt}, B. and {Ehrhardt}, T. and {Eichmann}, B. and {Euler}, S. and {Evenson}, P.~A. and {Fadiran}, O. and {Fahey}, S. and {Fazely}, A.~R. and {Fedynitch}, A. and {Feintzeig}, J. and {Felde}, J. and {Filimonov}, K. and {Finley}, C. and {Fischer-Wasels}, T. and {Flis}, S. and {Fuchs}, T. and {Gaisser}, T.~K. and {Gaior}, R. and {Gallagher}, J. and {Gerhardt}, L. and {Ghorbani}, K. and {Gier}, D. and {Gladstone}, L. and {Glagla}, M. and {Gl{\"u}senkamp}, T. and {Goldschmidt}, A. and {Golup}, G. and {Gonzalez}, J.~G. and {Goodman}, J.~A. and {G{\'o}ra}, D. and {Grant}, D. and {Gretskov}, P. and {Groh}, J.~C. and {Gross}, A. and {Ha}, C. and {Haack}, C. and {Haj Ismail}, A. and {Hallgren}, A. and {Halzen}, F. and {Hansmann}, B. and {Hanson}, K. and {Hebecker}, D. and {Heereman}, D. and {Helbing}, K. and {Hellauer}, R. and {Hellwig}, D. and {Hickford}, S. and {Hignight}, J. and {Hill}, G.~C. and {Hoffman}, K.~D. and {Hoffmann}, R. and {Holzapfel}, K. and {Homeier}, A. and {Hoshina}, K. and {Huang}, F. and {Huber}, M. and {Huelsnitz}, W. and {Hulth}, P.~O. and {Hultqvist}, K. and {In}, S. and {Ishihara}, A. and {Jacobi}, E. and {Japaridze}, G.~S. and {Jero}, K. and {Jurkovic}, M. and {Kaminsky}, B. and {Kappes}, A. and {Karg}, T. and {Karle}, A. and {Kauer}, M. and {Keivani}, A. and {Kelley}, J.~L. and {Kemp}, J. and {Kheirandish}, A. and {Kiryluk}, J. and {Kl{\"a}s}, J. and {Klein}, S.~R. and {Kohnen}, G. and {Kolanoski}, H. and {Konietz}, R. and {Koob}, A. and {K{\"o}pke}, L. and {Kopper}, C. and {Kopper}, S. and {Koskinen}, D.~J. and {Kowalski}, M. and {Krings}, K. and {Kroll}, G. and {Kroll}, M. and {Kunnen}, J. and {Kurahashi}, N. and {Kuwabara}, T. and {Labare}, M. and {Lanfranchi}, J.~L. and {Larson}, M.~J. and {Lesiak-Bzdak}, M. and {Leuermann}, M. and {Leuner}, J. and {L{\"u}nemann}, J. and {Madsen}, J. and {Maggi}, G. and {Mahn}, K.~B.~M. and {Maruyama}, R. and {Mase}, K. and {Matis}, H.~S. and {Maunu}, R. and {McNally}, F. and {Meagher}, K. and {Medici}, M. and {Meli}, A. and {Menne}, T. and {Merino}, G. and {Meures}, T. and {Miarecki}, S. and {Middell}, E. and {Middlemas}, E. and {Miller}, J. and {Mohrmann}, L. and {Montaruli}, T. and {Morse}, R. and {Nahnhauer}, R. and {Naumann}, U. and {Niederhausen}, H. and {Nowicki}, S.~C. and {Nygren}, D.~R. and {Obertacke}, A. and {Olivas}, A. and {Omairat}, A. and {O'Murchadha}, A. and {Palczewski}, T. and {Paul}, L. and {Pepper}, J.~A. and {P{\'e}rez de los Heros}, C. and {Pfendner}, C. and {Pieloth}, D. and {Pinat}, E. and {Posselt}, J. and {Price}, P.~B. and {Przybylski}, G.~T. and {P{\"u}tz}, J. and {Quinnan}, M. and {R{\"a}del}, L. and {Rameez}, M. and {Rawlins}, K. and {Redl}, P. and {Reimann}, R. and {Relich}, M. and {Resconi}, E. and {Rhode}, W. and {Richman}, M. and {Richter}, S. and {Riedel}, B. and {Robertson}, S. and {Rongen}, M. and {Rott}, C. and {Ruhe}, T. and {Ruzybayev}, B. and {Ryckbosch}, D. and {Saba}, S.~M. and {Sabbatini}, L. and {Sander}, H. -G. and {Sandrock}, A. and {Sandroos}, J. and {Sarkar}, S. and {Schatto}, K. and {Scheriau}, F. and {Schimp}, M. and {Schmidt}, T. and {Schmitz}, M. and {Schoenen}, S. and {Sch{\"o}neberg}, S. and {Sch{\"o}nwald}, A. and {Schukraft}, A. and {Schulte}, L. and {Seckel}, D. and {Seunarine}, S. and {Shanidze}, R. and {Smith}, M.~W.~E. and {Soldin}, D. and {Spiczak}, G.~M. and {Spiering}, C. and {Stahlberg}, M. and {Stamatikos}, M. and {Stanev}, T. and {Stanisha}, N.~A. and {Stasik}, A. and {Stezelberger}, T. and {Stokstad}, R.~G. and {St{\"o}ssl}, A. and {Strahler}, E.~A. and {Str{\"o}m}, R. and {Strotjohann}, N.~L. and {Sullivan}, G.~W. and {Sutherland}, M. and {Taavola}, H. and {Taboada}, I. and {Ter-Antonyan}, S. and {Terliuk}, A. and {Te{\v{s}}i{\'c}}, G. and {Tilav}, S. and {Toale}, P.~A. and {Tobin}, M.~N. and {Tosi}, D. and {Tselengidou}, M. and {Unger}, E. and {Usner}, M. and {Vallecorsa}, S. and {Vandenbroucke}, J. and {van Eijndhoven}, N. and {Vanheule}, S. and {van Santen}, J. and {Veenkamp}, J. and {Vehring}, M. and {Voge}, M. and {Vraeghe}, M. and {Walck}, C. and {Wallace}, A. and {Wallraff}, M. and {Wandkowsky}, N. and {Weaver}, Ch. and {Wendt}, C. and {Westerhoff}, S. and {Whelan}, B.~J. and {Whitehorn}, N. and {Wichary}, C. and {Wiebe}, K. and {Wiebusch}, C.~H. and {Wille}, L. and {Williams}, D.~R. and {Wissing}, H. and {Wolf}, M. and {Wood}, T.~R. and {Woschnagg}, K. and {Xu}, D.~L. and {Xu}, X.~W. and {Xu}, Y. and {Yanez}, J.~P. and {Yodh}, G. and {Yoshida}, S. and {Zarzhitsky}, P. and {Zoll}, M. and {IceCube Collaboration}},
        title = "{A Combined Maximum-likelihood Analysis of the High-energy Astrophysical Neutrino Flux Measured with IceCube}",
      journal = {The Astrophysics Journal},
         year = 2015,
        month = aug,
       volume = {809},
       number = {1},
          eid = {98},
        pages = {98},
          doi = {10.1088/0004-637X/809/1/98},
archivePrefix = {arXiv},
       eprint = {1507.03991},
 primaryClass = {astro-ph.HE},
       adsurl = {https://ui.adsabs.harvard.edu/abs/2015ApJ...809...98A}
}

@article{IceCube2017,
      author         = "Ahlers, Markus and Halzen, Francis",
      title          = "{IceCube: Neutrinos and Multimessenger Astronomy}",
      journal        = "PTEP",
      volume         = "2017",
      year           = "2017",
      number         = "12",
      pages          = "12A105",
      doi            = "10.1093/ptep/ptx021",
      SLACcitation   = "%%CITATION = INSPIRE-1648668;%%"
}

@INPROCEEDINGS{Domokos1,
   author = {{Domokos}, G. and {Kovesi-Domokos}, S.},
    title = "{Observation of UHE neutrino interactions from outer space}",
booktitle = {Workshop on Observing Giant Cosmic Ray Air Showers From >10(20) eV Particles From Space},
     year = 1998,
   series = {American Institute of Physics Conference Series},
   volume = 433,
   eprint = {hep-ph/9801362},
   editor = {{Krizmanic}, J.~F. and {Ormes}, J.~F. and {Streitmatter}, R.~E.
	},
    month = jun,
    pages = {390-393},
      doi = {10.1063/1.56127},
   adsurl = {http://adsabs.harvard.edu/abs/1998AIPC..433..390D}
}

@ARTICLE{Domokos2,
   author = {{Domokos}, G. and {Kovesi-Domokos}, S.},
    title = "{Observation of Ultrahigh Energy Neutrino Interactions by Orbiting Detectors}",
  journal = {ArXiv High Energy Physics - Phenomenology e-prints},
   eprint = {hep-ph/9805221},
     year = 1998,
    month = may,
   adsurl = {http://adsabs.harvard.edu/abs/1998hep.ph....5221D}
}

@ARTICLE{OWL,
       author = {{Stecker}, F.~W. and {Krizmanic}, J.~F. and {Barbier}, L.~M. and {Loh}, E. and {Mitchell}, J.~W. and {Sokolsky}, P. and {Streitmatter}, R.~E.},
    title = "{Observing the Ultrahigh Energy Universe with OWL Eyes}",
  journal = {Nuclear Physics B Proceedings Supplements},
   eprint = {astro-ph/0408162},
     year = 2004,
    month = nov,
   volume = 136,
    pages = {433-438},
      doi = {10.1016/j.nuclphysbps.2004.10.027},
   adsurl = {http://adsabs.harvard.edu/abs/2004NuPhS.136..433S}
}

@ARTICLE{Halzen:1998be,
   author = {{Halzen}, F. and {Saltzberg}, D.},
    title = "{Tau Neutrino Appearance with a 1000 Megaparsec Baseline}",
  journal = {Physical Review Letters},
   eprint = {hep-ph/9804354},
     year = 1998,
    month = nov,
   volume = 81,
    pages = {4305-4308},
      doi = {10.1103/PhysRevLett.81.4305},
   adsurl = {http://adsabs.harvard.edu/abs/1998PhRvL..81.4305H}
}

@ARTICLE{Feng:2001ue,
   author = {{Feng}, J.~L. and {Fisher}, P. and {Wilczek}, F. and {Yu}, T.~M.
	},
    title = "{Observability of Earth-Skimming Ultrahigh Energy Neutrinos}",
  journal = {Physical Review Letters},
   eprint = {hep-ph/0105067},
     year = 2002,
    month = apr,
   volume = 88,
   number = 16,
      eid = {161102},
    pages = {161102},
      doi = {10.1103/PhysRevLett.88.161102},
   adsurl = {http://adsabs.harvard.edu/abs/2002PhRvL..88p1102F}
}

@INPROCEEDINGS{Gaisser-Hillas-1977-ICRC-15-8-353,
   author = {{Gaisser}, T.~K. and {Hillas}, A.~M.},
   title = "{ Reliability of the method of constant intensity cuts for reconstructing the average development of vertical showers}",
   booktitle = {Proceedings of the 15th ICRC (Plovdiv)},
   year = 1977,
   volume = 8,
   pages = {353-357}
}

@ARTICLE{2020PhRvD.101b3012A,
       author = {{Anchordoqui}, Luis A. and {Bergman}, Douglas R. and {Bertaina}, Mario E. and {Fenu}, Francesco and {Krizmanic}, John F. and {Liberatore}, Alessandro and {Olinto}, Angela V. and {Reno}, Mary Hall and {Sarazin}, Fred and {Shinozaki}, Kenji and {Soriano}, Jorge F. and {Ulrich}, Ralf and {Unger}, Michael and {Venters}, Tonia M. and {Wiencke}, Lawrence},
        title = "{Performance and science reach of the Probe of Extreme Multimessenger Astrophysics for ultrahigh-energy particles}",
      journal = {Physical Review D},
         year = 2020,
        month = jan,
       volume = {101},
       number = {2},
          eid = {023012},
        pages = {023012},
          doi = {10.1103/PhysRevD.101.023012},
       adsurl = {https://ui.adsabs.harvard.edu/abs/2020PhRvD.101b3012A}
}

@ARTICLE{2018AdSpR..62.2902A,
       author = {{Aartsen}, M.~G. and {Ackermann}, M. and {Adams}, J. and {Aguilar}, J.~A. and {Ahlers}, M. and {Ahrens}, M. and {Altmann}, D. and {Andeen}, K. and {Anderson}, T. and {Ansseau}, I. and {Anton}, G. and {Archinger}, M. and {Arg{\"u}elles}, C. and {Auffenberg}, J. and {Axani}, S. and {Bai}, X. and {Barwick}, S.~W. and {Baum}, V. and {Bay}, R. and {Beatty}, J.~J. and {Becker Tjus}, J. and {Becker}, K. -H. and {BenZvi}, S. and {Berley}, D. and {Bernardini}, E. and {Bernhard}, A. and {Besson}, D.~Z. and {Binder}, G. and {Bindig}, D. and {Bissok}, M. and {Blaufuss}, E. and {Blot}, S. and {Bohm}, C. and {B{\"o}rner}, M. and {Bos}, F. and {Bose}, D. and {B{\"o}ser}, S. and {Botner}, O. and {Braun}, J. and {Brayeur}, L. and {Bretz}, H. -P. and {Bron}, S. and {Burgman}, A. and {Carver}, T. and {Casier}, M. and {Cheung}, E. and {Chirkin}, D. and {Christov}, A. and {Clark}, K. and {Classen}, L. and {Coenders}, S. and {Collin}, G.~H. and {Conrad}, J.~M. and {Cowen}, D.~F. and {Cross}, R. and {Day}, M. and {de Andr{\'e}}, J.~P.~A.~M. and {De Clercq}, C. and {del Pino Rosendo}, E. and {Dembinski}, H. and {De Ridder}, S. and {Desiati}, P. and {de Vries}, K.~D. and {de Wasseige}, G. and {de With}, M. and {DeYoung}, T. and {D{\'\i}az-V{\'e}lez}, J.~C. and {di Lorenzo}, V. and {Dujmovic}, H. and {Dumm}, J.~P. and {Dunkman}, M. and {Eberhardt}, B. and {Ehrhardt}, T. and {Eichmann}, B. and {Eller}, P. and {Euler}, S. and {Evenson}, P.~A. and {Fahey}, S. and {Fazely}, A.~R. and {Feintzeig}, J. and {Felde}, J. and {Filimonov}, K. and {Finley}, C. and {Flis}, S. and {F{\"o}sig}, C. -C. and {Franckowiak}, A. and {Friedman}, E. and {Fuchs}, T. and {Gaisser}, T.~K. and {Gallagher}, J. and {Gerhardt}, L. and {Ghorbani}, K. and {Giang}, W. and {Gladstone}, L. and {Glauch}, T. and {Gl{\"u}senkamp}, T. and {Goldschmidt}, A. and {Gonzalez}, J.~G. and {Grant}, D. and {Griffith}, Z. and {Haack}, C. and {Hallgren}, A. and {Halzen}, F. and {Hansen}, E. and {Hansmann}, T. and {Hanson}, K. and {Hebecker}, D. and {Heereman}, D. and {Helbing}, K. and {Hellauer}, R. and {Hickford}, S. and {Hignight}, J. and {Hill}, G.~C. and {Hoffman}, K.~D. and {Hoffmann}, R. and {Holzapfel}, K. and {Hoshina}, K. and {Huang}, F. and {Huber}, M. and {Hultqvist}, K. and {In}, S. and {Ishihara}, A. and {Jacobi}, E. and {Japaridze}, G.~S. and {Jeong}, M. and {Jero}, K. and {Jones}, B.~J.~P. and {Jurkovic}, M. and {Kang}, W. and {Kappes}, A. and {Karg}, T. and {Karle}, A. and {Katz}, U. and {Kauer}, M. and {Keivani}, A. and {Kelley}, J.~L. and {Kheirandish}, A. and {Kim}, J. and {Kim}, M. and {Kintscher}, T. and {Kiryluk}, J. and {Kittler}, T. and {Klein}, S.~R. and {Kohnen}, G. and {Koirala}, R. and {Kolanoski}, H. and {Konietz}, R. and {K{\"o}pke}, L. and {Kopper}, C. and {Kopper}, S. and {Koskinen}, D.~J. and {Kowalski}, M. and {Krings}, K. and {Kroll}, M. and {Kr{\"u}ckl}, G. and {Kr{\"u}ger}, C. and {Kunnen}, J. and {Kunwar}, S. and {Kurahashi}, N. and {Kuwabara}, T. and {Labare}, M. and {Lanfranchi}, J.~L. and {Larson}, M.~J. and {Lauber}, F. and {Lennarz}, D. and {Lesiak-Bzdak}, M. and {Leuermann}, M. and {Lu}, L. and {L{\"u}nemann}, J. and {Madsen}, J. and {Maggi}, G. and {Mahn}, K.~B.~M. and {Mancina}, S. and {Mandelartz}, M. and {Maruyama}, R. and {Mase}, K. and {Maunu}, R. and {McNally}, F. and {Meagher}, K. and {Medici}, M. and {Meier}, M. and {Meli}, A. and {Menne}, T. and {Merino}, G. and {Meures}, T. and {Miarecki}, S. and {Montaruli}, T. and {Moulai}, M. and {Nahnhauer}, R. and {Naumann}, U. and {Neer}, G. and {Niederhausen}, H. and {Nowicki}, S.~C. and {Nygren}, D.~R. and {Obertacke Pollmann}, A. and {Olivas}, A. and {O'Murchadha}, A. and {Palczewski}, T. and {Pandya}, H. and {Pankova}, D.~V. and {Peiffer}, P. and {Penek}, {\"O}. and {Pepper}, J.~A. and {P{\'e}rez de los Heros}, C. and {Pieloth}, D. and {Pinat}, E. and {Price}, P.~B. and {Przybylski}, G.~T. and {Quinnan}, M. and {Raab}, C. and {R{\"a}del}, L. and {Rameez}, M. and {Rawlins}, K. and {Reimann}, R. and {Relethford}, B. and {Relich}, M. and {Resconi}, E. and {Rhode}, W. and {Richman}, M. and {Riedel}, B. and {Robertson}, S. and {Rongen}, M. and {Rott}, C. and {Ruhe}, T. and {Ryckbosch}, D. and {Rysewyk}, D. and {Sabbatini}, L. and {Sanchez Herrera}, S.~E. and {Sandrock}, A. and {Sandroos}, J. and {Sarkar}, S. and {Satalecka}, K. and {Schlunder}, P. and {Schmidt}, T. and {Schoenen}, S. and {Sch{\"o}neberg}, S. and {Schumacher}, L. and {Seckel}, D. and {Seunarine}, S. and {Soldin}, D. and {Song}, M. and {Spiczak}, G.~M. and {Spiering}, C. and {Stanev}, T. and {Stasik}, A. and {Stettner}, J. and {Steuer}, A. and {Stezelberger}, T. and {Stokstad}, R.~G. and {St{\"o}{\ss}l}, A. and {Str{\"o}m}, R. and {Strotjohann}, N.~L. and {Sullivan}, G.~W. and {Sutherland}, M. and {Taavola}, H. and {Taboada}, I. and {Tatar}, J. and {Tenholt}, F. and {Ter-Antonyan}, S. and {Terliuk}, A. and {Te{\v{s}}i{\'c}}, G. and {Tilav}, S. and {Toale}, P.~A. and {Tobin}, M.~N. and {Toscano}, S. and {Tosi}, D. and {Tselengidou}, M. and {Turcati}, A. and {Unger}, E. and {Usner}, M. and {Vandenbroucke}, J. and {van Eijndhoven}, N. and {Vanheule}, S. and {van Rossem}, M. and {van Santen}, J. and {Veenkamp}, J. and {Vehring}, M. and {Voge}, M. and {Vogel}, E. and {Vraeghe}, M. and {Walck}, C. and {Wallace}, A. and {Wallraff}, M. and {Wandkowsky}, N. and {Weaver}, Ch. and {Weiss}, M.~J. and {Wendt}, C. and {Westerhoff}, S. and {Whelan}, B.~J. and {Wickmann}, S. and {Wiebe}, K. and {Wiebusch}, C.~H. and {Wille}, L. and {Williams}, D.~R. and {Wills}, L. and {Wolf}, M. and {Wood}, T.~R. and {Woolsey}, E. and {Woschnagg}, K. and {Xu}, D.~L. and {Xu}, X.~W. and {Xu}, Y. and {Yanez}, J.~P. and {Yodh}, G. and {Yoshida}, S. and {Zoll}, M.},
        title = "{Astrophysical neutrinos and cosmic rays observed by IceCube}",
      journal = {Advances in Space Research},
         year = 2018,
        month = nov,
       volume = {62},
       number = {10},
        pages = {2902-2930},
          doi = {10.1016/j.asr.2017.05.030},
archivePrefix = {arXiv},
       eprint = {1701.03731},
 primaryClass = {astro-ph.HE},
       adsurl = {https://ui.adsabs.harvard.edu/abs/2018AdSpR..62.2902A}
}

@ARTICLE{2018Sci...361.1378I,
       author = {{IceCube Collaboration} and {Aartsen}, M.~G. and {Ackermann}, M. and {Adams}, J. and {Aguilar}, J.~A. and {Ahlers}, M. and {Ahrens}, M. and {Al Samarai}, I. and {Altmann}, D. and {Andeen}, K. and {Anderson}, T. and {Ansseau}, I. and {Anton}, G. and {Arg{\"u}elles}, C. and {Auffenberg}, J. and {Axani}, S. and {Bagherpour}, H. and {Bai}, X. and {Barron}, J.~P. and {Barwick}, S.~W. and {Baum}, V. and {Bay}, R. and {Beatty}, J.~J. and {Becker Tjus}, J. and {Becker}, K. -H. and {BenZvi}, S. and {Berley}, D. and {Bernardini}, E. and {Besson}, D.~Z. and {Binder}, G. and {Bindig}, D. and {Blaufuss}, E. and {Blot}, S. and {Bohm}, C. and {B{\"o}rner}, M. and {Bos}, F. and {B{\"o}ser}, S. and {Botner}, O. and {Bourbeau}, E. and {Bourbeau}, J. and {Bradascio}, F. and {Braun}, J. and {Brenzke}, M. and {Bretz}, H. -P. and {Bron}, S. and {Brostean-Kaiser}, J. and {Burgman}, A. and {Busse}, R.~S. and {Carver}, T. and {Cheung}, E. and {Chirkin}, D. and {Christov}, A. and {Clark}, K. and {Classen}, L. and {Coenders}, S. and {Collin}, G.~H. and {Conrad}, J.~M. and {Coppin}, P. and {Correa}, P. and {Cowen}, D.~F. and {Cross}, R. and {Dave}, P. and {Day}, M. and {de Andr{\'e}}, J.~P.~A.~M. and {De Clercq}, C. and {DeLaunay}, J.~J. and {Dembinski}, H. and {De Ridder}, S. and {Desiati}, P. and {de Vries}, K.~D. and {de Wasseige}, G. and {de With}, M. and {DeYoung}, T. and {D{\'\i}az-V{\'e}lez}, J.~C. and {di Lorenzo}, V. and {Dujmovic}, H. and {Dumm}, J.~P. and {Dunkman}, M. and {Dvorak}, E. and {Eberhardt}, B. and {Ehrhardt}, T. and {Eichmann}, B. and {Eller}, P. and {Evenson}, P.~A. and {Fahey}, S. and {Fazely}, A.~R. and {Felde}, J. and {Filimonov}, K. and {Finley}, C. and {Flis}, S. and {Franckowiak}, A. and {Friedman}, E. and {Fritz}, A. and {Gaisser}, T.~K. and {Gallagher}, J. and {Gerhardt}, L. and {Ghorbani}, K. and {Glauch}, T. and {Gl{\"u}senkamp}, T. and {Goldschmidt}, A. and {Gonzalez}, J.~G. and {Grant}, D. and {Griffith}, Z. and {Haack}, C. and {Hallgren}, A. and {Halzen}, F. and {Hanson}, K. and {Hebecker}, D. and {Heereman}, D. and {Helbing}, K. and {Hellauer}, R. and {Hickford}, S. and {Hignight}, J. and {Hill}, G.~C. and {Hoffman}, K.~D. and {Hoffmann}, R. and {Hoinka}, T. and {Hokanson-Fasig}, B. and {Hoshina}, K. and {Huang}, F. and {Huber}, M. and {Hultqvist}, K. and {H{\"u}nnefeld}, M. and {Hussain}, R. and {In}, S. and {Iovine}, N. and {Ishihara}, A. and {Jacobi}, E. and {Japaridze}, G.~S. and {Jeong}, M. and {Jero}, K. and {Jones}, B.~J.~P. and {Kalaczynski}, P. and {Kang}, W. and {Kappes}, A. and {Kappesser}, D. and {Karg}, T. and {Karle}, A. and {Katz}, U. and {Kauer}, M. and {Keivani}, A. and {Kelley}, J.~L. and {Kheirandish}, A. and {Kim}, J. and {Kim}, M. and {Kintscher}, T. and {Kiryluk}, J. and {Kittler}, T. and {Klein}, S.~R. and {Koirala}, R. and {Kolanoski}, H. and {K{\"o}pke}, L. and {Kopper}, C. and {Kopper}, S. and {Koschinsky}, J.~P. and {Koskinen}, D.~J. and {Kowalski}, M. and {Krings}, K. and {Kroll}, M. and {Kr{\"u}ckl}, G. and {Kunwar}, S. and {Kurahashi}, N. and {Kuwabara}, T. and {Kyriacou}, A. and {Labare}, M. and {Lanfranchi}, J.~L. and {Larson}, M.~J. and {Lauber}, F. and {Leonard}, K. and {Lesiak-Bzdak}, M. and {Leuermann}, M. and {Liu}, Q.~R. and {Lozano Mariscal}, C.~J. and {Lu}, L. and {L{\"u}nemann}, J. and {Luszczak}, W. and {Madsen}, J. and {Maggi}, G. and {Mahn}, K.~B.~M. and {Mancina}, S. and {Maruyama}, R. and {Mase}, K. and {Maunu}, R. and {Meagher}, K. and {Medici}, M. and {Meier}, M. and {Menne}, T. and {Merino}, G. and {Meures}, T. and {Miarecki}, S. and {Micallef}, J. and {Moment{\'e}}, G. and {Montaruli}, T. and {Moore}, R.~W. and {S} and {Morse}, R. and {Moulai}, M. and {Nahnhauer}, R. and {Nakarmi}, P. and {Naumann}, U. and {Neer}, G. and {Niederhausen}, H. and {Nowicki}, S.~C. and {Nygren}, D.~R. and {Obertacke Pollmann}, A. and {Olivas}, A. and {O'Murchadha}, A. and {O'Sullivan}, E. and {Palczewski}, T. and {Pandya}, H. and {Pankova}, D.~V. and {Peiffer}, P. and {Pepper}, J.~A. and {P{\'e}rez de los Heros}, C. and {Pieloth}, D. and {Pinat}, E. and {Plum}, M. and {Price}, P.~B. and {Przybylski}, G.~T. and {Raab}, C. and {R{\"a}del}, L. and {Rameez}, M. and {Rauch}, L. and {Rawlins}, K. and {Rea}, I.~C. and {Reimann}, R. and {Relethford}, B. and {Relich}, M. and {Resconi}, E. and {Rhode}, W. and {Richman}, M. and {Robertson}, S. and {Rongen}, M. and {Rott}, C. and {Ruhe}, T. and {Ryckbosch}, D. and {Rysewyk}, D. and {Safa}, I. and {S{\"a}lzer}, T. and {Sanchez Herrera}, S.~E. and {Sandrock}, A. and {Sandroos}, J. and {Santander}, M. and {Sarkar}, S. and {Sarkar}, S. and {Satalecka}, K. and {Schlunder}, P. and {Schmidt}, T. and {Schneider}, A. and {Schoenen}, S. and {Sch{\"o}neberg}, S. and {Schumacher}, L. and {Sclafani}, S. and {Seckel}, D. and {Seunarine}, S. and {Soedingrekso}, J. and {Soldin}, D. and {Song}, M. and {Spiczak}, G.~M. and {Spiering}, C. and {Stachurska}, J. and {Stamatikos}, M. and {Stanev}, T. and {Stasik}, A. and {Stein}, R. and {Stettner}, J. and {Steuer}, A. and {Stezelberger}, T. and {Stokstad}, R.~G. and {St{\"o}{\ss}l}, A. and {Strotjohann}, N.~L. and {Stuttard}, T. and {Sullivan}, G.~W. and {Sutherland}, M. and {Taboada}, I. and {Tatar}, J. and {Tenholt}, F. and {Ter-Antonyan}, S. and {Terliuk}, A. and {Tilav}, S. and {Toale}, P.~A. and {Tobin}, M.~N. and {Toennis}, C. and {Toscano}, S. and {Tosi}, D. and {Tselengidou}, M. and {Tung}, C.~F. and {Turcati}, A. and {Turley}, C.~F. and {Ty}, B. and {Unger}, E. and {Usner}, M. and {Vandenbroucke}, J. and {Van Driessche}, W. and {van Eijk}, D. and {van Eijndhoven}, N. and {Vanheule}, S. and {van Santen}, J. and {Vogel}, E. and {Vraeghe}, M. and {Walck}, C. and {Wallace}, A. and {Wallraff}, M. and {Wandler}, F.~D. and {Wandkowsky}, N. and {Waza}, A. and {Weaver}, C. and {Weiss}, M.~J. and {Wendt}, C. and {Werthebach}, J. and {Westerhoff}, S. and {Whelan}, B.~J. and {Whitehorn}, N. and {Wiebe}, K. and {Wiebusch}, C.~H. and {Wille}, L. and {Williams}, D.~R. and {Wills}, L. and {Wolf}, M. and {Wood}, J. and {Wood}, T.~R. and {Woschnagg}, K. and {Xu}, D.~L. and {Xu}, X.~W. and {Xu}, Y. and {Yanez}, J.~P. and {Yodh}, G. and {Yoshida}, S. and {Yuan}, T. and {Fermi-LAT Collaboration} and {Abdollahi}, S. and {Ajello}, M. and {Angioni}, R. and {Baldini}, L. and {Ballet}, J. and {Barbiellini}, G. and {Bastieri}, D. and {Bechtol}, K. and {Bellazzini}, R. and {Berenji}, B. and {Bissaldi}, E. and {Blandford}, R.~D. and {Bonino}, R. and {Bottacini}, E. and {Bregeon}, J. and {Bruel}, P. and {Buehler}, R. and {Burnett}, T.~H. and {Burns}, E. and {Buson}, S. and {Cameron}, R.~A. and {Caputo}, R. and {Caraveo}, P.~A. and {Cavazzuti}, E. and {Charles}, E. and {Chen}, S. and {Cheung}, C.~C. and {Chiang}, J. and {Chiaro}, G. and {Ciprini}, S. and {Cohen-Tanugi}, J. and {Conrad}, J. and {Costantin}, D. and {Cutini}, S. and {D'Ammando}, F. and {de Palma}, F. and {Digel}, S.~W. and {Di Lalla}, N. and {Di Mauro}, M. and {Di Venere}, L. and {Dom{\'\i}nguez}, A. and {Favuzzi}, C. and {Franckowiak}, A. and {Fukazawa}, Y. and {Funk}, S. and {Fusco}, P. and {Gargano}, F. and {Gasparrini}, D. and {Giglietto}, N. and {Giomi}, M. and {Giommi}, P. and {Giordano}, F. and {Giroletti}, M. and {Glanzman}, T. and {Green}, D. and {Grenier}, I.~A. and {Grondin}, M. -H. and {Guiriec}, S. and {Harding}, A.~K. and {Hayashida}, M. and {Hays}, E. and {Hewitt}, J.~W. and {Horan}, D. and {J{\'o}hannesson}, G. and {Kadler}, M. and {Kensei}, S. and {Kocevski}, D. and {Krauss}, F. and {Kreter}, M. and {Kuss}, M. and {La Mura}, G. and {Larsson}, S. and {Latronico}, L. and {Lemoine-Goumard}, M. and {Li}, J. and {Longo}, F. and {Loparco}, F. and {Lovellette}, M.~N. and {Lubrano}, P. and {Magill}, J.~D. and {Maldera}, S. and {Malyshev}, D. and {Manfreda}, A. and {Mazziotta}, M.~N. and {McEnery}, J.~E. and {Meyer}, M. and {Michelson}, P.~F. and {Mizuno}, T. and {Monzani}, M.~E. and {Morselli}, A. and {Moskalenko}, I.~V. and {Negro}, M. and {Nuss}, E. and {Ojha}, R. and {Omodei}, N. and {Orienti}, M. and {Orlando}, E. and {Palatiello}, M. and {Paliya}, V.~S. and {Perkins}, J.~S. and {Persic}, M. and {Pesce-Rollins}, M. and {Piron}, F. and {Porter}, T.~A. and {Principe}, G. and {Rain{\`o}}, S. and {Rando}, R. and {Rani}, B. and {Razzano}, M. and {Razzaque}, S. and {Reimer}, A. and {Reimer}, O. and {Renault-Tinacci}, N. and {Ritz}, S. and {Rochester}, L.~S. and {Saz Parkinson}, P.~M. and {Sgr{\`o}}, C. and {Siskind}, E.~J. and {Spandre}, G. and {Spinelli}, P. and {Suson}, D.~J. and {Tajima}, H. and {Takahashi}, M. and {Tanaka}, Y. and {Thayer}, J.~B. and {Thompson}, D.~J. and {Tibaldo}, L. and {Torres}, D.~F. and {Torresi}, E. and {Tosti}, G. and {Troja}, E. and {Valverde}, J. and {Vianello}, G. and {Vogel}, M. and {Wood}, K. and {Wood}, M. and {Zaharijas}, G. and {MAGIC Collaboration} and {Ahnen}, M.~L. and {Ansoldi}, S. and {Antonelli}, L.~A. and {Arcaro}, C. and {Baack}, D. and {Babi{\'c}}, A. and {Banerjee}, B. and {Bangale}, P. and {Barres de Almeida}, U. and {Barrio}, J.~A. and {Becerra Gonz{\'a}lez}, J. and {Bednarek}, W. and {Bernardini}, E. and {Berti}, A. and {Bhattacharyya}, W. and {Biland}, A. and {Blanch}, O. and {Bonnoli}, G. and {Carosi}, A. and {Carosi}, R. and {Ceribella}, G. and {Chatterjee}, A. and {Colak}, S.~M. and {Colin}, P. and {Colombo}, E. and {Contreras}, J.~L. and {Cortina}, J. and {Covino}, S. and {Cumani}, P. and {Da Vela}, P. and {Dazzi}, F. and {De Angelis}, A. and {De Lotto}, B. and {Delfino}, M. and {Delgado}, J. and {Di Pierro}, F. and {Dom{\'\i}nguez}, A. and {Dominis Prester}, D. and {Dorner}, D. and {Doro}, M. and {Einecke}, S. and {Elsaesser}, D. and {Fallah Ramazani}, V. and {Fern{\'a}ndez-Barral}, A. and {Fidalgo}, D. and {Foffano}, L. and {Pfrang}, K. and {Fonseca}, M.~V. and {Font}, L. and {Franceschini}, A. and {Fruck}, C. and {Galindo}, D. and {Gallozzi}, S. and {Garc{\'\i}a L{\'o}pez}, R.~J. and {Garczarczyk}, M. and {Gaug}, M. and {Giammaria}, P. and {Godinovi{\'c}}, N. and {Gora}, D. and {Guberman}, D. and {Hadasch}, D. and {Hahn}, A. and {Hassan}, T. and {Hayashida}, M. and {Herrera}, J. and {Hose}, J. and {Hrupec}, D. and {Inoue}, S. and {Ishio}, K. and {Konno}, Y. and {Kubo}, H. and {Kushida}, J. and {Lelas}, D. and {Lindfors}, E. and {Lombardi}, S. and {Longo}, F. and {L{\'o}pez}, M. and {Maggio}, C. and {Majumdar}, P. and {Makariev}, M. and {Maneva}, G. and {Manganaro}, M. and {Mannheim}, K. and {Maraschi}, L. and {Mariotti}, M. and {Mart{\'\i}nez}, M. and {Masuda}, S. and {Mazin}, D. and {Minev}, M. and {M}, J.~M. and {Mirzoyan}, R. and {Moralejo}, A. and {Moreno}, V. and {Moretti}, E. and {Nagayoshi}, T. and {Neustroev}, V. and {Niedzwiecki}, A. and {Nievas Rosillo}, M. and {Nigro}, C. and {Nilsson}, K. and {Ninci}, D. and {Nishijima}, K. and {Noda}, K. and {Nogu{\'e}s}, L. and {Paiano}, S. and {Palacio}, J. and {Paneque}, D. and {Paoletti}, R. and {Paredes}, J.~M. and {Pedaletti}, G. and {Peresano}, M. and {Persic}, M. and {Prada Moroni}, P.~G. and {Prandini}, E. and {Puljak}, I. and {Rodriguez Garcia}, J. and {Reichardt}, I. and {Rhode}, W. and {Rib{\'o}}, M. and {Rico}, J. and {Righi}, C. and {Rugliancich}, A. and {Saito}, T. and {Satalecka}, K. and {Schweizer}, T. and {Sitarek}, J. and {{\v{S}}nidaric {\textasciiacute}}, I. and {Sobczynska}, D. and {Stamerra}, A. and {Strzys}, M. and {Suri{\'c}}, T. and {Takahashi}, M. and {Tavecchio}, F. and {Temnikov}, P. and {Terzi{\'c}}, T. and {Teshima}, M. and {Torres-Alb{\`a}}, N. and {Treves}, A. and {Tsujimoto}, S. and {Vanzo}, G. and {Vazquez Acosta}, M. and {Vovk}, I. and {Ward}, J.~E. and {Will}, M. and {S} and {Zaric {\textasciiacute}}, D. and {AGILE Team} and {Lucarelli}, F. and {Tavani}, M. and {Piano}, G. and {Donnarumma}, I. and {Pittori}, C. and {Verrecchia}, F. and {Barbiellini}, G. and {Bulgarelli}, A. and {Caraveo}, P. and {Cattaneo}, P.~W. and {Colafrancesco}, S. and {Costa}, E. and {Di Cocco}, G. and {Ferrari}, A. and {Gianotti}, F. and {Giuliani}, A. and {Lipari}, P. and {Mereghetti}, S. and {Morselli}, A. and {Pacciani}, L. and {Paoletti}, F. and {Parmiggiani}, N. and {Pellizzoni}, A. and {Picozza}, P. and {Pilia}, M. and {Rappoldi}, A. and {Trois}, A. and {Vercellone}, S. and {Vittorini}, V. and {ASAS-SN Team} and {Stanek}, A. Franckowiak K.~Z. and {Kochanek}, C.~S. and {Beacom}, J.~F. and {Thompson}, T.~A. and {Holoien}, T.~W. -S. and {Dong}, S. and {Prieto}, J.~L. and {Shappee}, B.~J. and {Holmbo}, S. and {HAWC Collaboration} and {Abeysekara}, A.~U. and {Albert}, A. and {Alfaro}, R. and {Alvarez}, C. and {Arceo}, R. and {Arteaga-Vel{\'a}zquez}, J.~C. and {Avila Rojas}, D. and {Ayala Solares}, H.~A. and {Becerril}, A. and {Belmont-Moreno}, E. and {Bernal}, A. and {Caballero-Mora}, K.~S. and {Capistr{\'a}n}, T. and {Carrami{\~n}ana}, A. and {Casanova}, S. and {Castillo}, M. and {Cotti}, U. and {Cotzomi}, J. and {Couti{\~n}o de Le{\'o}n}, S. and {De Le{\'o}n}, C. and {De la Fuente}, E. and {Diaz Hernandez}, R. and {Dichiara}, S. and {Dingus}, B.~L. and {DuVernois}, M.~A. and {D{\'\i}az-V{\'e}lez}, J.~C. and {Ellsworth}, R.~W. and {Engel}, K. and {Fiorino}, D.~W. and {Fleischhack}, H. and {Fraija}, N. and {Garc{\'\i}a-Gonz{\'a}lez}, J.~A. and {Garfias}, F. and {Gonz{\'a}lez Mu{\~n}oz}, A. and {Gonz{\'a}lez}, M.~M. and {Goodman}, J.~A. and {Hampel-Arias}, Z. and {Harding}, J.~P. and {Hernandez}, S. and {Hona}, B. and {Hueyotl-Zahuantitla}, F. and {Hui}, C.~M. and {H{\"u}ntemeyer}, P. and {Iriarte}, A. and {Jardin-Blicq}, A. and {Joshi}, V. and {Kaufmann}, S. and {Kunde}, G.~J. and {Lara}, A. and {Lauer}, R.~J. and {Lee}, W.~H. and {Lennarz}, D. and {Le{\'o}n Vargas}, H. and {Linnemann}, J.~T. and {Longinotti}, A.~L. and {Luis-Raya}, G. and {Luna-Garc{\'\i}a}, R. and {Malone}, K. and {Marinelli}, S.~S. and {Martinez}, O. and {Martinez-Castellanos}, I. and {Mart{\'\i}nez-Castro}, J. and {Mart{\'\i}nez-Huerta}, H. and {Matthews}, J.~A. and {Miranda-Romagnoli}, P. and {Moreno}, E. and {Mostaf{\'a}}, M. and {Nayerhoda}, A. and {Nellen}, L. and {Newbold}, M. and {Nisa}, M.~U. and {Noriega-Papaqui}, R. and {Pelayo}, R. and {Pretz}, J. and {P{\'e}rez-P{\'e}rez}, E.~G. and {Ren}, Z. and {Rho}, C.~D. and {Rivi{\`e}re}, C. and {Rosa-Gonz{\'a}lez}, D. and {Rosenberg}, M. and {Ruiz-Velasco}, E. and {Ruiz-Velasco}, E. and {Salesa Greus}, F. and {Sandoval}, A. and {Schneider}, M. and {Schoorlemmer}, H. and {Sinnis}, G. and {Smith}, A.~J. and {Springer}, R.~W. and {Surajbali}, P. and {Tibolla}, O. and {Tollefson}, K. and {Torres}, I. and {Villase{\~n}or}, L. and {Weisgarber}, T. and {Werner}, F. and {Yapici}, T. and {Gaurang}, Y. and {Zepeda}, A. and {Zhou}, H. and {{\'A}lvarez}, J.~D. and {H.~E.~S.~S. Collaboration} and {Abdalla}, H. and {Ang{\"u}ner}, E.~O. and {Armand}, C. and {Backes}, M. and {Becherini}, Y. and {Berge}, D. and {B{\"o}ttcher}, M. and {Boisson}, C. and {Bolmont}, J. and {Bonnefoy}, S. and {Bordas}, P. and {Brun}, F. and {B{\"u}chele}, M. and {Bulik}, T. and {Caroff}, S. and {Carosi}, A. and {Casanova}, S. and {Cerruti}, M. and {Chakraborty}, N. and {Chandra}, S. and {Chen}, A. and {Colafrancesco}, S. and {Davids}, I.~D. and {Deil}, C. and {Devin}, J. and {Djannati-Ata{\"\i}}, A. and {Egberts}, K. and {Emery}, G. and {Eschbach}, S. and {Fiasson}, A. and {Fontaine}, G. and {Funk}, S. and {F{\"u}{\ss}ling}, M. and {Gallant}, Y.~A. and {Gat{\'e}}, F. and {Giavitto}, G. and {Glawion}, D. and {Glicenstein}, J.~F. and {Gottschall}, D. and {Grondin}, M. -H. and {Haupt}, M. and {Henri}, G. and {Hinton}, J.~A. and {Hoischen}, C. and {Holch}, T.~L. and {Huber}, D. and {Jamrozy}, M. and {Jankowsky}, D. and {Jankowsky}, F. and {Jouvin}, L. and {Jung-Richardt}, I. and {Kerszberg}, D. and {Kh{\'e}lifi}, B. and {King}, J. and {Klepser}, S. and {Kluz {\textasciiacute}niak}, W. and {Komin}, Nu. and {Kraus}, M. and {Lefaucheur}, J. and {Lemi{\`e}re}, A. and {Lemoine-Goumard}, M. and {Lenain}, J. -P. and {Leser}, E. and {Lohse}, T. and {L{\'o}pez-Coto}, R. and {Lorentz}, M. and {Lypova}, I. and {Marandon}, V. and {Guillem Mart{\'\i}-Devesa}, G. and {Maurin}, G. and {Mitchell}, A.~M.~W. and {Moderski}, R. and {Mohamed}, M. and {Mohrmann}, L. and {Moulin}, E. and {Murach}, T. and {de Naurois}, M. and {Niederwanger}, F. and {Niemiec}, J. and {Oakes}, L. and {O'Brien}, P. and {Ohm}, S. and {Ostrowski}, M. and {Oya}, I. and {Panter}, M. and {Parsons}, R.~D. and {Perennes}, C. and {Piel}, Q. and {Pita}, S. and {Poireau}, V. and {Priyana Noel}, A. and {Prokoph}, H. and {P{\"u}hlhofer}, G. and {Quirrenbach}, A. and {Raab}, S. and {Rauth}, R. and {Renaud}, M. and {Rieger}, F. and {Rinchiuso}, L. and {Romoli}, C. and {Rowell}, G. and {Rudak}, B. and {Sasaki}, D.~A. and {Sanchez}, M. and {Schlickeiser}, R. and {Sch{\"u}ssler}, F. and {Schulz}, A. and {Schwanke}, U. and {Seglar-Arroyo}, M. and {Shafi}, N. and {Simoni}, R. and {Sol}, H. and {Stegmann}, C. and {Steppa}, C. and {Tavernier}, T. and {Taylor}, A.~M. and {Tiziani}, D. and {Trichard}, C. and {Tsirou}, M. and {van Eldik}, C. and {van Rensburg}, C. and {van Soelen}, B. and {Veh}, J. and {Vincent}, P. and {Voisin}, F. and {Wagner}, S.~J. and {Wagner}, R.~M. and {Wierzcholska}, A. and {Zanin}, R. and {Zdziarski}, A.~A. and {Zech}, A. and {Ziegler}, A. and {Zorn}, J. and {{\.Z}ywucka}, N. and {INTEGRAL Team} and {Savchenko}, V. and {Ferrigno}, C. and {Bazzano}, A. and {Diehl}, R. and {Kuulkers}, E. and {Laurent}, P. and {Mereghetti}, S. and {Natalucci}, L. and {Panessa}, F. and {Rodi}, J. and {Ubertini}, P. and {Kanata}, Kiso and Subaru Observing Teams and {Morokuma}, T. and {Ohta}, K. and {Tanaka}, Y.~T. and {Mori}, H. and {Yamanaka}, M. and {Kawabata}, K.~S. and {Utsumi}, Y. and {Nakaoka}, T. and {Kawabata}, M. and {Nagashima}, H. and {Yoshida}, M. and {Matsuoka}, Y. and {Itoh}, R. and {Kapteyn Team} and {Keel}, W. and {Liverpool Telescope Team} and {Copperwheat}, C. and {Steele}, I. and {Swift/NuSTAR Team} and {Cenko}, S.~B. and {Cowen}, D.~F. and {DeLaunay}, J.~J. and {Evans}, P.~A. and {Fox}, D.~B. and {Keivani}, A. and {Kennea}, J.~A. and {Marshall}, F.~E. and {Osborne}, J.~P. and {Santander}, M. and {Tohuvavohu}, A. and {Turley}, C.~F. and {VERITAS Collaboration} and {Abeysekara}, A.~U. and {Archer}, A. and {Benbow}, W. and {Bird}, R. and {Brill}, A. and {Brose}, R. and {Buchovecky}, M. and {Buckley}, J.~H. and {Bugaev}, V. and {Christiansen}, J.~L. and {Connolly}, M.~P. and {Cui}, W. and {Daniel}, M.~K. and {Errando}, M. and {Falcone}, A. and {Feng}, Q. and {Finley}, J.~P. and {Fortson}, L. and {Furniss}, A. and {Gueta}, O. and {H{\"u}tten}, M. and {Hervet}, O. and {Hughes}, G. and {Humensky}, T.~B. and {Johnson}, C.~A. and {Kaaret}, P. and {Kar}, P. and {Kelley-Hoskins}, N. and {Kertzman}, M. and {Kieda}, D. and {Krause}, M. and {Krennrich}, F. and {Kumar}, S. and {Lang}, M.~J. and {Lin}, T.~T.~Y. and {Maier}, G. and {McArthur}, S. and {Moriarty}, P. and {Mukherjee}, R. and {Nieto}, D. and {O'Brien}, S. and {Ong}, R.~A. and {Otte}, A.~N. and {Park}, N. and {Petrashyk}, A. and {Pohl}, M. and {Popkow}, A. and {Pueschel}, S.~E. and {Quinn}, J. and {Ragan}, K. and {Reynolds}, P.~T. and {Richards}, G.~T. and {Roache}, E. and {Rulten}, C. and {Sadeh}, I. and {Santander}, M. and {Scott}, S.~S. and {Sembroski}, G.~H. and {Shahinyan}, K. and {Sushch}, I. and {Tr{\'e}panier}, S. and {Tyler}, J. and {Vassiliev}, V.~V. and {Wakely}, S.~P. and {Weinstein}, A. and {Wells}, R.~M. and {Wilcox}, P. and {Wilhelm}, A. and {Williams}, D.~A. and {Zitzer}, B. and {VLA/B Team} and {Tetarenko}, A.~J. and {Kimball}, A.~E. and {Miller-Jones}, J.~C.~A. and {Sivakoff}, G.~R.},
        title = "{Multimessenger observations of a flaring blazar coincident with high-energy neutrino IceCube-170922A}",
      journal = {Science},
         year = 2018,
        month = jul,
       volume = {361},
       number = {6398},
          eid = {eaat1378},
        pages = {eaat1378},
          doi = {10.1126/science.aat1378},
archivePrefix = {arXiv},
       eprint = {1807.08816},
 primaryClass = {astro-ph.HE},
       adsurl = {https://ui.adsabs.harvard.edu/abs/2018Sci...361.1378I}
}

@ARTICLE{2018Sci...361..147I,
       author = {{IceCube Collaboration} and {Aartsen}, M.~G. and {Ackermann}, M. and {Adams}, J. and {Aguilar}, J.~A. and {Ahlers}, M. and {Ahrens}, M. and {Samarai}, I. Al and {Altmann}, D. and {Andeen}, K. and {Anderson}, T. and {Ansseau}, I. and {Anton}, G. and {Arg{\"u}elles}, C. and {Arsioli}, B. and {Auffenberg}, J. and {Axani}, S. and {Bagherpour}, H. and {Bai}, X. and {Barron}, J.~P. and {Barwick}, S.~W. and {Baum}, V. and {Bay}, R. and {Beatty}, J.~J. and {Becker Tjus}, J. and {Becker}, K. -H. and {BenZvi}, S. and {Berley}, D. and {Bernardini}, E. and {Besson}, D.~Z. and {Binder}, G. and {Bindig}, D. and {Blaufuss}, E. and {Blot}, S. and {Bohm}, C. and {B{\"o}rner}, M. and {Bos}, F. and {B{\"o}ser}, S. and {Botner}, O. and {Bourbeau}, E. and {Bourbeau}, J. and {Bradascio}, F. and {Braun}, J. and {Brenzke}, M. and {Bretz}, H. -P. and {Bron}, S. and {Brostean-Kaiser}, J. and {Burgman}, A. and {Busse}, R.~S. and {Carver}, T. and {Cheung}, E. and {Chirkin}, D. and {Christov}, A. and {Clark}, K. and {Classen}, L. and {Coenders}, S. and {Collin}, G.~H. and {Conrad}, J.~M. and {Coppin}, P. and {Correa}, P. and {Cowen}, D.~F. and {Cross}, R. and {Dave}, P. and {Day}, M. and {de Andr{\'e}}, J.~P.~A.~M. and {De Clercq}, C. and {DeLaunay}, J.~J. and {Dembinski}, H. and {DeRidder}, S. and {Desiati}, P. and {de Vries}, K.~D. and {de Wasseige}, G. and {de With}, M. and {DeYoung}, T. and {D{\'\i}az-V{\'e}lez}, J.~C. and {di Lorenzo}, V. and {Dujmovic}, H. and {Dumm}, J.~P. and {Dunkman}, M. and {Dvorak}, E. and {Eberhardt}, B. and {Ehrhardt}, T. and {Eichmann}, B. and {Eller}, P. and {Evenson}, P.~A. and {Fahey}, S. and {Fazely}, A.~R. and {Felde}, J. and {Filimonov}, K. and {Finley}, C. and {Flis}, S. and {Franckowiak}, A. and {Friedman}, E. and {Fritz}, A. and {Gaisser}, T.~K. and {Gallagher}, J. and {Gerhardt}, L. and {Ghorbani}, K. and {Giommi}, P. and {Glauch}, T. and {Gl{\"u}senkamp}, T. and {Goldschmidt}, A. and {Gonzalez}, J.~G. and {Grant}, D. and {Griffith}, Z. and {Haack}, C. and {Hallgren}, A. and {Halzen}, F. and {Hanson}, K. and {Hebecker}, D. and {Heereman}, D. and {Helbing}, K. and {Hellauer}, R. and {Hickford}, S. and {Hignight}, J. and {Hill}, G.~C. and {Hoffman}, K.~D. and {Hoffmann}, R. and {Hoinka}, T. and {Hokanson-Fasig}, B. and {Hoshina}, K. and {Huang}, F. and {Huber}, M. and {Hultqvist}, K. and {H{\"u}nnefeld}, M. and {Hussain}, R. and {In}, S. and {Iovine}, N. and {Ishihara}, A. and {Jacobi}, E. and {Japaridze}, G.~S. and {Jeong}, M. and {Jero}, K. and {Jones}, B.~J.~P. and {Kalaczynski}, P. and {Kang}, W. and {Kappes}, A. and {Kappesser}, D. and {Karg}, T. and {Karle}, A. and {Katz}, U. and {Kauer}, M. and {Keivani}, A. and {Kelley}, J.~L. and {Kheirandish}, A. and {Kim}, J. and {Kim}, M. and {Kintscher}, T. and {Kiryluk}, J. and {Kittler}, T. and {Klein}, S.~R. and {Koirala}, R. and {Kolanoski}, H. and {K{\"o}pke}, L. and {Kopper}, C. and {Kopper}, S. and {Koschinsky}, J.~P. and {Koskinen}, D.~J. and {Kowalski}, M. and {Krammer}, B. and {Krings}, K. and {Kroll}, M. and {Kr{\"u}ckl}, G. and {Kunwar}, S. and {Kurahashi}, N. and {Kuwabara}, T. and {Kyriacou}, A. and {Labare}, M. and {Lanfranchi}, J.~L. and {Larson}, M.~J. and {Lauber}, F. and {Leonard}, K. and {Lesiak-Bzdak}, M. and {Leuermann}, M. and {Liu}, Q.~R. and {Lozano Mariscal}, C.~J. and {Lu}, L. and {L{\"u}nemann}, J. and {Luszczak}, W. and {Madsen}, J. and {Maggi}, G. and {Mahn}, K.~B.~M. and {Mancina}, S. and {Maruyama}, R. and {Mase}, K. and {Maunu}, R. and {Meagher}, K. and {Medici}, M. and {Meier}, M. and {Menne}, T. and {Merino}, G. and {Meures}, T. and {Miarecki}, S. and {Micallef}, J. and {Moment{\'e}}, G. and {Montaruli}, T. and {Moore}, R.~W. and {Morse}, R. and {Moulai}, M. and {Nahnhauer}, R. and {Nakarmi}, P. and {Naumann}, U. and {Neer}, G. and {Niederhausen}, H. and {Nowicki}, S.~C. and {Nygren}, D.~R. and {Obertacke Pollmann}, A. and {Olivas}, A. and {O'Murchadha}, A. and {O'Sullivan}, E. and {Padovani}, P. and {Palczewski}, T. and {Pandya}, H. and {Pankova}, D.~V. and {Peiffer}, P. and {Pepper}, J.~A. and {P{\'e}rez de los Heros}, C. and {Pieloth}, D. and {Pinat}, E. and {Plum}, M. and {Price}, P.~B. and {Przybylski}, G.~T. and {Raab}, C. and {R{\"a}del}, L. and {Rameez}, M. and {Rawlins}, K. and {Rea}, I.~C. and {Reimann}, R. and {Relethford}, B. and {Relich}, M. and {Resconi}, E. and {Rhode}, W. and {Richman}, M. and {Robertson}, S. and {Rongen}, M. and {Rott}, C. and {Ruhe}, T. and {Ryckbosch}, D. and {Rysewyk}, D. and {Safa}, I. and {Sahakyan}, N. and {S{\"a}lzer}, T. and {Sanchez Herrera}, S.~E. and {Sandrock}, A. and {Sandroos}, J. and {Santander}, M. and {Sarkar}, S. and {Sarkar}, S. and {Satalecka}, K. and {Schlunder}, P. and {Schmidt}, T. and {Schneider}, A. and {Schoenen}, S. and {Sch{\"o}neberg}, S. and {Schumacher}, L. and {Sclafani}, S. and {Seckel}, D. and {Seunarine}, S. and {Soedingrekso}, J. and {Soldin}, D. and {Song}, M. and {Spiczak}, G.~M. and {Spiering}, C. and {Stachurska}, J. and {Stamatikos}, M. and {Stanev}, T. and {Stasik}, A. and {Stettner}, J. and {Steuer}, A. and {Stezelberger}, T. and {Stokstad}, R.~G. and {St{\"o}{\ss}l}, A. and {Strotjohann}, N.~L. and {Stuttard}, T. and {Sullivan}, G.~W. and {Sutherland}, M. and {Taboada}, I. and {Tatar}, J. and {Tenholt}, F. and {Ter-Antonyan}, S. and {Terliuk}, A. and {Tilav}, S. and {Toale}, P.~A. and {Tobin}, M.~N. and {Toennis}, C. and {Toscano}, S. and {Tosi}, D. and {Tselengidou}, M. and {Tung}, C.~F. and {Turcati}, A. and {Turley}, C.~F. and {Ty}, B. and {Unger}, E. and {Usner}, M. and {Vandenbroucke}, J. and {Van Driessche}, W. and {van Eijk}, D. and {van Eijndhoven}, N. and {Vanheule}, S. and {van Santen}, J. and {Vogel}, E. and {Vraeghe}, M. and {Walck}, C. and {Wallace}, A. and {Wallraff}, M. and {Wandler}, F.~D. and {Wandkowsky}, N. and {Waza}, A. and {Weaver}, C. and {Weiss}, M.~J. and {Wendt}, C. and {Werthebach}, J. and {Westerhoff}, S. and {Whelan}, B.~J. and {Whitehorn}, N. and {Wiebe}, K. and {Wiebusch}, C.~H. and {Wille}, L. and {Williams}, D.~R. and {Wills}, L. and {Wolf}, M. and {Wood}, J. and {Wood}, T.~R. and {Woschnagg}, K. and {Xu}, D.~L. and {Xu}, X.~W. and {Xu}, Y. and {Yanez}, J.~P. and {Yodh}, G. and {Yoshida}, S. and {Yuan}, T.},
        title = "{Neutrino emission from the direction of the blazar TXS 0506+056 prior to the IceCube-170922A alert}",
      journal = {Science},
         year = 2018,
        month = jul,
       volume = {361},
       number = {6398},
        pages = {147-151},
          doi = {10.1126/science.aat2890},
archivePrefix = {arXiv},
       eprint = {1807.08794},
 primaryClass = {astro-ph.HE},
       adsurl = {https://ui.adsabs.harvard.edu/abs/2018Sci...361..147I}
}

@ARTICLE{2013APh....44...76A,
       author = {{Adams}, J.~H. and {Ahmad}, S. and {Albert}, J. -N. and {Allard}, D. and {Ambrosio}, M. and {Anchordoqui}, L. and {Anzalone}, A. and {Arai}, Y. and {Aramo}, C. and {Asano}, K. and {Ave}, M. and {Barrillon}, P. and {Batsch}, T. and {Bayer}, J. and {Belenguer}, T. and {Bellotti}, R. and {Berlind}, A.~A. and {Bertaina}, M. and {Biermann}, P.~L. and {Biktemerova}, S. and {Blaksley}, C. and {B{\l}e{\c{c}}ki}, J. and {Blin-Bondil}, S. and {Bl{\"u}mer}, J. and {Bobik}, P. and {Bogomilov}, M. and {Bonamente}, M. and {Briggs}, M.~S. and {Briz}, S. and {Bruno}, A. and {Cafagna}, F. and {Campana}, D. and {Capdevielle}, J. -N. and {Caruso}, R. and {Casolino}, M. and {Cassardo}, C. and {Castellini}, G. and {Catalano}, O. and {Cellino}, A. and {Chikawa}, M. and {Christl}, M.~J. and {Connaughton}, V. and {Cort{\'e}s}, J.~F. and {Crawford}, H.~J. and {Cremonini}, R. and {Csorna}, S. and {D'Olivo}, J.~C. and {Dagoret-Campagne}, S. and {de Castro}, A.~J. and {De Donato}, C. and {de la Taille}, C. and {del Peral}, L. and {Dell'Oro}, A. and {De Pascale}, M.~P. and {Di Martino}, M. and {Distratis}, G. and {Dupieux}, M. and {Ebersoldt}, A. and {Ebisuzaki}, T. and {Engel}, R. and {Falk}, S. and {Fang}, K. and {Fenu}, F. and {Fern{\'a}ndez-G{\'o}mez}, I. and {Ferrarese}, S. and {Franceschi}, A. and {Fujimoto}, J. and {Galeotti}, P. and {Garipov}, G. and {Geary}, J. and {Giaccari}, U.~G. and {Giraudo}, G. and {Gonchar}, M. and {Gonz{\'a}lez Alvarado}, C. and {Gorodetzky}, P. and {Guarino}, F. and {Guzm{\'a}n}, A. and {Hachisu}, Y. and {Harlov}, B. and {Haungs}, A. and {Hern{\'a}ndez Carretero}, J. and {Higashide}, K. and {Iguchi}, T. and {Ikeda}, H. and {Inoue}, N. and {Inoue}, S. and {Insolia}, A. and {Isgr{\`o}}, F. and {Itow}, Y. and {Joven}, E. and {Judd}, E.~G. and {Jung}, A. and {Kajino}, F. and {Kajino}, T. and {Kaneko}, I. and {Karadzhov}, Y. and {Karczmarczyk}, J. and {Katahira}, K. and {Kawai}, K. and {Kawasaki}, Y. and {Keilhauer}, B. and {Khrenov}, B.~A. and {Kim}, Jeong-Sook and {Kim}, Soon-Wook and {Kim}, Sug-Whan and {Kleifges}, M. and {Klimov}, P.~A. and {Ko}, S.~H. and {Kolev}, D. and {Kreykenbohm}, I. and {Kudela}, K. and {Kurihara}, Y. and {Kuznetsov}, E. and {La Rosa}, G. and {Lee}, J. and {Licandro}, J. and {Lim}, H. and {L{\'o}pez}, F. and {Maccarone}, M.~C. and {Mannheim}, K. and {Marcelli}, L. and {Marini}, A. and {Martin-Chassard}, G. and {Martinez}, O. and {Masciantonio}, G. and {Mase}, K. and {Matev}, R. and {Maurissen}, A. and {Medina-Tanco}, G. and {Mernik}, T. and {Miyamoto}, H. and {Miyazaki}, Y. and {Mizumoto}, Y. and {Modestino}, G. and {Monnier-Ragaigne}, D. and {Morales de los R{\'\i}os}, J.~A. and {Mot}, B. and {Murakami}, T. and {Nagano}, M. and {Nagata}, M. and {Nagataki}, S. and {Nakamura}, T. and {Nam}, J.~W. and {Nam}, S. and {Nam}, K. and {Napolitano}, T. and {Naumov}, D. and {Neronov}, A. and {Nomoto}, K. and {Ogawa}, T. and {Ohmori}, H. and {Olinto}, A.~V. and {Orlea{\'n}ski}, P. and {Osteria}, G. and {Pacheco}, N. and {Panasyuk}, M.~I. and {Parizot}, E. and {Park}, I.~H. and {Pastircak}, B. and {Patzak}, T. and {Paul}, T. and {Pennypacker}, C. and {Peter}, T. and {Picozza}, P. and {Pollini}, A. and {Prieto}, H. and {Reardon}, P. and {Reina}, M. and {Reyes}, M. and {Ricci}, M. and {Rodr{\'\i}guez}, I. and {Rodr{\'\i}guez Fr{\'\i}as}, M.~D. and {Ronga}, F. and {Rothkaehl}, H. and {Roudil}, G. and {Rusinov}, I. and {Rybczy{\'n}ski}, M. and {Sabau}, M.~D. and {S{\'a}ez Cano}, G. and {Saito}, A. and {Sakaki}, N. and {Sakata}, M. and {Salazar}, H. and {S{\'a}nchez}, S. and {Santangelo}, A. and {Santiago Cr{\'u}z}, L. and {Sanz Palomino}, M. and {Saprykin}, O. and {Sarazin}, F. and {Sato}, H. and {Sato}, M. and {Schanz}, T. and {Schieler}, H. and {Scotti}, V. and {Scuderi}, M. and {Segreto}, A. and {Selmane}, S. and {Semikoz}, D. and {Serra}, M. and {Sharakin}, S. and {Shibata}, T. and {Shimizu}, H.~M. and {Shinozaki}, K. and {Shirahama}, T. and {Siemieniec-Ozieb{\l}o}, G. and {Silva L{\'o}pez}, H.~H. and {Sledd}, J. and {S{\l}omi{\'n}ska}, K. and {Sobey}, A. and {Sugiyama}, T. and {Supanitsky}, D. and {Suzuki}, M. and {Szabelska}, B. and {Szabelski}, J. and {Tajima}, F. and {Tajima}, N. and {Tajima}, T. and {Takahashi}, Y. and {Takami}, H. and {Takeda}, M. and {Takizawa}, Y. and {Tenzer}, C. and {Tibolla}, O. and {Tkachev}, L. and {Tomida}, T. and {Tone}, N. and {Trillaud}, F. and {Tsenov}, R. and {Tsuno}, K. and {Tymieniecka}, T. and {Uchihori}, Y. and {Vaduvescu}, O. and {Vald{\'e}s-Galicia}, J.~F. and {Vallania}, P. and {Valore}, L. and {Vankova}, G. and {Vigorito}, C. and {Villase{\~n}or}, L. and {von Ballmoos}, P. and {Wada}, S. and {Watanabe}, J. and {Watanabe}, S. and {Watts}, J. and {Weber}, M. and {Weiler}, T.~J. and {Wibig}, T. and {Wiencke}, L. and {Wille}, M. and {Wilms}, J. and {W{\l}odarczyk}, Z. and {Yamamoto}, T. and {Yamamoto}, Y. and {Yang}, J. and {Yano}, H. and {Yashin}, I.~V. and {Yonetoku}, D. and {Yoshida}, K. and {Yoshida}, S. and {Young}, R. and {Zamora}, A. and {Zuccaro Marchi}, A.},
        title = "{An evaluation of the exposure in nadir observation of the JEM-EUSO mission}",
      journal = {Astroparticle Physics},
         year = 2013,
        month = apr,
       volume = {44},
        pages = {76-90},
          doi = {10.1016/j.astropartphys.2013.01.008},
archivePrefix = {arXiv},
       eprint = {1305.2478},
 primaryClass = {astro-ph.HE},
       adsurl = {https://ui.adsabs.harvard.edu/abs/2013APh....44...76A}
}

@INPROCEEDINGS{1981BAAS...13R.883L,
       author = {{Linsley}, J.},
        title = "{Detection of High Energy Cosmic Neutrinos by means of Atmosphere Fluorescence}",
    booktitle = {Bulletin of the American Astronomical Society},
         year = 1981,
       volume = {13},
        month = mar,
        pages = {883},
       adsurl = {https://ui.adsabs.harvard.edu/abs/1981BAAS...13R.883L}
}

@INPROCEEDINGS{1981ICRC....8..145B,
       author = {{Benson}, R. and {Linsley}, J.},
        title = "{Satellite Observation of Cosmic-Ray Air Showers}",
    booktitle = {International Cosmic Ray Conference},
         year = 1981,
       series = {International Cosmic Ray Conference},
       volume = {8},
        month = jan,
        pages = {145},
       adsurl = {https://ui.adsabs.harvard.edu/abs/1981ICRC....8..145B}
}

@INPROCEEDINGS{1981ICRC....7..125L,
       author = {{Linsley}, J.},
        title = "{Methods of Detecting and Identifying Cosmic-Neutrinos with Energy > 10(18) EV}",
    booktitle = {International Cosmic Ray Conference},
         year = 1981,
       series = {International Cosmic Ray Conference},
       volume = {7},
        month = jan,
        pages = {125},
       adsurl = {https://ui.adsabs.harvard.edu/abs/1981ICRC....7..125L}
}

@INPROCEEDINGS{1980BAAS...12Q.818B,
       author = {{Benson}, R. and {Linsley}, J.},
        title = "{Satellite Observation of Cosmic-Ray Air Showers}",
    booktitle = {Bulletin of the American Astronomical Society},
         year = 1980,
       volume = {12},
        month = sep,
        pages = {818},
       adsurl = {https://ui.adsabs.harvard.edu/abs/1980BAAS...12Q.818B}
}

@INPROCEEDINGS{1995ICRC....3..595T,
       author = {{Takahashi}, Y.},
        title = "{Maximum-energy Auger Air Shower Satellite (MASS) for Observing Cosmic Rays in the Energy Region 10$^{19-22}$ eV}",
    booktitle = {International Cosmic Ray Conference},
         year = 1995,
       series = {International Cosmic Ray Conference},
       volume = {3},
        month = jan,
        pages = {595},
       adsurl = {https://ui.adsabs.harvard.edu/abs/1995ICRC....3..595T}
}

@INPROCEEDINGS{1982srca.conf..169L,
       author = {{Linsley}, J.},
        title = "{Neutrino astronomy using SOCRAS, a satellite observatory for cosmic ray air showers.}",
    booktitle = {Southwest Regional Conference for Astronomy and Astrophysics, 7},
         year = 1982,
       volume = {7},
        month = jan,
        pages = {169-178},
       adsurl = {https://ui.adsabs.harvard.edu/abs/1982srca.conf..169L}
}

@ARTICLE{2017SSRv..212.1687K,
       author = {{Klimov}, P.~A. and {Panasyuk}, M.~I. and {Khrenov}, B.~A. and {Garipov}, G.~K. and {Kalmykov}, N.~N. and {Petrov}, V.~L. and {Sharakin}, S.~A. and {Shirokov}, A.~V. and {Yashin}, I.~V. and {Zotov}, M.~Y. and {Biktemerova}, S.~V. and {Grinyuk}, A.~A. and {Grebenyuk}, V.~M. and {Lavrova}, M.~V. and {Tkachev}, L.~G. and {Tkachenko}, A.~V. and {Park}, I.~H. and {Lee}, J. and {Jeong}, S. and {Martinez}, O. and {Salazar}, H. and {Ponce}, E. and {Saprykin}, O.~A. and {Botvinko}, A.~A. and {Senkovsky}, A.~N. and {Puchkov}, A.~E.},
        title = "{The TUS Detector of Extreme Energy Cosmic Rays on Board the Lomonosov Satellite}",
      journal = {Space Science Reviews},
         year = 2017,
        month = nov,
       volume = {212},
       number = {3-4},
        pages = {1687-1703},
          doi = {10.1007/s11214-017-0403-3},
archivePrefix = {arXiv},
       eprint = {1706.04976},
 primaryClass = {astro-ph.IM},
       adsurl = {https://ui.adsabs.harvard.edu/abs/2017SSRv..212.1687K}
}

@ARTICLE{2020JCAP...03..033K,
       author = {{Khrenov}, B.~A. and {Garipov}, G.~K. and {Kaznacheeva}, M.~A. and {Klimov}, P.~A. and {Panasyuk}, M.~I. and {Petrov}, V.~L. and {Sharakin}, S.~A. and {Shirokov}, A.~V. and {Yashin}, I.~V. and {Zotov}, M. Yu. and {Grinyuk}, A.~A. and {Grebenyuk}, V.~M. and {Lavrova}, M.~V. and {Tkachev}, L.~G. and {Tkachenko}, A.~V. and {Saprykin}, O.~A. and {Botvinko}, A.~A. and {Senkovsky}, A.~N. and {Puchkov}, A.~E. and {Bertaina}, M. and {Golzio}, A.},
        title = "{An extensive-air-shower-like event registered with the TUS orbital detector}",
      journal = {Journal of Cosmology and Astroparticle Physics},
         year = 2020,
        month = mar,
       volume = {2020},
       number = {3},
          eid = {033},
        pages = {033},
          doi = {10.1088/1475-7516/2020/03/033},
archivePrefix = {arXiv},
       eprint = {1907.06028},
 primaryClass = {astro-ph.HE},
       adsurl = {https://ui.adsabs.harvard.edu/abs/2020JCAP...03..033K}
}

@article{Klimov:2017dul,
    author = "Klimov, Pavel",
    collaboration = "Lomonosov-UHECR/TLE",
    title = "{Ultra-high energy cosmic ray detector TUS: preliminary results of the first year of measurements}",
    doi = "10.22323/1.301.1098",
    journal = "PoS",
    volume = "ICRC2017",
    pages = "1098",
    year = "2018"
}

@ARTICLE{2021NatAs.tmp...34S,
       author = {{Stein}, Robert and {Velzen}, Sjoert van and {Kowalski}, Marek and {Franckowiak}, Anna and {Gezari}, Suvi and {Miller-Jones}, James C.~A. and {Frederick}, Sara and {Sfaradi}, Itai and {Bietenholz}, Michael F. and {Horesh}, Assaf and {Fender}, Rob and {Garrappa}, Simone and {Ahumada}, Tom{\'a}s and {Andreoni}, Igor and {Belicki}, Justin and {Bellm}, Eric C. and {B{\"o}ttcher}, Markus and {Brinnel}, Valery and {Burruss}, Rick and {Cenko}, S. Bradley and {Coughlin}, Michael W. and {Cunningham}, Virginia and {Drake}, Andrew and {Farrar}, Glennys R. and {Feeney}, Michael and {Foley}, Ryan J. and {Gal-Yam}, Avishay and {Golkhou}, V. Zach and {Goobar}, Ariel and {Graham}, Matthew J. and {Hammerstein}, Erica and {Helou}, George and {Hung}, Tiara and {Kasliwal}, Mansi M. and {Kilpatrick}, Charles D. and {Kong}, Albert K.~H. and {Kupfer}, Thomas and {Laher}, Russ R. and {Mahabal}, Ashish A. and {Masci}, Frank J. and {Necker}, Jannis and {Nordin}, Jakob and {Perley}, Daniel A. and {Rigault}, Mickael and {Reusch}, Simeon and {Rodriguez}, Hector and {Rojas-Bravo}, C{\'e}sar and {Rusholme}, Ben and {Shupe}, David L. and {Singer}, Leo P. and {Sollerman}, Jesper and {Soumagnac}, Maayane T. and {Stern}, Daniel and {Taggart}, Kirsty and {van Santen}, Jakob and {Ward}, Charlotte and {Woudt}, Patrick and {Yao}, Yuhan},
        title = "{A tidal disruption event coincident with a high-energy neutrino}",
      journal = {Nature Astronomy},
         year = 2021,
        month = feb,
          doi = {10.1038/s41550-020-01295-8},
       adsurl = {https://ui.adsabs.harvard.edu/abs/2021NatAs.tmp...34S}
}

@INPROCEEDINGS{Krizmanic2011,
   author = {{Krizmanic}, J. and {Mitchell}, J.W.},
    title = "{The Potential of Spaced-based High-Energy Neutrino Measurements via the Airshower Cherenkov Signal}",
  booktitle = {Proceedings of the 32nd ICRC (Beijing)},
archivePrefix = "arXiv",
   eprint = {1109.6544},
 primaryClass = "astro-ph.HE",
     year = 2011,
   volume = 4,
    pages = {316},
      doi = {10.7529/ICRC2011/V04/1331},
   adsurl = {http://adsabs.harvard.edu/abs/2011ICRC....4..316K}
}

@ARTICLE{CHANT,
       author = {{Neronov}, Andrii and {Semikoz}, Dmitri V. and {Anchordoqui}, Luis A. and {Adams}, James H. and {Olinto}, Angela V.},
    title = "{Sensitivity of a proposed space-based Cherenkov astrophysical-neutrino telescope}",
    journal  = "Phys. Rev.",
    volume  = "D95",
archivePrefix = "arXiv",
   eprint = {1606.03629},
 primaryClass = "astro-ph.IM",
     year = 2017,
    month = jan,
   number = 2,
      eid = {023004},
    pages = {023004},
      doi = {10.1103/PhysRevD.95.023004},
   adsurl = {http://adsabs.harvard.edu/abs/2017PhRvD..95b3004N}
}

@ARTICLE{2009APh....32...10A,
       author = {{Anita Collaboration} and {Gorham}, P.~W. and {Allison}, P. and {Barwick}, S.~W. and {Beatty}, J.~J. and {Besson}, D.~Z. and {Binns}, W.~R. and {Chen}, C. and {Chen}, P. and {Clem}, J.~M. and {Connolly}, A. and {Dowkontt}, P.~F. and {Duvernois}, M.~A. and {Field}, R.~C. and {Goldstein}, D. and {Goodhue}, A. and {Hast}, C. and {Hebert}, C.~L. and {Hoover}, S. and {Israel}, M.~H. and {Kowalski}, J. and {Learned}, J.~G. and {Liewer}, K.~M. and {Link}, J.~T. and {Lusczek}, E. and {Matsuno}, S. and {Mercurio}, B.~C. and {Miki}, C. and {Mio{\v{c}}inovi{\'c}}, P. and {Nam}, J. and {Naudet}, C.~J. and {Nichol}, R.~J. and {Palladino}, K. and {Reil}, K. and {Romero-Wolf}, A. and {Rosen}, M. and {Ruckman}, L. and {Saltzberg}, D. and {Seckel}, D. and {Varner}, G.~S. and {Walz}, D. and {Wang}, Y. and {Williams}, C. and {Wu}, F. and {ANITA Collaboration}},
        title = "{The Antarctic Impulsive Transient Antenna ultra-high energy neutrino detector: Design, performance, and sensitivity for the 2006-2007 balloon flight}",
      journal = {Astroparticle Physics},
         year = 2009,
        month = aug,
       volume = {32},
       number = {1},
        pages = {10-41},
          doi = {10.1016/j.astropartphys.2009.05.003},
archivePrefix = {arXiv},
       eprint = {0812.1920},
 primaryClass = {astro-ph},
       adsurl = {https://ui.adsabs.harvard.edu/abs/2009APh....32...10A}
}

@ARTICLE{2010PhRvD..82b2004G,
       author = {{Gorham}, P.~W. and {Allison}, P. and {Baughman}, B.~M. and {Beatty}, J.~J. and {Belov}, K. and {Besson}, D.~Z. and {Bevan}, S. and {Binns}, W.~R. and {Chen}, C. and {Chen}, P. and {Clem}, J.~M. and {Connolly}, A. and {Detrixhe}, M. and {de Marco}, D. and {Dowkontt}, P.~F. and {Duvernois}, M. and {Grashorn}, E.~W. and {Hill}, B. and {Hoover}, S. and {Huang}, M. and {Israel}, M.~H. and {Javaid}, A. and {Liewer}, K.~M. and {Matsuno}, S. and {Mercurio}, B.~C. and {Miki}, C. and {Mottram}, M. and {Nam}, J. and {Nichol}, R.~J. and {Palladino}, K. and {Romero-Wolf}, A. and {Ruckman}, L. and {Saltzberg}, D. and {Seckel}, D. and {Varner}, G.~S. and {Vieregg}, A.~G. and {Wang}, Y. and {ANITA Collaboration}},
        title = "{Observational constraints on the ultrahigh energy cosmic neutrino flux from the second flight of the ANITA experiment}",
      journal = {Phys. Rev. D},
         year = 2010,
        month = jul,
       volume = {82},
       number = {2},
          eid = {022004},
        pages = {022004},
          doi = {10.1103/PhysRevD.82.022004},
archivePrefix = {arXiv},
       eprint = {1003.2961},
 primaryClass = {astro-ph.HE},
       adsurl = {https://ui.adsabs.harvard.edu/abs/2010PhRvD..82b2004G}
}

@INPROCEEDINGS{2019EPJWC.21601009V,
       author = {{Vieregg}, Abigail},
        title = "{Results from the third flight of ANITA}",
    booktitle = {European Physical Journal Web of Conferences},
         year = 2019,
       series = {European Physical Journal Web of Conferences},
       volume = {216},
        month = jun,
          eid = {01009},
        pages = {01009},
          doi = {10.1051/epjconf/201921601009},
       adsurl = {https://ui.adsabs.harvard.edu/abs/2019EPJWC.21601009V}
}

@ARTICLE{2019PhRvD..99l2001G,
       author = {{Gorham}, P.~W. and {Allison}, P. and {Banerjee}, O. and {Batten}, L. and {Beatty}, J.~J. and {Belov}, K. and {Besson}, D.~Z. and {Binns}, W.~R. and {Bugaev}, V. and {Cao}, P. and {Chen}, C.~H. and {Chen}, P. and {Chen}, Y. and {Clem}, J.~M. and {Connolly}, A. and {Cremonesi}, L. and {Dailey}, B. and {Deaconu}, C. and {Dowkontt}, P.~F. and {Fox}, B.~D. and {Gordon}, J.~W.~H. and {Hast}, C. and {Hill}, B. and {Hsu}, S.~Y. and {Huang}, J.~J. and {Hughes}, K. and {Hupe}, R. and {Israel}, M.~H. and {Liewer}, K.~M. and {Liu}, T.~C. and {Ludwig}, A.~B. and {Macchiarulo}, L. and {Matsuno}, S. and {McBride}, K. and {Miki}, C. and {Mulrey}, K. and {Nam}, J. and {Naudet}, C. and {Nichol}, R.~J. and {Novikov}, A. and {Oberla}, E. and {Prohira}, S. and {Rauch}, B.~F. and {Ripa}, J. and {Roberts}, J.~M. and {Romero-Wolf}, A. and {Rotter}, B. and {Russell}, J.~W. and {Saltzberg}, D. and {Seckel}, D. and {Schoorlemmer}, H. and {Shiao}, J. and {Stafford}, S. and {Stockham}, J. and {Stockham}, M. and {Strutt}, B. and {Sutherland}, M.~S. and {Varner}, G.~S. and {Vieregg}, A.~G. and {Wang}, N. and {Wang}, S.~H. and {Wissel}, S.~A. and {Anita Collaboration}},
        title = "{Constraints on the ultrahigh-energy cosmic neutrino flux from the fourth flight of ANITA}",
      journal = {Phys. Rev. D},
         year = 2019,
        month = jun,
       volume = {99},
       number = {12},
          eid = {122001},
        pages = {122001},
          doi = {10.1103/PhysRevD.99.122001},
archivePrefix = {arXiv},
       eprint = {1902.04005},
 primaryClass = {astro-ph.HE},
       adsurl = {https://ui.adsabs.harvard.edu/abs/2019PhRvD..99l2001G}
}

@article{Askaryan:1962hbi,
    author = "Askar'yan, G. A.",
    title = "{Excess negative charge of an electron-photon shower and its coherent radio emission}",
    journal = "Zh. Eksp. Teor. Fiz.",
    volume = "41",
    pages = "616--618",
    year = "1961"
}

@ARTICLE{ANITAtd,
  author = {Barwick, S. W. and Beatty, J. J. and Besson, D. Z. and Binns, W. R. and Cai, B. and Clem, J. M. and Connolly, A. and Cowen, D. F. and Dowkontt, P. F. and DuVernois, M. A. and Evenson, P. A. and Goldstein, D. and Gorham, P. W. and Hebert, C. L. and Israel, M. H. and Learned, J. G. and Liewer, K. M. and Link, J. T. and Matsuno, S. and Mio\ifmmode \check{c}\else \v{c}\fi{}inovi\ifmmode \acute{c}\else \'{c}\fi{}, P. and Nam, J. and Naudet, C. J. and Nichol, R. and Palladino, K. and Rosen, M. and Saltzberg, D. and Seckel, D. and Silvestri, A. and Stokes, B. T. and Varner, G. S. and Wu, F.},
    title = "{Constraints on Cosmic Neutrino Fluxes from the Antarctic Impulsive Transient Antenna Experiment}",
  journal = {Physical Review Letters},
   eprint = {astro-ph/0512265},
     year = 2006,
    month = may,
   volume = 96,
   number = 17,
      eid = {171101},
    pages = {171101},
      doi = {10.1103/PhysRevLett.96.171101},
   adsurl = {http://adsabs.harvard.edu/abs/2006PhRvL..96q1101B}
}

@ARTICLE{2019PhRvD..99f3011R,
       author = {{Romero-Wolf}, A. and {Wissel}, S.~A. and {Schoorlemmer}, H. and {Carvalho}, W.~R. and {Alvarez-Mu{\~n}iz}, J. and {Zas}, E. and {Allison}, P. and {Banerjee}, O. and {Batten}, L. and {Beatty}, J.~J. and {Bechtol}, K. and {Belov}, K. and {Besson}, D.~Z. and {Binns}, W.~R. and {Bugaev}, V. and {Cao}, P. and {Chen}, C.~C. and {Chen}, C.~H. and {Chen}, P. and {Clem}, J.~M. and {Connolly}, A. and {Cremonesi}, L. and {Dailey}, B. and {Deaconu}, C. and {Dowkontt}, P.~F. and {Fox}, B.~D. and {Gordon}, J.~W.~H. and {Gorham}, P.~W. and {Hast}, C. and {Hill}, B. and {Hsu}, S.~Y. and {Huang}, J.~J. and {Hughes}, K. and {Hupe}, R. and {Israel}, M.~H. and {Liewer}, K.~M. and {Liu}, T.~C. and {Ludwig}, A.~B. and {Macchiarulo}, L. and {Matsuno}, S. and {Miki}, C. and {Mulrey}, K. and {Nam}, J. and {Naudet}, C. and {Nichol}, R.~J. and {Novikov}, A. and {Oberla}, E. and {Prohira}, S. and {Rauch}, B.~F. and {Roberts}, J.~M. and {Rotter}, B. and {Russell}, J.~W. and {Saltzberg}, D. and {Seckel}, D. and {Shiao}, J. and {Stafford}, S. and {Stockham}, J. and {Stockham}, M. and {Strutt}, B. and {Sutherland}, M.~S. and {Varner}, G.~S. and {Vieregg}, A.~G. and {Wang}, S.~H.},
        title = "{Comprehensive analysis of anomalous ANITA events disfavors a diffuse tau-neutrino flux origin}",
      journal = {Phys. Rev. D},
         year = 2019,
        month = mar,
       volume = {99},
       number = {6},
          eid = {063011},
        pages = {063011},
          doi = {10.1103/PhysRevD.99.063011},
archivePrefix = {arXiv},
       eprint = {1811.07261},
 primaryClass = {astro-ph.HE},
       adsurl = {https://ui.adsabs.harvard.edu/abs/2019PhRvD..99f3011R}
}

@ARTICLE{2018PhRvL.121p1102G,
       author = {{Gorham}, P.~W. and {Rotter}, B. and {Allison}, P. and {Banerjee}, O. and {Batten}, L. and {Beatty}, J.~J. and {Bechtol}, K. and {Belov}, K. and {Besson}, D.~Z. and {Binns}, W.~R. and {Bugaev}, V. and {Cao}, P. and {Chen}, C.~C. and {Chen}, C.~H. and {Chen}, P. and {Clem}, J.~M. and {Connolly}, A. and {Cremonesi}, L. and {Dailey}, B. and {Deaconu}, C. and {Dowkontt}, P.~F. and {Fox}, B.~D. and {Gordon}, J.~W.~H. and {Hast}, C. and {Hill}, B. and {Hughes}, K. and {Huang}, J.~J. and {Hupe}, R. and {Israel}, M.~H. and {Javaid}, A. and {Lam}, J. and {Liewer}, K.~M. and {Lin}, S.~Y. and {Liu}, T.~C. and {Ludwig}, A. and {Macchiarulo}, L. and {Matsuno}, S. and {Miki}, C. and {Mulrey}, K. and {Nam}, J. and {Naudet}, C.~J. and {Nichol}, R.~J. and {Novikov}, A. and {Oberla}, E. and {Olmedo}, M. and {Prechelt}, R. and {Prohira}, S. and {Rauch}, B.~F. and {Roberts}, J.~M. and {Romero-Wolf}, A. and {Russell}, J.~W. and {Saltzberg}, D. and {Seckel}, D. and {Schoorlemmer}, H. and {Shiao}, J. and {Stafford}, S. and {Stockham}, J. and {Stockham}, M. and {Strutt}, B. and {Varner}, G.~S. and {Vieregg}, A.~G. and {Wang}, S.~H. and {Wissel}, S.~A.},
        title = "{Observation of an Unusual Upward-Going Cosmic-Ray-like Event in the Third Flight of ANITA}",
      journal = {Physical Review Letters},
         year = 2018,
        month = oct,
       volume = {121},
       number = {16},
          eid = {161102},
        pages = {161102},
          doi = {10.1103/PhysRevLett.121.161102},
archivePrefix = {arXiv},
       eprint = {1803.05088},
 primaryClass = {astro-ph.HE},
       adsurl = {https://ui.adsabs.harvard.edu/abs/2018PhRvL.121p1102G}
}

@ARTICLE{2021PhRvL.126g1103G,
       author = {{Gorham}, P.~W. and {Ludwig}, A. and {Deaconu}, C. and {Cao}, P. and {Allison}, P. and {Banerjee}, O. and {Batten}, L. and {Bhattacharya}, D. and {Beatty}, J.~J. and {Belov}, K. and {Binns}, W.~R. and {Bugaev}, V. and {Chen}, C.~H. and {Chen}, P. and {Chen}, Y. and {Clem}, J.~M. and {Cremonesi}, L. and {Dailey}, B. and {Dowkontt}, P.~F. and {Fox}, B.~D. and {Gordon}, J.~W.~H. and {Hast}, C. and {Hill}, B. and {Hsu}, S.~Y. and {Huang}, J.~J. and {Hughes}, K. and {Hupe}, R. and {Israel}, M.~H. and {Liu}, T.~C. and {Macchiarulo}, L. and {Matsuno}, S. and {McBride}, K. and {Miki}, C. and {Nam}, J. and {Naudet}, C.~J. and {Nichol}, R.~J. and {Novikov}, A. and {Oberla}, E. and {Olmedo}, M. and {Prechelt}, R. and {Rauch}, B.~F. and {Roberts}, J.~M. and {Romero-Wolf}, A. and {Rotter}, B. and {Russell}, J.~W. and {Saltzberg}, D. and {Seckel}, D. and {Schoorlemmer}, H. and {Shiao}, J. and {Stafford}, S. and {Stockham}, J. and {Stockham}, M. and {Strutt}, B. and {Sutherland}, M.~S. and {Varner}, G.~S. and {Vieregg}, A.~G. and {Wang}, S.~H. and {Wissel}, S.~A.},
        title = "{Unusual Near-Horizon Cosmic-Ray-like Events Observed by ANITA-IV}",
      journal = {Physical Review Letters},
         year = 2021,
        month = feb,
       volume = {126},
       number = {7},
          eid = {071103},
        pages = {071103},
          doi = {10.1103/PhysRevLett.126.071103},
       adsurl = {https://ui.adsabs.harvard.edu/abs/2021PhRvL.126g1103G}
}

@ARTICLE{2019NIMPA.936..237O,
       author = {{Osteria}, G. and {Scotti}, V. and {JEM-EUSO Collaboration}},
        title = "{EUSO-SPB: In-flight performance}",
      journal = {Nuclear Instruments and Methods in Physics Research A},
         year = 2019,
        month = aug,
       volume = {936},
        pages = {237-238},
          doi = {10.1016/j.nima.2018.10.213},
       adsurl = {https://ui.adsabs.harvard.edu/abs/2019NIMPA.936..237O}
}

@INPROCEEDINGS{2017ICRC...35..445B,
       author = {{Bertaina}, M.~E. and {von Ballmoos}, P. and {JEM-EUSO Collaboration}},
        title = "{Results of the EUSO-Balloon flight}",
    booktitle = {35th International Cosmic Ray Conference (Busan)},
         year = 2017,
       series = {International Cosmic Ray Conference},
       volume = {301},
        month = jan,
          eid = {445},
        pages = {445},
       adsurl = {https://ui.adsabs.harvard.edu/abs/2017ICRC...35..445B}
}

@ARTICLE{2020NIMPA.95862164S,
       author = {{Scotti}, V. and {Osteria}, G. and {JEM-EUSO Collaboration}},
        title = "{The EUSO-SPB2 mission}",
      journal = {Nuclear Instruments and Methods in Physics Research A},
         year = 2020,
        month = apr,
       volume = {958},
          eid = {162164},
        pages = {162164},
          doi = {10.1016/j.nima.2019.05.005},
       adsurl = {https://ui.adsabs.harvard.edu/abs/2020NIMPA.95862164S}
}

@article{PhysRev.118.316,
  title = {Resonant Scattering of Antineutrinos},
  author = {Glashow, Sheldon L.},
  journal = {Phys. Rev.},
  volume = {118},
  issue = {1},
  pages = {316--317},
  numpages = {0},
  year = {1960},
  month = {Apr},
  publisher = {American Physical Society},
  doi = {10.1103/PhysRev.118.316},
  url = {https://link.aps.org/doi/10.1103/PhysRev.118.316}
}

@PHDTHESIS{1990PhDT........94K,
       author = {{Krizmanic}, John Francis},
        title = "{a Search for the Oscillation of Muon Antineutrinos to Electron Antineutrinos Using the AGS Wide Band Beam.}",
       school = {THE JOHNS HOPKINS UNIVERSITY.},
         year = 1990,
        month = jan,
       adsurl = {https://ui.adsabs.harvard.edu/abs/1990PhDT........94K}
}

@article{Block:2014kza,
      author         = "Block, Martin M. and Durand, Loyal and Ha, Phuoc",
      title          = "{Connection of the virtual $\gamma^*p$ cross section of
                        $ep$ deep inelastic scattering to real $\gamma p$ scattering, and
                        the implications for $\nu N$ and $ep$ total cross sections}",
      journal        = "Phys. Rev.",
      volume         = "D89",
      year           = "2014",
      number         = "9",
      pages          = "094027",
      doi            = "10.1103/PhysRevD.89.094027",
      eprint         = "1404.4530",
      archivePrefix  = "arXiv",
      primaryClass   = "hep-ph",
      SLACcitation   = "%%CITATION = ARXIV:1404.4530;%%"
}

@article{Block:2013mia,
      author         = "Block, Martin M. and Durand, Loyal and Ha, Phuoc and
                        McKay, Douglas W.",
      title          = "{Implications of a Froissart bound saturation of $\gamma^*p$
                        deep inelastic scattering. I. Quark distributions at ultra
                        small x}",
      journal        = "Phys. Rev.",
      volume         = "D88",
      year           = "2013",
      number         = "1",
      pages          = "014006",
      doi            = "10.1103/PhysRevD.88.014006",
      eprint         = "1302.6119",
      archivePrefix  = "arXiv",
      primaryClass   = "hep-ph",
      SLACcitation   = "%%CITATION = ARXIV:1302.6119;%%"
}

@article{PhysRevD.58.093009,
  title = {Neutrino interactions at ultrahigh energies},
  author = {Gandhi, Raj and Quigg, Chris and Reno, Mary Hall and Sarcevic, Ina},
  journal = {Phys. Rev. D},
  volume = {58},
  issue = {9},
  pages = {093009},
  numpages = {15},
  year = {1998},
  month = {Sep},
  publisher = {American Physical Society},
  doi = {10.1103/PhysRevD.58.093009},
  url = {https://link.aps.org/doi/10.1103/PhysRevD.58.093009}
}

@article{PhysRevD.83.113009,
  title = {Calculation of high energy neutrino-nucleon cross sections and uncertainties using the Martin-Stirling-Thorne-Watt parton distribution functions and implications for future experiments},
  author = {Connolly, Amy and Thorne, Robert S. and Waters, David},
  journal = {Phys. Rev. D},
  volume = {83},
  issue = {11},
  pages = {113009},
  numpages = {13},
  year = {2011},
  month = {Jun},
  publisher = {American Physical Society},
  doi = {10.1103/PhysRevD.83.113009},
  url = {https://link.aps.org/doi/10.1103/PhysRevD.83.113009}
}

@unpublished{NeuBACH,
  title = "{Study of Neutrino-Induced Airshower Detection Using a Balloon-Borne Instrument}",
  author = {Streitmatter, Robert E. (PI) and Mitchell, John W. and Krizmanic, John F. and others},
  note = {NASA APRA Proposal},
  year = {2001},
 }

@INPROCEEDINGS{2018LPICo2063.3002S,
       author = {{Solomey}, N. and {Barghouty}, N. and {Christl}, M. and {Johnson}, L. and {Meyer}, H.},
        title = "{Deep-Space Test of a Neutrino Detector}",
    booktitle = {Deep Space Gateway Concept Science Workshop},
         year = 2018,
       volume = {2063},
        month = feb,
          eid = {3002},
        pages = {3002},
       adsurl = {https://ui.adsabs.harvard.edu/abs/2018LPICo2063.3002S}
}

@ARTICLE{2016AdSpR..58..464R,
       author = {{Ryabov}, V.~A. and {Chechin}, V.~A. and {Gusev}, G.~A. and {Maung}, K.~T.},
        title = "{Prospects for ultrahigh-energy particle observation based on the lunar orbital LORD space experiment}",
      journal = {Advances in Space Research},
         year = 2016,
        month = aug,
       volume = {58},
       number = {3},
        pages = {464-474},
          doi = {10.1016/j.asr.2016.04.030},
       adsurl = {https://ui.adsabs.harvard.edu/abs/2016AdSpR..58..464R}
}

@ARTICLE{2011JHEP...08..042C,
       author = {{Cooper-Sarkar}, Amanda and {Mertsch}, Philipp and {Sarkar}, Subir},
        title = "{The high energy neutrino cross-section in the Standard Model and its uncertainty}",
      journal = {Journal of High Energy Physics},
         year = 2011,
        month = aug,
       volume = {2011},
          eid = {42},
        pages = {42},
          doi = {10.1007/JHEP08(2011)042},
archivePrefix = {arXiv},
       eprint = {1106.3723},
 primaryClass = {hep-ph},
       adsurl = {https://ui.adsabs.harvard.edu/abs/2011JHEP...08..042C}
}

@ARTICLE{2021PhRvD.103a4013H,
       author = {{Hou}, Tie-Jiun and {Gao}, Jun and {Hobbs}, T.~J. and {Xie}, Keping and {Dulat}, Sayipjamal and {Guzzi}, Marco and {Huston}, Joey and {Nadolsky}, Pavel and {Pumplin}, Jon and {Schmidt}, Carl and {Sitiwaldi}, Ibrahim and {Stump}, Daniel and {Yuan}, C. -P.},
        title = "{New CTEQ global analysis of quantum chromodynamics with high-precision data from the LHC}",
      journal = {Phys. Rev. D},
         year = 2021,
        month = jan,
       volume = {103},
       number = {1},
          eid = {014013},
        pages = {014013},
          doi = {10.1103/PhysRevD.103.014013},
       adsurl = {https://ui.adsabs.harvard.edu/abs/2021PhRvD.103a4013H}
}

@misc{Reno21,
  note = {Figure generated by Mary Hall Reno (University of Iowa), private communication},
 }

@misc{AustinPlot,
  note = {Figure generated by Austin Cummings (PennState), private communication},
 }

@misc{ToniV21,
  note = {Figure generated by Toni Venters (NASA/GSFC), private communication},
 }

@misc{Evoli,
 author = {{Evoli}, Carmelo.},
 note = {The Cosmic-Ray Energy Spectrum. Zenodo. https://doi.org/10.5281/zenodo.4396125},
 year= {2020}
 }

@ARTICLE{CONEX,
   author = {{Pierog}, T. and {Alekseeva}, M.~K. and {Bergmann}, T. and {Chernatkin}, V. and 
	{Engel}, R. and {Heck}, D. and {Kalmykov}, N.~N. and {Moyon}, J. and 
	{Ostapchenko}, S. and {Thouw}, T. and {Werner}, K.},
    title = "{First results of fast one-dimensional hybrid simulation of EAS using CONEX}",
  journal = {Nuclear Physics B Proceedings Supplements},
   eprint = {astro-ph/0411260},
     year = 2006,
    month = jan,
   volume = 151,
    pages = {159-162},
      doi = {10.1016/j.nuclphysbps.2005.07.029},
   adsurl = {http://adsabs.harvard.edu/abs/2006NuPhS.151..159P}
}

@article{PhysRev.93.646.2,
  title = {Influence of the Earth's Magnetic Field on the Extensive Air Showers},
  author = {Cocconi, Giuseppe},
  journal = {Phys. Rev.},
  volume = {93},
  issue = {3},
  pages = {646--647},
  numpages = {0},
  year = {1954},
  month = {Feb},
  publisher = {American Physical Society},
  doi = {10.1103/PhysRev.93.646.2},
  url = {https://link.aps.org/doi/10.1103/PhysRev.93.646.2}
}

@article{PhysRev.95.1705.4,
  title = {Influence of the Earth's Magnetic Field on the Extensive Air Showers},
  author = {Cocconi, Giuseppe},
  journal = {Phys. Rev.},
  volume = {95},
  issue = {6},
  pages = {1705--1706},
  numpages = {0},
  year = {1954},
  month = {Sep},
  publisher = {American Physical Society},
  doi = {10.1103/PhysRev.95.1705.4},
  url = {https://link.aps.org/doi/10.1103/PhysRev.95.1705.4}
}

@ARTICLE{Alvarez-Muniz_2012,
   author = {{Alvarez-Mu{\~n}iz}, J. and {Carvalho}, W.~R. and {Zas}, E.},
    title = "{Monte Carlo simulations of radio pulses in atmospheric showers using ZHAireS}",
  journal = {Astroparticle Physics},
archivePrefix = "arXiv",
   eprint = {1107.1189},
 primaryClass = "astro-ph.HE",
     year = 2012,
    month = jan,
   volume = 35,
    pages = {325-341},
      doi = {10.1016/j.astropartphys.2011.10.005},
   adsurl = {http://adsabs.harvard.edu/abs/2012APh....35..325A}
}

@ARTICLE{Alvarez-Muniz_2012b,
   author = {{Alvarez-Mu{\~n}iz}, J. and {Carvalho}, Jr., W.~R. and {Romero-Wolf}, A. and 
	{Tueros}, M. and {Zas}, E.},
    title = "{Coherent radiation from extensive air showers in the ultrahigh frequency band}",
      journal        = "Phys. Rev.",
      volume         = "D86",
archivePrefix = "arXiv",
   eprint = {1208.0951},
 primaryClass = "astro-ph.HE",
     year = 2012,
    month = dec,
     number = 12,
      eid = {123007},
    pages = {123007},
      doi = {10.1103/PhysRevD.86.123007},
   adsurl = {http://adsabs.harvard.edu/abs/2012PhRvD..86l3007A}
}

@ARTICLE{Askaryan_1962,
   author = {{Askaryan}, G.~A.},
    title = "{Excess Negative Charge of the Electron-Photon Shower and Coherent Radiation Originating 
from It. Radio Recording of Showers Under the Ground and on the Moon}",
  journal = {JETP},
     year = 1962,
   volume = 14,
    pages = {441}
}

@Article{IGRF_2015,
author="Th{\'e}bault, Erwan and Finlay, C.C. and Beggan, C.D.  and others",
title="International Geomagnetic Reference Field: the 12th generation",
journal="Earth, Planets and Space",
year="2015",
month="May",
day="27",
volume="67",
number="1",
pages="79",
issn="1880-5981",
doi="10.1186/s40623-015-0228-9",
url="https://doi.org/10.1186/s40623-015-0228-9"
}

@ARTICLE{Buitink_2016,
       author = {{Buitink}, S. and {Corstanje}, A. and {Falcke}, H. and {H{\"o}randel}, J.~R. and {Huege}, T. and {Nelles}, A. and {Rachen}, J.~P. and {Rossetto}, L. and {Schellart}, P. and {Scholten}, O. and {Ter Veen}, S. and {Thoudam}, S. and {Trinh}, T.~N.~G. and {Anderson}, J. and {Asgekar}, A. and {Avruch}, I.~M. and {Bell}, M.~E. and {Bentum}, M.~J. and {Bernardi}, G. and {Best}, P. and {Bonafede}, A. and {Breitling}, F. and {Broderick}, J.~W. and {Brouw}, W.~N. and {Br{\"u}ggen}, M. and {Butcher}, H.~R. and {Carbone}, D. and {Ciardi}, B. and {Conway}, J.~E. and {de Gasperin}, F. and {de Geus}, E. and {Deller}, A. and {Dettmar}, R. -J. and {van Diepen}, G. and {Duscha}, S. and {Eisl{\"o}ffel}, J. and {Engels}, D. and {Enriquez}, J.~E. and {Fallows}, R.~A. and {Fender}, R. and {Ferrari}, C. and {Frieswijk}, W. and {Garrett}, M.~A. and {Grie{\ss}meier}, J.~M. and {Gunst}, A.~W. and {van Haarlem}, M.~P. and {Hassall}, T.~E. and {Heald}, G. and {Hessels}, J.~W.~T. and {Hoeft}, M. and {Horneffer}, A. and {Iacobelli}, M. and {Intema}, H. and {Juette}, E. and {Karastergiou}, A. and {Kondratiev}, V.~I. and {Kramer}, M. and {Kuniyoshi}, M. and {Kuper}, G. and {van Leeuwen}, J. and {Loose}, G.~M. and {Maat}, P. and {Mann}, G. and {Markoff}, S. and {McFadden}, R. and {McKay-Bukowski}, D. and {McKean}, J.~P. and {Mevius}, M. and {Mulcahy}, D.~D. and {Munk}, H. and {Norden}, M.~J. and {Orru}, E. and {Paas}, H. and {Pandey-Pommier}, M. and {Pandey}, V.~N. and {Pietka}, M. and {Pizzo}, R. and {Polatidis}, A.~G. and {Reich}, W. and {R{\"o}ttgering}, H.~J.~A. and {Scaife}, A.~M.~M. and {Schwarz}, D.~J. and {Serylak}, M. and {Sluman}, J. and {Smirnov}, O. and {Stappers}, B.~W. and {Steinmetz}, M. and {Stewart}, A. and {Swinbank}, J. and {Tagger}, M. and {Tang}, Y. and {Tasse}, C. and {Toribio}, M.~C. and {Vermeulen}, R. and {Vocks}, C. and {Vogt}, C. and {van Weeren}, R.~J. and {Wijers}, R.~A.~M.~J. and {Wijnholds}, S.~J. and {Wise}, M.~W. and {Wucknitz}, O. and {Yatawatta}, S. and {Zarka}, P. and {Zensus}, J.~A.},
       title = "{A large light-mass component of cosmic rays at 10$^{17}$-10$^{17.5}$ electronvolts from radio observations}",
  journal = {Nature},
archivePrefix = "arXiv",
   eprint = {1603.01594},
 primaryClass = "astro-ph.HE",
     year = 2016,
    month = mar,
   volume = 531,
    pages = {70-73},
      doi = {10.1038/nature16976},
   adsurl = {http://adsabs.harvard.edu/abs/2016Natur.531...70B}
}

@ARTICLE{Hoover_2010,
      author = {{Hoover}, S. and {Nam}, J. and {Gorham}, P.~W. and {Grashorn}, E. and {Allison}, P. and {Barwick}, S.~W. and {Beatty}, J.~J. and {Belov}, K. and {Besson}, D.~Z. and {Binns}, W.~R. and {Chen}, C. and {Chen}, P. and {Clem}, J.~M. and {Connolly}, A. and {Dowkontt}, P.~F. and {Duvernois}, M.~A. and {Field}, R.~C. and {Goldstein}, D. and {Vieregg}, A.~G. and {Hast}, C. and {Israel}, M.~H. and {Javaid}, A. and {Kowalski}, J. and {Learned}, J.~G. and {Liewer}, K.~M. and {Link}, J.~T. and {Lusczek}, E. and {Matsuno}, S. and {Mercurio}, B.~C. and {Miki}, C. and {Mio{\v{c}}inovi{\'c}}, P. and {Naudet}, C.~J. and {Ng}, J. and {Nichol}, R.~J. and {Palladino}, K. and {Reil}, K. and {Romero-Wolf}, A. and {Rosen}, M. and {Ruckman}, L. and {Saltzberg}, D. and {Seckel}, D. and {Varner}, G.~S. and {Walz}, D. and {Wu}, F.},
    title = "{Observation of Ultrahigh-Energy Cosmic Rays with the ANITA Balloon-Borne Radio Interferometer}",
  journal = {Physical Review Letters},
archivePrefix = "arXiv",
   eprint = {1005.0035},
 primaryClass = "astro-ph.HE",
     year = 2010,
    month = oct,
   volume = 105,
   number = 15,
      eid = {151101},
    pages = {151101},
      doi = {10.1103/PhysRevLett.105.151101},
   adsurl = {http://adsabs.harvard.edu/abs/2010PhRvL.105o1101H}
}

@ARTICLE{Schoorlemmer_2016,
       author = {{Schoorlemmer}, H. and {Belov}, K. and {Romero-Wolf}, A. and {Garc{\'\i}a-Fern{\'a}ndez}, D. and {Bugaev}, V. and {Wissel}, S.~A. and {Allison}, P. and {Alvarez-Mu{\~n}iz}, J. and {Barwick}, S.~W. and {Beatty}, J.~J. and {Besson}, D.~Z. and {Binns}, W.~R. and {Carvalho}, W.~R., Jr. and {Chen}, C. and {Chen}, P. and {Clem}, J.~M. and {Connolly}, A. and {Dowkontt}, P.~F. and {DuVernois}, M.~A. and {Field}, R.~C. and {Goldstein}, D. and {Gorham}, P.~W. and {Hast}, C. and {Huege}, T. and {Heber}, C.~L. and {Hoover}, S. and {Israel}, M.~H. and {Javaid}, A. and {Kowalski}, J. and {Lam}, J. and {Learned}, J.~G. and {Link}, J.~T. and {Lusczek}, E. and {Matsuno}, S. and {Mercurio}, B.~C. and {Miki}, C. and {Mio{\v{c}}inovi{\'c}}, P. and {Mulrey}, K. and {Nam}, J. and {Naudet}, C.~J. and {Ng}, J. and {Nichol}, R.~J. and {Palladino}, K. and {Rauch}, B.~F. and {Roberts}, J. and {Reil}, K. and {Rotter}, B. and {Rosen}, M. and {Ruckman}, L. and {Saltzberg}, D. and {Seckel}, D. and {Urdaneta}, D. and {Varner}, G.~S. and {Vieregg}, A.~G. and {Walz}, D. and {Wu}, F. and {Zas}, E.},
       title = "{Energy and flux measurements of ultra-high energy cosmic rays observed during the first ANITA flight}",
  journal = {Astroparticle Physics},
archivePrefix = "arXiv",
   eprint = {1506.05396},
 primaryClass = "astro-ph.HE",
     year = 2016,
    month = apr,
   volume = 77,
    pages = {32-43},
      doi = {10.1016/j.astropartphys.2016.01.001},
   adsurl = {http://adsabs.harvard.edu/abs/2016APh....77...32S}
}

@ARTICLE{Gorham_2016,
       author = {{Gorham}, P.~W. and {Nam}, J. and {Romero-Wolf}, A. and {Hoover}, S. and {Allison}, P. and {Banerjee}, O. and {Beatty}, J.~J. and {Belov}, K. and {Besson}, D.~Z. and {Binns}, W.~R. and {Bugaev}, V. and {Cao}, P. and {Chen}, C. and {Chen}, P. and {Clem}, J.~M. and {Connolly}, A. and {Dailey}, B. and {Deaconu}, C. and {Cremonesi}, L. and {Dowkontt}, P.~F. and {Duvernois}, M.~A. and {Field}, R.~C. and {Fox}, B.~D. and {Goldstein}, D. and {Gordon}, J. and {Hast}, C. and {Hebert}, C.~L. and {Hill}, B. and {Hughes}, K. and {Hupe}, R. and {Israel}, M.~H. and {Javaid}, A. and {Kowalski}, J. and {Lam}, J. and {Learned}, J.~G. and {Liewer}, K.~M. and {Liu}, T.~C. and {Link}, J.~T. and {Lusczek}, E. and {Matsuno}, S. and {Mercurio}, B.~C. and {Miki}, C. and {Mio{\v{c}}inovi{\'c}}, P. and {Mottram}, M. and {Mulrey}, K. and {Naudet}, C.~J. and {Ng}, J. and {Nichol}, R.~J. and {Palladino}, K. and {Rauch}, B.~F. and {Reil}, K. and {Roberts}, J. and {Rosen}, M. and {Rotter}, B. and {Russell}, J. and {Ruckman}, L. and {Saltzberg}, D. and {Seckel}, D. and {Schoorlemmer}, H. and {Stafford}, S. and {Stockham}, J. and {Stockham}, M. and {Strutt}, B. and {Tatem}, K. and {Varner}, G.~S. and {Vieregg}, A.~G. and {Walz}, D. and {Wissel}, S.~A. and {Wu}, F. and {Anita Collaboration}},
       journal = {Physical Review Letters},
archivePrefix = "arXiv",
   eprint = {1603.05218},
 primaryClass = "astro-ph.HE",
     year = 2016,
    month = aug,
   volume = 117,
   number = 7,
      eid = {071101},
    pages = {071101},
      doi = {10.1103/PhysRevLett.117.071101},
   adsurl = {http://adsabs.harvard.edu/abs/2016PhRvL.117g1101G}
}

@INPROCEEDINGS{Romero-Wolf_2017,
author = {Romero-Wolf, Andres and Gorham, PW and Nam, Jae-Hwan and Hoover, S and Allison, P and Banerjee, O and Batten, Luke and Beatty, J. and Belov, K and Besson, DZ and Binns, W. and Bugaev, V and Cao, P and Chen, Chulin and Chen, Pengchong and Clem, JM and Connolly, A and Dailey, B and Deaconu, Cosmin and Zas, E.},
   title = "{Upward-Pointing Cosmic-Ray-like Events Observed with ANITA}",
   booktitle = {Proceedings of the 35th ICRC (Busan)},
   year = 2017,
   number = {PoS(ICRC2017)935}
}

@ARTICLE{Aab_2016_radio_CR_energy,
       author = {{Aab}, A. and {Abreu}, P. and {Aglietta}, M. and {Ahn}, E.~J. and {Al Samarai}, I. and {Albuquerque}, I.~F.~M. and {Allekotte}, I. and {Allison}, P. and {Almela}, A. and {Alvarez Castillo}, J. and {Alvarez-Mu{\~n}iz}, J. and {Alves Batista}, R. and {Ambrosio}, M. and {Aminaei}, A. and {Anastasi}, G.~A. and {Anchordoqui}, L. and {Andringa}, S. and {Aramo}, C. and {Arqueros}, F. and {Arsene}, N. and {Asorey}, H. and {Assis}, P. and {Aublin}, J. and {Avila}, G. and {Awal}, N. and {Badescu}, A.~M. and {Baus}, C. and {Beatty}, J.~J. and {Becker}, K.~H. and {Bellido}, J.~A. and {Berat}, C. and {Bertaina}, M.~E. and {Bertou}, X. and {Biermann}, P.~L. and {Billoir}, P. and {Blaess}, S.~G. and {Blanco}, A. and {Blanco}, M. and {Blazek}, J. and {Bleve}, C. and {Bl{\"u}mer}, H. and {Boh{\'a}{\v{c}}ov{\'a}}, M. and {Boncioli}, D. and {Bonifazi}, C. and {Borodai}, N. and {Brack}, J. and {Brancus}, I. and {Bretz}, T. and {Bridgeman}, A. and {Brogueira}, P. and {Buchholz}, P. and {Bueno}, A. and {Buitink}, S. and {Buscemi}, M. and {Caballero-Mora}, K.~S. and {Caccianiga}, B. and {Caccianiga}, L. and {Candusso}, M. and {Caramete}, L. and {Caruso}, R. and {Castellina}, A. and {Cataldi}, G. and {Cazon}, L. and {Cester}, R. and {Chavez}, A.~G. and {Chiavassa}, A. and {Chinellato}, J.~A. and {Chudoba}, J. and {Cilmo}, M. and {Clay}, R.~W. and {Cocciolo}, G. and {Colalillo}, R. and {Coleman}, A. and {Collica}, L. and {Coluccia}, M.~R. and {Concei{\c{c}}{\~a}o}, R. and {Contreras}, F. and {Cooper}, M.~J. and {Cordier}, A. and {Coutu}, S. and {Covault}, C.~E. and {Cronin}, J. and {Dallier}, R. and {Daniel}, B. and {Dasso}, S. and {Daumiller}, K. and {Dawson}, B.~R. and {de Almeida}, R.~M. and {de Jong}, S.~J. and {De Mauro}, G. and {de Mello Neto}, J.~R.~T. and {De Mitri}, I. and {de Oliveira}, J. and {de Souza}, V. and {del Peral}, L. and {Deligny}, O. and {Dhital}, N. and {Di Giulio}, C. and {Di Matteo}, A. and {Diaz}, J.~C. and {D{\'\i}az Castro}, M.~L. and {Diogo}, F. and {Dobrigkeit}, C. and {Docters}, W. and {D'Olivo}, J.~C. and {Dorofeev}, A. and {Dorosti Hasankiadeh}, Q. and {dos Anjos}, R.~C. and {Dova}, M.~T. and {Ebr}, J. and {Engel}, R. and {Erdmann}, M. and {Erfani}, M. and {Escobar}, C.~O. and {Espadanal}, J. and {Etchegoyen}, A. and {Falcke}, H. and {Fang}, K. and {Farrar}, G. and {Fauth}, A.~C. and {Fazzini}, N. and {Ferguson}, A.~P. and {Fick}, B. and {Figueira}, J.~M. and {Filevich}, A. and {Filip{\v{c}}i{\v{c}}}, A. and {Fratu}, O. and {Freire}, M.~M. and {Fujii}, T. and {Garc{\'\i}a}, B. and {Garcia-Gamez}, D. and {Garcia-Pinto}, D. and {Gate}, F. and {Gemmeke}, H. and {Gherghel-Lascu}, A. and {Ghia}, P.~L. and {Giaccari}, U. and {Giammarchi}, M. and {Giller}, M. and {G{\l}as}, D. and {Glaser}, C. and {Glass}, H. and {Golup}, G. and {G{\'o}mez Berisso}, M. and {G{\'o}mez Vitale}, P.~F. and {Gonz{\'a}lez}, N. and {Gookin}, B. and {Gordon}, J. and {Gorgi}, A. and {Gorham}, P. and {Gouffon}, P. and {Griffith}, N. and {Grillo}, A.~F. and {Grubb}, T.~D. and {Guarino}, F. and {Guedes}, G.~P. and {Hampel}, M.~R. and {Hansen}, P. and {Harari}, D. and {Harrison}, T.~A. and {Hartmann}, S. and {Harton}, J.~L. and {Haungs}, A. and {Hebbeker}, T. and {Heck}, D. and {Heimann}, P. and {Herve}, A.~E. and {Hill}, G.~C. and {Hojvat}, C. and {Hollon}, N. and {Holt}, E. and {Homola}, P. and {H{\"o}randel}, J.~R. and {Horvath}, P. and {Hrabovsk{\'y}}, M. and {Huber}, D. and {Huege}, T. and {Insolia}, A. and {Isar}, P.~G. and {Jandt}, I. and {Jansen}, S. and {Jarne}, C. and {Johnsen}, J.~A. and {Josebachuili}, M. and {K{\"a}{\"a}p{\"a}}, A. and {Kambeitz}, O. and {Kampert}, K.~H. and {Kasper}, P. and {Katkov}, I. and {Keilhauer}, B. and {Kemp}, E. and {Kieckhafer}, R.~M. and {Klages}, H.~O. and {Kleifges}, M. and {Kleinfeller}, J. and {Krause}, R. and {Krohm}, N. and {Kuempel}, D. and {Kukec Mezek}, G. and {Kunka}, N. and {Kuotb Awad}, A.~W. and {LaHurd}, D. and {Latronico}, L. and {Lauer}, R. and {Lauscher}, M. and {Lautridou}, P. and {Le Coz}, S. and {Lebrun}, D. and {Lebrun}, P. and {Leigui de Oliveira}, M.~A. and {Letessier-Selvon}, A. and {Lhenry-Yvon}, I. and {Link}, K. and {Lopes}, L. and {L{\'o}pez}, R. and {L{\'o}pez Casado}, A. and {Louedec}, K. and {Lucero}, A. and {Malacari}, M. and {Mallamaci}, M. and {Maller}, J. and {Mandat}, D. and {Mantsch}, P. and {Mariazzi}, A.~G. and {Marin}, V. and {Mari{\c{s}}}, I.~C. and {Marsella}, G. and {Martello}, D. and {Martinez}, H. and {Mart{\'\i}nez Bravo}, O. and {Martraire}, D. and {Mas{\'\i}as Meza}, J.~J. and {Mathes}, H.~J. and {Mathys}, S. and {Matthews}, J. and {Matthews}, J.~A.~J. and {Matthiae}, G. and {Maurizio}, D. and {Mayotte}, E. and {Mazur}, P.~O. and {Medina}, C. and {Medina-Tanco}, G. and {Meissner}, R. and {Mello}, V.~B.~B. and {Melo}, D. and {Menshikov}, A. and {Messina}, S. and {Micheletti}, M.~I. and {Middendorf}, L. and {Minaya}, I.~A. and {Miramonti}, L. and {Mitrica}, B. and {Molina-Bueno}, L. and {Mollerach}, S. and {Montanet}, F. and {Morello}, C. and {Mostaf{\'a}}, M. and {Moura}, C.~A. and {Muller}, M.~A. and {M{\"u}ller}, G. and {M{\"u}ller}, S. and {Navas}, S. and {Necesal}, P. and {Nellen}, L. and {Nelles}, A. and {Neuser}, J. and {Nguyen}, P.~H. and {Niculescu-Oglinzanu}, M. and {Niechciol}, M. and {Niemietz}, L. and {Niggemann}, T. and {Nitz}, D. and {Nosek}, D. and {Novotny}, V. and {No{\v{z}}ka}, L. and {N{\'u}{\~n}ez}, L.~A. and {Ochilo}, L. and {Oikonomou}, F. and {Olinto}, A. and {Pacheco}, N. and {Pakk Selmi-Dei}, D. and {Palatka}, M. and {Pallotta}, J. and {Papenbreer}, P. and {Parente}, G. and {Parra}, A. and {Paul}, T. and {Pech}, M. and {P{\k{e}}kala}, J. and {Pelayo}, R. and {Pepe}, I.~M. and {Perrone}, L. and {Petermann}, E. and {Peters}, C. and {Petrera}, S. and {Petrov}, Y. and {Phuntsok}, J. and {Piegaia}, R. and {Pierog}, T. and {Pieroni}, P. and {Pimenta}, M. and {Pirronello}, V. and {Platino}, M. and {Plum}, M. and {Porcelli}, A. and {Porowski}, C. and {Prado}, R.~R. and {Privitera}, P. and {Prouza}, M. and {Quel}, E.~J. and {Querchfeld}, S. and {Quinn}, S. and {Rautenberg}, J. and {Ravel}, O. and {Ravignani}, D. and {Reinert}, D. and {Revenu}, B. and {Ridky}, J. and {Risse}, M. and {Ristori}, P. and {Rizi}, V. and {Rodrigues de Carvalho}, W. and {Rodriguez Rojo}, J. and {Rodr{\'\i}guez-Fr{\'\i}as}, M.~D. and {Rogozin}, D. and {Rosado}, J. and {Roth}, M. and {Roulet}, E. and {Rovero}, A.~C. and {Saffi}, S.~J. and {Saftoiu}, A. and {Salazar}, H. and {Saleh}, A. and {Salesa Greus}, F. and {Salina}, G. and {Sanabria Gomez}, J.~D. and {S{\'a}nchez}, F. and {Sanchez-Lucas}, P. and {Santos}, E. and {Santos}, E.~M. and {Sarazin}, F. and {Sarkar}, B. and {Sarmento}, R. and {Sarmiento-Cano}, C. and {Sato}, R. and {Scarso}, C. and {Schauer}, M. and {Scherini}, V. and {Schieler}, H. and {Schmidt}, D. and {Scholten}, O. and {Schoorlemmer}, H. and {Schov{\'a}nek}, P. and {Schr{\"o}der}, F.~G. and {Schulz}, A. and {Schulz}, J. and {Schumacher}, J. and {Sciutto}, S.~J. and {Segreto}, A. and {Settimo}, M. and {Shadkam}, A. and {Shellard}, R.~C. and {Sigl}, G. and {Sima}, O. and {{\'S}mia{\l}kowski}, A. and {{\v{S}}m{\'\i}da}, R. and {Snow}, G.~R. and {Sommers}, P. and {Sonntag}, S. and {Sorokin}, J. and {Squartini}, R. and {Srivastava}, Y.~N. and {Stanca}, D. and {Stani{\v{c}}}, S. and {Stapleton}, J. and {Stasielak}, J. and {Stephan}, M. and {Stutz}, A. and {Suarez}, F. and {Suarez Dur{\'a}n}, M. and {Suomij{\"a}rvi}, T. and {Supanitsky}, A.~D. and {Sutherland}, M.~S. and {Swain}, J. and {Szadkowski}, Z. and {Taborda}, O.~A. and {Tapia}, A. and {Tepe}, A. and {Theodoro}, V.~M. and {Timmermans}, C. and {Todero Peixoto}, C.~J. and {Toma}, G. and {Tomankova}, L. and {Tom{\'e}}, B. and {Tonachini}, A. and {Torralba Elipe}, G. and {Torres Machado}, D. and {Travnicek}, P. and {Trini}, M. and {Ulrich}, R. and {Unger}, M. and {Urban}, M. and {Vald{\'e}s Galicia}, J.~F. and {Vali{\~n}o}, I. and {Valore}, L. and {van Aar}, G. and {van Bodegom}, P. and {van den Berg}, A.~M. and {van Velzen}, S. and {van Vliet}, A. and {Varela}, E. and {Vargas C{\'a}rdenas}, B. and {Varner}, G. and {Vasquez}, R. and {V{\'a}zquez}, J.~R. and {V{\'a}zquez}, R.~A. and {Veberi{\v{c}}}, D. and {Verzi}, V. and {Vicha}, J. and {Videla}, M. and {Villase{\~n}or}, L. and {Vlcek}, B. and {Vorobiov}, S. and {Wahlberg}, H. and {Wainberg}, O. and {Walz}, D. and {Watson}, A.~A. and {Weber}, M. and {Weidenhaupt}, K. and {Weindl}, A. and {Welling}, C. and {Werner}, F. and {Widom}, A. and {Wiencke}, L. and {Wilczy{\'n}ski}, H. and {Winchen}, T. and {Wittkowski}, D. and {Wundheiler}, B. and {Wykes}, S. and {Yang}, L. and {Yapici}, T. and {Yushkov}, A. and {Zas}, E. and {Zavrtanik}, D. and {Zavrtanik}, M. and {Zepeda}, A. and {Zimmermann}, B. and {Ziolkowski}, M. and {Zuccarello}, F. and {Pierre Auger Collaboration}},
       title = "{Energy estimation of cosmic rays with the Engineering Radio Array of the Pierre Auger Observatory}",
    journal  = "Phys. Rev.",
    volume  = "D93",
archivePrefix = "arXiv",
   eprint = {1508.04267},
 primaryClass = "astro-ph.HE",
     year = 2016,
    month = jun,
   number = 12,
      eid = {122005},
    pages = {122005},
      doi = {10.1103/PhysRevD.93.122005},
   adsurl = {http://adsabs.harvard.edu/abs/2016PhRvD..93l2005A}
}

@ARTICLE{2020SCPMA..6319501A,
       author = {{{\'A}lvarez-Mu{\~n}iz}, Jaime and {Alves Batista}, Rafael and {Balagopal V.}, Aswathi and {Bolmont}, Julien and {Bustamante}, Mauricio and {Carvalho}, Washington and {Charrier}, Didier and {Cognard}, Isma{\"e}l and {Decoene}, Valentin and {Denton}, Peter B. and {De Jong}, Sijbrand and {De Vries}, Krijn D. and {Engel}, Ralph and {Fang}, Ke and {Finley}, Chad and {Gabici}, Stefano and {Gou}, QuanBu and {Gu}, JunHua and {Gu{\'e}pin}, Claire and {Hu}, HongBo and {Huang}, Yan and {Kotera}, Kumiko and {Le Coz}, Sandra and {Lenain}, Jean-Philippe and {L{\"u}}, GuoLiang and {Martineau-Huynh}, Olivier and {Mostaf{\'a}}, Miguel and {Mottez}, Fabrice and {Murase}, Kohta and {Niess}, Valentin and {Oikonomou}, Foteini and {Pierog}, Tanguy and {Qian}, XiangLi and {Qin}, Bo and {Ran}, Duan and {Renault-Tinacci}, Nicolas and {Roth}, Markus and {Schr{\"o}der}, Frank G. and {Sch{\"u}ssler}, Fabian and {Tasse}, Cyril and {Timmermans}, Charles and {Tueros}, Mat{\'\i}as and {Wu}, XiangPing and {Zarka}, Philippe and {Zech}, Andreas and {Zhang}, B. Theodore and {Zhang}, JianLi and {Zhang}, Yi and {Zheng}, Qian and {Zilles}, Anne},
        title = "{The Giant Radio Array for Neutrino Detection (GRAND): Science and design}",
      journal = {Science China Physics, Mechanics, and Astronomy},
         year = 2020,
        month = jan,
       volume = {63},
       number = {1},
          eid = {219501},
        pages = {219501},
          doi = {10.1007/s11433-018-9385-7},
archivePrefix = {arXiv},
       eprint = {1810.09994},
 primaryClass = {astro-ph.HE},
       adsurl = {https://ui.adsabs.harvard.edu/abs/2020SCPMA..6319501A}
}

@ARTICLE{Romero-Wolf_2013,
   author = {{Romero-Wolf}, A. and {Gorham}, P. and {Liewer}, K. and {Booth}, J. and 
	{Duren}, R.},
    title = "{Concept and Analysis of a Satellite for Space-based Radio Detection of Ultra-high Energy Cosmic Rays}",
  journal = {ArXiv e-prints},
archivePrefix = "arXiv",
   eprint = {1302.1263},
 primaryClass = "astro-ph.IM",
     year = 2013,
    month = feb,
   adsurl = {http://adsabs.harvard.edu/abs/2013arXiv1302.1263R}
}

@Book{Thomson_Moran_Swenson_2017,
  author =	 {{Thomson}, A. R. and {Moran}, J. M. and {Swenson Jr.}, G. W.},
  title = 	 {Interferometry and Synthesis in Radio Astronomy, Third Edition},
  publisher = 	 {Springer},
  year = 	 2017
}

@Article{Nerling-2006-APP-24-421,
       author = {{Nerling}, F. and {Bl{\"u}mer}, J. and {Engel}, R. and {Risse}, M.},
        title = "{Universality of electron distributions in high-energy air showers{\textemdash}Description of Cherenkov light production}",
      journal = {Astroparticle Physics},
         year = 2006,
        month = jan,
       volume = {24},
       number = {6},
        pages = {421-437},
          doi = {10.1016/j.astropartphys.2005.09.002},
archivePrefix = {arXiv},
       eprint = {astro-ph/0506729},
 primaryClass = {astro-ph},
       adsurl = {https://ui.adsabs.harvard.edu/abs/2006APh....24..421N}
}

@Article{Giller-2004-JPG-30-97,
       author = {{Giller}, M. and {Wieczorek}, G. and {Kacperczyk}, A. and {Stojek}, H. and {Tkaczyk}, W.},
        title = "{Energy spectra of electrons in the extensive air showers of ultra-high energy}",
      journal = {Journal of Physics G Nuclear Physics},
         year = 2004,
        month = feb,
       volume = {30},
       number = {2},
        pages = {97-105},
          doi = {10.1088/0954-3899/30/2/009},
       adsurl = {https://ui.adsabs.harvard.edu/abs/2004JPhG...30...97G}
}

@Article{Lafebre-2009-APP-31-243,
       author = {{Lafebre}, S. and {Engel}, R. and {Falcke}, H. and {H{\"o}randel}, J. and {Huege}, T. and {Kuijpers}, J. and {Ulrich}, R.},
        title = "{Universality of electron-positron distributions in extensive air showers}",
      journal = {Astroparticle Physics},
         year = 2009,
        month = apr,
       volume = {31},
       number = {3},
        pages = {243-254},
          doi = {10.1016/j.astropartphys.2009.02.002},
archivePrefix = {arXiv},
       eprint = {0902.0548},
 primaryClass = {astro-ph.HE},
       adsurl = {https://ui.adsabs.harvard.edu/abs/2009APh....31..243L}
}

@InProceedings{Bergman-2013-ICRC-33-0983,
  author = 	 {Douglas Bergman},
  title = 	 {An Efficient Technique for the Reconstruction of
                  Extensive Air Showers using Non-Imaging Cherenkov
                  Measurements},
  booktitle = {Proceedings of the 33rd ICRC (Rio de Janeiro)},
  year = 	 2013,
  number = 	 00983,
  
}

@ARTICLE{Hillas1,
   author = {{Hillas}, A.~M.},
   title = "{The sensitivity of Cerenkov radiation pulses to the longitudinal development of cosmic-ray showers}",
   journal = {J. Phys. G: Nucl. Phys.},
   year = 1982,
   volume = 8,
   pages = {1475 - 1492}
}

@ARTICLE{Hillas2,
   author = {{Hillas}, A.~M.},
   title = "{Angular and energy distributions of charged particles in electron-photon cascades in air}",
   journal = {J. Phys. G: Nucl. Phys.},
   year = 1982,
   volume = 8,
   pages = {1461 - 1473}
}

@INPROCEEDINGS{2019ICRC...36..936K,
       author = {{Krizmanic}, J.~F. and {Akaike}, Y. and {Bergman}, D. and {Eser}, J. and
         {Patel}, S. and {Romero-Wolf}, A. and {Reno}, M.~H. and
         {Sarazin}, F. and {Venters}, T. and {Anchordoqui}, L.~A. and
         {Mackovjak}, S. and {Olinto}, A. and {Wiencke}, L. and {Wissel}, S. and
         {Reustle}, A.},
        title = "{nuSpaceSim: A Comprehensive Neutrino Simulation Package for Space-based \& Suborbital Experiments}",
    booktitle = {36th International Cosmic Ray Conference (ICRC2019)},
         year = 2019,
       series = {International Cosmic Ray Conference},
       volume = {36},
        month = jul,
          eid = {936},
        pages = {936},
       adsurl = {https://ui.adsabs.harvard.edu/abs/2019ICRC...36..936K}
}

@INPROCEEDINGS{2021ICRC...37..1205K,
       author = {{Krizmanic}, J.~F. and {Akaike}, Y. and {Anchordoqui}, L.~A.  and {Bergman}, D. and {Buckland}, I. and {Cummings}, A. and {Eser}, J. and {Guepin}, C. and {Mackovjak}, S. and {Olinto}, A. and {Paul}, T. and  
         {Patel}, S. and {Reustle}, A. and {Romero-Wolf}, A. and {Reno}, M.~H. and
         {Sarazin}, F. and {Venters}, T. and and
          {Wiencke}, L. and {Wissel}, S.},
        title = "{nuSpaceSim: A Comprehensive Simulation for the Modeling of Optical and Radio Signals from Extensive Air Showers Induced by Cosmic Neutrinos for Space-based Experiments}",
    booktitle = {37th International Cosmic Ray Conference (ICRC2021)},
         year = 2021,
       series = {International Cosmic Ray Conference},
       volume = {37},
        month = jul,
          eid = {1205},
        pages = {1205},
}

@INPROCEEDINGS{2021ICRC...37..1203P,
       author = {{Patel}, S. and {Reno}, M.~H. and {nuSpaceSim Collaboration}},
        title = "{Monte Carlo simulations of neutrino and charged lepton propagation in the Earth with nuPyProp}",
    booktitle = {37th International Cosmic Ray Conference (ICRC2021)},
         year = 2021,
       series = {International Cosmic Ray Conference},
       volume = {37},
        month = jul,
          eid = {1203},
        pages = {1203},
}

@INPROCEEDINGS{2021ICRC...37..316K,
       author = {{Klimov}, P. and {Sharakin}, S. and {Zotov}, M. and {Bertaina}, M. and. {Fenu}, F. and {Lomonosov-UHECR/TLE Collaboration}},
        title = "{Main results of the TUS experiment on board the Lomonosov satellite}",
    booktitle = {37th International Cosmic Ray Conference (ICRC2021)},
         year = 2021,
       series = {International Cosmic Ray Conference},
       volume = {37},
        month = jul,
          eid = {316},
        pages = {316},
}

@INPROCEEDINGS{2021ICRC...37..1031P,
       author = {{Romero-Wolf}, A. and {nuSpaceSim Collaboration}},
        title = "{Radio Simulations of Upgoing Extensive Air Showers Observed from Low-Earth Orbit}",
    booktitle = {37th International Cosmic Ray Conference (ICRC2021)},
         year = 2021,
       series = {International Cosmic Ray Conference},
       volume = {37},
        month = jul,
          eid = {1031},
        pages = {1031},
}

@INPROCEEDINGS{2021ICRC...37..0387P,
       author = {{Pont}, B. and {Auger Collaboration}},
        title = "{The depth of the shower maximum of air showers measured with AERA}",
    booktitle = {37th International Cosmic Ray Conference (ICRC2021)},
         year = 2021,
       series = {International Cosmic Ray Conference},
       volume = {37},
        month = jul,
          eid = {0387},
        pages = {0387},
}

@article{OTTE2019283,
title = {Locating the avalanche structure and the origin of breakdown generating charge carriers in silicon photomultipliers by using the bias dependent breakdown probability},
journal = {Nuclear Instruments and Methods in Physics Research Section A: Accelerators, Spectrometers, Detectors and Associated Equipment},
volume = {916},
pages = {283-289},
year = {2019},
issn = {0168-9002},
doi = {https://doi.org/10.1016/j.nima.2018.11.086},
url = {https://www.sciencedirect.com/science/article/pii/S0168900218317133},
author = {Adam Nepomuk Otte and Thanh Nguyen and Joel Stansbury}
}

@article{allm,
    author = "Abramowicz, H. and Levy, A.",
    title = "{The ALLM parameterization of sigma(tot)(gamma* p): An Update}",
    eprint = "hep-ph/9712415",
    archivePrefix = "arXiv",
    reportNumber = "DESY-97-251",
    month = "12",
    year = "1997"
}

@article{reno_beta,
	title = {Propagation of muons and taus at high energies},
	author = {Iyer Dutta, S. and Reno, M. H. and Sarcevic, I. and Seckel, D.},
	journal = {Phys. Rev. D},
	volume = {63},
	issue = {9},
	pages = {094020},
	numpages = {10},
	year = {2001},
	month = {Apr},
	publisher = {American Physical Society},
	doi = {10.1103/PhysRevD.63.094020},
	url = {https://link.aps.org/doi/10.1103/PhysRevD.63.094020}
}

@article{hls_param,
  title = {Quark mass effects in high energy neutrino-nucleon scattering},
  author = {Jeong, Yu Seon and Reno, Mary Hall},
  journal = {Phys. Rev. D},
  volume = {81},
  issue = {11},
  pages = {114012},
  numpages = {8},
  year = {2010},
  month = {Jun},
  publisher = {American Physical Society},
  doi = {10.1103/PhysRevD.81.114012},
  url = {https://link.aps.org/doi/10.1103/PhysRevD.81.114012}
}

@article{cteq18_nlo,
  author = {Hou, Tie-Jiun and Gao, Jun and Hobbs, T. J. and Xie, Keping and Dulat, Sayipjamal and Guzzi, Marco and Huston, Joey and Nadolsky, Pavel and Pumplin, Jon and Schmidt, Carl and Sitiwaldi, Ibrahim and Stump, Daniel and Yuan, C.-P.},
    title = "{New CTEQ global analysis of quantum chromodynamics with high-precision data from the LHC}",
    eprint = "1912.10053",
    archivePrefix = "arXiv",
    primaryClass = "hep-ph",
    reportNumber = "MSUHEP-19-025, PITT-PACC-1911, SMU-HEP-19-03",
    doi = "10.1103/PhysRevD.103.014013",
    journal = "Phys. Rev. D",
    volume = "103",
    number = "1",
    pages = "014013",
    year = "2021"
}

@article{PREM,
	author = {Adam M. Dziewonski and Don L. Anderson},
	doi = {10.1016/0031-9201(81)90046-7},
	journal = {Physics of the Earth and Planetary Interiors},
	number = {4},
	pages = {297–356},
	title = {Preliminary reference Earth model},
	url = {https://www.sciencedirect.com/science/article/pii/0031920181900467},
	volume = {25},
	year = {1981}
}

@article{Landau:1953um,
    author = "Landau, L. D. and Pomeranchuk, I.",
    title = "{Limits of applicability of the theory of bremsstrahlung electrons and pair production at high-energies}",
    journal = "Dokl. Akad. Nauk Ser. Fiz.",
    volume = "92",
    pages = "535--536",
    year = "1953"
}

@ARTICLE{1956PhRv..103.1811M,
       author = {{Migdal}, A.~B.},
        title = "{Bremsstrahlung and Pair Production in Condensed Media at High Energies}",
      journal = {Physical Review},
         year = 1956,
        month = sep,
       volume = {103},
       number = {6},
        pages = {1811-1820},
          doi = {10.1103/PhysRev.103.1811},
       adsurl = {https://ui.adsabs.harvard.edu/abs/1956PhRv..103.1811M}
}

@ARTICLE{2020arXiv201002892A,
       author = {{Abarr}, Q. and {Allison}, P. and {Ammerman Yebra}, J. and {Alvarez-Mu{\~n}iz}, J. and {Beatty}, J.~J. and {Besson}, D.~Z. and {Chen}, P. and {Chen}, Y. and {Cheng}, X. and {Clem}, J.~M. and {Connolly}, A. and {Cremonesi}, L. and {Deaconu}, C. and {Flaherty}, J. and {Frikken}, D. and {Gorham}, P.~W. and {Hast}, C. and {Hornhuber}, C. and {Huang}, J.~J. and {Hughes}, K. and {Hynous}, A. and {Ku}, Y. and {Kuo}, C. -Y. and {Liu}, T.~C. and {Martin}, Z. and {Miki}, C. and {Nam}, J. and {Nichol}, R.~J. and {Nishimura}, K. and {Novikov}, A. and {Nozdrina}, A. and {Oberla}, E. and {Prohira}, S. and {Prechelt}, R. and {Rauch}, B.~F. and {Roberts}, J.~M. and {Romero-Wolf}, A. and {Russell}, J.~W. and {Seckel}, D. and {Shiao}, J. and {Smith}, D. and {Southall}, D. and {Varner}, G.~S. and {Vieregg}, A.~G. and {Wang}, S. -H. and {Wang}, Y. -H. and {Wissel}, S.~A. and {Young}, R. and {Zas}, E. and {Zeolla}, A.},
        title = "{The Payload for Ultrahigh Energy Observations (PUEO): A White Paper}",
      journal = {arXiv e-prints},
         year = 2020,
        month = oct,
          eid = {arXiv:2010.02892},
        pages = {arXiv:2010.02892},
archivePrefix = {arXiv},
       eprint = {2010.02892},
 primaryClass = {astro-ph.IM},
       adsurl = {https://ui.adsabs.harvard.edu/abs/2020arXiv201002892A}
}

@ARTICLE{2020PhRvD.102l3013V,
       author = {{Venters}, Tonia M. and {Reno}, Mary Hall and {Krizmanic}, John F. and {Anchordoqui}, Luis A. and {Gu{\'e}pin}, Claire and {Olinto}, Angela V.},
        title = "{POEMMA's target-of-opportunity sensitivity to cosmic neutrino transient sources}",
      journal = {Phys. Rev. D},
         year = 2020,
        month = dec,
       volume = {102},
       number = {12},
          eid = {123013},
        pages = {123013},
          doi = {10.1103/PhysRevD.102.123013},
archivePrefix = {arXiv},
       eprint = {1906.07209},
 primaryClass = {astro-ph.HE},
       adsurl = {https://ui.adsabs.harvard.edu/abs/2020PhRvD.102l3013V}
}

@ARTICLE{2019PhRvD.100f3010R,
       author = {{Reno}, Mary Hall and {Krizmanic}, John F. and {Venters}, Tonia M.},
        title = "{Cosmic tau neutrino detection via Cherenkov signals from air showers from Earth-emerging taus}",
      journal = {Phys. Rev. D},
         year = 2019,
        month = sep,
       volume = {100},
       number = {6},
          eid = {063010},
        pages = {063010},
          doi = {10.1103/PhysRevD.100.063010},
       adsurl = {https://ui.adsabs.harvard.edu/abs/2019PhRvD.100f3010R}
}

@ARTICLE{2020arXiv200805690A,
       author = {{ANITA Collaboration} and {Gorham}, P.~W. and {Ludwig}, A. and {Deaconu}, C. and {Cao}, P. and {Allison}, P. and {Banerjee}, O. and {Batten}, L. and {Bhattacharya}, D. and {Beatty}, J.~J. and {Belov}, K. and {Binns}, W.~R. and {Bugaev}, V. and {Chen}, C.~H. and {Chen}, P. and {Chen}, Y. and {Clem}, J.~M. and {Cremonesi}, L. and {Dailey}, B. and {Dowkontt}, P.~F. and {Fox}, B.~D. and {Gordon}, J.~W.~H. and {Hast}, C. and {Hill}, B. and {Hsu}, S.~Y. and {Huang}, J.~J. and {Hughes}, K. and {Hupe}, R. and {Israel}, M.~H. and {Liu}, T.~C. and {Macchiarulo}, L. and {Matsuno}, S. and {McBride}, K. and {Miki}, C. and {Nam}, J. and {Naudet}, C.~J. and {Nichol}, R.~J. and {Novikov}, A. and {Oberla}, E. and {Olmedo}, M. and {Prechelt}, R. and {Prohira}, S. and {Rauch}, B.~F. and {Roberts}, J.~M. and {Romero-Wolf}, A. and {Rotter}, B. and {Russell}, J.~W. and {Saltzberg}, D. and {Seckel}, D. and {Schoorlemmer}, H. and {Shiao}, J. and {Stafford}, S. and {Stockham}, J. and {Stockham}, M. and {Strutt}, B. and {Sutherland}, M.~S. and {Varner}, G.~S. and {Vieregg}, A.~G. and {Wang}, S.~H. and {Wissel}, S.~A.},
        title = "{Unusual Near-horizon Cosmic-ray-like Events Observed by ANITA-IV}",
      journal = {arXiv e-prints},
         year = 2020,
        month = aug,
          eid = {arXiv:2008.05690},
        pages = {arXiv:2008.05690},
archivePrefix = {arXiv},
       eprint = {2008.05690},
 primaryClass = {astro-ph.HE},
       adsurl = {https://ui.adsabs.harvard.edu/abs/2020arXiv200805690A}
}

@ARTICLE{2021ApJS..253...36B,
       author = {{Bacholle}, S. and {Barrillon}, P. and {Battisti}, M. and {Belov}, A. and {Bertaina}, M. and {Bisconti}, F. and {Blaksley}, C. and {Blin-Bondil}, S. and {Cafagna}, F. and {Cambi{\`e}}, G. and {Capel}, F. and {Casolino}, M. and {Crisconio}, M. and {Churilo}, I. and {Cotto}, G. and {de la Taille}, C. and {Djakonow}, A. and {Ebisuzaki}, T. and {Fenu}, F. and {Franceschi}, A. and {Fuglesang}, C. and {Gorodetzky}, P. and {Haungs}, A. and {Kajino}, F. and {Kasuga}, H. and {Khrenov}, B. and {Klimov}, P. and {Kochepasov}, S. and {Kuznetsov}, V. and {Marcelli}, L. and {Marsza{\l}}, W. and {Mignone}, M. and {Mascetti}, G. and {Miyamoto}, H. and {Murashov}, A. and {Napolitano}, T. and {Olinto}, A.~V. and {Ohmori}, H. and {Osteria}, G. and {Panasyuk}, M. and {Porfilio}, M. and {Poroshin}, A. and {Parizot}, E. and {Picozza}, P. and {Piotrowski}, L.~W. and {Plebaniak}, Z. and {Pr{\'e}v{\^o}t}, G. and {Przybylak}, M. and {Reali}, E. and {Ricci}, M. and {Sakaki}, N. and {Shinozaki}, K. and {Szabelski}, J. and {Takizawa}, Y. and {Turriziani}, S. and {Tra{\"\i}che}, M. and {Valentini}, G. and {Wada}, S. and {Wiencke}, L. and {Yashin}, I. and {Zuccaro-Marchi}, A.},
        title = "{Mini-EUSO Mission to Study Earth UV Emissions on board the ISS}",
      journal = {The Astrophysical Journal Supplement Series},
         year = 2021,
        month = apr,
       volume = {253},
       number = {2},
          eid = {36},
        pages = {36},
          doi = {10.3847/1538-4365/abd93d},
archivePrefix = {arXiv},
       eprint = {2010.01937},
 primaryClass = {astro-ph.IM},
       adsurl = {https://ui.adsabs.harvard.edu/abs/2021ApJS..253...36B}
}

@ARTICLE{2004PhRvD..69a3008L,
       author = {{Lehtinen}, Nikolai G. and {Gorham}, Peter W. and {Jacobson}, Abram R. and {Roussel-Dupr{\'e}}, Robert A.},
        title = "{FORTE satellite constraints on ultrahigh energy cosmic particle fluxes}",
      journal = {Phys. Rev. D},
         year = 2004,
        month = jan,
       volume = {69},
       number = {1},
          eid = {013008},
        pages = {013008},
          doi = {10.1103/PhysRevD.69.013008},
archivePrefix = {arXiv},
       eprint = {astro-ph/0309656},
 primaryClass = {astro-ph},
       adsurl = {https://ui.adsabs.harvard.edu/abs/2004PhRvD..69a3008L}
}

@ARTICLE{2021PhRvD.103d3017C,
       author = {{Cummings}, A.~L. and {Aloisio}, R. and {Krizmanic}, J.~F.},
        title = "{Modeling of the tau and muon neutrino-induced optical Cherenkov signals from upward-moving extensive air showers}",
      journal = {Phys. Rev. D.},
         year = 2021,
        month = feb,
       volume = {103},
       number = {4},
          eid = {043017},
        pages = {043017},
          doi = {10.1103/PhysRevD.103.043017},
archivePrefix = {arXiv},
       eprint = {2011.09869},
 primaryClass = {astro-ph.HE},
       adsurl = {https://ui.adsabs.harvard.edu/abs/2021PhRvD.103d3017C}
}

@ARTICLE{2021arXiv210503255C,
       author = {{Cummings}, Austin and {Aloisio}, Roberto and {Eser}, Johannes and {Krizmanic}, John},
        title = "{Modeling the Optical Cherenkov Signals by Cosmic Ray Extensive Air Showers Directly Observed from Sub-Orbital and Orbital Altitudes}",
      journal = {arXiv e-prints},
         year = 2021,
        month = may,
          eid = {arXiv:2105.03255},
        pages = {arXiv:2105.03255},
archivePrefix = {arXiv},
       eprint = {2105.03255},
 primaryClass = {astro-ph.IM},
       adsurl = {https://ui.adsabs.harvard.edu/abs/2021arXiv210503255C}
}

@ARTICLE{2021NIMPA.98564614K,
       author = {{Krizmanic}, John F. and {POEMMA Collaboration<Sup></Sup>}},
        title = "{Space-based extensive air shower optical Cherenkov and fluorescence measurements using SiPM detectors in context of POEMMA}",
      journal = {Nuclear Instruments and Methods in Physics Research A},
         year = 2021,
        month = jan,
       volume = {985},
          eid = {164614},
        pages = {164614},
          doi = {10.1016/j.nima.2020.164614},
archivePrefix = {arXiv},
       eprint = {2008.04984},
 primaryClass = {astro-ph.IM},
       adsurl = {https://ui.adsabs.harvard.edu/abs/2021NIMPA.98564614K}
}

@misc{Fermi,
      note         = {https://fermi.gsfc.nasa.gov/}
}

@ARTICLE{2009ApJ...697.1071A,
       author = {{Atwood}, W.~B. and {Abdo}, A.~A. and {Ackermann}, M. and {Althouse}, W. and {Anderson}, B. and {Axelsson}, M. and {Baldini}, L. and {Ballet}, J. and {Band}, D.~L. and {Barbiellini}, G. and {Bartelt}, J. and {Bastieri}, D. and {Baughman}, B.~M. and {Bechtol}, K. and {B{\'e}d{\'e}r{\`e}de}, D. and {Bellardi}, F. and {Bellazzini}, R. and {Berenji}, B. and {Bignami}, G.~F. and {Bisello}, D. and {Bissaldi}, E. and {Blandford}, R.~D. and {Bloom}, E.~D. and {Bogart}, J.~R. and {Bonamente}, E. and {Bonnell}, J. and {Borgland}, A.~W. and {Bouvier}, A. and {Bregeon}, J. and {Brez}, A. and {Brigida}, M. and {Bruel}, P. and {Burnett}, T.~H. and {Busetto}, G. and {Caliandro}, G.~A. and {Cameron}, R.~A. and {Caraveo}, P.~A. and {Carius}, S. and {Carlson}, P. and {Casandjian}, J.~M. and {Cavazzuti}, E. and {Ceccanti}, M. and {Cecchi}, C. and {Charles}, E. and {Chekhtman}, A. and {Cheung}, C.~C. and {Chiang}, J. and {Chipaux}, R. and {Cillis}, A.~N. and {Ciprini}, S. and {Claus}, R. and {Cohen-Tanugi}, J. and {Condamoor}, S. and {Conrad}, J. and {Corbet}, R. and {Corucci}, L. and {Costamante}, L. and {Cutini}, S. and {Davis}, D.~S. and {Decotigny}, D. and {DeKlotz}, M. and {Dermer}, C.~D. and {de Angelis}, A. and {Digel}, S.~W. and {do Couto e Silva}, E. and {Drell}, P.~S. and {Dubois}, R. and {Dumora}, D. and {Edmonds}, Y. and {Fabiani}, D. and {Farnier}, C. and {Favuzzi}, C. and {Flath}, D.~L. and {Fleury}, P. and {Focke}, W.~B. and {Funk}, S. and {Fusco}, P. and {Gargano}, F. and {Gasparrini}, D. and {Gehrels}, N. and {Gentit}, F. -X. and {Germani}, S. and {Giebels}, B. and {Giglietto}, N. and {Giommi}, P. and {Giordano}, F. and {Glanzman}, T. and {Godfrey}, G. and {Grenier}, I.~A. and {Grondin}, M. -H. and {Grove}, J.~E. and {Guillemot}, L. and {Guiriec}, S. and {Haller}, G. and {Harding}, A.~K. and {Hart}, P.~A. and {Hays}, E. and {Healey}, S.~E. and {Hirayama}, M. and {Hjalmarsdotter}, L. and {Horn}, R. and {Hughes}, R.~E. and {J{\'o}hannesson}, G. and {Johansson}, G. and {Johnson}, A.~S. and {Johnson}, R.~P. and {Johnson}, T.~J. and {Johnson}, W.~N. and {Kamae}, T. and {Katagiri}, H. and {Kataoka}, J. and {Kavelaars}, A. and {Kawai}, N. and {Kelly}, H. and {Kerr}, M. and {Klamra}, W. and {Kn{\"o}dlseder}, J. and {Kocian}, M.~L. and {Komin}, N. and {Kuehn}, F. and {Kuss}, M. and {Landriu}, D. and {Latronico}, L. and {Lee}, B. and {Lee}, S. -H. and {Lemoine-Goumard}, M. and {Lionetto}, A.~M. and {Longo}, F. and {Loparco}, F. and {Lott}, B. and {Lovellette}, M.~N. and {Lubrano}, P. and {Madejski}, G.~M. and {Makeev}, A. and {Marangelli}, B. and {Massai}, M.~M. and {Mazziotta}, M.~N. and {McEnery}, J.~E. and {Menon}, N. and {Meurer}, C. and {Michelson}, P.~F. and {Minuti}, M. and {Mirizzi}, N. and {Mitthumsiri}, W. and {Mizuno}, T. and {Moiseev}, A.~A. and {Monte}, C. and {Monzani}, M.~E. and {Moretti}, E. and {Morselli}, A. and {Moskalenko}, I.~V. and {Murgia}, S. and {Nakamori}, T. and {Nishino}, S. and {Nolan}, P.~L. and {Norris}, J.~P. and {Nuss}, E. and {Ohno}, M. and {Ohsugi}, T. and {Omodei}, N. and {Orlando}, E. and {Ormes}, J.~F. and {Paccagnella}, A. and {Paneque}, D. and {Panetta}, J.~H. and {Parent}, D. and {Pearce}, M. and {Pepe}, M. and {Perazzo}, A. and {Pesce-Rollins}, M. and {Picozza}, P. and {Pieri}, L. and {Pinchera}, M. and {Piron}, F. and {Porter}, T.~A. and {Poupard}, L. and {Rain{\`o}}, S. and {Rando}, R. and {Rapposelli}, E. and {Razzano}, M. and {Reimer}, A. and {Reimer}, O. and {Reposeur}, T. and {Reyes}, L.~C. and {Ritz}, S. and {Rochester}, L.~S. and {Rodriguez}, A.~Y. and {Romani}, R.~W. and {Roth}, M. and {Russell}, J.~J. and {Ryde}, F. and {Sabatini}, S. and {Sadrozinski}, H.~F. -W. and {Sanchez}, D. and {Sander}, A. and {Sapozhnikov}, L. and {Parkinson}, P.~M. Saz and {Scargle}, J.~D. and {Schalk}, T.~L. and {Scolieri}, G. and {Sgr{\`o}}, C. and {Share}, G.~H. and {Shaw}, M. and {Shimokawabe}, T. and {Shrader}, C. and {Sierpowska-Bartosik}, A. and {Siskind}, E.~J. and {Smith}, D.~A. and {Smith}, P.~D. and {Spandre}, G. and {Spinelli}, P. and {Starck}, J. -L. and {Stephens}, T.~E. and {Strickman}, M.~S. and {Strong}, A.~W. and {Suson}, D.~J. and {Tajima}, H. and {Takahashi}, H. and {Takahashi}, T. and {Tanaka}, T. and {Tenze}, A. and {Tether}, S. and {Thayer}, J.~B. and {Thayer}, J.~G. and {Thompson}, D.~J. and {Tibaldo}, L. and {Tibolla}, O. and {Torres}, D.~F. and {Tosti}, G. and {Tramacere}, A. and {Turri}, M. and {Usher}, T.~L. and {Vilchez}, N. and {Vitale}, V. and {Wang}, P. and {Watters}, K. and {Winer}, B.~L. and {Wood}, K.~S. and {Ylinen}, T. and {Ziegler}, M.},
        title = "{The Large Area Telescope on the Fermi Gamma-Ray Space Telescope Mission}",
      journal = {The Astrophysical Journal},
         year = 2009,
        month = jun,
       volume = {697},
       number = {2},
        pages = {1071-1102},
          doi = {10.1088/0004-637X/697/2/1071},
archivePrefix = {arXiv},
       eprint = {0902.1089},
 primaryClass = {astro-ph.IM},
       adsurl = {https://ui.adsabs.harvard.edu/abs/2009ApJ...697.1071A}
}

@misc{PDG2,
      note         = {see Particle Data Group, Lepton Property Summary Tables, https://pdg.lbl.gov}
}

@misc{PDG,
      note         = {see Particle Data Group, ``Particle Detectors for Non-Accelerator Physics'', https://pdg.lbl.gov}
}

@ARTICLE{2021JCAP...06..007P,
       author = {{POEMMA Collaboration} and {Olinto}, A.~V. and {Krizmanic}, J. and {Adams}, J.~H. and {Aloisio}, R. and {Anchordoqui}, L.~A. and {Anzalone}, A. and {Bagheri}, M. and {Barghini}, D. and {Battisti}, M. and {Bergman}, D.~R. and {Bertaina}, M.~E. and {Bertone}, P.~F. and {Bisconti}, F. and {Bustamante}, M. and {Cafagna}, F. and {Caruso}, R. and {Casolino}, M. and {{\v{C}}ern{\'y}}, K. and {Christl}, M.~J. and {Cummings}, A.~L. and {de Mitri}, I. and {Diesing}, R. and {Engel}, R. and {Eser}, J. and {Fang}, K. and {Fenu}, F. and {Filippatos}, G. and {Gazda}, E. and {Guepin}, C. and {Haungs}, A. and {Hays}, E.~A. and {Judd}, E.~G. and {Klimov}, P. and {Kungel}, V. and {Kuznetsov}, E. and {Mackovjak}, {\v{S}}. and {Mand{\'a}t}, D. and {Marcelli}, L. and {McEnery}, J. and {Medina-Tanco}, G. and {Merenda}, K. -D. and {Meyer}, S.~S. and {Mitchell}, J.~W. and {Miyamoto}, H. and {Nachtman}, J.~M. and {Neronov}, A. and {Oikonomou}, F. and {Onel}, Y. and {Osteria}, G. and {Otte}, A.~N. and {Parizot}, E. and {Paul}, T. and {Pech}, M. and {Perkins}, J.~S. and {Picozza}, P. and {Piotrowski}, L.~W. and {Plebaniak}, Z. and {Pr{\'e}v{\^o}t}, G. and {Reardon}, P. and {Reno}, M.~H. and {Ricci}, M. and {Romero Matamala}, O. and {Sarazin}, F. and {Schov{\'a}nek}, P. and {Scotti}, V. and {Shinozaki}, K. and {Soriano}, J.~F. and {Stecker}, F. and {Takizawa}, Y. and {Ulrich}, R. and {Unger}, M. and {Venters}, T.~M. and {Wiencke}, L. and {Winn}, D. and {Young}, R.~M. and {Zotov}, M.},
        title = "{The POEMMA (Probe of Extreme Multi-Messenger Astrophysics) observatory}",
      journal = {Journal of Cosmology and Astroparticle Physics},
         year = 2021,
        month = jun,
       volume = {2021},
       number = {6},
          eid = {007},
        pages = {007},
          doi = {10.1088/1475-7516/2021/06/007},
       adsurl = {https://ui.adsabs.harvard.edu/abs/2021JCAP...06..007P}
}

@ARTICLE{Meier,
   author = {{Meier}, R.~R.},
    title = "{Ultraviolet spectroscopy and remote sensing of the upper atmosphere}",
  journal = {Space Science Reviews},
     year = 1991,
   volume = 58,
    pages = {1-185},
      doi = {10.1007/BF01206000},
   adsurl = {http://adsabs.harvard.edu/abs/1991SSRv...58....1M}
}

@ARTICLE{Leinert,
       author = {{Leinert}, Ch. and {Bowyer}, S. and {Haikala}, L.~K. and {Hanner}, M.~S. and {Hauser}, M.~G. and {Levasseur-Regourd}, A. -Ch. and {Mann}, I. and {Mattila}, K. and {Reach}, W.~T. and {Schlosser}, W. and {Staude}, H.~J. and {Toller}, G.~N. and {Weiland}, J.~L. and {Weinberg}, J.~L. and {Witt}, A.~N.},
    title = "{The 1997 reference of diffuse night sky brightness}",
  journal = {Astron. Astrophys. Suppl. Ser.},
     year = 1998,
    month = jan,
   volume = 127,
    pages = {1-99},
      doi = {10.1051/aas:1998105},
   adsurl = {http://adsabs.harvard.edu/abs/1998A%26AS..127....1L}
}

@ARTICLE{2006JGRA..11112307C,
       author = {{Cosby}, Philip C. and {Sharpee}, Brian D. and {Slanger}, Tom G. and {Huestis}, David L. and {Hanuschik}, Reinhard W.},
        title = "{High-resolution terrestrial nightglow emission line atlas from UVES/VLT: Positions, intensities, and identifications for 2808 lines at 314-1043 nm}",
      journal = {Journal of Geophysical Research (Space Physics)},
         year = "2006",
        month = "Dec",
       volume = {111},
       number = {A12},
          eid = {A12307},
        pages = {A12307},
          doi = {10.1029/2006JA012023},
       adsurl = {https://ui.adsabs.harvard.edu/abs/2006JGRA..11112307C}
}

@ARTICLE{2003A&A...407.1157H,
       author = {{Hanuschik}, R.~W.},
        title = "{A flux-calibrated, high-resolution atlas of optical sky emission from UVES}",
      journal = {Astronomy and Astrophysics},
         year = "2003",
        month = "Sep",
       volume = {407},
        pages = {1157-1164},
          doi = {10.1051/0004-6361:20030885},
       adsurl = {https://ui.adsabs.harvard.edu/abs/2003A&A...407.1157H}
}

@ARTICLE{Roach,
   author = {{Roach}, F.~E. and {Meinel}, A.~B.},
    title = "{~~}",
  journal = {ApJ},
     year = 1955,
   volume = 122,
    pages = {530}
}

@article{Kimura_2017,
	doi = {10.3847/2041-8213/aa8d14},
	url = {https://doi.org/10.3847/2041-8213/aa8d14},
	year = 2017,
	month = {oct},
	publisher = {American Astronomical Society},
	volume = {848},
	number = {1},
	pages = {L4},
	author = {Shigeo S. Kimura and Kohta Murase and Peter M{\'{e}}sz{\'{a}}ros and Kenta Kiuchi},
	title = {High-energy Neutrino Emission from Short Gamma-Ray Bursts: Prospects for Coincident Detection with Gravitational Waves},
	journal = {The Astrophysical Journal}
}

@article{FangMetzger_2017,
	doi = {10.3847/1538-4357/aa8b6a},
	url = {https://doi.org/10.3847/1538-4357/aa8b6a},
	year = 2017,
	month = {nov},
	publisher = {American Astronomical Society},
	volume = {849},
	number = {2},
	pages = {153},
	author = {Ke Fang and Brian D. Metzger},
	title = {High-energy Neutrinos from Millisecond Magnetars Formed from the Merger of Binary Neutron Stars},
	journal = {The Astrophysical Journal}
}

@ARTICLE{2005APh....22..439B,
       author = {{Barbier}, Louis M. and {Smith}, Robert and {Murphy}, Scott and {Christian}, Eric R. and {Farley}, Rodger and {Krizmanic}, John F. and {Mitchell}, John W. and {Streitmatter}, Robert E. and {Loh}, Eugene C. and {Stochaj}, Stephen},
        title = "{NIGHTGLOW: an instrument to measure the Earth{\textquoteright}s nighttime ultraviolet glow{\textemdash}results from the first engineering flight}",
      journal = {Astroparticle Physics},
         year = 2005,
        month = jan,
       volume = {22},
       number = {5-6},
        pages = {439-449},
          doi = {10.1016/j.astropartphys.2004.10.002},
       adsurl = {https://ui.adsabs.harvard.edu/abs/2005APh....22..439B}
}

@PHDTHESIS{1967PhDT........28B,
       author = {{Bunner}, Alan Newton},
        title = "{Cosmic Ray Detection by Atmospheric Fluorescence.}",
       school = {CORNELL UNIVERSITY.},
         year = 1967,
        month = jan,
       adsurl = {https://ui.adsabs.harvard.edu/abs/1967PhDT........28B}
}

\end{document}